\documentclass{article}
\usepackage{graphicx}
\usepackage{graphics}
\usepackage[mathscr]{eucal}
\usepackage{amssymb}
\usepackage{amsmath}
\usepackage{xspace}
\usepackage{listings}
\usepackage{jhephacked}
\usepackage{longtable}

\let\origthelstnumber\thelstnumber
\makeatletter
\newcommand*\Suppressnumber{%
  \lst@AddToHook{OnNewLine}{%
    \let\thelstnumber\relax%
     \advance\c@lstnumber-\@ne\relax%
    }%
}

\newcommand*\Reactivatenumber{%
  \lst@AddToHook{OnNewLine}{%
   \let\thelstnumber\origthelstnumber%
   \advance\c@lstnumber\@ne\relax}%
}

\definecolor{gray}{rgb}{0.8,0.8,0.8}

\newcommand{\La}{\mathfrak{L}}
\newcommand{\DRbar}{{\ensuremath{\overline{\mathrm{DR}}}}}
\newcommand{\MSbar}{{\ensuremath{\overline{\mathrm{MS}}}}}
\newcommand\SARAH{{\tt SARAH}\xspace}
\newcommand{\SARAHv}[1]{\SARAH~\texttt{#1}\xspace}
\newcommand\MG{{\tt MadGraph}\xspace}
\newcommand\ME{{\tt MadEvent}\xspace}
\newcommand\MA{{\tt MadAnalysis}\xspace}
\newcommand{\MGv}[1]{\MG~{\texttt{#1}}\xspace}
\newcommand\FeynArts{{\tt FeynArts}\xspace}

\newcommand\FlavorKit{{\tt FlavorKit}\xspace}
\newcommand\FormCalc{{\tt FormCalc}\xspace}
\newcommand\CalcHep{{\tt CalcHep}\xspace}
\newcommand\CompHep{{\tt CompHep}\xspace}
\newcommand\PreSARAH{{\tt PreSARAH}\xspace}
\newcommand\WHIZARD{{\tt WHIZARD}\xspace}
\newcommand\OMEGA{{\tt O'Mega}\xspace}
\newcommand\Sherpa{{\tt Sherpa}\xspace}
\newcommand\SPheno{{\tt SPheno}\xspace}
\newcommand\Herwig{{\tt Herwig++}\xspace}
\newcommand\FlexibleSUSY{{\tt FlexibleSUSY}\xspace}
\newcommand\Vevacious{{\tt Vevacious}\xspace}
\newcommand\MO{{\tt MicrOmegas}\xspace}
\newcommand\HB{{\tt HiggsBounds}\xspace}
\newcommand\HS{{\tt HiggsSignals}\xspace}
\newcommand\SSP{{\tt SSP}\xspace}
\newcommand\UFO{{\tt UFO}\xspace}

\newcommand\Susyno{{\tt Susyno}\xspace}
\newcommand\Mathematica{{\tt Mathematica}\xspace}

\newcommand{\Fortran}{\texttt{Fortran}\xspace}

\newcommand{\sign}{\text{sign}\xspace}

\newcommand{\BLSSM}{\textit{B-L}-SSM\xspace}

\def\gsim{\raise0.3ex\hbox{$\;>$\kern-0.75em\raise-1.1ex\hbox{$\sim\;$}}}

\font\btt=rm-lmtk10

\lstdefinestyle{mathematica}{
        basicstyle=\ttfamily\mdseries,
	language=bash,
	frame=false,
	xleftmargin=.25in}   

\lstdefinestyle{terminal}{
	language=bash,
	frame=lines,
	xleftmargin=.5in,
        numbers=none}
        
\lstdefinestyle{file}{
        basicstyle=\ttfamily\mdseries,
	language=bash,
	frame=shadowbox,
        numbers=left,   
        numberstyle=\tiny} 
        
\lstdefinestyle{mssm}{
        basicstyle=\ttfamily\mdseries,
	language=bash,
	frame=shadowbox,
        numbers=left,  
        stringstyle=\color{gray},  
        title=\hspace{14.5cm}MSSM,
        numberstyle=\tiny}     
        
\lstdefinestyle{blssm}{
        basicstyle=\ttfamily\mdseries,
	language=bash,
	frame=shadowbox,
        title=\hspace{14.2cm}\BLSSM,	
        numbers=left,   
        numberstyle=\tiny}
        
\lstdefinestyle{both}{
        basicstyle=\ttfamily\mdseries,
	language=bash,
	frame=shadowbox,
        title=\hspace{13cm}{MSSM,\,\BLSSM},		
        numbers=left,   
        numberstyle=\tiny}

\begin{document}

\preprint{CERN-TH-2015-051}

\title{Exploring new models in all detail with \SARAH}

\author[a]{Florian Staub}
\affiliation[a]{Theory Division, CERN, 1211 Geneva 23, Switzerland}
\emailAdd{florian.staub@cern.ch} 

\abstract{I give about overview of the features the \Mathematica package \SARAH provides to study new models.
In general, \SARAH can handle a wide range of models beyond the MSSM 
coming with additional chiral superfields, extra gauge groups, or distinctive features like 
Dirac gaugino masses. All of these models can be implemented in a compact form in 
\SARAH and are easy to use: \SARAH extracts all analytical properties of the given model like
two-loop renormalization group equations, tadpole equations, mass matrices and 
vertices. Also one- and two-loop corrections to tadpoles and self-energies can be obtained.  
For numerical calculations \SARAH can be interfaced 
to other tools to get the mass spectrum, to check flavour or dark matter constraints, and to test the vacuum 
stability, or to perform collider studies. In particular, the interface to \SPheno allows a 
precise prediction of the Higgs mass in a given model comparable to MSSM precision 
by incorporating the important two-loop corrections.
I show in great detail at the example of the \BLSSM  how \SARAH together with \SPheno,
\HB/\HS, \FlavorKit, \Vevacious, \CalcHep, \MO, \WHIZARD, and \MG can be used to 
study all phenomenological aspects of a model. Even if I concentrate in this manuscript on the analysis of supersymmetric models most features are also 
available in the  non-supersymmetric case.  }

\maketitle

\section{Introduction}
Supersymmetry (SUSY) has been the top candidate for beyond standard model (BSM) physics since many years \cite{Ramond:1971gb,Wess:1974tw,Volkov:1973ix}. This has many reasons. SUSY solves the hierarchy problem of the standard model (SM) \cite{Weinberg:1975gm,Weinberg:1979bn}, provides a dark matter candidate \cite{Goldberg:1983nd, Ellis:1983ew,Drees:1992am}, leads to gauge coupling unification \cite{Dimopoulos:1981yj,Ibanez:1981yh,Marciano:1981un,Einhorn:1981sx,Amaldi:1991cn,Langacker:1991an,Ellis:1990wk} and gives an explanation for electroweak symmetry breaking (EWSB) \cite{Martin:2001vx,Ibanez:1982fr}. Before the LHC has turned on, the main focus has been on the minimal supersymmetric extensions of the SM, the MSSM. The 105 additional parameters of this model, mainly located in the SUSY breaking sector, can be constrained by assuming a fundamental, grand unified theory (GUT) and a specific mechanism for SUSY breaking \cite{Witten:1981nf,Witten:1982df,Hall:1983iz,Nilles:1983ge,AlvarezGaume:1983gj,Kane:1993td,Dine:1995ag,Dine:1994vc,Giudice:1998bp,Randall:1998uk}. In these cases often four or five free parameters are left and the model becomes very predictive. 
However, the negative results from SUSY searches at the LHC\footnote{see for instance Refs.~\cite{Craig:2013cxa} for an overview of SUSY searches} as well as the measured Higgs mass of about 125~GeV \cite{Chatrchyan:2012ufa,Aad:2012tfa} put large pressure on the simplest scenarios. Wide regions of the parameter space, which had been considered as natural before LHC has started, have been ruled out. This has caused more interest in non-minimal SUSY models. Beyond-MSSM model can provide many advantages compared the MSSM not only addressing the two issues mentioned so far. A more complete list of good reasons to take a look on extensions of the MSSM is:
\begin{itemize}
 \item {\bf Naturalness}: the Higgs mass in SUSY is not a free parameter like in the SM. In the MSSM the tree level mass is bounded from above by $m_h^{(T)} < m_Z$. Thus, about one third of the Higgs mass has to be generated radiatively to explain the observation. Heavy SUSY masses are needed to gain sufficiently large loop corrections, i.e. a soft version of the hierarchy problem appears again. The need for large loop corrections gets significantly
 softened if $F$- or $D$-terms are present which already give a push to the tree-level mass \cite{Ellwanger:2009dp,Ellwanger:2006rm,Ma:2011ea,Zhang:2008jm,Hirsch:2011hg}.
 \item {\bf SUSY searches}: the negative results from all SUSY searches at the LHC have put impressive limits on the Sparticle masses. 
 However, the different searches are based on certain assumptions like a stable, neutral and colourless lightest SUSY particle (LSP), a sufficiently  mass splitting between the SUSY states, and so on. As soon as these conditions are no longer given like in models with broken $R$-parity, the limits become much weaker \cite{Dreiner:2012gx,Bhattacherjee:2013gr,Kim:2014eva}. Also 
 in scenarios with compressed spectra where SUSY states are nearly degenerated, the strong limits do often not apply \cite{Giudice:2010wb, Gori:2013ala,Han:2013usa, Schwaller:2013baa,Baer:2014cua,Han:2014kaa,Bramante:2014dza,Han:2014xoa,Baer:2014kya,Gori:2014oua,Bramante:2014tba,Han:2015lma}. 
 \item {\bf Neutrino data}: there is an overwhelming experimental evidence that neutrinos have masses and do mix among each other, see Ref.~\cite{Gonzalez-Garcia:2014bfa} and references therein. However, neutrino masses are not incorporated in the MSSM. To do that, either one of the different seesaw mechanisms can be utilised or $R$-parity 
 must be broken to allow a neutrino--neutralino mixing \cite{Hisano:1998wn,Rossi:2002zb,Buckley:2006nv,Hirsch:2008gh,Hirsch:2008dy,Borzumati:2009hu,Esteves:2009vg,Esteves:2010ff,%
 Malinsky:2005bi,Abada:2012cq,BhupalDev:2012ru,Abada:2014kba}.
 \item {\bf Strong CP-problem}: the strong CP problems remains not only an open question in the SM but also in the MSSM. In principle, for both models the same solution exists to explain the smallness of the $\Theta$ term in QCD: the presence of a broken Peccei-Quinn (PQ)
 symmetry \cite{Peccei:1977hh}. In its supersymmetric version PQ models not only predict an axion but also an axino which could be another DM candidate \cite{Covi:2001nw, Kim:2001sh, Covi:2009pq,  Choi:2011yf, Bae:2011iw, Bae:2011jb, Strumia:2010aa}. In general, the phenomenological aspects of axion--axino models are often even richer, in particular if the DSFZ version is considered \cite{Dine:1981rt, Zhitnitsky:1980tq}: the minimal, self-consistent supersymmetric DSFZ-axion model needs in total three additional superfields compared to the MSSM \cite{Dreiner:2014eda}.
 \item {\bf $\mu$-problem}: the superpotential of the MSSM involves one parameter with dimension mass: the $\mu$-term. This term is not protected by any symmetry, i.e. the natural values would be either exactly 0 or $O(M_{GUT})$. However, both extreme values are ruled out by phenomenological considerations. The optimal size of this parameter would be comparable to the electroweak scale. This could be obtained if the $\mu$-term is actually not a fundamental parameter but generated dynamically. For instance, in singlet extensions an effective $\mu$-term appears as consequence of SUSY breaking and is therefore naturally $O(M_{SUSY})$  \cite{Kim:1983dt,Ellwanger:2009dp}.
 \item {\bf Top-Down approach}: starting with a GUT or String theory it is not necessarily clear that only the gauge sector and particle content of the MSSM is present at the low scale. Realistic UV completions come often with many additional matter close to the SUSY scale. In many cases also additional neutral and even charged gauge bosons are predicted \cite{Athron:2009ue,Athron:2009bs,Arbelaez:2013hr,Arbelaez:2013nga}.
 \item {\bf $R$-symmetry}: if one considers $R$-symmetric models, Majorana gaugino masses are forbidden. To give masses to the gauginos in these models, a coupling to a chiral superfield in the adjoint representation is needed. This gives rise to Dirac masses for the gauginos which are in agreement with $R$-symmetry 
\cite{Fayet:1978qc,Polchinski:1982an,Hall:1990hq,Fox:2002bu,Nelson:2002ca,Antoniadis:2005em,Antoniadis:2006uj,%
Amigo:2008rc,Plehn:2008ae,Benakli:2008pg,Belanger:2009wf,Benakli:2009mk,Choi:2009ue,Benakli:2010gi,Choi:2010gc,%
Carpenter:2010as,Kribs:2010md,Abel:2011dc,Davies:2011mp,Benakli:2011vb,Benakli:2011kz,Kalinowski:2011zz,Frugiuele:2011mh,%
Itoyama:2011zi,Rehermann:2011ax,Bertuzzo:2012su,Davies:2012vu,Argurio:2012cd,Fok:2012fb,Argurio:2012bi,Frugiuele:2012pe,%
Frugiuele:2012kp,Benakli:2012cy,Itoyama:2013sn,Chakraborty:2013gea,Csaki:2013fla,Itoyama:2013vxa,Beauchesne:2014pra,%
Benakli:2014daa,Bertuzzo:2014bwa,Alves:2015kia}. Dirac gauginos are also attractive because they can weaken LHC search bounds \cite{Heikinheimo:2011fk,Kribs:2012gx,Alves:2013wra}, and flavour constraints \cite{Kribs:2007ac,Fok:2012me,Dudas:2013gga}. 
\end{itemize}

Despite the large variety and flexibility of SUSY, many dedicated public computer tools like {\tt SoftSUSY} \cite{Allanach:2001kg,Allanach:2009bv,Allanach:2014nba}, {\tt SPheno} \cite{Porod:2003um,Porod:2011nf}, {\tt Suspect} \cite{Djouadi:2002ze}, {\tt Isajet} \cite{Paige:2003mg,Baer:2003xc,Baer:1999sp,Paige:1998xm,Paige:1998ux,Baer:1993ae} or {\tt FeynHiggs} \cite{Heinemeyer:1998yj,Hahn:2009zz} are restricted to the simplest realization of SUSY, the MSSM, or small extensions of it. Therefore, more generic tools are needed to allow to study non-minimal SUSY models with the same precision as the MSSM. This precision is needed to confront also these models with the strong limits from SUSY searches, flavour observables, dark matter observations, and Higgs measurements.
The most powerful tool in this direction is the \Mathematica package \SARAH \cite{Staub:2008uz,Staub:2009bi,Staub:2010jh,Staub:2012pb,Staub:2013tta}. \SARAH has been optimized for an easy, fast 
and exhaustive study of non-minimal SUSY models. While the first version of \SARAH has been focused on the derivation of 
tree-level properties of a SUSY model, i.e.  mass matrices and vertices, and interfacing this information to Monte-Carlo (MC) tools, with the second version of \SARAH the calculation of one-loop self-energies as well as two-loop renormalization group equations (RGEs) has been automatized. With version {\tt 3}, \SARAH became the first 'spectrum-generator-generator': all analytical information derived by \SARAH can be exported to \Fortran code which provides a fully-fledged spectrum generator based on \SPheno. This functionality has been later extended by the \FlavorKit \cite{Porod:2014xia} interface which allows a modular implementation of new flavour observables based on the tools \FeynArts/\FormCalc\ -- \SARAH\ -- \SPheno. Also different methods to calculate the two-loop corrections to the Higgs states in a non-minimal model are available with \SPheno modules generated by \SARAH today: the radiative contributions to CP even scalar masses at the two-loop level can be obtained by either using the effective potential approach \cite{Goodsell:2014bna} based on generic results given in Ref.~\cite{Martin:2001vx}, or a fully diagrammatic calculation \cite{Goodsell:2015ira}. Both calculations provide Higgs masses with a precision which is otherwise just available for the MSSM. 
Beginning with \SARAHv{4}, the package is no longer restricted to SUSY models but can handle also a general, renormalizable quantum field theory and provides nearly the same features as for SUSY models. Today, \SARAH can be used for SUSY and non-SUSY models to write model files for
\CalcHep/\CompHep \cite{Pukhov:2004ca,Boos:1994xb}, \FeynArts/\FormCalc \cite{Hahn:2000kx,Hahn:2009bf}, \WHIZARD/\OMEGA \cite{Kilian:2007gr,Moretti:2001zz} as well as in the \UFO format \cite{Degrande:2011ua} which can be handled for instance by 
\MGv{5} \cite{Alwall:2011uj}, {\tt GoSam} \cite{Cullen:2011ac}, {\tt Herwig++} \cite{Gieseke:2003hm,Gieseke:2006ga,Bellm:2013lba}, and {\tt Sherpa} \cite{Gleisberg:2003xi,Gleisberg:2008ta,Hoche:2014kca}. 
The modules created by \SARAH for \SPheno calculate the full one-loop and partially two-loop corrected mass spectrum, branching ratios 
and decays widths of all states, and many flavour and precision observables. Also an easy link to \HB \cite{Bechtle:2008jh,Bechtle:2011sb} and \HS \cite{Bechtle:2013xfa} exists.  Another possibility to get a tailor made spectrum generator for a non-minimal SUSY model based on \SARAH is the tool {\tt FlexibleSUSY} \cite{Athron:2014yba}. Finally, \SARAH can also produce model files for \Vevacious \cite{Camargo-Molina:2013qva}. 
The combination \SARAH--\SPheno--\Vevacious provides the possibility to find the global minimum of the one-loop effective potential of a given model and parameter point. \\

The range of models covered by \SARAH is very broad. \SARAH and its different interfaces have been successfully used 
to study many different SUSY scenarios: singlet extensions
with and without CP violation \cite{Stal:2011cz,Ender:2011qh,Aparicio:2012vk,Graf:2012hh,SchmidtHoberg:2012yy,SchmidtHoberg:2012ip,Ross:2012nr,Kaminska:2013mya,Binjonaid:2014oga,Kaminska:2014wia,Muhlleitner:2014vsa}, 
triplet extensions \cite{Arina:2014xya,Bandyopadhyay:2014vma},
models with $R$-parity violation \cite{List:2013dga,Dreiner:2012mx,Dreiner:2013jta,Biswas:2014gga,Allanach:2014lca,Chamoun:2014eda,Dreiner:2014lqa}, different kinds of seesaw 
mechanisms \cite{Esteves:2010ff,Abada:2011mg,Abada:2012cq,Hirsch:2012yv,Banerjee:2013fga,Krauss:2013gya,BhupalDev:2012ru,DeRomeri:2012qd,Abada:2014kba,Boucenna:2015zwa}, 
models with extended gauge sectors at intermediate scales \cite{Esteves:2010si,Esteves:2011gk,DeRomeri:2011ie,Krauss:2013jva}
or the SUSY scale \cite{O'Leary:2011yq,Hirsch:2011hg,Hirsch:2012kv,Frank:2014bma,Brooijmans:2014eja,Corcella:2014lha}, models with
Dirac Gauginos \cite{Frugiuele:2012pe,Benakli:2012cy,Busbridge:2014sha,Benakli:2014cia,Diessner:2014ksa} or vectorlike states \cite{Lalak:2015xea}, and even more
exotic extensions \cite{Athron:2012sq,Alves:2012fx,Bharucha:2013ela,Ding:2015wma}. In addition, \SARAH can be also very useful to perform studies in the context of the MSSM which can't be done with any other public tool out of the box. That's the case for instance, if new SUSY breaking mechanisms should be considered \cite{Brummer:2013upa,Ding:2013pya,Louis:2014pia,Ding:2014bqa,Kyae:2014aka,Abel:2014fka,Un:2014afa,Fichet:2015oha} or if the presence of charge and colour breaking minima should be checked \cite{Camargo-Molina:2014pwa,Chattopadhyay:2014gfa}. For the NMSSM, despite the presence of specialized tools like 
{\tt NMSSMTools} \cite{Ellwanger:2006rn}, {\tt SoftSUSY} \cite{Allanach:2013kza} or {\tt NMSSMCalc} \cite{Baglio:2013iia}, the \SPheno version created by \SARAH is the only code providing two-loop corrections beyond $O(\alpha_S(\alpha_t+\alpha_b))$ not relying on MSSM approximations \cite{Goodsell:2014pla}. Also the full one-loop corrections to all SUSY states in the NMSSM have first been derived with \SARAH \cite{Staub:2010ty}.

This paper is organized as follows: in the next section an overview about the models supported by \SARAH is given.
In sec.~\ref{sec:analytical}, I'll discuss the possible analytical calculations which can be done with \SARAH, and list the possible output of the derived information for further evaluation. The main part of this manuscript is a detailed example how \SARAH can be used to study all phenomenological aspects of a model. That's done in sections~\ref{sec:example_implementation}--\ref{sec:example_scans}: in sec.~\ref{sec:example_implementation} the implementation of the \BLSSM in \SARAH is described, in sec.~\ref{sec:example_mathematica} it is discussed how the model can be understood at the analytical level in \Mathematica. The \SPheno output with all its features is presented in sec.~\ref{sec:example_spheno}. In sec.~\ref{sec:example_tools} I'll show how other tools can be used together with \SARAH and \SPheno to study for instance the dark matter and collider phenomenological. In  sec.~\ref{sec:example_scans} different possibilities to perform parameter scans are presented. I summarize in sec.~\ref{sec:summary}.  Throughout the paper and in the given examples I will focus mainly on SUSY models, but many statements apply one-to-one also to non-SUSY models.
\section{Models}
\label{sec:models}

\subsection{Input needed by \SARAH to define a model}
\SARAH is optimized for the handling of a wide range of SUSY models. The basic idea of \SARAH was to give the user the possibility to implement models in an easy, compact and straightforward way. Most tasks to get the Lagrangian are fully  automatized: it is sufficient to define just the fundamental properties of the model. That means, that the {\bf necessary input} to completely define the gauge eigenstates with all their interactions are:
\begin{enumerate}
 \item Global symmetries
 \item Gauge symmetries
 \item Chiral superfields
 \item Superpotential
\end{enumerate}
That means that \SARAH automatizes many steps to derive the Lagrangian from that input:
\begin{enumerate}
 \item All interactions of matter fermions and the $F$-terms are derived from the superpotential
 \item All vector boson and gaugino interactions as well as $D$-terms are derived from gauge invariance
 \item All gauge fixing terms are derived by demanding that scalar--vector mixing vanishes in the kinetic terms
 \item All ghost interactions are derived from the gauge fixing terms
 \item All soft-breaking masses for scalars and gauginos as well as the soft-breaking counterparts to the superpotential couplings are added automatically 
\end{enumerate} 

Of course, the Lagrangian of the gauge eigenstates is not the final aim. Usually one is interested in the mass eigenstates after gauge symmetry breaking. To perform the necessary rotations to the new  eigenstates, the user has to give some more information: 
\begin{enumerate}
 \item Definition of the fields which get a vacuum expectation value (VEV) to break gauge symmetries
 \item Definition of what vector bosons, scalars and fermions mix among each other
\end{enumerate}
Using this information, all necessary re-definitions and fields rotations are done by \SARAH. Also the gauge fixing terms are derived for the new eigenstates and the ghost interactions are added. For all eigenstates plenty of information can be derived by \SARAH as explained in sec.~\ref{sec:analytical}. Before coming to that, I'll give more details what kind of models and what features are supported by \SARAH.

\subsection{Supported models and features}
\label{sec:supported_models}
As we have seen in the introduction, there are many possibilities to go beyond the widely studied MSSM. Each approach modifies the on or the other sector of the model. In general, possible changes compared to the MSSM are:  (i) using other global symmetries to extent the set of allowed couplings, (ii) adding chiral superfields, (iii) extending the gauge sector, (iv) giving VEVs to other particle than only the Higgs doublets, (v) adding Dirac masses for gauginos, (vi) considering non-canonical terms like non-holomorphic soft SUSY breaking interactions or Fayet-Iliopoulos $D$-terms. All of these roads can in principle be gone by \SARAH and I'll briefly discuss what is possible in the different sectors and which steps are done by \SARAH to get the Lagrangian. Of course, extending the gauge sector or adding Dirac masses to gauginos comes inevitable with an extended matter sector as well. Thus, often several new effects appear together and can be covered by \SARAH.

\subsubsection{Global symmetries} \SARAH can handle an arbitrary number of global symmetries which are either $Z_N$ or $U(1)$ symmetries. Also a continuous $R$-symmetry $U(1)_R$ is possible.  Global symmetries are used in \SARAH mainly for three different purposes: first, they help to constrain the allowed couplings in the superpotential. However,  \SARAH doesn't strictly forbid terms in the superpotential which violate a global symmetry. \SARAH only prints a warning to point out the potential contradiction. The reason is that such a term might be included on purpose to explain its tininess. 
Global symmetries can also affect the soft-breaking terms written down by \SARAH. \SARAH always tries to generate the most general Lagrangian and includes also soft-masses of the form $m^2 \phi_i \phi_j^*$ for two scalars $\phi_i$,$\phi_j$ with identical charges. However, these terms are dropped if they are forbidden by a global symmetry. By the same consideration, Dirac gaugino mass terms are written down or not. Finally, global symmetries are crucial for the output of model files for \MO to calculate the relic density. For this output at least one unbroken discrete global symmetry must be present. \\
By modifying the global symmetries one can already go beyond the MSSM without changing the particle content: choosing a $Z_3$ (Baryon triality) instead of $R$-Parity \cite{Ibanez:1991hv,Ibanez:1991pr,Banks:1991xj,Dreiner:2005rd,Dreiner:2006xw}, lepton number violating terms would be allowed while the proton is still stable. \SARAH comes not only with $R$-parity violating models based on Baryon triality, but also a variant for Baryon number violation but conserved Lepton number is included.

\subsubsection{Gauge sector}
\paragraph*{Gauge groups} The gauge sector of a SUSY model in \SARAH is fixed by defining a set of vector superfields. \SARAH is not restricted to three vector superfields like in the MSSM, but many more gauge groups can be defined. To improve the power in dealing with gauge groups, \SARAH has linked routines from the \Mathematica package \Susyno \cite{Fonseca:2011sy}. \SARAH together with \Susyno take care of all group-theoretical calculations: the Dynkin and Casimir invariants are calculated, and the needed representation matrices as well as Clebsch-Gordan coefficients are derived. This is not only done for $U(1)$ and $SU(N)$ gauge groups, but also $SO(N)$, $Sp(2N)$ and expectational groups can be used.  For all Abelian groups also a GUT normalization can be given. This factor comes usually from considerations about the embedding of a model in a greater symmetry group like $SU(5)$ or $SO(10)$. If a GUT normalization is defined for a group, it will be used in the calculation of the RGEs. The soft-breaking terms for a gaugino $\lambda$ of a gauge group $A$ are usually included as
\begin{equation}
\label{eq:SoftVector}
\La_{SB,\lambda_A} = \frac{1}{2} \lambda_A^a \lambda_A^a M_A + h.c. 
\end{equation}

\paragraph*{Gauge interactions} With the definition of the vector superfields already the self-interactions of vector bosons as well as the interactions between vector bosons and gauginos are fixed. Those are taken to be
\begin{equation}
\label{eq:LagVS}
\La = - \frac{1}{4} F^{A,a}_{\mu\nu} F^{A,\mu\nu a} - i \lambda_A^{\dagger a } \bar{\sigma}^\mu D_\mu \lambda_A^a
\end{equation}
I'm using here and in the following capital letters $A,B$ to label the gauge groups and small letter $a,b,c$ to label the generators, vector bosons and gauginos of a particular gauge group. The field strength tensor is defined as
\begin{equation}
\label{eq:FieldStrength}
F_{\mu\nu}^{A,a} = \partial_\mu V^{A,a}_\nu - \partial_\nu V^{A,a}_\mu + g_A f_A^{abc} V^{A,b}_\mu V^{A,c}_\nu \thickspace ,
\end{equation}
and the covariant derivative is
\begin{equation}
D_\mu \lambda_A^a = \partial_\mu \lambda_A^a + g_A f^{abc} A_\mu^b \lambda^c
\thickspace .
\end{equation}
Here, $f_A^{abc}$ is the structure constant of the gauge group $A$. Plugging eq.~(\ref{eq:FieldStrength}) in the first term of eq.~(\ref{eq:LagVS}) leads to self-interactions of three and four gauge bosons.  In general, the procedure to obtain the Lagrangian from the vector and chiral superfields is very similar to Ref.~\cite{Martin:1997ns}. Interested readers might check this reference for more details.

\paragraph*{Gauge interactions of matter fields} Vector superfields usually don't come alone but also matter fields are present. I'm going to discuss the possibilities to define chiral superfields in sec.~\ref{sec:mattersector}. Here, I assume that a number of chiral superfields are present and I want to discuss the gauge interactions which are taken into account for those. 
First, the $D$-terms stemming from the auxiliary component of the superfield are calculated. These terms cause four scalar interactions and read
\begin{equation}
\label{eq:Dterms}
\La_{D_A} = \frac{1}{2} g^2_A \sum_{i,j} |(\phi_i^* T_{A r}^a \phi_j)|^2
\end{equation}
Here, the sum is over all scalars $i,j$ in the model, $T_{A r}^a$ are the generators of the gauge group $A$ for a irreducible representation $r$. For Abelian groups $T_{A r}^a$ simplify to the charges  $Q_\phi^A$ of the different fields.  In addition, Abelian gauge groups can come also with another feature: a Fayet-Iliopoulos $D$-term \cite{Fayet:1974jb}:
\begin{equation}
\La_{FI,A} = \xi_A \frac{g_A}{2} \sum_{i} (\phi_i^* Q^A_\phi \phi_i)
\end{equation}
This term can optionally be included in \SARAH for any $U(1)$.\\

The other gauge--matter interactions are those stemming from the kinetic terms:
\begin{equation}
\La_{kin} = - D^\mu \phi^{*i} D_\mu \phi_i  - i \psi^{\dagger i }
\bar{\sigma}^\mu D_\mu \psi_i 
\end{equation}
with covariant derivatives $D_\mu \equiv \partial_\mu - i g_A V^{A,a}_\mu (T_{Ar}^a )$.
The SUSY counterparts of these interactions are those between gauginos and matter fermions and scalars:
\begin{equation}
\La_{GFS} = - \sqrt{2} g_A (\phi_i^* T_{Ar}^a \psi_j) \lambda_A^a + \mbox{h.c.} \thickspace .
\end{equation} 

\paragraph*{Gauge kinetic mixing} The terms mentioned so far cover all gauge interactions which are possible in the MSSM. These are derived for any other SUSY model in exactly the same way. However, there is another subtlety which arises if more than one Abelian gauge group is present. In that case 
\begin{equation}
\label{eq:off}
\La = -\frac{1}{4} \kappa F_{\mu\nu}^A F^{B,\mu\nu} \hspace{1cm} A\neq B
\end{equation}
are allowed for field strength tensors $F^{\mu\nu}$ of two different Abelian groups $A$, $B$  \cite{Holdom:1985ag}. $\kappa$ is in general a $n \times n$ matrix if $n$ Abelian groups are present. \SARAH fully includes the effect of kinetic mixing independent of the number of Abelian groups. For this purpose \SARAH is not working with field strength interactions like eq.~(\ref{eq:off}) but performs a rotation to bring the field strength in a diagonal form. That's done by a redefinition of the vector $\Upsilon$ carrying all gauge fields $V^\mu_X$:
\begin{equation}
\label{eq:kappaRot}
 \Upsilon \to \sqrt{\kappa} \Upsilon
\end{equation}
This rotation has an impact on the interactions of the gauge bosons with matter fields. In general, the interaction of a particle $\phi$ with all gauge fields can be expressed by
\begin{equation}
\Theta_\phi^T \tilde{G}  \Upsilon
\end{equation}
$\Theta_\phi$ is a vector containing the charges $Q^x_\phi$ of $\phi$ under all $U(1)$ groups $x$ and $\tilde{G}$ is a $n \times n$ diagonal matrix carrying the gauge couplings of the different groups. After the rotation according to eq.~(\ref{eq:kappaRot}) the interaction part can be expressed by
\begin{equation}
\label{eq:QGA}
\Theta_\phi^T G  \Upsilon
\end{equation}
with a general $n \times n$ matrix $G$ which is no longer diagonal. In that way, the effect of gauge kinetic mixing has been absorbed in 'off-diagonal' gauge couplings. That means the covariant derivative in \SARAH reads
\begin{equation}
D_\mu \phi = \left(\partial_\mu - i \sum_{x,y} Q_\phi^x g_{xy}   V^\mu_y \right)\phi \hspace{1cm}
\end{equation}
$x,y$ are running over all $U(1)$ groups, and $g_{xy}$ are the entries of the matrix $G$. Gauge-kinetic mixing is not only included in the interactions with vector bosons, but also in the derivation of the $D$-terms. Therefore, the $D$-terms for the Abelian sector in \SARAH read
\begin{equation}
\label{eq:Dabelian}
\La_{D,U(1)} = \sum_{ij} (\phi_i^*\phi_i) (G^T \Theta_{\phi_i}) (G \Theta_{\phi_j}) (\phi_j^* \phi_j)
\end{equation}
while the non-Abelian $D$-terms keep the standard form eq.~(\ref{eq:Dterms}).
Finally, also 'off-diagonal' gaugino masses are introduced. The soft-breaking part of the Lagrangian reads then
\begin{equation}
\label{eq:SoftKinM}
\La_{SB,\lambda,U(1)} \supset \sum_{xy} \frac{1}{2} \lambda_x \lambda_y M_{xy} + h.c. 
\end{equation}
\SARAH takes the off-diagonal gaugino masses to be symmetric: $M_{xy} = M_{yx}$.

\subsubsection{Gauge fixing sector} All terms written down so far lead to a Lagrangian which is invariant under a
general gauge transformation. To break this invariance one can add 'gauge fixing' terms to the Lagrangian. The general form of these terms is 
\begin{equation}
\La_{GF} = - \frac{1}{2} |\mathscr{F}_A^a|^2 \thickspace .
\end{equation} 
Here, $\mathscr{F}_A^a$ is usually a function involving partial derivatives of gauge bosons $V_\mu^{A,a}$. \SARAH uses $R_\xi$ gauge. That means that for an unbroken gauge symmetry, the gauge fixing terms are 
\begin{equation}
\La_{GF} = - \frac{1}{2 R_{\xi_A}} \left|\partial^\mu V_\mu^{A,a} \right|^2 
\thickspace.
\end{equation}
For broken symmetries, the gauge fixings terms are chosen in a way that the mixing terms between vector bosons and scalars disappears from the
Lagrangian.  This generates usually terms of the form 
\begin{equation}
\label{GFewsb}
\La_{GF, R_\xi} =  - \frac{1}{2 R_{\xi_A}} \left| \partial^\mu V^{A}_\mu + R_{\xi_{A}} M_A G^A \right|^2 
\end{equation}
Here, $G^A$ is the Goldstone boson of the  vector boson $V^A_\mu$ with mass $M_A$.  From the gauge fixing part, the interactions of ghost fields 
$\bar{\eta}^a_A$  are derived by
\begin{equation}
\La_{Ghost} = - \bar{\eta}^a_A (\delta \mathscr{F}_A^a) \thickspace. 
\end{equation}
Here, $\delta$ assigns the operator for a BRST transformation. All steps to get the gauge fixing parts and the ghost interactions are completely done automatically by \SARAH and adjusted to the gauge groups in the model. 

\subsubsection{Matter sector} 
\label{sec:mattersector}
There can be up to 99 chiral superfields in a single SUSY model in \SARAH. All superfields can come with an arbitrary number of generations and can transform as any irreducible representation with respect to the defined gauge groups. In the handling of non-fundamental fields under a symmetry, \SARAH distinguishes if the corresponding symmetry gets broken or not: for unbroken symmetries it is convenient to work with fields which transform as vector under the symmetry with the appropriate length. For instance, a {\bf 6} under $SU(3)_c$ is taken to be
\begin{equation}
 \phi_\alpha \hspace{1cm} \alpha=1,2, \dots 6
\end{equation}
I.e. it carries one charge index. In contrast, non-fundamental fields under a broken gauge symmetry are represented by tensor products of the fundamental representation. For instance, a {\bf 3} under $SU(2)_L$ is taken to be 
\begin{equation}
 \phi_{ab} \hspace{1cm} a,b=1,2
\end{equation}
Thus, the triplet can be given as usual as $2 \times 2$ matrix. \\
For Abelian gauge groups one can not only define charges for superfields which are real numbers, but also variables can be used for that. All interactions are then expressed keeping these charges as free parameter.\\

For all chiral superfield \SARAH adds the soft-breaking masses. For fields appearing in $N$ generations, these are treated as hermitian $N \times N$ matrices. As written above, also soft-terms mixing two scalars are included if allowed by all symmetries. Hence, the soft-breaking mass terms read in general
\begin{equation}
\label{eq:LSBphi} 
\La_{SB,\phi}  = \sum_{ij} \tilde{\delta}_{ij} \phi_i^\dagger m^2_{ij} \phi_j + h.c.
\end{equation}
Note, $i,j$ label different scalar fields, generation indices are not shown. $\tilde{\delta}_{ij}$ is 1, if fields $\phi_i$ and $\phi_j$ have exactly the same transformation properties under all local and global symmetries, and otherwise 0. 

\subsubsection{Models with Dirac Gauginos}
Another feature which became popular in the last years are models with Dirac gauginos. In these models mass terms $m^{\hat \phi_i A}_D \lambda_A \psi_i $ between gauginos $\lambda_A$ and a fermionic component $\psi_i$ of the chiral superfield $\hat \phi_i$ in the adjoint representation of the gauge group $A$ are present. In addition, also new $D$-terms are introduced in these models \cite{Benakli:2011vb}. Thus, the new terms in the Lagrangian are
\begin{equation}
\La_{DG} = - m^{\hat \phi_i A}_D \lambda^a_A \psi_i + \sqrt{2} m^{\hat \phi_i A}_D \phi_i D_A 
\end{equation}
$D_A$ is the auxiliary component of the vector superfield of the group $A$. To allow for Dirac mass terms, these models come always with an extended matter sector: to generate Dirac mass terms for all MSSM gauginos at least one singlet, one triplet under $SU(2)$ and one octet under $SU(3)$ must be added. Furthermore, models with Dirac gauginos generate also new structures in the RGEs \cite{Goodsell:2012fm}. All of this is fully supported in \SARAH. \\
If Dirac masses for gauginos are explicitly turned on in \SARAH, it will check for all allowed combinations of vector and chiral superfields which can generate Dirac masses and which are consistent with all symmetries. For instance, in models with several gauge singlets, the bino might even get several Dirac mass terms. 

\subsubsection{Superpotential, soft-terms and non-canonical interactions}
The matter interactions in SUSY models are usually fixed by the superpotential and the soft-SUSY breaking terms. \SARAH fully supports all renormalizable terms in the superpotential
\begin{equation}
\label{eq:W}
 W = c_L L_i \hat \phi_i + c_M M^{ij} \hat \phi_i \hat \phi_j + c_T Y^{ijk} \hat \phi_i \hat \phi_j \hat \phi_k 
\end{equation}
and generates the corresponding soft-breaking terms 
\begin{equation}
L_{SB,W} =  c_L t_i  \phi_i + c_M B^{ij}  \phi_i \phi_j + c_T T^{ijk}  \phi_i  \phi_j  \phi_k + h.c.
\end{equation}
$c_L$, $c_M$, $c_T$ are real coefficients. All parameters are treated by default in the most general way by taking them as complex tensors of appropriate order and dimension. If identical fields are involved in the same coupling, \SARAH derives also the symmetry properties for the parameter.\\
As discussed below, \SARAH can also handle to some extent non-renormalizable terms with four superfields in the superpotential
\begin{equation}
\label{eq:WW}
W_{NR} = c_W W^{ijkl} \hat \phi_i \hat \phi_j \hat \phi_k \hat \phi_l 
\end{equation}
From the superpotential, all the $F$-terms
\begin{equation}
|F|^2 = \sum_i \left|\frac{\partial \tilde{W}}{\partial  \phi_i}\right|^2
\end{equation}
and interactions of matter fermions 
\begin{equation}
\La_Y = - \frac{1}{2} \frac{\partial^2 \tilde{W}}{\partial \phi_i \partial \phi_j}  \psi_i \psi_j + \mbox{h.c.} \thickspace, 
\end{equation}
are derived. Here $\tilde{W}$ is the superpotential $W$ with all superfields $\hat \phi_i$ replaced by their scalar component $\phi_i$. $\psi_i$ is the fermionic component of that superfield. \\
Usually, the $F$- and $D$-terms and the soft-breaking terms for chiral and vector superfields fix the full scalar potential of the model. However, in some cases also non-canonical terms should be studied. These are for instance non-holomorphic soft-terms 
\begin{equation}
\La_{SB,NH} = \tilde{T}^{ijk} \phi_i \phi_j \phi^*_k
\end{equation}
Those can be added as well and they are taken into account in the calculation of the vertices and masses and as consequence also in all loop calculations. However, they are not included in the calculation of the RGEs because of the lack of generic results in literature.

\subsubsection{Symmetry breaking and VEVs} All gauge symmetries can also be broken. This is in general done by decomposing a complex scalar into its real components and a VEV:
\begin{equation}
S_i \to \frac{1}{\sqrt{2}}(v_i + \phi_i +  i \sigma_i)
\end{equation}
Assigning a VEV to a scalar is not restricted to colourless and neutral particles. Also models with spontaneous colour or charge breaking (CCB) can be studied with \SARAH. Also explicit CP violation in the Higgs sector is possible. There are two possibilities to define that. Either a 
complex phase is added 
\begin{equation}
S_i \to \frac{1}{\sqrt{2}} e^{i\eta}(v_i + \phi_i +  i \sigma_i)
\end{equation}
or a VEV for the CP odd component is defined
\begin{equation}
S_i \to \frac{1}{\sqrt{2}} (v^R_i + \phi_i +  i (v^I_i + \sigma_i))
\end{equation}
Both options are possible in \SARAH, even if the first one might often be preferred. \\

In the case of an extended gauge sector also additional gauge bosons are present. Depending on the quantum numbers of the states which get a VEV these gauge bosons might mix with the SM ones. Also this mixing is fully supported by \SARAH. There is no restriction if the additional gauge bosons are ultra-light (dark photons) or much heavier ($Z'$, $W'$-bosons). 

\subsubsection{Mixing in matter sector} Mixing between gauge eigenstates to new mass eigenstate appears not only in the gauge but also in the matter sector.   In general the mixing is induced via bilinear terms in the Lagrangian between gauge eigenstates. These bilinear terms can either be a consequence of gauge symmetry breaking or they can correspond to bilinear superpotential- or soft-terms. In general, four kinds of bilinear terms can show up in the matter part of the Lagrangian:
\begin{equation}
\label{eq:lagMasses}
 \La = - m^{ij}_C \phi_i^* \phi_j - \frac{1}{2} m^{ij}_R \varphi_i \varphi_j  - \frac{1}{2} m^{ij}_M \Psi^0_i \Psi^0_j - m_D^{ij} \Psi^1_i \Psi^2_j
\end{equation}
Here, $\phi$, $\varphi$, $\Psi^x$ ($x=0,1,2$) are vectors whose components are gauge eigenstates. $\phi$ are complex and $\varphi$ are real scalars, $\Psi^0$, $\Psi_1$ and $\Psi_2$ are Weyl spinors. The rotation of complex scalars $\phi$ to mass eigenstates $\bar \phi$ happens via an unitary matrix $U$ which diagonalizes the matrix $m_C$. For real scalars the rotation is done via a real matrix $Z$ which diagonalizes $m_R$:
\begin{eqnarray}
& \bar \phi = U \phi \hspace{1cm}   M_C^{diag} = U m_C U^\dagger & \\
& \bar \varphi = Z \varphi \hspace{1cm}   M_R^{diag} = Z m_R Z^T & 
\end{eqnarray}
We have to distinguish for fermions if the bilinear terms are symmetric or not. In the symmetric case the gauge eigenstates are  rotated to Majorana fermions. The mass matrix $m_M$ is then diagonalized by one unitary matrix. In the second case, two unitary matrices are needed to transform $\Psi_1$ and $\Psi_2$ differently. This results in Dirac fermions. Both matrices together diagonalize the mass matrix $m_D$. 
\begin{eqnarray}
& \bar  \Psi^0 = N  \Psi^0 \hspace{1cm}   M_M^{diag} = N^* m_M N^{-1} & \\
& \bar  \Psi^1 = V  \Psi^1\,,\quad  \bar \Psi^2 = U \Psi^2  \hspace{1cm}   M_D^{diag} = U^* m_D V^{-1} & 
\end{eqnarray}
There is no restriction in \SARAH how many states do mix. The most extreme case is the one with spontaneous charge, colour and CP violation where all fermions, scalars and vector bosons mix among each other. This results in a huge mass matrix which would be derived by \SARAH. Phenomenological more relevant models can still have a neutralino sector mixing seven to ten states. That's done without any problem with \SARAH. 
Information about the calculation of the mass matrices in \SARAH are given in sec.~\ref{sec:calc_massmatrix}.  

\subsubsection{Superheavy particles}
Extensions of the MSSM can not only be present at the SUSY scale but also appear at much higher scales. These superheavy states have then only indirect effects on the SUSY phenomenology compared to the MSSM: they alter the RGE evolution and give a different prediction for the SUSY parameters. In addition, they can also induce higher dimensional operators which are important. \SARAH provides features to explore models with superheavy states: it is possible to change stepwise the set of RGEs which is used to run the parameters numerically with \SPheno. In addition, the most important thresholds are included at the scale $M_T$ at which the fields of mass $M$ are integrated out. These are the corrections to the gauge couplings and gaugino masses \cite{Hall:1980kf}
\begin{eqnarray}
\label{eq:shift1}
 g_A & \rightarrow & g_A \left( 1\pm \frac{1}{16 \pi^2} g_A^2 S^A(r) \ln\left(\frac{M^2}{M_T^2}\right)\right)  \thickspace ,\\
\label{eq:shift2}
 M_A & \rightarrow & M_A \left( 1\pm \frac{1}{16 \pi^2} g_A^2 S^A(r) \ln\left(\frac{M^2}{M_T^2}\right)\right) \thickspace .
\end{eqnarray}
$S^A(r)$ is the Dynkin index of a superfield transforming as representation $r$ with respect to the gauge group $A$. When evaluating the RGEs from the low to the high scale the contribution is positive, when running down, it is negative. Eqs.~(\ref{eq:shift1})--(\ref{eq:shift2}) assume that the mass splitting between the components of the chiral superfield integrated out is negligible. That's often a good approximation for very heavy states. Nevertheless, \SARAH can also take into account the mass splitting among the components if necessary. \\
Also higher dimensional operators can be initialized which give rise to terms like eq.~(\ref{eq:WW}). However, those are only partially supported in \SARAH. That means that only the RGEs are calculated for these terms and the resulting interactions between two fermions and two scalars are included in the Lagrangian. The six scalar interactions are not taken into account. This approach is for instance sufficient to work with the Weinberg operator necessary for neutrino masses \cite{Weinberg:1979sa,Weinberg:1980bf}.

\subsection{Checks of implemented models}
\label{sec:checks}
After the initialization of a model \SARAH provides functions to check the (self-) consistency of this model. The function {\tt CheckModel} performs the following checks:

\paragraph*{Causes the particle content gauge anomalies?} Gauge anomalies are caused by triangle diagrams with three external gauge bosons and internal fermions \cite{Adler:1969er}. The corresponding conditions for all $SU(N)_A$ groups to be anomaly free are
\begin{eqnarray}
   \sum_i \mbox{Tr}\left[T_{Ar}^a(\psi_i) T_{Ar}^a(\psi_i) T_{Ar}^a(\psi_i)\right] = 0 
\end{eqnarray}
Again, $T_{Ar}^a(\psi_i)$ are the generators for a fermion $\psi_i$ transforming as irreducible representation $r$ under the gauge group $SU(N)_A$. The sum is taken over all chiral superfields. In the Abelian sector several conditions have to be fulfilled depending on the number of $U(1)$ gauge groups
\begin{eqnarray}
 U(1)_A^3 &:& \sum_i (Q^A_{\psi_i})^3  = 0 \\
 U(1)_A\times U(1)_B^2 &:& \thinspace  \sum_i  Q^A_{\psi_i} (Q^B_{\psi_i})^2 = 0   \\
 U(1)_A\times U(1)_B\times U(1)_C &:& \thinspace  \sum_i  Q^A_{\psi_i} Q^B_{\psi_i} Q^C_{\psi_i}= 0
\end{eqnarray}
The mixed condition involving Abelian and non-Abelian groups is
\begin{equation}
 U(1)_A\times SU(N)^2_B  :  \sum_i Q^A_{\psi_i}\, \mbox{Tr}\left[T_{Br}^a(\psi_i) T_{Br}^a(\psi_i)\right] = 0  
\end{equation}
Finally, conditions involving gravity $\mathfrak{G}$ are
\begin{eqnarray}
 \mathfrak{G} \times U(1)_A^2 &:& \sum_i (Q^A_{\psi_i})^2  = 0 \\
 \mathfrak{G} \times U(1)_A\times U(1)_B &:& \thinspace  \sum_i  Q^A_{\psi_i} Q^B_{\psi_i} = 0   \\
 \mathfrak{G}^2 \times U(1)_A &:& \sum_i Q^A_{\psi_i}  = 0 
\end{eqnarray}
If one if these conditions is not-fulfilled a warning is printed by \SARAH. If some $U(1)$ charges were defined as variable, the conditions on these variables for anomaly cancellation are printed.
 \paragraph*{Leads the particle content to the Witten anomaly?} \SARAH checks that there is an even number of \(SU(2)\) doublets. This is the necessary for a model in order to be free of the Witten anomaly \cite{Witten:1982fp}
 \paragraph*{Are all terms in the (super)potential in agreement with global and local symmetries?} As mentioned above, \SARAH doesn't forbid to include terms in the superpotential which violate global or gauge symmetries. However, it prints a warning if this happens. 
 \paragraph*{Are there other terms allowed in the (super)potential by global and local symmetries?} \SARAH will print a list of renormalizable terms which are allowed by all symmetries but which haven't been included in the model file. 
 \paragraph*{Are all unbroken gauge groups respected?} \SARAH checks what gauge symmetries remain unbroken and if the definition of all rotations in the matter sector and of the Dirac spinors are consistent with that. 
 \paragraph*{Are there terms in the Lagrangian of the mass eigenstates which can cause additional mixing between fields?} If in the final Lagrangian bilinear terms between different matter eigenstates are present this means that not the entire mixing of states has been taken into account. \SARAH checks if those terms are present and returns a warning showing the involved fields and the non-vanishing coefficients. 
 \paragraph*{Are all mass matrices irreducible?} If mass matrices are block diagonal, a mixing has been assumed which is actually not there. In that case \SARAH will point this out.
 \paragraph*{Are the properties of all particles and parameters defined correctly?} These are formal checks about the implementation of a model. It is checked for instance, if the number of PDGs fits to the number of generations for each particle class, if \LaTeX\ names are defined for all particles and parameters, if the position in a Les Houches spectrum file are defined for all parameters, etc. Not all of these warnings have to be addressed by the user. Especially, if he/she is not interested in the output which would fail because of missing definitions.

\section{Calculations and output}
\label{sec:analytical}
\SARAH can perform in its natural \Mathematica environment many calculations for a model on the analytical level. For an exhaustive numerical analysis usually one of the dedicated interfaces  to other tools is the best approach. I give in this section an overview what \SARAH calculates itself, and how that information is linked to other codes. 

\subsection{Renormalization group equations}
\label{sec:RGEs}
\SARAH calculates the SUSY RGEs at the one- and two-loop level. In general, the $\beta$-function of a parameter $c$ is parametrized by
\begin{equation}
\frac{d}{dt} c \equiv \beta_c = \frac{1}{16 \pi^2} \beta^{(1)}_c + \frac{1}{(16 \pi^2)} \beta^{(2)}_c 
\end{equation}
$\beta^{(1)}_c$, $\beta^{(2)}_c$ are the coefficients at one- and two-loop level. For the gauge couplings the generic one-loop expression is rather simple and reads
\begin{equation}
\label{eq:BetaG}
\beta_{g_A}^{(1)} = g_A^3 (S(R) - 3 C(G)) 
\end{equation}
$S(R)$ is the Dynkin index for the gauge group summed over all chiral superfields charged under that group, and $C(G)$ is the Casimir of the adjoint representation of the group. The two-loop expressions are more complicated and are skipped here. They are for instance given in Ref.~\cite{Martin:1993zk}.\\

The starting point for the calculation of the RGEs for the superpotential terms in \SARAH are the anomalous dimensions $\gamma$ for all superfields. These can be also parametrized by
\begin{equation}
 \gamma_{\hat \phi_i \hat \phi_j} = \frac{1}{16\pi^2} \gamma^{(1)}_{\hat \phi_i \hat \phi_i} + \frac{1}{(16 \pi^2)^2} \gamma^{(2)}_{\hat \phi_i \hat \phi_j}
\end{equation}
I want to stress again that $i,j$ are not generation indices but label the different fields. Generic formula for the one- and two-loop coefficients $\gamma^{(1)}$, $\gamma^{(2)}$ are given in Ref.~\cite{Martin:1993zk} as well. \SARAH includes the case of an anomalous dimension matrix with off-diagonal entries, i.e. $\hat \phi_i \neq \hat \phi_j$. That's for instance necessary in models with vector like quarks where the superpotential reads
\begin{equation}
W \supset Y_u \hat u \hat q \hat H_u + Y_U \hat U \hat q \hat H_u + M_U \hat U \hat \hat{\bar{{U}}}
\end{equation}
$\gamma_{\hat u \hat U}$ is not vanishing but receives already at one-loop contributions $\propto Y_u Y_U$. \\

From the anomalous dimensions it is straightforward to get the $\beta$-functions of the superpotential terms: for a generic superpotential of the form 
eqs.~(\ref{eq:W}) and (\ref{eq:WW}) the coefficients $\beta^{(x)}$ are given by
\begin{align}
\beta^{(x)}_{L^i} \sim & L^a \gamma^{(x)}_{a \hat \phi_i} \\
\beta^{(x)}_{M^{ij}} \sim & M^{ia} \gamma^{(x)}_{a \hat \phi_j} +  (j\leftrightarrow i) \\
\beta^{(x)}_{Y^{ijk}} \sim & Y^{ija} \gamma^{(x)}_{a \hat \phi_k} + (k\leftrightarrow i) + (k\leftrightarrow j) \\
\beta^{(x)}_{W^{ijkl}} \sim & W^{ijka} \gamma^{(x)}_{a \hat \phi_l} + (l\leftrightarrow i) + (l\leftrightarrow j) + (l\leftrightarrow k) 
\end{align}
up to constant coefficients. In the soft-breaking sector \SARAH includes also all standard terms of the form
\begin{equation}
- \La_{SB} = t^i \phi_i + \frac{1}{2} B^{ij} \phi_i \phi_j + \frac{1}{3!} T^{ijk} \phi_i \phi_j \phi_k + \frac{1}{4!} Q^{ijkl} \phi_i \phi_j \phi_k \phi_l + \frac{1}{2} (m^2)^j_i  \phi^{* i} \phi_j - \frac{1}{2} M \lambda \lambda
\end{equation}
The generic expressions for $B$'s, $T$'s, $m^2$'s and $M$'s up to two-loop are given again in Ref.~\cite{Martin:1993zk} which is used by \SARAH. The $\beta$-function for the linear soft-term $t$ is calculated using Ref.~\cite{Yamada:1994id}. For the quartic soft-term $Q$ the approach of Ref.~\cite{Jack:1997eh} is adopted. In this approach $\bar{\gamma}$ is defined by
\begin{equation}
\bar{\gamma}^{(x)}_{\hat \phi_i \hat \phi_j} = \left(M_A g_A^2 \frac{\partial }{\partial g_A^2} - T^{lmn} \frac{\partial}{\partial Y^{lmn}} \right) \gamma^{(x)}_{\hat \phi_i \hat \phi_j}
\end{equation}
The $\beta$-functions for $Q$ can then expressed by $\gamma$ and $\bar{\gamma}$:
\begin{equation}
\beta^{(x)}_{Q_{ijkl}} = \left[Q^{ijka} \gamma^{(x)}_{a \hat \phi_l} +  2 W^{ijka} \bar{\gamma}^{(x)}_{a \hat \phi_l} \right] + (l\leftrightarrow i) + (l\leftrightarrow j) + (l\leftrightarrow k) 
\end{equation}
In principle, the same approach can also be used for $B$ and $T$ terms as long as no gauge singlet exists in the model. Because of this restriction, \SARAH uses the more general expressions of Ref.~\cite{Martin:1993zk}. \\

The running of the Fayet-Iliopoulos $D$-term $\xi$ receives two contributions:
\begin{equation}
\beta_{\xi_A}^{(x)} = \frac{\beta_{g_A}^{(x)}}{g_A} \xi_A + \beta^{(x)}_{\hat \xi_A}
\end{equation}
The first part is already fixed by the running of the gauge coupling of the Abelian group, the second part, $\beta^{(x)}_{\hat \xi}$, is known even to three loops \cite{Jack:1999zs,Jack:2000nm}. \SARAH has implemented the one- and two-loop results which are rather simple: 
\begin{align}
\beta^{(1)}_{\hat \xi_A} =& 2 g_A \sum_i (Q^A_{\phi_i} m_{\phi_i \phi_i}^2) \equiv \sigma_{1,A}\\
\beta^{(1)}_{\hat \xi_A} =& - 4 g_A \sum_{ij} (Q^A_{\phi_i}m^2_{\phi_i \phi_j} \gamma^{(1)}_{\hat \phi_j \hat \phi_i}) \equiv \sigma_{3,A} 
\end{align}
$\sigma_1$ and $\sigma_3$ are traces which are also used to express the $\beta$-functions of the soft-scalar masses at one- and two-loop, see for instance  Ref.~\cite{Martin:1993zk}. \\

Finally, the $\beta$-functions for the gaugino mass parameters are 
\begin{equation}
\frac{d}{dt} \equiv \beta_M = \frac{1}{16\pi^2} \beta_M^{(1)} + \frac{1}{(16 \pi^2)} \beta_M^{(2)}
\end{equation}
where the expressions for $\beta_M^{(x)}$ are also given in Refs.~\cite{Martin:1993zk,Jack:1997pa,Yamada:1994id}. $\beta_M^{(1)}$ has actually a rather simple form similar to the one of the gauge couplings. One finds 
\begin{equation}
\label{eq:BetaM}
\beta_{M_A}^{(1)} = 2 g_A^2 \left(S(R) - 3 C(G)\right) M_A
\end{equation}
Therefore, the running of the gaugino masses are strongly correlated with the one of the gauge couplings. Thus, for a GUT model the hierarchy of the running gaugino masses is the same as the one for the gauge couplings. \\

The expressions presented in the early works of Refs.~\cite{Martin:1993zk,Jack:1997pa,Yamada:1994id} did actually not cover all possibilities and are not sufficient for any possible SUSY models which can be implemented in \SARAH. Therefore, \SARAH has implemented also some more results from literature which became available in the last few years. In the case of several $U(1)$'s, gauge-kinetic mixing can arise if the groups are not orthogonal. Substitution rules to translate the results of Ref.~\cite{Martin:1993zk} to those including gauge kinetic mixing where presented in Ref.~\cite{Fonseca:2011vn} and have been implemented in \SARAH \footnote{Another method to deal with gauge-kinetic mixing was proposed in Ref.~\cite{Braam:2011xh}}. For instance, to include gauge-kinetic mixing in the running of the gauge couplings and gaugino masses eqs.~(\ref{eq:BetaG}) and (\ref{eq:BetaM}) can be used together with the substitutions
\begin{align}
g_A^3 S(R) \to  & G \sum_{\hat \phi} V_{\hat \phi} V_{\hat \phi}^T \\
g_A^2 M_A S(R) \to  & M \sum_{\hat \phi} V_{\hat \phi} V_{\hat \phi}^T + \sum_{\hat \phi} V_{\hat \phi} V_{\hat \phi}^T M
\end{align}
Here, $G$ and $M$ are matrices carrying the gauge couplings and gaugino masses of all $U(1)$ groups, see also sec.~\ref{sec:supported_models}, and I introduced $V_{\hat \phi}= G^T Q_{\hat \phi}$. The sums are running over all chiral superfields ${\hat \phi}$. Also for all other terms involving gauge couplings and gaugino masses appearing in the $\beta$ functions similar rules are presented in Ref.~\cite{Fonseca:2011vn} which are used by \SARAH. \\

Furthermore, also the changes in the RGEs in the presence of Dirac gaugino mass terms are known today at the two-loop level, see Ref.~\cite{Goodsell:2012fm}. \SARAH makes use of Ref.~\cite{Goodsell:2012fm} to obtain the $\beta$-functions for the new mass parameters as well as to include new contribution to the RGEs of tadpole terms in presence of Dirac gauginos. The $\beta$ functions of a Dirac mass terms $m_D^{\hat \phi \lambda}\Psi \lambda_i$ are related to the anomalous dimension of the involved chiral superfield $\hat \phi$, whose fermionic component is $\Psi$, and to the running of the corresponding gauge coupling:
\begin{equation}
\beta_{m^{\hat \phi A}_D} = \gamma_{\hat \phi a} m_D^{a A} + \frac{\beta_{g_A}}{g_A} m_D^{\hat \phi A}
\end{equation}
The tadpole term receives two new contributions from Fayet-Iliopoulos terms discussed above and terms mimicking $B$ insertions
\begin{equation}
\beta^{(x)}_{t,DG} = \beta^{(x)}_t + \beta^{(x)}_{\hat \xi} + \beta^{(x)}_D
\end{equation}
Thus, the only missing piece is $\beta^{(x)}_D$ which is also calculated by \SARAH up to two-loop based on Ref.~\cite{Goodsell:2012fm}. \\

Finally, the set of SUSY RGEs is completed by using the results of Refs.~\cite{Sperling:2013eva,Sperling:2013xqa} to get the gauge dependence in the running of the VEVs. As consequence, the $\beta$-functions for the VEVs consist of two parts which are calculated independently by \SARAH
\begin{equation}
\beta^{(x)}_{v_\phi} = (\gamma^{S,(x)}_{\phi a}  + \hat{\gamma}^{S,(x)}_{\phi a})v_a
\end{equation}
$\gamma^S$ is the anomalous dimension of the scalar $\phi$ which receives the VEV $v_\phi$. The gauge dependent parts which vanish in Landau gauge are absorbed in $\hat{\gamma}^S$. All details about this calculation and the generic results for $\hat{\gamma}^{S,(x)}$  are given in Refs.~\cite{Sperling:2013eva,Sperling:2013xqa}.\\

I want to mention that \SARAH provides the same accuracy also for the RGEs for a non-SUSY model by making use of the generic results of Refs.~\cite{Machacek:1983tz,Machacek:1983fi,Machacek:1984zw,Luo:2002ti}. These results are completed by \cite{Fonseca:2013bua} to cover gauge kinetic mixing and again by Refs.~\cite{Sperling:2013eva,Sperling:2013xqa} to include the gauge-dependence of the running VEVs also in the non-SUSY case.

\paragraph*{Output} The RGEs calculated by \SARAH are outputted in different formats: (i) they are written in the internal \SARAH format in the output directory, (ii) they are included in the \LaTeX\ output in a much more readable format, (iii) they are exported into a format which can be used together with {\tt NDSolve} of \Mathematica to solve the RGEs numerically within \Mathematica, (iv) they are exported into \Fortran code which is used by {\btt SPheno}.

\subsection{Tadpole equations}
During the evaluation of a model, \SARAH calculates 'on the fly' all minimum conditions of the tree-level potential, the so called tadpole equations. In the case of no CP violation, in which complex scalars are decomposed as
\begin{equation}
S_i \to \frac{1}{\sqrt{2}}(v_i + \phi_i +  i \sigma_i) \,,
\end{equation}
the expressions 
\begin{equation}
0 = \frac{\partial V}{\partial \phi_i} \equiv T_i
\end{equation}
are calculated. These are equivalent to $\frac{\partial V}{\partial v_i}$. For models with CP violation in the Higgs sector, i.e. where either complex phases appear between the real scalars or where the VEVs have an imaginary part, \SARAH calculates the minimum conditions with respect to the CP-even and CP-odd components:
\begin{equation}
0 =  \frac{\partial V}{\partial \phi_i} \equiv T_{\phi_i} \,,\hspace{1cm} 0 =  \frac{\partial V}{\partial \sigma_i}   \equiv T_{\sigma_i}
\end{equation}
The set of all tadpole equations is in this case $T_i = \{T_{\phi_i},T_{\sigma_i}\}$.

\paragraph*{Output} The tadpole equations are exported into \LaTeX\ format as well as in \Fortran code used by {\btt SPheno}. This ensures that all parameter points evaluated by \SPheno are at least sitting at a local minimum of the scalar potential. Moreover, the tadpole equations are included in the model files for {\btt \Vevacious} which is used to find all possible solutions of them with respect to the different VEVs. 

\subsection{Masses and mass matrices}
\label{sec:calc_massmatrix}
\SARAH uses the definition of the rotations defined in the model file to calculate the mass matrices for particles which mix. 
The mass matrices for scalars are calculated by
 \begin{equation}
   M^S_{ij} = \frac{-\partial^2  \La}{\partial {\phi}_i \partial {\phi}_j^*} 
 \end{equation}
where $\phi$ can be either real or complex, i.e. the resulting $M^S$ corresponds to $m_C$ or $m_R$ of eq.~(\ref{eq:lagMasses}). In the mass matrices of states which include Goldstone bosons also the $R_\xi$ dependent terms are included.\\

The mass matrices for fermions are calculated as
\begin{equation}
  M^F_{ij} = \frac{-\partial^2  \La}{\partial {\psi}^x_i {\psi}^y_j} 
\end{equation}
with $x=y=0$ for Majorana fermions, and $x=1$, $y=2$ for Dirac fermions. 
 
\SARAH calculates for all states which are rotated to mass eigenstates the mass matrices during the evaluation of a model. In addition, it checks if there are also particles where gauge and mass eigenstates are identical. In that case, it calculates also the expressions for the masses of these states. 
 
\paragraph*{Output} The tree-level masses and mass matrices are also exported to \LaTeX\ files as well as to \Fortran code for {\btt \SPheno}. In addition, they are used in the {\btt Vevacious} output to enable the calculation of the one-loop effective potential. The mass matrices can also be exported to the {\btt CalcHep} model files if the user wants to calculate the masses internally with \CalcHep instead of using them as input.

\subsection{Vertices}
Vertices are not automatically calculated during the initialization of a model like this is done for mass matrices and tadpole equations. However, the calculation can be started very easily. In general, \SARAH is optimized for the extraction of three- and four-point interactions with renormalizable operators.  That means, usually only the following generic interactions are taken into account in the calculations: interactions of two fermions or two ghosts with one scalar or vector bosons ({\tt FFS}, {\tt FFV}, {\tt GGS}, {\tt GGV}), interactions of three or four scalars or vector bosons ({\tt SSS}, {\tt SSSS}, {\tt VVV}, {\tt VVVV}), as well as interactions of two scalars with one or two vector bosons ({\tt SSV}, {\tt SSVV}) or two vector bosons with one scalar ({\tt SVV}).  \\
In this context, vertices not involving fermions are calculated by
\begin{align}
V(\eta_a,\eta_b,\eta_c) \equiv & i \frac{\partial^3 \La}{\partial \eta_a \partial \eta_b \partial \eta_c} = C \Gamma \\
V(\eta_a,\eta_b,\eta_c, \eta_d) \equiv= & i \frac{\partial^4 \La}{\partial \eta_a \partial \eta_b \partial \eta_c \partial \eta_d} = C \Gamma
\end{align}
Here, $\eta_i$ are either scalars, vector bosons, or ghosts. The results are expressed by a coefficient $C$ which is a Lorentz invariant and a Lorentz factor $\Gamma$ which involves $\gamma_\mu$, $p_\mu$, or  $g^{\mu\nu}$. Vertices for Dirac fermions are first expressed in terms of Weyl fermions. The two vertices are then calculated separately. Taking two Dirac fermions $F_a = (\Psi^1_a, \Psi^{2*}_a)$, $F_b = (\Psi^1_b, \Psi^{2*}_b)$ and distinguishing the two cases for fermion--vector and fermion--scalar couplings, the vertices are calculated and expressed by
\begin{align}
V(\bar F_a,F_b,V_c) &= \{V(\Psi_a^{1 *},\Psi_b^1,V_c), V(\Psi_a^2,\Psi_b^{2*},V_c)\} \equiv \{C^L \gamma_\mu P_L, C^R \gamma_\mu P_R\} \\
V(\bar F_a,F_b,S_c) &= \{V(\Psi_a^2,\Psi_b^1,S_c), V(\Psi_a^{1*},\Psi_b^{2*},V_c)\} \equiv \{C^L P_L, C^R P_R\}
\end{align}
Here, the polarization operators $P_{L,R}$ are used. \\

The user can either calculate specific vertices for a particular set of external states or call functions that \SARAH derives all existing interactions from the Lagrangian. The first option might be useful to check the exact structure of single vertices, while the second one is needed to get all vertices to write model files for other tools. 

\paragraph*{Output} The vertices are exported into many different formats. They are saved in the \SARAH internal format and they can be written to \LaTeX\ files. The main purpose is the export into formats which can be used with other tools. \SARAH writes model files for {\btt FeynArts}, {\btt WHIZARD/OMEGA}, {\btt CalcHep/CompHep} as well as in the {\btt UFO} format. The {\btt UFO} format is supported by {\btt MadGraph}, {\btt Herwigg+} and {\btt Sherpa}. Thus, by the output of the vertices into these different format, \SARAH provides an implementation of a given model in a wide range of HEP tools. In addition, \SARAH generates also \Fortran code to implement all vertices in {\btt SPheno}. 

\subsection{One- and two-loop corrections to tadpoles and self-energies}
\label{sec:loopcorrections}
\subsubsection{One-loop corrections}
\label{sec:OneLoopSelf}
\begin{figure}[hbt]
\centering
\includegraphics[width=1.0\linewidth]{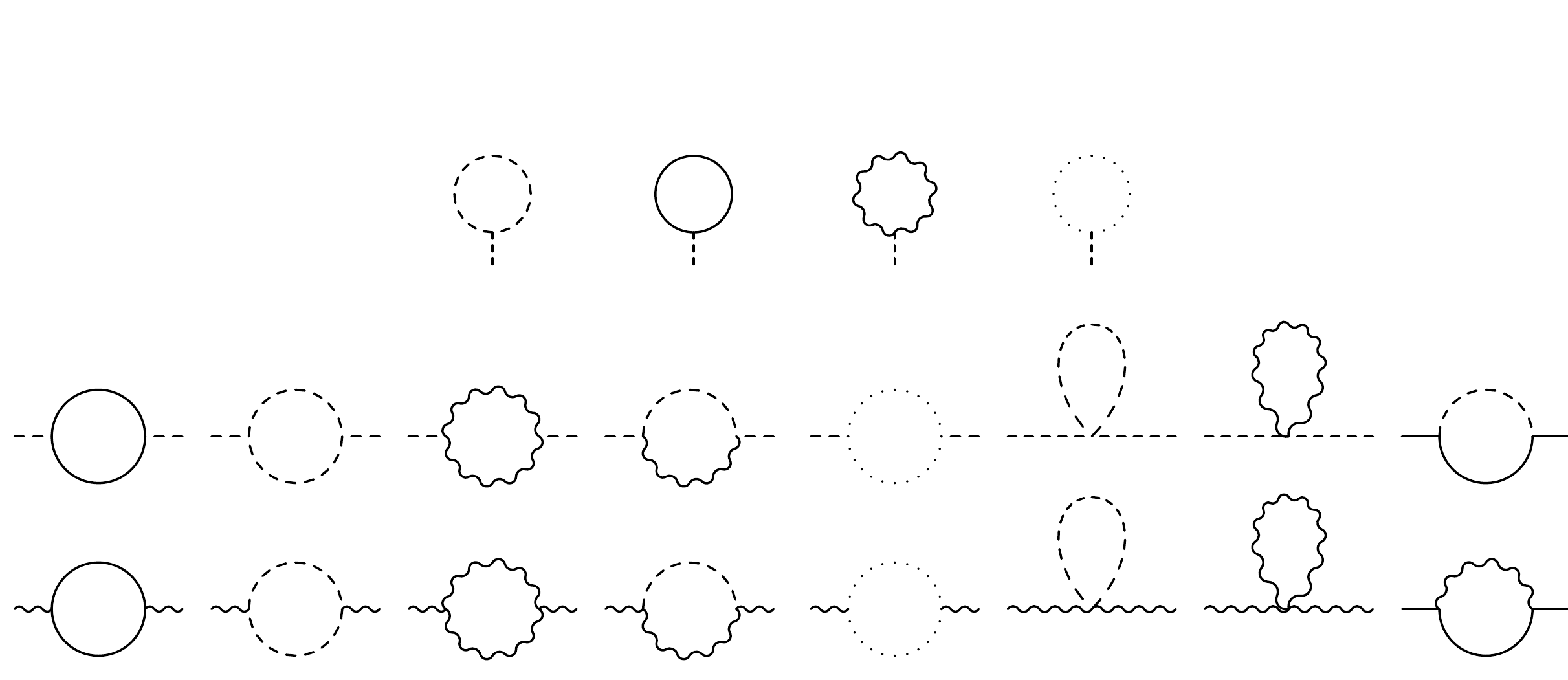} 
\caption{Generic diagrams included by \SARAH to calculate one-loop tadpoles and self-energies.}
\label{fig:1loopDiagrams}
\end{figure}

\SARAH calculates the analytical expressions for the one-loop corrections to the tadpoles and the one-loop self-energies for all particles. For states which are a mixture of several gauge eigenstates, the self-energy matrices are calculated. For doing that, \SARAH is working with gauge eigenstates as external particles but uses mass eigenstates in the loop. The calculations are performed in $\DRbar$-scheme using 't Hooft gauge. In the case of non-SUSY models \SARAH switches to $\MSbar$-scheme. This approach is a generalization of the procedure applied in Ref.~\cite{Pierce:1996zz} to the MSSM.  In this context, the following results are obtained:
\begin{itemize}
 \item The self-energies $\Pi$ of scalars and scalar mass matrices
 \item The self-energies $\Sigma^L$, $\Sigma^R$, $\Sigma^S$ for fermions and fermion mass matrices
 \item The transversal self-energy $\Pi^T$ of massive vector bosons
\end{itemize}
The approach to calculate the loop corrections is as follows: all possible generic diagrams at the one-loop level shown in Fig.~\ref{fig:1loopDiagrams} are included in \SARAH. Each generic amplitude is parametrized by
\begin{equation}
\label{eq:genericloop}
\mathscr{M} = \text{Symmetry} \times \text{Colour} \times \text{Couplings} \times \text{Loop-Function} 
\end{equation}
Here 'Symmetry' and 'Colour' are real factors. The loop-functions are expressed by standard Passarino-Veltman  integrals $A_0$ and $B_0$ and some related functions: $B_1$, $B_{22}$, $F_0$, $G_0$, $H_0$, $\bar{B}_{22}$ as defined in Ref.~\cite{Pierce:1996zz}. \\

As first step to get the loop corrections, \SARAH generates all possible Feynman diagrams with all field combinations possible in the considered model. 
%
The second step is to match these diagrams to the generic expressions. All calculations are done without any assumption and always the most general case is taken. For instance, the generic expression for a purely scalar contribution to the scalar self-energy reads
\begin{equation}
\mathscr{M}_{SSS} = c_S \times c_F \times |c|^2 B_0(p^2, m_{S_1}^2, m_{S_2}^2)
\end{equation}
In the case of an external charged Higgs $\phi^+ = ((H_d^-)^*, H_u^+)$ together with down- and up-squarks in the loop the correction to the charged Higgs mass matrix becomes
\begin{equation}
\mathscr{M}_{\phi^+_a \tilde{u} \tilde{d}^*} = 3 \times \sum_{i=1}^6 \sum_{j=1}^6 |c(\phi^+_a \tilde{u}_i \tilde{d}^*_j)|^2 B_0(p^2,m_{\tilde u_i}^2,m_{\tilde d_j}^2)
\end{equation}
$c(\phi^+_a \tilde{u}_i \tilde{d}^*_j)$ is the charged Higgs-sdown-sup vertex where the rotation matrix of the charged Higgs are replaced by the identity matrix to get the projection on the gauge eigenstates. One can see that all possible combinations of internal generations are included, i.e. also effects like flavour mixing are completely covered. Also the entire $p^2$ dependence is kept. 

\paragraph*{Output} The one-loop expressions are saved in the \SARAH internal \Mathematica format and can be included in the \LaTeX\ output. In addition, all self-energies and one-tadpoles are exported into \Fortran code for {\btt SPheno}. This enable \SPheno to calculate the loop-corrected masses for all particles  as discussed below.

\subsubsection{Two-loop corrections}
It is even possible to go beyond one-loop with \SARAH and to calculate two-loop contributions to the self-energies of real scalars. There are two equivalent approaches implemented in the \SPheno interface of \SARAH to perform these calculations: an effective potential approach, and a diagrammatic approach with vanishing external momenta. Because of the very complicated form of the results there is no output of the corresponding expressions in the \Mathematica or \LaTeX\ format but the results are just included in the \Fortran code for numerical evaluation. I'll discuss both calculations a bit more.  
\begin{figure}[hbt]
\centering
 \includegraphics[width=0.5\linewidth]{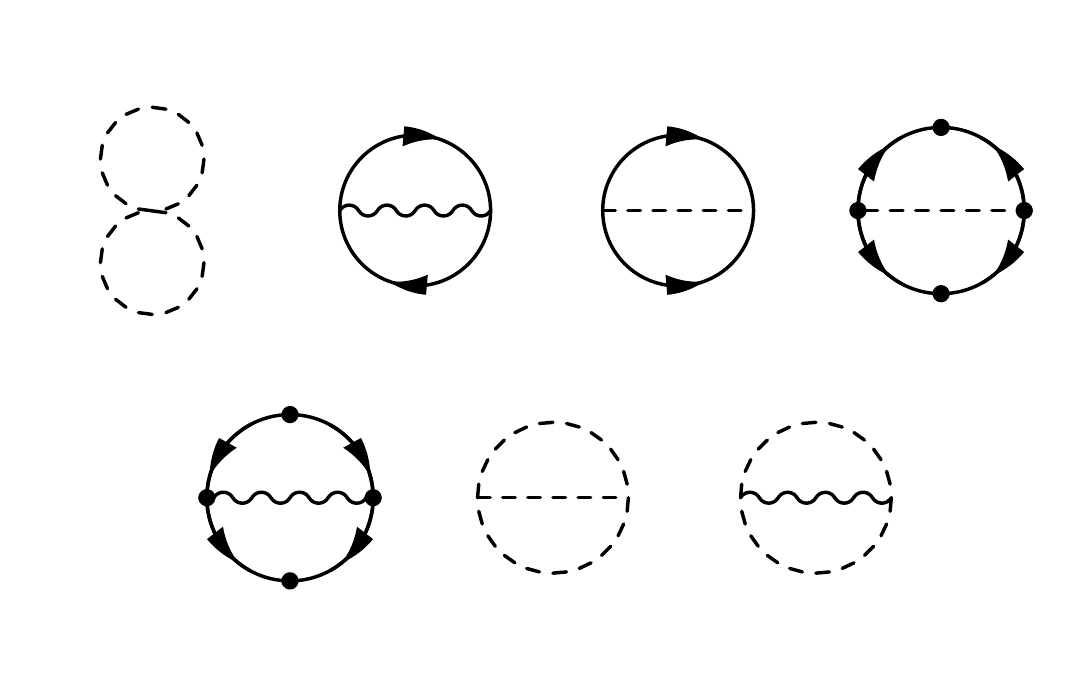}
\caption{Generic diagrams included to calculate the effective potential at the two-loop level. These are the diagram which don't vanish in the gaugeless limit.}
\label{fig:2loopEffPot}
\end{figure}
\paragraph{Effective potential calculation} The first calculation of the two-loop self-energies is based on the effective potential approach.  The starting point of the calculation are the generic results for the two-loop effective potential given in Ref.~\cite{Martin:2001vx}. These have been translated to four component notation and were implemented in \SARAH. When \SARAH creates the \SPheno output it writes down the amplitude for all two-loop diagrams which don't vanish in the gaugeless limit. This limit means that contributions from broken gauge groups are ignored. The remaining generic diagrams which are included are shown in Fig.~\ref{fig:2loopEffPot}. Using these diagrams includes for instance all two-loop contributions which are also taken into account in the MSSM. To get the values for the two-loop self-energies and two-loop tadpoles, the derivatives of the potential with respect to the VEVs are taken numerically as proposed in Ref.~\cite{Martin:2002wn}. There are two possibilities for this derivation implemented in \SARAH/\SPheno: (i)  a fully numerical procedure which takes the derivative of the full effective potential with respect to the VEVs. A semi-analytical derivation which takes analytical the derivative of the loop functions with respect to involved masses, but derives the masses and coupling numerically with respect to the VEVs. More details about both methods and the numerical differences are given in Ref.~\cite{Goodsell:2014bna}.

\begin{figure}[hbt]
 \includegraphics[width=1.0\linewidth]{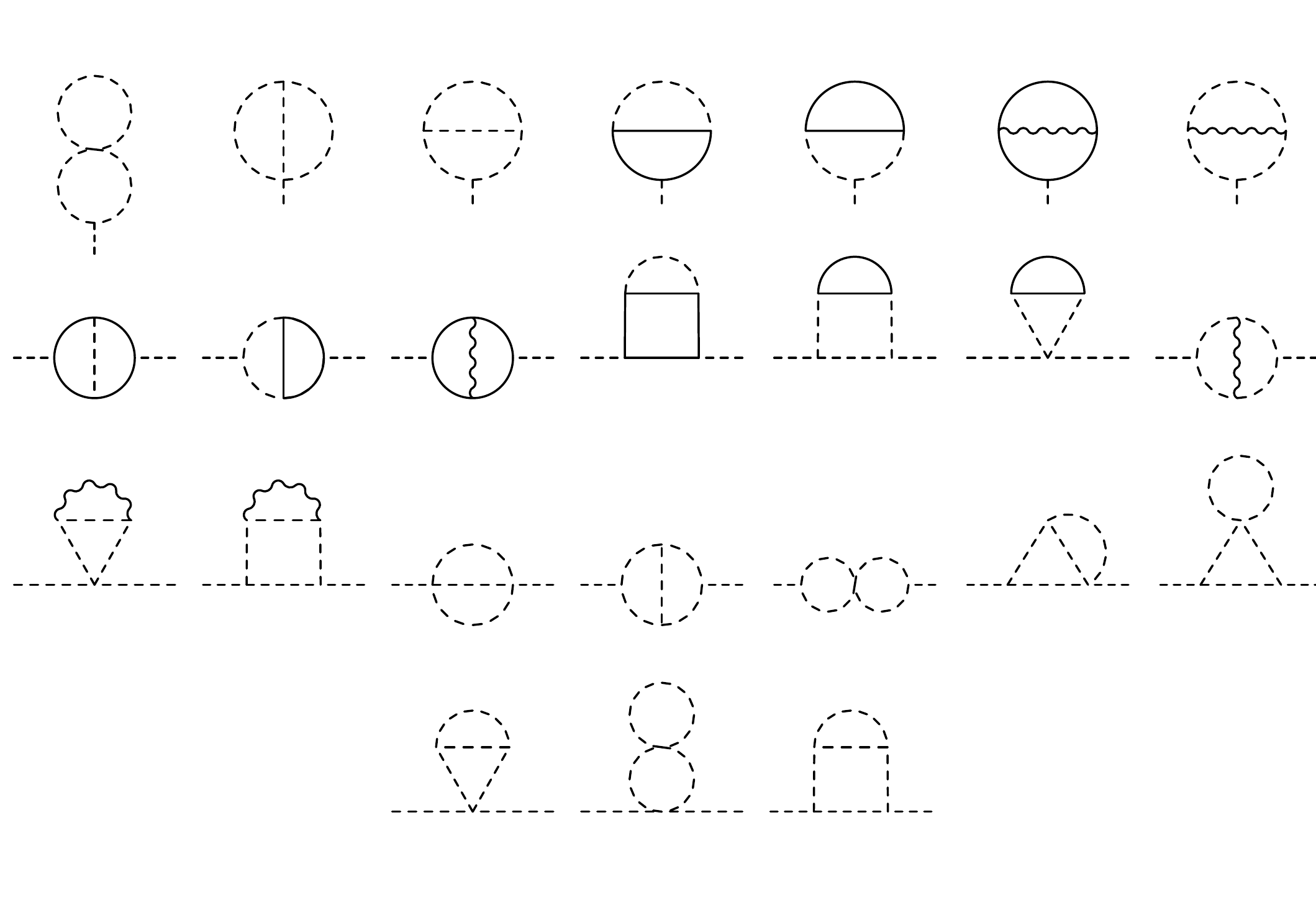}
 \caption{Generic diagrams included to calculate the two two-loop corrections to tadpoles and scalars. These are the diagram which don't vanish in the gaugeless.}
\label{fig:2loopDiagram}
\end{figure}

\paragraph*{Diagrammatic calculation} A fully diagrammatic calculation for two-loop contributions to scalar self-energies with \SARAH--\SPheno became available with Ref.~\cite{Goodsell:2015ira}. In this setup a set of generic expressions first derived in  Ref.~\cite{Goodsell:2015ira} is used. All two-loop diagrams shown in Fig.~\ref{fig:2loopDiagram} are included in the limit $p^2=0$. These are again the diagrams which don't vanish in general in the gaugeless limit. The results of Ref.~\cite{Goodsell:2015ira} have the advantage that the expressions which are derived from the effective potential are much simpler than taking the limit $p^2\to0$ in other two-loop functions available in literature \cite{Martin:2005eg}. The diagrammatic method gives completely equivalent results to the effective potential calculation but is usually numerically more robust.

\paragraph*{The need for both calculations} Since both calculations are based on a completely independent implementation and use a different approach they are very useful to perform cross checks. For the MSSM and NMSSM both calculations reproduce exactly the results obtained by widely used routines based on Refs.~\cite{Brignole:2001jy,Degrassi:2001yf,Brignole:2002bz,Dedes:2002dy,Dedes:2003km,Degrassi:2009yq}.  However, for non-minimal SUSY models are no references available to compare with. Thus, the only possibility to cross check the results is within \SPheno and comparing the two different methods. 

\paragraph*{Output} The two-loop expressions for the effective potential, the tadpoles and the self-energies are just exported to {\btt SPheno} at the moment to calculate the loop corrected mass spectrum. 

\subsection{Loop corrected mass spectrum}
The information about the one- and two-loop corrections to the one- and two-point functions introduced in sec.~\ref{sec:loopcorrections} can be used to calculate the loop corrected mass spectrum. Sticking to approach of Ref.~\cite{Pierce:1996zz}, the renormalized mass matrices (or masses) are related to the tree-level mass matrices (or masses) and the self-energies as follows.

\subsubsection{Loop corrected masses}
\paragraph*{Real scalars} For a real scalar $\phi$, the one-loop, and in some cases also two-loop, self-energies are calculated by \SPheno. The loop corrected mass matrix squared $m_\phi^{2,(L)}$ is related to the tree-level mass matrix squared $m_\phi^{2,(T)}$ and the self-energies via
\begin{equation}
\label{eq:RealScalarLoop}
 m_\phi^{2,(L)}(p^2) = m^{2,(T)}_\phi -  \Re(\Pi^{(1L)}_{\phi}(p^2)) -  \Re(\Pi^{(2L)}_{\phi}(0))
\end{equation}
The one-shell condition for the eigenvalue $M^2_{\phi_i}(p^2)$ of the loop corrected mass matrix $m_\phi^{2,(L)}(p^2)$ reads 
\begin{equation}
\mathrm{Det}\left[ p^2_i \mathbf{1} - M^2_{\phi_i}(p^2) \right] = 0,
\label{eq:propagator}
\end{equation}
A stable solution of eq.~(\ref{eq:propagator}) for each eigenvalue $M^2_{\phi_i}(p^2=M^2_{\phi_i})$ is usually just found via an iterative procedure. In this approach one has to be careful how $m_\phi^{2,(T)}$ is defined: this is the tree-level mass matrix where the parameters are taken at the minimum of the effective potential evaluated at the same loop-level at which the self-energies are known. The physical masses are associated with the eigenvalues $M^2_{\phi_i}(p^2=M^2_{\phi_i})$. In general, for each eigenvalue the rotation matrix is slightly different because of the $p^2$ dependence of the self-energies. The convention by \SARAH and \SPheno is that the rotation matrix of the lightest eigenvalue is used in all further calculations and the output. 

\paragraph*{Complex scalars} For a complex scalar $\eta$ the one-loop corrected mass matrix squared is related to the tree-level mass and the one-loop self-energy via
\begin{equation}
	m_\eta^{2,(1L)}(p^2) = m_\eta^{(T)} - \Pi^{(1L)}_{\eta}(p^2_i) ,
\end{equation}
The same on-shell condition, eq.~(\ref{eq:propagator}),  as for real scalars is used. 

\paragraph*{Vector bosons} For vector bosons we have similar simple expressions as for scalar. The one-loop masses of real or complex vector bosons $V$ are given by
\begin{equation}
 m^{2,(1L)}_{V} = m^{2,(T)}_{V} - \Re(\Pi^{T,(1L)}_{V}(p^2)) 
\end{equation}

\paragraph*{Majorana fermions} The one-loop mass matrix of a Majorana  fermion $\chi$ is related
to the tree-level mass matrix $m_\chi^{(T)}$ and the different parts of the self-energies by
\begin{eqnarray}
m_\chi^{(1L)} (p^2) &=& m_\chi^{(T)} - \frac{1}{2} \bigg[ \Sigma^\chi_S(p^2) + \Sigma^{\chi,T}_S(p^2)
 + \left(\Sigma^{\chi,T}_L(p^2)+   \Sigma^\chi_R(p^2)\right) m_\chi^{(T)}
 \nonumber \\
&& \hspace{16mm}
+ m_\chi^{(T)} \left(\Sigma^{\chi,T}_R(p^2) +  \Sigma^\chi_L(p^2) \right)
 \bigg] 
\end{eqnarray}
Note, $(T)$ is used to assign tree-level values while $T$ denotes a transposition. Eq.~(\ref{eq:propagator}) can also be used for fermions by taking the eigenvalues of $m_\chi^{2,(1L)}= m_\chi^{(1L)*}  m_\chi^{(1L)}$. 

\paragraph*{Dirac fermions} For a Dirac fermion $\Psi$ one obtains the one-loop corrected mass matrix via
\begin{eqnarray}
\label{eq:DiracLoop}
m_\Psi^{(1L)}(p^2) =  m_\Psi^{(T)} - \Sigma^+_S(p^2)  - \Sigma^+_R(p^2) m_\Psi^{(T)} - m_\Psi^{(T)} \Sigma^+_L(p^2) .
\end{eqnarray}
Here, the eigenvalues of $(m_\Psi^{(1L)})^\dagger m_\Psi^{(1L)}$ are used in eq.~(\ref{eq:propagator}) to get the pole masses.

\subsubsection{Renormalization procedure} I have explained so far how \SPheno does calculate the one- and two-loop self-energies and how these are related to the loop corrected masses. Now, it is time to put this in a more global picture by describing step-by-step the entire renormalization procedure \SPheno uses:
\begin{enumerate}
 \item Everything starts with calculating the running parameters at the renormalization scale $Q=M_{SUSY}$ from the given input parameters. The  parameters can be given either directly at $M_{SUSY}$ as input or they are fixed by some GUT conditions and a RGE running is performed. $M_{SUSY}$ itself can be either be a fixed value or can be dynamically chosen. It is common to choose the geometric mean of the stop masses because this usually minimizes the scale dependence of the Higgs mass prediction. 
 \item Not all parameters are fixed by the input but some parameters are kept free. These parameters are arranged in a way that all further calculations are done at the minimum of the potential. For this purpose the tadpole equations $T_i$ are solved at tree-level with respect to these free parameters.
 \item As soon as all running parameters are known at the SUSY scale, they are used to calculate the tree-level mass spectrum. 
 \item The tree-level masses are used to calculate the self-energies of the $Z$-boson, $\Pi^T_{Z}(M^{(p),2}_Z)$ where $M^{(p)}_Z$ is the pole mass.
 \item $M^{(p)}_Z$, $\Pi^T_{Z}(M^{(p),2}_Z)$ are used to get the tree-level value of the electroweak VEV $v$. $v$ and the running value of $\tan\beta$ are used to get tree-level VEVs $v_u$, $v_d$. Note, in this step it is assumed that always two Higgs doublets are present in SUSY models which give mass to up- and down quark as well as leptons and gauge bosons. 
 \item Now, all tree-level parameters are known and the tree-level masses and rotation matrices are re-calculated using the obtained values.
 \item Tree-level masses, rotation matrices and parameter are used to get all vertices at tree-level. The vertices and masses are then plugged in the expressions for the one- and two-loop corrections to the tadpoles  $\delta t_i^{(x)}$ ($x=1,2$). The conditions to work at the minimum of the effective potential are
\begin{equation}
\label{eq:TadLoop}
T_i + \delta t_i^{(1)} + \delta  t_i^{(2)} \equiv 0.
\end{equation}
These equations are again solved for the same set of parameters as at tree-level. 
\item The self-energies for all particles are calculated at the highest available loop-level as explained above. Note, these calculations involve purely tree-level parameters but not the ones obtained from eq.~(\ref{eq:TadLoop}).
\item Eqs.~(\ref{eq:RealScalarLoop})--(\ref{eq:DiracLoop}) are used to get the loop corrected mass matrices for all particle. Now, the parameters coming from loop corrected tadpoles are used to express the tree-level mass matrices. All calculations are iterated until the on-shell condition is satisfied for all masses. 
\end{enumerate}

\subsubsection{Thresholds} 
\label{sec:thresholds}
So far, I haven't mentioned another subtlety: in general, the running SM parameters depend on the SUSY masses. The reason are the thresholds to match the running parameters to the measured ones. These thresholds change, when the mass spectrum changes. Therefore, the above procedure is iterated until the entire loop corrected mass spectrum has converged. The calculation of the thresholds is also dynamically adjusted by \SARAH to include all new physics contributions. The general procedure to obtain the running gauge and Yukawa at $M_Z$ is as follows:
\begin{enumerate}
 \item The first step is the calculation of $\alpha^{\DRbar}(M_Z)$, $\alpha_S^{\DRbar}(M_Z)$ via
\begin{align}
  \alpha^{\DRbar}(M_Z) &= \frac{\alpha^{(5),\overline{\text{MS}}}(M_Z)}{1 - \Delta\alpha^{\text{SM}}(M_Z) - \Delta\alpha^{\text{NP}}(M_Z)} ,\\
  \alpha_S^{\DRbar}(M_Z) &= \frac{\alpha_S^{(5),\overline{\text{MS}}}(M_Z)}{1 - \Delta\alpha_S^{\text{SM}}(M_Z)    - \Delta\alpha_S^{\text{NP}}(M_Z)} 
\end{align}
Here, $\alpha_S^{(5),\overline{\text{MS}}}$ and $\alpha^{(5),\overline{\text{MS}}}$ are taken as input and receive corrections from the top loops as well as form new physics (NP):
\begin{eqnarray}
&  \Delta\alpha^{\text{SM}}(\mu) = \frac{\alpha}{2\pi} \left(\frac{1}{3}- \frac{16}{9} \log{\frac{m_t}{\mu}} \right), \hspace{1cm}
  \Delta\alpha^{\text{NP}}(\mu) =\frac{\alpha}{2\pi} \left( -\sum_i c_i \log{\frac{m_i}{\mu}} \right), &\\
&  \Delta\alpha_S^{\text{SM}}(\mu) = \frac{\alpha_\text{s}}{2\pi} \left( -\frac{2}{3} \log{\frac{m_t}{\mu}} \right), \hspace{1cm}
  \Delta\alpha_S^{\text{NP}}(\mu) = \frac{\alpha_S}{2\pi}\left( \frac{1}{2}-\sum_i c_i \log{\frac{m_i}{\mu}} \right) .&
\end{eqnarray}
The sum runs over all particles $i$ which are not present in the SM and which are either charged or coloured. The coefficients $c_i$ depends on the charge respectively colour representation, the generic type of the particle (scalar, fermion, vector), and the degrees of freedom of the particle (real/complex boson respectively Majorana/Dirac fermion). 
\item The next step is the calculation of the running Weinberg angle $\sin\Theta^{\DRbar}$ and electroweak VEV $v$. For that the one-loop corrections $\delta M_Z^2$ and $\delta M_W^2$ to the $Z$- and $W$-mass are needed. And an iterative procedure is applied with   $\Theta^{\DRbar}_W = \Theta_W^{\text{SM}}$ in the first iteration together with:
\begin{align}
v^2 =& (M_Z^2 + \delta M_Z^2) \frac{(1- \sin^2\Theta^{\DRbar}_W)\sin^2\Theta^\DRbar_W}{\pi \alpha^{\DRbar}} \\
sin^2\Theta^{\DRbar}_W =&  \frac{1}{2} - \sqrt{\frac{1}{4} - \frac{\pi \alpha^{\DRbar}}{\sqrt{2} M_Z^2 G_F (1-\delta_r)}}
\end{align}
Here, $G_F$ is the Fermi constant and $\delta_r$ is defined by
\begin{equation}
\delta_r = \rho \frac{\delta M_W^2}{M_W^2} - \frac{\delta M_Z^2}{M_Z^2} + \delta_{VB}
\end{equation}
where $\delta_{VB}$ are the corrections to the muon decay $\mu \to e \nu_i \bar \nu_j$ which are calculated at one-loop as well. The $\rho$ parameter is calculated also at full one-loop and the known two-loop SM corrections are added.
\item With the obtained information, the running gauge couplings at $M_Z$ are given by
\begin{eqnarray}
&  g_1^{\DRbar}(M_Z) =
  \frac{\sqrt{4 \pi \alpha^{\DRbar}(M_Z)}}{\cos\theta_{W}^{\DRbar}(M_Z)}, \hspace{0.5cm}
  g_2^{\DRbar}(M_Z) =
  \frac{\sqrt{4\pi \alpha^{\DRbar}(M_Z)}}{\sin\theta_{W}^{\DRbar}(M_Z)}, \hspace{0.5cm}
  g_3^{\DRbar}(M_Z) = \sqrt{4 \pi \alpha_S^{\DRbar}(M_Z)} &
\end{eqnarray}
\item The running Yukawa couplings are also calculated in an iterative way. The starting point are the running fermion masses in \DRbar\ obtained from the pole masses given as input:
\begin{align}
m_{l,\mu,\tau}^{\DRbar,\text{SM}} =& m_{l,\mu,\tau} \left(1 - \frac{3}{128 \pi^2}(g^{\DRbar,2}_1+g^{\DRbar,2}_2)\right) \\
m_{d,s,b}^{\DRbar,\text{SM}} =& m_{d,s,b} \left(1- \frac{\alpha^\DRbar_S}{3\pi}-\frac{23 \alpha_S^{\DRbar,2}}{72 \pi^2} + \frac{3}{128 \pi^2} g^{\DRbar,2}_2 - \frac{13}{1152 \pi^2} g^{\DRbar,2}_1\right)\\
m_{u,c}^{\DRbar,\text{SM}} =& m_{u,c} \left(1- \frac{\alpha^\DRbar_S}{3\pi}-\frac{23 \alpha^{\DRbar,2}_S}{72 \pi^2} + \frac{3}{128 \pi^2} g^{\DRbar,2}_2 - \frac{7}{1152 \pi^2} g^{\DRbar,2}_1\right)\\
m^{\DRbar,\text{SM}}_t =& m_t \left[1 + \frac{1}{16\pi^2} \left(\Delta m_t^{(1),qcd} +\Delta m_t^{(2),qcd} + \Delta m_t^{(1),ew}\right)\right]
\end{align}
with 
\begin{align}
 \Delta m_t^{(1),qcd} &= -\frac{16 \pi \alpha_S^\DRbar }{3} \left(5 + 3 \log\frac{M_Z^2}{m_t^2} \right) \\
 \Delta m_t^{(2),qcd} &=  -\frac{64 \pi^2 \alpha_S^{\DRbar,2} }{3} \left(\frac{1}{24}+\frac{2011}{384\pi^2}+\frac{\ln2}{12}-\frac{\zeta(3)}{8\pi^2}+\frac{123}{32\pi^2} \log\frac{M_Z^2}{m_t^2}+\frac{33}{32\pi^2} \left(\log\frac{M_Z^2}{m_t^2}\right)^2 \right)\\
 \Delta m_t^{(1),ew} &= - \frac{4}{9} g_2^{\DRbar,2} \sin^2 \Theta^\DRbar_W \left(5 + 3 \log\frac{M_Z^2}{m_t^2}\right)
\end{align}
The two-loop parts are taken from Ref.~\cite{Avdeev:1997sz,Bednyakov:2002sf}. 
The masses are matched to the eigenvalues of the loop-corrected fermion mass matrices calculated as
\begin{equation}
m_f^{(1L)}(p^2_i) =  m_f^{(T)} - \tilde{\Sigma}^{+}_S(p^2_i)
 - \tilde{\Sigma}^{+}_R(p^2_i) m_f^{(T)} - m_f^{(T)} \tilde{\Sigma}^{+}_L(p^2_i) 
\end{equation}
Here, the pure QCD and QED corrections are dropped in the self-energies $\tilde \Sigma $. Inverting this relation to get the running tree-level mass matrix $m_f^{(T)}$ leads to  $Y_d^{\DRbar}$, $Y_u^{\DRbar}$, $Y_e^{\DRbar}$. Since the self-energies depend also one the Yukawa matrices, this calculation has to be iterated until a stable point is reached. Optionally, also the constraint that the CKM matrix is reproduced can be included in the matching. 
\end{enumerate}

\paragraph*{Output} The calculation of the loop corrected mass spectrum and the thresholds is included in the {\btt SPheno} output.

\subsection{Decays and branching ratios}
The calculation of decays widths and branching ratios can be done by using the interface between \SARAH and \SPheno. 
\SPheno modules created by \SARAH calculate all two-body decays for SUSY and Higgs states as well as for additional gauge bosons. In addition, the three-body decays of a fermion into three other fermions and of a scalar into another scalar and two fermions are included. \\
In the Higgs sector, possible decays into two SUSY particles, leptons and massive gauge bosons are calculated at tree-level. For two quarks in the final state the dominant QCD corrections due to gluons are included \cite{Spira:1995rr}. The loop induced decays into two photons and gluons are fully calculated at leading-order (LO) with the dominant next-to-leading-order corrections known from the MSSM. For the LO contributions all charged and coloured states in the given model are included in the loop, i.e. new contributions rising in a model beyond the MSSM are fully covered at one-loop. 
In addition, in the Higgs decays also final states with off-shell gauge bosons ($Z Z^*$, $W W^*$) are included. The only missing piece is the $\gamma Z$ channel. The corresponding loops are not yet derived by \SARAH and the partial width is set to zero. \\

In contrast to other spectrum generators, \SPheno modules by \SARAH perform a RGE running to the mass scale of the decaying particle. This should give a more accurate prediction for the decay width and branching ratios. However, the user can also turn off this running and use always the parameters as calculated at $M_{SUSY}$ in all decays as this is done by other codes. 

\paragraph*{Output} All necessary routines to calculate the two- and three-body decays are included by default in the \Fortran output for {\btt SPheno}. 

\subsection{Higgs coupling ratios}
With the discovery of the Higgs boson at the LHC and the precise measurements of its mass and couplings to other particles, a new era of high energy physics has started. Today, many SUSY models have not only be confronted with the exclusion limits from direct searches, but they have also to reproduce the Higgs properties correctly. The agreement with respect to the mass can be easily read off a spectrum file. For the rates this is usually not so easy. One can parametrize how 'SM-like' the couplings of a particular scalar are by considering the ratio 
\begin{equation}
 \label{eq:HiggsCouplingRatios}
r^{\phi XY} = (c_{\phi XY}^{SUSY}/c_{h XY}^{SM})^2 \, .
\end{equation}
Here, $c_{\phi XY}^{SUSY}$ is the calculated coupling between a scalar $\phi$ and two SM particles $X$ and $Y$ for a particular parameter point in a particular model. This coupling is normalized to the SM expectation for the same interaction. Nowadays, all $r^{\phi XY}$ are constrained to be rather close to 1 if $\phi$ should be associated with the SM Higgs. \SARAH uses the information which is already available from the calculation of the decays to obtain also values for $r^{\phi XY}$. Of course, also here the $\gamma Z$ channel is missing and $r^{\phi \gamma Z}$ is therefore put always to 0. 

\paragraph*{Output} All necessary routines to calculate the Higgs coupling ratios are included by default in the \Fortran output for {\btt SPheno}. 

\subsection{Flavour and precision observables}
\begin{figure}[hbt]
\centering
\includegraphics[width=\linewidth]{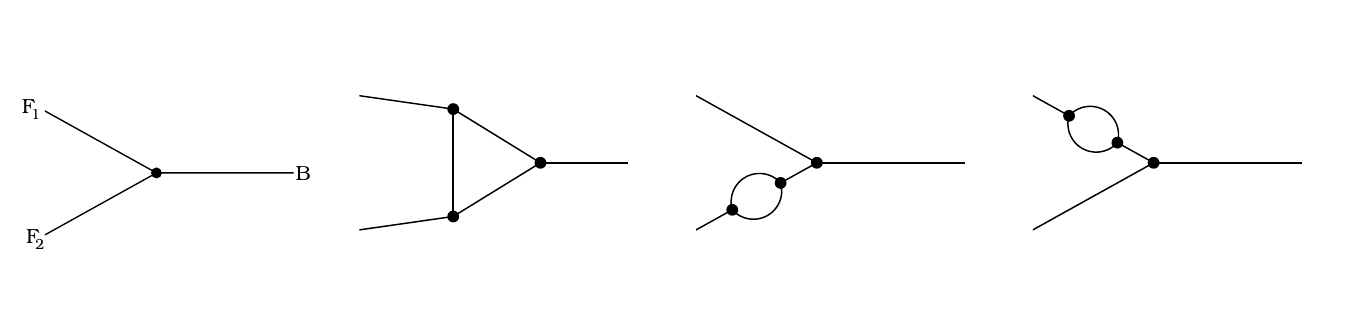}
\caption{All topologies considered by \FlavorKit to calculate the
  Wilson coefficients of 2-fermion-1-boson operators. All possible generic
  combinations of the internal fields are taken into account.}
\label{fig:topologies2F}
\end{figure}
\begin{figure}[hbt]
\centering
\includegraphics[width=\linewidth]{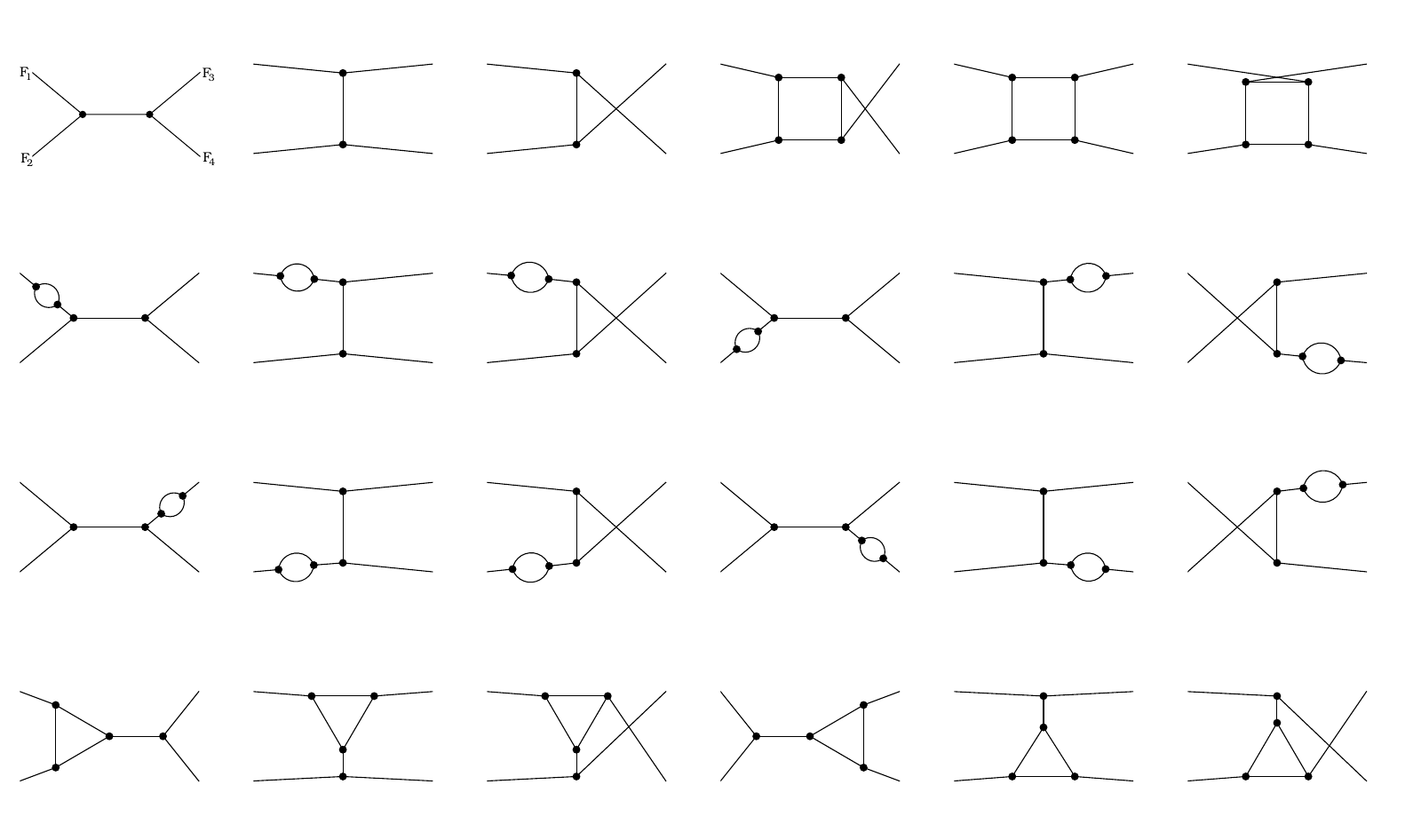}
\caption{All tree topologies considered by \FlavorKit to calculate the Wilson coefficients of 4-fermion operators. All possible generic
  combinations of the internal fields are taken into account.}
\label{fig:topologies4Ftree}
\end{figure}
Constraints for new physics scenarios come not only from direct searches and the Higgs mass observation but also from the measurement of processes which happen only very rarely in the SM and/or which are known to a very high accuracy. These are in particular flavour violation observables. When using the \SPheno output of \SARAH, routines for the calculation of many quark and lepton flavour (QFV, LFV) observables are generated. 

\paragraph*{Lepton flavour violation} The radiative decays of a lepton into another lepton and a photon (Br($l_i \to l_j \gamma$)), and the purely leptonic three-body decays of leptons are included (Br($l \to 3l'$)). Also flavour violating decays of the $Z$-boson (Br($Z\to l l'$)) are tested by \SARAH/\SPheno. Moreover, there are also semi-hadronic observables in the output: $\mu$--$e$ conversion in nuclei (CR($\mu-e,N$)) where the considered nuclei are ($N$=Al, Ti, Sr, Sb, Au, Pb) as well as decays of $\tau$'s into pseudo-scalars, Br($\tau\to l+P$)  with  ($P$=$\pi$, $\eta$,$\eta'$). \\

\paragraph*{Quark flavour violation} The radiative $B$-decay Br($B\to X_s\gamma$), and a set of $B$-decays stemming from four-fermion operators are calculated: Br($B_{s,d}^0 \to l \bar{l}$), Br($B \to s l \bar{l}$), Br($B \to q \nu\nu$),  Br($B \to K \mu \bar{\mu}$). Also Koan decays are considered (Br($K \to \mu \nu$),  Br($K^+ \to \pi^+ \nu\nu$), Br($K_L \to \pi^0 \nu\nu$)) as well as CP observables ($\Delta M_{B_s,B_d}$, $\Delta M_K$, $\epsilon_K$). Finally, some decays which take place already at tree-level are included, namely Br($B\to l \nu$) and Br($D_s \to l \nu$). \\

The approach in \SARAH  to generate the routines to calculate all these observables  is similar to the approach used for loop calculations needed for radiative corrections to the masses: generic formulas for all possible Feynman diagrams which contribute to the Wilson coefficients \footnote{for simplicity, I'll use here 'Wilson coefficients' also for the coefficients of LFV operators which are more commonly called 'form factors'.} of widely used dimension 5 and 6 operators are implemented in \SARAH.  I show in Figs.~\ref{fig:topologies2F} and \ref{fig:topologies4Ftree} only the topologies which are considered because these are already many. For each topology the amplitudes with all possible generic insertions are included. Today, these implementations are mainly based on the \FlavorKit functionality discussed below. \SARAH generates all possible one-loop diagrams and uses the generic expression to get their amplitudes. In this context, not only all possible particles in the loop are included but also all different propagators for penguin diagrams are considered. Thus, not only photonic penguins which are often considered to be dominant in many processes are taken into account. Also all Higgs, $Z$ and -- if existing -- $Z'$ penguins are generated.   After the calculation of the Wilson coefficient, these  are then combined to calculate the observables. This can easily be done by using expressions from literature which are usually model independent. \\

With the development of the \FlavorKit \cite{Porod:2014xia} interface all information to calculate flavour observables is no longer hard-coded in \SARAH but provided by external files. This makes it possible for the user to extent the list of flavour observables when necessary. The \FlavorKit is an automatization of the procedure presented in Ref.~\cite{Dreiner:2012dh} to implement $B_{s,d}^0 \to l \bar l$ in \SARAH and \SPheno. Users interested in the internal calculation might take a look at these two references. \\
 
Also some other observables are calculated by the combination \SARAH--\SPheno which are  measured with high precision: electric dipole moments (EDMs), anomalous magnetic moments of leptons ($(g-2)_l$) and $\delta\rho$
\begin{equation}
 \delta\rho = \frac{\Pi^T_{ZZ}(M_Z^2)}{M_Z^2} - \frac{\Pi^T_{WW}(M_W^2)}{M_W^2}
\end{equation}

\paragraph*{Output} When generating {\btt \SPheno} code with \SARAH, the above listed flavour and precision observables are included in the \Fortran code. In addition, \SARAH writes also \LaTeX\ files with all contribution to the Wilson coefficients from any possible diagram.

\subsection{Fine-Tuning}
A measure for the electroweak fine-tuning was introduced in Refs.~\cite{Ellis:1986yg, Barbieri:1987fn} 
\begin{equation} 
\label{eq:measure}
\Delta_{FT} \equiv \max {\text{Abs}}\big[\Delta _{\alpha}\big],\qquad \Delta _{\alpha}\equiv \frac{\partial \ln
  M_Z^{2}}{\partial \ln \alpha} = \frac{\alpha}{M_Z^2}\frac{\partial M_Z^2}{\partial \alpha} \;.
\end{equation}
$\alpha$ is a set of independent parameters. $\Delta_\alpha^{-1}$ gives an estimate of the accuracy to which the parameter $\alpha$ must be tuned to get the correct electroweak breaking scale \cite{Ghilencea:2012qk}. Using this definition the fine-tuning of a given models depends on the choice what parameters are considered as fundamental and at which scale they are defined. The approach by \SARAH is that it takes by default the scale at which the SUSY breaking parameters are set. This corresponds in models where SUSY is broken by gravity to the scale of grand unification (GUT scale), while for models with gauge mediated SUSY breaking (GMSB) the messenger scale would be used. For simplicity, I call both $M_{Boundary}$. The choice of the set of parameters $\alpha$ is made by user. \\

To calculate the fine-tuning in practice, an iteration of the RGEs between $M_{SUSY}$ and$M_{Boundary}$ happens using the full two-loop RGEs. In each iteration one of the fundamental parameters is slightly varied and the running parameters at $M_{SUSY}$ are calculated. These parameters are used to solve the tadpole equations numerically with respect to all VEVs and to re-calculate the $Z$-boson mass. To give an even more accurate estimate, also one-loop corrections to the $Z$ mass stemming from $\Pi^T_Z$ can be included. 

\paragraph*{Output} A fine-tuning calculation is optionally included in the \Fortran output for {\btt SPheno}.

\subsection{Summary}
\SARAH derives a lot of information about a given model. This information can be used in different interfaces to study a model in all detail. In general,  one can get (i) \LaTeX\ files, (ii) a spectrum generator based on \SPheno, (iii) model files for different HEP tools. \\

\paragraph*{\LaTeX} All analytical information derived about a model can be exported to 
\LaTeX\ files. These files provide in a human readable format the following information: (i) list of all superfields as well as component fields for all eigenstates; (ii) the superpotential and important parts of the Lagrangian like soft-breaking and gauge fixing terms added by \SARAH; (iii) all mass matrices and tadpole equations; (iv) the full two-loop RGEs; (v) analytical expressions for the one-loop self energies and tadpoles; (vi) all interactions and the corresponding Feynman diagrams; (vii) details about the implementation in \SARAH. Separated files are also generated for the flavour observables showing all contributing diagrams with their amplitudes.
 
 \paragraph*{Spectrum Generator} \SARAHv{3} has been the first `spectrum--generator--generator': using the derived information about the 
mass matrices, tadpole equations, vertices, loop corrections and RGEs for the given model \SARAH writes Fortran source code
for \SPheno.  Based on this code the user gets a fully functional spectrum generator for a new model. 
The features of a spectrum generator created in this way are:
\begin{enumerate}
 \item {\bf RGE running}: the full two-loop RGEs are included
 \item {\bf Precise mass spectrum calculation}: \SPheno modules created by \SARAH include the one-loop corrections to all SUSY particles. For Higgs states the full one-loop and in addition dominant two-loop corrections are included.  
 \item {\bf Calculation of decays}: \SPheno calculates all two-body decays for SUSY and Higgs states. In addition, the three body decays of a fermion into three other fermions and the three-body decays of scalar into another scalar and a pair of fermions are included. 
 \item {\bf \FlavorKit interface}:  \SPheno modules calculate out of the box many flavour observables for a given model. 
 \item {\bf Calculation of precision observables}: \SPheno does also calculate $\delta\rho$, electromagnetic dipole moments as well as anomalous magnetic moments of leptons.
 \item {\bf Output for \HB and \HS}: \SPheno generates all necessary files with the Higgs properties (masses, widths, couplings to SM states) which are needed to run \HB and \HS.
 \item {\bf Estimate of the Fine-Tuning}: \SPheno modules can calculate the electroweak fine-tuning with respect to a set of defined parameters
\end{enumerate}
 
 \paragraph*{Model files} Especially the vertex lists can be exported into several formats to create model files for \FeynArts/\FormCalc, \CalcHep/\CompHep, \WHIZARD/\OMEGA as well as for \MG, \Herwig or \Sherpa based on the \UFO format. Also model files for \Vevacious can be generated which include the tree-level potential as well as the mass matrices to generate the one-loop effective potential.

\section{Example -- Part I: The \BLSSM and its implementation in \SARAH}
\label{sec:example_implementation}
I'll discuss in this section and the subsequent ones the implementation of the \BLSSM in \SARAH and how all phenomenological aspects of this model can be studied. The \BLSSM is for sure not the simplest extension of the MSSM, but it provides many interesting features. There are some subtleties in the implementation which won't show up in singlet extensions for instance. I hope that the examples presented in the following sections show how even such a complicated model can be studied with a very high precision without too much effort. Applying the same methodology to other SUSY or non-SUSY models should be straightforward. However, I can't discuss here all topics which might be interesting and useful for some models. In particular, I won't show how models with threshold scales are implemented. \SARAH supports thresholds where heavy particles get integrated out and where either the gauge symmetries do change or not. Users interested in that topic might take a look at the manual as well as the implementations of seesaw models \footnote{models with a threshold scale where heavy superfields are integrated out  are  {\tt Seesaw1}, {\tt Seesaw2}, {\tt Seesaw3}} or left-right symmetric models \footnote{a model with threshold scales where the gauge group changes is the left-right symmetric model called {\tt Omega} in \SARAH}. A brief summary of the general approach to include thresholds is also given in Appendix~\ref{app:thresholds}. 

In the first part of this section I'll give a short introduction to the \BLSSM, before I come to the tools which are going to be used. The implementation of the \BLSSM in \SARAH is discussed in sec.~\ref{sec:example_model_sarah} and how it is evaluated is shown in sec.~\ref{sec:example_running}. The next section explains what can be done with the model using just \Mathematica. It is shown how the \LaTeX\ output is generated, how mass matrices and tadpole equations can be handled with \Mathematica and how RGEs are calculated and solved. Sec.~\ref{sec:example_spheno} is about the interface between \SARAH and \SPheno. The mass spectrum calculation is explained, and what else can be obtained with \SPheno: decays and branching ratios, flavour and precision observables and the fine-tuning. Afterwards, we include more tools in our study in sec.~\ref{sec:example_tools}: \HB/\HS to test Higgs constraints, \Vevacious to check the vacuum stability, \MO to calculate the dark matter relic density and direct detections rates, \WHIZARD/\OMEGA to produce monojet events and \MG to make a simple dilepton analysis. Sec.~\ref{sec:example_scans} is about parameter scans and how the different tools can be interfaced. 

\subsection{The model}
Supersymmetric models with an additional $U(1)_{B-L}$ gauge symmetry at the TeV scale have recently received considerable attention: they can explain neutrino data, they might help to understand the origin of $R$-parity and its possible spontaneous violation~\cite{Khalil:2007dr,Barger:2008wn,FileviezPerez:2010ek,CamargoMolina:2012hv,Marshall:2014kea}, and they have a rich phenomenology \cite{Elsayed:2012ec}. In the $R$-parity conserving case, these models come with a new Higgs which mixes with the MSSM one \cite{O'Leary:2011yq}, they provide new dark matter candidates \cite{Basso:2012gz}, and can have an impact on the Higgs decays \cite{Basso:2012tr}. In both cases of broken and unbroken $R$-parity these models have interesting consequences for LHC searches \cite{FileviezPerez:2011kd,Khalil:2012gs,Krauss:2012ku,Perez:2013kla}.

\subsubsection{Particle content and superpotential}
\label{sec:BLSSMparticle}
We study the minimal supersymmetric model where the SM gauge sector is extended by a $U(1)_{B-L}$
\begin{equation}
 \mathscr{G} = U(1)_Y \otimes SU(2)_L \otimes  SU(3)_c  \otimes  U(1)_{B-L}
\end{equation}
and where $R$-parity is not broken by sneutrino VEVs. This model is called the \BLSSM. In this model the matter sector of the MSSM is extended by three generations of right-handed neutrino superfields $\hat \nu^c$ and two fields which are responsible for $B-L$ breaking, $\hat \eta$ and $\hat \bar{\eta}$. These fields carry lepton number 2 and are called 'bileptons'. The chiral superfields and their quantum numbers are summarized in Tab.~\ref{tab:cSF}. 
\begin{table} 
\centering
\begin{tabular}{|c|c|c|c|c|c|} 
\hline \hline 
Superfield & Spin 0 & Spin $1/2\) & Generations & $(U(1)_Y\otimes\,SU(2)_L\otimes\, SU(3)_C\otimes\, U(1)_{B-L})\) \\ 
\hline 
$\hat{Q}$ & $\tilde{Q}$ & $Q$ & 3 & $1/6\, \otimes \, {\bf 2}\, \otimes \, {\bf 3}\, \otimes \, 1/6$ \\ 
$\hat{D}$ & $\tilde{d}^c$ & $d^c$ & 3 & $1/3\, \otimes \, {\bf 1}\, \otimes \, {\bf \overline{3}}\, \otimes \, -1/6$ \\ 
$\hat{U}$ & $\tilde{u}^c$ & $u^c$ & 3 & $-2/3\, \otimes \, {\bf 1}\, \otimes \, {\bf \overline{3}}\, \otimes \, -1/6$ \\ 
$\hat{L}$ & $\tilde{L}$ & $L$ & 3 & $-1/2\, \otimes \, {\bf 2}\, \otimes \, {\bf 1}\, \otimes \, -1/2$ \\ 
$\hat{E}$ & $\tilde{e}^c$ & $e^c$ & 3  & $1\, \otimes \, {\bf 1}\, \otimes \, {\bf 1}\, \otimes \, 1/2$ \\ 
$\hat{\nu}$ & $\tilde{\nu}^c$ & $\nu^c$ & 3  & $0\, \otimes \, {\bf 1}\, \otimes \, {\bf 1}\, \otimes \, 1/2$ \\ 
$\hat{H}_d$ & $H_d$ & $\tilde{H}_d$ & 1  & $-1/2\, \otimes \, {\bf 2}\, \otimes \, {\bf 1}\, \otimes \, 0$ \\ 
$\hat{H}_u$ & $H_u$ & $\tilde{H}_u$ & 1  & $1/2\, \otimes \, {\bf 2}\, \otimes \, {\bf 1}\, \otimes \, 0$ \\ 
$\hat{\eta}$ & $\eta$ & $\tilde{\eta}$ & 1  & $0\, \otimes \, {\bf 1}\, \otimes \, {\bf 1}\, \otimes \, -1$ \\ 
$\hat{\bar{\eta}}$ & $\bar{\eta}$ & $\tilde{\bar{\eta}}$ & 1  & $0\, \otimes \, {\bf 1}\, \otimes \, {\bf 1}\, \otimes \, 1$ \\ 
\hline \hline
\end{tabular} 
\caption{Chiral superfields of the \BLSSM and their quantum numbers under $U(1)_Y \otimes SU(2)_L \otimes  SU(3)_c  \otimes U(1)_Y U(1)_{B-L}$.}
\label{tab:cSF}
\end{table}
The superpotential of the \BLSSM is given by
\begin{align} 
\label{eq:Sup}
W = & \, Y^{ij}_u\,\hat{U}_i\,\hat{Q}_j\,\hat{H}_u\, - Y_d^{ij} \,\hat{D}_i\,\hat{Q}_j\,\hat{H}_d\, - Y^{ij}_e \,\hat{E}_i\,\hat{L}_j\,\hat{H}_d\,+\mu\,\hat{H}_u\,\hat{H}_d\, \nonumber \\
& +Y^{ij}_{\nu}\,\hat{L}_i\,\hat{H}_u\,\hat{\nu}_j\,- \mu' \,\hat{\eta}\,\hat{\bar{\eta}}\, +Y^{ij}_x\,\hat{\nu}_i\,\hat{\eta}\,\hat{\nu}_j\,
\end{align} 
Here, $i,j$ are generation indices and we suppressed all colour and isospin indices. The first line is identical to the MSSM, and all new terms are collected in the second line of eq.~(\ref{eq:Sup}). After $B-L$ breaking a Majorana mass term $Y_x \langle \eta \rangle $ for the right-handed neutrinos is generated. This term causes a mass splitting between the CP even and odd parts of the complex sneutrinos. \\

The additional soft-breaking terms compared to the MSSM are 
\begin{align}
\nonumber \La_{SB,B-L-SSM} = \,\,& \La_{SB,MSSM}   - \lambda_{\tilde{B}} \lambda_{\tilde{B}'} {M}_{B B'}  - \frac{1}{2} \lambda_{\tilde{B}'} \lambda_{\tilde{B}'} {M}_{B'} \nonumber \\
& - m_{\eta}^2 |\eta|^2 - m_{\bar{\eta}}^2 |\bar{\eta}|^2  - {m_{\nu,ij}^{2}} (\tilde{\nu}_i^c)^* \tilde{\nu}_j^c - \eta \bar{\eta} B_{\mu'} + T^{ij}_{\nu}  H_u \tilde{\nu}_i^c \tilde{L}_j + T^{ij}_{x} \eta \tilde{\nu}_i^c \tilde{\nu}_j^c 
\end{align}
The electroweak and $B-L$ symmetry are broken to $U(1)_{em}$ by the following VEVs of Higgs states and bileptons:
\begin{align} 
H_d^0 = & \, \frac{1}{\sqrt{2}} \left(i \sigma_{d} + v_d  +  \phi_{d} \right), \hspace{1cm}
H_u^0 = \, \frac{1}{\sqrt{2}} \left(i \sigma_{u} + v_u  +  \phi_{u} \right)\\ 
\eta 
= & \, \frac{1}{\sqrt{2}} \left(i \sigma_\eta + v_{\eta} +  \phi_{\eta} \right), \hspace{1cm}
\bar{\eta}
= \, \frac{1}{\sqrt{2}} \left(i \sigma_{\bar{\eta}} + v_{\bar{\eta}}  +  \phi_{\bar{\eta}} \right)
\end{align} 
In analogy to the MSSM definition  $\tan\beta = \frac{v_u}{v_d}$, we call the ratio of the two bilepton VEVs $\tan\beta' = \frac{v_{\eta}}{v_{\bar{\eta}}}$. 
As mentioned above, the Majorana mass term causes a mass splitting between the CP even and odd parts of the right sneutrinos. This makes it necessary to define
\begin{equation}
\tilde{\nu}^i_R =  \, \frac{1}{\sqrt{2}} \left(i \sigma^i_R +  \phi^i_R \right),
\hspace{1cm}
\tilde{\nu}^i_L = \, \frac{1}{\sqrt{2}} \left(i \sigma^i_{L}  +  \phi^i_{L} \right)
\end{equation}
Here, $i=1,2,3$ is the generation index of the sneutrinos.

\subsubsection{Gauge kinetic mixing}
\label{subsec:kineticmixing}
The particle content of the \BLSSM gives rise to gauge-kinetic mixing because the matrix 
\begin{equation}
\gamma_{AB} = \frac{1}{16 \pi^2} \sum_\phi Q_\phi^A Q_\phi^B 
\end{equation}
is not diagonal. The indices $A$ and $B$ run over all $U(1)$ groups, and $\phi$ over all superfields. For the particle content of Tab.~\ref{tab:cSF} we find
\begin{equation}
\label{eq:gammaMatrix}
\gamma = \frac{1}{16 \pi^2} N \left( \begin{array}{cc} 11 & 4 \\
 4 & 6 \end{array} \right) N.
\end{equation}
$N$ contains the GUT normalization  of the two Abelian gauge groups. We will take  $\sqrt{\frac{3}{5}}$ for $U(1)_Y$ and $\sqrt{\frac{3}{2}}$ for $U(1)_{B-L}$. From eq.~(\ref{eq:gammaMatrix}) we see that even if gauge-kinetic mixing vanishes at one scale, it is induced again by RGE running and must be included therefore.  

However, as long as the two Abelian gauge groups are unbroken, we can make a change of basis. This freedom is used go to a basis where electroweak  precision data is respected in a simple way:  by choosing a triangle form of the gauge coupling matrix, the bilepton contributions to the $Z$ mass vanish:
\begin{equation}
\left(\begin{array}{cc} g_{YY} & g_{YB} \\ g_{BY} & g_{BB}   \end{array} \right) \to  \left(\begin{array}{cc} g_1 & \tilde{g} \\ 0 & g_{BL} \end{array} \right)
\end{equation}
The gauge couplings are related by  \cite{Chankowski:2006jk}:
\begin{eqnarray}
\label{eq:Triangle1}
& g_1 =  \frac{g_{YY} g_{B B} - g_{Y B} g_{B Y}}{\sqrt{g_{B B}^2 + g_{B Y}^2}}  \hspace{1cm}
 \tilde{g}  =  \frac{g_{Y B} g_{B B} + g_{B Y} g_{YY}}{\sqrt{g_{B B}^2 + g_{B Y}^2}}   \hspace{1cm}
  g_{BL} =  \sqrt{g_{B B}^2 + g_{B Y}^2}  &
\end{eqnarray}

\subsubsection{Mass eigenstates}
\label{sec:BLSSMmixing}
After electroweak and $B-L$ breaking and also because of gauge-kinetic mixing there is a mixing between the neutral SUSY particles from the MSSM and the new sector. 

\paragraph*{Neutral gauge bosons} In the gauge sector, the neutral gauge bosons $B$, $W^3$ and $B'$ mix. This gives three mass eigenstates $\gamma$, $Z$, $Z'$ which are related by a rotation matrix which can be expressed by two angles $\Theta$ and $\Theta'$
\begin{align} 
\label{eq:rotZZp}
\left(\begin{array}{c} 
B\\ 
W\\ 
{B'}\end{array} \right) 
 = & \,\left( 
\begin{array}{ccc} 
\cos\Theta_W & -\cos{\Theta'}_W \sin\Theta_W &  \sin\Theta_W \sin{\Theta'}_W \\ 
\sin\Theta_W &  \cos\Theta_W \cos{\Theta'}_W & -\cos\Theta_W \sin{\Theta'}_W \\ 
 0 & \sin{\Theta'}_W  & \cos{\Theta'}_W \end{array} 
\right)
\left(\begin{array}{c} 
\gamma\\ 
Z\\ 
{Z'}\end{array} \right) 
\end{align}
The entire mixing between the $B-L$ and the SM gauge boson comes from $\Theta'$ which can be approximated by  \cite{Basso:2010jm}
\begin{equation}
\label{eq:ThetaWP}
\tan 2 {\Theta'}_W \simeq \frac{2 \tilde{g} \sqrt{g_1^2 + g_2^2}}{\tilde{g}^2 + 16 \left(\frac{x}{v}\right)^2 g_{BL}^2 -g_2^2 - g_1^2}
\end{equation}
with $v=\sqrt{v_d^2+v_u^2}$ and $x=\sqrt{v_\eta^2 + v_{\bar \eta}^2}$. 

\paragraph*{Neutral Higgs bosons} In the Higgs sector, the CP even and odd parts of the doublets mix with the corresponding CP eigenstates of the bileptons. This leads to four physical scalar Higgs particles
\begin{equation}
 (\phi_d, \phi_u, \phi_\eta, \phi_{\bar \eta}) \to (h_1, h_2, h_3,h_4)
\end{equation}
and four pseudo-scalars. Two pseudo-scalars are physical, the other two are Goldstone bosons of the $Z$ and $Z'$.
\begin{equation}
 (\sigma_d, \sigma_u, \sigma_\eta, \sigma_{\bar \eta}) \to (G^Z, G^{Z'}, A^h_1, A^h_2)
\end{equation}

\paragraph*{Neutralinos} There are seven neutralinos in the model which are an admixture of the three gauginos, the two Higgsinos and the two bileptinos. 
\begin{equation}
\left(\lambda_{\tilde{B}}, \tilde{W}^0, \tilde{H}_d^0, \tilde{H}_u^0, \lambda_{\tilde{B}{}'}, \tilde{\eta},  \tilde{\bar{\eta}}\right) \to (\tilde \chi_1, \dots, \tilde \chi^0_7)
\end{equation}

\paragraph*{Neutrinos and sneutrinos} There are six Majorana neutrinos which are a superposition of the left- and right neutrino gauge eigenstates
\begin{equation}
(\nu_L^i, \nu_R^i) \to \nu_j \hspace{1cm} i=1,2,3; \,\, j=1,\dots 6  
\end{equation}
Similarly, the sneutrinos are admixtures of left- and right sneutrino gauge eigenstates. Because of the mass splitting due to the Majorana mass term the CP-even and odd parts mix separately, and there are 12 real mass eigenstates:
\begin{align}
(\phi_L^i, \phi_R^i) \to & \tilde \nu^R_j \hspace{1cm} i=1,2,3; \,\, j=1,\dots 6    \\
(\sigma_L^i, \sigma_R^i) \to & \tilde \nu^I_j \hspace{1cm} i=1,2,3; \,\, j=1,\dots 6    \\
\end{align}

\subsection{Constrained model}
\label{sec:CBLSSM}
The \BLSSM has 55 additional parameters, including all phases, compared to the general MSSM. Therefore, an organizing principle to relate these parameter to reduce the number of free parameters is often helpful.  We choose a CMSSM-like setup inspired by minimal supergravity. We set the boundary conditions at the GUT scale which is taken to be 
\begin{equation}
 g^{GUT}_1 = g^{GUT}_2 = g^{GUT}_{BL} \thickspace .
\end{equation}
We assume that at the GUT gauge-kinetic mixing is absent, but gets induced below that scale. This gives as additional conditions
\begin{equation}
g_{YB}^{GUT} = g_{BY}^{GUT} = 0 \,.
\end{equation}
Thus, at the GUT scale the relations $g^{GUT}_1=g^{GUT}_{YY}$ and $g^{GUT}_{BL} = g^{GUT}_{BB}$ hold. \\
In minimal supergravity all scalar and gaugino soft masses unify at the GUT scale, i.e. there are only two free parameters to fix all soft masses: $m_0$, $M_{1/2}$. The boundary conditions at the GUT scale are:
\begin{align}
\label{eq:boundary1}
 m^2_0 \equiv & m^2_{H_d} = m^2_{H_u} = m^2_{\eta} = m^2_{\bar{\eta}} \\
 \label{eq:boundary2}
m^2_0 {\bf 1} \equiv & m_D^2  =  m_U^2  = m_Q^2 
= m_E^2 = m_L^2  = m_{\nu}^2  \\
\label{eq:boundary3}
 M_{1/2} \equiv & M_1 = M_2 = M_3 = M_{\tilde{B}'}
\end{align}
with the identity matrix {\bf 1}. 
We assume also that no gauge kinetic mixing is present in the gaugino sector at this scale, i.e.
\begin{equation}
 M_{B B'} =  0 
\end{equation}
For the trilinear soft-breaking terms  we use in analogy to the CMSSM the relation
\begin{align}
\label{eq:boundary4}
 T_i = A_0 Y_i, \hspace{1cm} i = e,d,u,x,\nu \thickspace . 
\end{align}
with a free parameter $A_0$. The remaining soft-terms $B_\mu$, $B_{\mu'}$ are not taken to be input, but the tadpole equations are, if not stated otherwise, solved with respect to $\mu, B_\mu, \mu'$ and $B_{\mu'}$. The advantage of this choice is that these parameters don't enter the RGE evaluation of all other terms and they can be considered independently. \\

To fix the gauge part of the $B-L$ sector, we need two more input parameters. These are the mass of the $Z'$ and the ratio of the bilepton VEVs. Finally, there are two more Yukawa-like matrices, $Y_\nu$ and $Y_x$. Thus, the full set of input parameter is
\begin{eqnarray}
\label{eq:BLinput}
& m_0, \thickspace M_{1/2},\thickspace A_0,\thickspace \tan\beta,\thickspace
 \tan\beta',\thickspace \sign(\mu),\thickspace \sign(\mu'),\thickspace M_{Z'},
 \thickspace  Y_x \thickspace \mbox{and} \thickspace Y_{\nu} . &
\end{eqnarray}
The masses of the light neutrinos in this model are proportional to $Y_{\nu}$. This gives strong constraints and we can often neglect $Y_\nu$ because the entries must be tiny. 

\subsection{Setup}
We want to study all aspects of the \BLSSM: the mass spectrum, decays, flavour observables, Higgs constraints, dark matter, vacuum stability and collider physics. For this purpose, we make not only use of \SARAH but also interface it to several other public tools. To simplify the following presentation, I'm going to assume that all the following tools are installed in the same directory {\tt \$PATH}:
\begin{enumerate}
 \item \SARAH\ {\tt 4.5.0} or newer in {\tt \$PATH/SARAH} \footnote{ {\tt http://www.hepforge.org/downloads/sarah} }
 \item \SPheno\  {\tt 3.3.6} or newer in {\tt \$PATH/SPHENO} \footnote{ {\tt http://www.hepforge.org/downloads/spheno } }
 \item \SSP\ {\tt 1.2.0} or newer in {\tt \$PATH/SSP} \footnote{{\tt http://www.hepforge.org/downloads/sarah}}
 \item \PreSARAH\ {\tt 1.0.3} or newer in {\tt \$PATH/PRESARAH} \footnote{{\tt http://www.hepforge.org/downloads/sarah}}
 \item \HB\  {\tt 4.1.0} or newer in {\tt \$PATH/HIGGSBOUNDS} \footnote{{\tt http://higgsbounds.hepforge.org/downloads.html}}
 \item \HS\  {\tt 1.2.0} or newer  in {\tt \$PATH/HIGGSSIGNALS} \footnote{{\tt http://higgsbounds.hepforge.org/downloads.html}}
 \item \CalcHep\ {\tt 3.4.6} or newer in {\tt \$PATH/CALCHEP} \footnote{{\tt http://theory.sinp.msu.ru/~pukhov/calchep.html}}
 \item \MO\ {\tt 4.1} or newer in {\tt \$PATH/MICROMEGAS} \footnote{{\tt http://lapth.cnrs.fr/micromegas/}}
 \item \WHIZARD\ {\tt 2.2.2} or newer in {\tt \$PATH/WHIZARD} \footnote{{\tt http://www.hepforge.org/downloads/whizard}}
 \item \MG\ {\tt 5.2.2.2} or newer in {\tt \$PATH/MADGRAPH} \footnote{{\tt https://launchpad.net/mg5amcnlo}}
 \item \Vevacious\ {\tt 1.1.0} or newer in {\tt \$PATH/VEVACIOUS} \footnote{{\tt http://www.hepforge.org/downloads/vevacious}}
 \item \MA\ {\tt 1.1.11} or newer in {\tt \$PATH/MADANALYSIS} \footnote{{\tt https://launchpad.net/madanalysis5}}
\end{enumerate}
It is possible to use the {\tt BSM Toolbox} scripts \cite{Staub:2011dp} from 
\begin{center}
{\tt http://sarah.hepforge.org/Toolbox.html}
\end{center}
to install most tools at once via the included {\tt configure} script. The {\tt Toolbox} contains also a script called {\tt butler} to implement a model in the different tools automatically based on the \SARAH implementation. This might be the convenient choice for the user. However, I'm going to use the explicit route and discuss the implementation in each tool in some detail. \\

\paragraph*{Installation of \SARAH} The installation of \SARAH is very simple: the package can be downloaded from 
\begin{center}
{\tt http://sarah.hepforg.org }
\end{center}
After copying the {\tt tar} file to the directory {\tt \$PATH}, it can be extracted 
\begin{lstlisting}[style=terminal]
$ cp [Download-Directory]/SARAH-X.Y.Z.tar.gz $PATH/
$ cd $PATH
$ tar -xf SARAH-X.Y.Z.tar.gz
$ ln -s SARAH SARAH-X.Y.Z
\end{lstlisting}
{\tt X.Y.Z} has to be replaced by the version which was downloaded. In the last line a symbolic link from the directory \SARAH to {\tt SARAH-X.Y.Z} is created. There is no compilation necessary, but \SARAH can directly be used with any \Mathematica version between 7 and 10.\\

\paragraph*{Installation of all other tools} For the installation of all other tools, please, check the manual or {\tt readme} files of these tools.

\subsection{Implementation of the $B-L$-SSM}
\label{sec:example_model_sarah}
I assume in the following that the \BLSSM was not be delivered with \SARAH and we have to implement it ourselves. 
All information about the model is saved in three different files: \verb"B-L-SSM.m", \verb"parameters.m" and \verb"particles.m" which have to be located in the subdirectory \verb"B-L-SSM" in the {\tt Model} directory of \SARAH. 
Only the first file, {\tt B-L-SSM.m}, is absolute necessary and contains all physical information about the model: the symmetries, particle content, superpotential and mixings. In \verb"parameters.m" properties of all parameters can be defined, e.g. \LaTeX\ name, Les Houches block and number, relations among parameters, real/complex, etc. In \verb"particles.m" additional information about particles are set: mass, width, electric charge, PDG, \LaTeX\ name, output name, and so on. The optional information in \verb"parameters.m" and \verb"particles.m" might be needed for the different outputs of \SARAH as discussed later. 

\subsubsection{Main file: {\tt B-L-SSM.m}}
Usually the easiest way to implement a new model in \SARAH is to start with an existing model file. For non-SUSY models the convenient choice is the model file for the SM, for SUSY models the one for the MSSM. Therefore, I'm going to discuss the differences in the model file of the \BLSSM compared to the MSSM one. As reference, the corresponding parts in the MSSM model file are shown as well. In addition, there are some specific flags listed in Appendix.~\ref{app:sarahflags} which can be used to turn on/off particular structures in the Lagrangian. Those are not needed for our purpose here but might be useful for other models.\\

At the very beginning of a model file, some additional information about the model implementation can be given: what's the name of the model (in \LaTeX\ syntax as well as string without any special character), who is the author of the model file and when was the last change:

\begin{lstlisting}[style=blssm,firstnumber=3]
Model`Name = "BLSSM";
Model`NameLaTeX ="B-L-SSM";
Model`Authors = "L.Basso, F.Staub";
Model`Date = "2012-09-01"; 
\end{lstlisting}

\paragraph{Global symmetries} \SARAH supports $Z_N$ as well as $U(1)$ global symmetries which are defined in the array {\tt Global}. Usually, one considers the MSSM with conserved $R$-parity. This discrete symmetry is defined via 
\begin{lstlisting}[style=both,firstnumber=18]
Global[[1]] = {Z[2],RParity};
\end{lstlisting}
First, the kind of the symmetry  is defined ({\tt Z[N]} with integer {\tt N}, or {\tt U[1]}). Note, $Z_N$ symmetries are always understood as multiplicative symmetries. The second entry gives a name to the symmetry. For the $U(1)$ there is one specific name which can be used to define $R$-symmetries: {\tt RSymmetry}. In that case, the $R$ charges of the SUSY coordinates are considered as well.  There are two possibilities to define the charges of SUSY fields with respect to a global symmetry: 
\begin{enumerate}
 \item  if in the definition of the vector or chiral superfields, which will be explained below, only one quantum number is given per superfield per global symmetry this number is used for the superfield itself but also for component fields. 
 \item if a list with three entries is given as charge for a vector or chiral superfield the following convention applies: for chiral superfields, the first entry is the charge for the superfield, the second one for the scalar  component, the third one for the fermionic component. For vector superfield, the second entry refers  to the gaugino, the third to the gauge boson. 
\end{enumerate}
With these conventions a suitable definition of the global symmetries for states with $R$-parity $\pm 1$ would be
\begin{lstlisting}[style=both,firstnumber=19]
RpM = {-1, -1,  1};
RpP = { 1,  1, -1};
\end{lstlisting}

In principle, the \BLSSM has no global symmetry but $R$-parity is just a remnant of the broken $U(1)_{B-L}$. However, it turns out to be helpful to keep the standard definition for $R$-parity: we can use this $Z_2$ to get the relic density with \MO. Sometimes, one uses also matter parities to have an additional $Z_2$ in models with gauged $B-L$ \cite{Dev:2009aw,BhupalDev:2010he}. 

\paragraph{Gauge symmetries} The next step to define a SUSY model is to fix the gauge sector. That's done by adding for each gauge group one entry to the array {\tt Gauge}. For SUSY models for each entry in {\tt Gauge} also the corresponding vector superfield are set automatically. 
For instance, the SM gauge group $\mathfrak{G}_{SM}=U(1)_Y \times SU(2)_L \times SU(3)_C$ is set via
\begin{lstlisting}[style=mssm,firstnumber=24]
(* Vector Superfields *)
Gauge[[1]]={B,   U[1], hypercharge, g1,  False, RpM};
Gauge[[2]]={WB, SU[2], left,        g2,  True,  RpM};
Gauge[[3]]={G,  SU[3], color,       g3,  False, RpM};
\end{lstlisting}
First, the names of the gauge fields are given\footnote{Gauge bosons get this name with a prefix {\tt V}, gauginos with a prefix {\tt f} and ghosts with the prefix {\tt g}. For instance, the gluon and gluino are called {\tt VG} and {\tt fG} by this definition and the Ghost {\tt gG}.}, the second entry defines the kind of the group. The third entry gives a name to the gauge group\footnote{The charge indices of non-Abelian groups start with the first three letters of the gauge group's name followed by an integer, i.e. {\tt col1} is used for colour indices.} and the fourth entry fixes the name of the corresponding gauge coupling. In the fifth entry it is defined if the group will be broken later. The last entry sets the global charges of the gauginos and vector bosons of the corresponding vector superfield using the conventions explained above: here, the vector superfields as well as its fermionic components get $R$-parity -1, while the spin-1 states have $R$-parity +1. \\

For the \BLSSM only a fourth entry is needed. We call the group {\tt Bp}, i.e. the vector boson will be named {\tt VBp} by \SARAH and the gaugino {\tt fBp}. For the gauge coupling we chose as name {\tt gBL}\footnote{Note, {\tt gBp} is not possible because this is already the name of the ghost!}. Thus, the definition of the \BLSSM gauge sector is
\begin{lstlisting}[style=blssm,firstnumber=24]
(* Vector Superfields *)
Gauge[[1]]={B,   U[1], hypercharge, g1,  False, RpM};
Gauge[[2]]={WB, SU[2], left,        g2,  True,  RpM};
Gauge[[3]]={G,  SU[3], color,       g3,  False, RpM};
Gauge[[4]]={Bp,  U[1], BminusL,     gBL, False, RpM};
\end{lstlisting}
Since this is the second $U(1)$ beside hypercharge, \SARAH will generate two off-diagonal gauge couplings {\tt g1BL} and {\tt gBL1} stemming from kinetic mixing, see also sec.~\ref{sec:models}.  The soft-masses for the gauginos which are added by \SARAH are {\tt MassB}, {\tt MassWB}, {\tt MassG} and {\tt MassBp}. And as consequence of gauge-kinetic mixing {\tt MassBBp} appears as well. 

\paragraph{Chiral superfields} Chiral superfields in \SARAH are defined via the array {\tt SuperFields}. The conventions are as follows: first, a name for the superfield is given, in the second entry the number of generations is set and in the third entry  the names for the isospin components are defined. Afterwards, the transformation under the different gauge groups  are given, the last entries set the charges under the global symmetries. To define the transformation with respect to Abelian groups the charge is given. For non-Abelian groups the dimension is given as integer, where conjugated representations are negative. If the dimension is not unique, also the Dynkin labels can be defined. The treatment of higher-dimensional representation is different for groups which get broken and those which stay unbroken
\begin{enumerate}
 \item Representation with respect to groups which get broken: in that case it is convenient to define higher-dimensional representations as tensor products of the fundamental representation. 
 \item Representation with respect to groups which get not broken: in that case the representation is treated as vector which the appropriate dimension. To do this \SARAH makes use of the generators as well as the Clebsch-Gordan-coefficient provided by \Susyno.
\end{enumerate}
We have to deal here only with doublets under a broken $SU(2)$, i.e. we need a vector of length two for the isospin components. \\

The particle content of the MSSM are the squark superfields {\tt q}, {\tt d} and {\tt u}, the lepton superfields {\tt l} and {\tt e} as well as the two superfields for the Higgs doublets {\tt Hd} and {\tt Hu}. We have {\tt 3} generations for all matter fields and the usual charges with respect to the SM gauge groups. These are defined by 
\begin{lstlisting}[style=mssm,firstnumber=31]
(* Chiral Superfields *)
SuperFields[[1]] = {q, 3, {uL,  dL},    1/6, 2, 3,  RpM};  
SuperFields[[2]] = {l, 3, {vL,  eL},   -1/2, 2, 1,  RpM};
SuperFields[[3]] = {Hd,1, {Hd0, Hdm},  -1/2, 2, 1,  RpP};
SuperFields[[4]] = {Hu,1, {Hup, Hu0},   1/2, 2, 1,  RpP};

SuperFields[[5]] = {d, 3, conj[dR],   1/3, 1, -3,  RpM};
SuperFields[[6]] = {u, 3, conj[uR],  -2/3, 1, -3,  RpM};
SuperFields[[7]] = {e, 3, conj[eR],     1, 1,  1,  RpM};
\end{lstlisting}
The \SARAH conventions for the component fields of a superfield are: scalars start with {\tt S} and fermions with {\tt F}, e.g. the left down-squark is called {\tt SdL} and the right lepton {\tt FeR}. The soft masses for the scalars are {\tt mq2}, {\tt ml2}, {\tt mHd2}, etc. \\

In the \BLSSM, we have first to define the $U(1)_{B-L}$ charge of all MSSM fields \SARAH includes a factor $\frac{1}{2}$ for all $U(1)$ charges as this is done usually by conventions for the hypercharge. Therefore, quark superfields have $B-L$ charge $\pm \frac{1}{6}$ and lepton superfields carry  $\pm \frac{1}{2}$. The other changes are the definition of the right sneutrino superfield {\tt vR} which comes in three generations and which is a gauge singlet under the SM gauge groups. The bileptons {\tt C1} and {\tt C2} are also SM singlets and come with $B-L$ charge $\pm 1$ according to the just mentioned conventions. 
\begin{lstlisting}[style=blssm,firstnumber=31]
(* Chiral Superfields *)
SuperFields[[1]] = {q, 3, {uL,  dL},    1/6, 2, 3,  1/6, RpM};  
SuperFields[[2]] = {l, 3, {vL,  eL},   -1/2, 2, 1, -1/2, RpM};
SuperFields[[3]] = {Hd,1, {Hd0, Hdm},  -1/2, 2, 1,    0, RpP};
SuperFields[[4]] = {Hu,1, {Hup, Hu0},   1/2, 2, 1,    0, RpP};

SuperFields[[5]] = {d, 3, conj[dR],   1/3, 1, -3, -1/6, RpM};
SuperFields[[6]] = {u, 3, conj[uR],  -2/3, 1, -3, -1/6, RpM};
SuperFields[[7]] = {e, 3, conj[eR],     1, 1,  1,  1/2, RpM};
SuperFields[[8]] = {vR,3, conj[vR],     0, 1,  1,  1/2, RpM};

SuperFields[[9]]  = {C1, 1, C10,  0, 1, 1, -1, RpP};
SuperFields[[10]] = {C2, 1, C20,  0, 1, 1,  1, RpP};
\end{lstlisting}
The soft-terms for the new superfields are {\tt mvR2}, {\tt mC12} and {\tt mC22}.

\paragraph{Superpotential} The superpotential of the MSSM consists of the Yukawa interaction {\tt Yu}, {\tt Yd}, {\tt Ye} as well as the $\mu$-term called \verb"\[Mu]". \SARAH will add the corresponding soft-terms called {\tt T[Yu]}, {\tt T[Yd]}, {\tt T[Ye]}, and \verb"B[\[Mu]]". The Yukawas as well as the trilinear soft-terms are treated as complex $3 \times 3$ matrices, $\mu$ and $B_\mu$ are complex by default. The MSSM superpotential is set in \SARAH by
\begin{lstlisting}[style=mssm,firstnumber=45]
(* Superpotential *) 
SuperPotential = Yu u.q.Hu - Yd d.q.Hd - Ye e.l.Hd + \[Mu] Hu.Hd;
\end{lstlisting}
We see here that we don't have to define the charge indices and the contraction of these. All of this is done automatically by \SARAH and the term {\tt Yu u.q.Hu} gets interpreted internally as
\begin{equation}
Y_u^{ij} \epsilon_{ab} \delta_{\alpha\beta} \hat u_i^\alpha \hat q_j^{\beta a} \hat H_u^b
\end{equation}

In the \BLSSM three more terms are present in the superpotential: the neutrino Yukawa coupling {\tt Yv}, the coupling between bileptons and right-neutrino {\tt Yn} and the $\mu'$-term called {\tt MuP}. The corresponding soft-terms added by \SARAH are {\tt T[Yv]}, {\tt T[Yn]}, and {\tt B[MuP]}. Also those are treated by default in the most general way. The full superpotential of the \BLSSM reads
\begin{lstlisting}[style=blssm,firstnumber=49]
(* Superpotential *) 
SuperPotential = Yu u.q.Hu - Yd d.q.Hd - Ye e.l.Hd +
              \[Mu] Hu.Hd + Yv vR.l.Hu - MuP C1.C2 + Yn vR.C1.vR;
\end{lstlisting}

\paragraph{Eigenstates} In the MSSM and the \BLSSM we have to deal with two sets of eigenstates. First, the gauge eigenstates which will always be present. These states are called by default {\tt GaugeES}. Second, we have the mass eigenstates after symmetry breaking which we call {\tt EWSB}. Both sets are defined in the array {\tt NameOfStates}. Sometimes, it might be useful to use also intermediate states to make step-wise rotations from the gauge to the final matter eigenstates. Therefore, the length of {\tt NameOfStates} is not restricted. 
\begin{lstlisting}[style=both,firstnumber=56]
NameOfStates={GaugeES, EWSB};
\end{lstlisting}

\paragraph{Electroweak and $B-L$ symmetry breaking} In the MSSM, EWSB happens when the neutral components of the Higgs doublets receive VEVs. In addition, the states are split in their CP even and odd parts. These definitions are done in the array {\tt DEFINITION[EWSB][VEVs]}. 
In the MSSM implementation in \SARAH the factors of $1/\sqrt{2}$ are chosen in a way that the electroweak (ew) VEV is about 246~GeV. 
\begin{lstlisting}[style=mssm,firstnumber=87]
(*--- VEVs ---- *)
DEFINITION[EWSB][VEVs]= 
{ {SHd0, {vd, 1/Sqrt[2]}, {sigmad, I/Sqrt[2]},{phid,1/Sqrt[2]}},
  {SHu0, {vu, 1/Sqrt[2]}, {sigmau, I/Sqrt[2]},{phiu,1/Sqrt[2]}}
};
\end{lstlisting} 

In the \BLSSM we have to extend {\tt DEFINITION[EWSB][VEVs]} to give also VEVs to bileptons. Here, we use the same normalization as for the ew VEVs. As mentioned in sec.~\ref{sec:BLSSMparticle}, also the sneutrino will split into their real and imaginary parts because of the Majorana mass term. We can also define this splitting in {\tt DEFINITION[EWSB][VEVs]}, and keep 0 for the VEVs.  We could study spontaneous $R$-parity violation by adding a non-zero VEV for these particles. 
\begin{lstlisting}[style=blssm,firstnumber=95]
(*--- VEVs ---- *)
DEFINITION[EWSB][VEVs]= 
{ {SHd0, {vd, 1/Sqrt[2]}, {sigmad, I/Sqrt[2]},{phid,1/Sqrt[2]}},
  {SHu0, {vu, 1/Sqrt[2]}, {sigmau, I/Sqrt[2]},{phiu,1/Sqrt[2]}},
  {SC10, {x1, 1/Sqrt[2]}, {sigma1, I/Sqrt[2]},{phi1,1/Sqrt[2]}},
  {SC20, {x2, 1/Sqrt[2]}, {sigma2, I/Sqrt[2]},{phi2,1/Sqrt[2]}},
  {SvL,  {0, 0},          {sigmaL, I/Sqrt[2]},{phiL,1/Sqrt[2]}},
  {SvR,  {0, 0},          {sigmaR, I/Sqrt[2]},{phiR,1/Sqrt[2]}}     
};
\end{lstlisting}

\paragraph{Rotations in gauge sector} The electroweak gauge bosons mix in the MSSM as they do in the SM after EWSB. This mixing is defined in the array {\tt DEFINITION[EWSB][GaugeSector]} and is encoded by three matrices {\tt ZZ}, {\tt ZW}, {\tt ZfW}. These matrices have a simple form: {\tt ZZ} is usually parametrized by the Weinberg angle, {\tt ZW} and {\tt ZfW} just involve constant factors. This parametrization can be defined later in parameters.m, see \ref{sec:parameters.m}. If no parametrization is given, \SARAH will treat them as general rotation matrix with the appropriate dimension\footnote{One has to be careful with the rotations of charged vector bosons like $W^+$. Here, the mass matrix in the basis $(W^1,W^2)$ is diagonal and the calculated rotation matrix from diagonalizing this matrix would be just the identity matrix. In that case it is inevitable to give the standard parametrization $Z^W=1/\sqrt{2}\left(\begin{array}{cc} 1 & 1 \\-i & i  \end{array}\right)$ in numerical studies as input. For neutral vector bosons such a problems won't show up. }
\begin{lstlisting}[style=mssm,firstnumber=77]
(*--- Gauge Sector ---- *)
DEFINITION[EWSB][GaugeSector] =
{ {{VB,VWB[3]},{VP,VZ},ZZ},
  {{VWB[1],VWB[2]},{VWm,conj[VWm]},ZW},
  {{fWB[1],fWB[2],fWB[3]},{fWm,fWp,fW0},ZfW}
};
\end{lstlisting}

In the \BLSSM the vector boson of the $U(1)_{B-L}$ mixes with the $B$-Boson and third component of the $W$-boson to three neutral states. We call the rotation matrix again {\tt ZZ} but use another parametrization based on eq.~(\ref{eq:rotZZp}) as shown in sec.~\ref{sec:parameters.m}. The mixing in the charged gauge boson and gaugino sector doesn't change compared to the MSSM.
\begin{lstlisting}[style=blssm,firstnumber=83]
(*--- Gauge Sector ---- *)
DEFINITION[EWSB][GaugeSector] =
{ {{VB,VWB[3],VBp},{VP,VZ,VZp},ZZ},
  {{VWB[1],VWB[2]},{VWm,conj[VWm]},ZW},
  {{fWB[1],fWB[2],fWB[3]},{fWm,fWp,fW0},ZfW}
};
\end{lstlisting}

\paragraph{Rotations in matter sector} All rotations in the matter sector of the MSSM are defined via the array {\tt DEFINITION[EWSB][MatterSector]}. First the involved gauge eigenstates are given, then the name for the mass eigenstates and the rotation matrices. 
\begin{lstlisting}[style=mssm,firstnumber=95]
(*--- Matter Sector ---- *)
DEFINITION[EWSB][MatterSector]= 
{    {{SdL, SdR}, {Sd, ZD}},
     {{SuL, SuR}, {Su, ZU}},
     {{SeL, SeR}, {Se, ZE}},
     {{SvL}, {Sv, ZV}},
     {{phid, phiu}, {hh, ZH}}, 
     {{sigmad, sigmau}, {Ah, ZA}},
     {{SHdm,conj[SHup]},{Hpm,ZP}},
     {{fB, fW0, FHd0, FHu0}, {L0, ZN}}, 
     {{{fWm, FHdm}, {fWp, FHup}}, {{Lm,UM}, {Lp,UP}}},
     {{{FeL},{conj[FeR]}},{{FEL,ZEL},{FER,ZER}}},
     {{{FdL},{conj[FdR]}},{{FDL,ZDL},{FDR,ZDR}}},
     {{{FuL},{conj[FuR]}},{{FUL,ZUL},{FUR,ZUR}}} 
       }; 
\end{lstlisting} 
Lines 97 and 98 set the mixing of the down- and up-squarks ({\tt Sd}, {\tt Su}), the next two lines the mixing for charged and neutral sleptons ({\tt Se}, {\tt Sv}). Afterwards the rotations of the CP-even, CP-odd and charged Higgs bosons follow ({\tt hh}, {\tt Ah}, {\tt Hpm}). Lines 104--108 are the mixing for the fermions: neutralinos ({\tt L0}), charginos ({\tt Lm}/{\tt Lp}), charged leptons  ({\tt FEL}/{\tt FEL}), down quarks  ({\tt FDL}/{\tt FDL}) and up quarks  ({\tt FUL}/{\tt FUR}). One sees the different conventions used for the mixing of Dirac fermions compared to Majorana fermions and scalars. For instance, the first line is interpreted as
\begin{align} 
\tilde{d}_{L,{i \alpha}} = \sum_{j}Z^{D,*}_{j i}\tilde{d}_{{j \alpha}}\,, \hspace{1cm} 
\tilde{d}_{R,{i \alpha}} = \sum_{j+3}Z^{D,*}_{j i}\tilde{d}_{{j \alpha}}
\end{align} 
Here $i=1,2,3$ is the generation index for the gauge eigenstates $\tilde{d}_L$, $\tilde{d}_R$, $j=1,\dots,6$ is the generation index for the mass eigenstates $\tilde d$, and $Z^D$ is the rotation matrix. $\alpha$ is a colour index.
In the case of charginos,  first the two basis vectors are defined, before the names for the mass eigenstates and the rotation matrices follow. Thus line 105 leads to the following relations:
\begin{align} 
& \tilde{W}^- = \sum_j U^{-*}_{j 1}\lambda^-_{{j}}\,, \hspace{1cm} 
\tilde{H}_d^- = \sum_j U^{-*}_{j 2}\lambda^-_{{j}} \,, \hspace{1cm} \\
& \tilde{W}^+ = \sum_j U^{+*}_{1 j}\lambda^+_{{j}}\,, \hspace{1cm} 
\tilde{H}_u^+ = \sum_j U^{+*}_{2 j}\lambda^+_{{j}}, 
\end{align} 
The two rotation matrices $U^-$ ({\tt Um}) and $U^+$ ({\tt Up}) diagonalize the chargino mass matrix. \\
A few modifications are necessary in the \BLSSM to include the additional mixing discussed in sec.~\ref{sec:BLSSMmixing}: the mixing of the sneutrinos has to be defined for CP even and odd states separately ({\tt SvIm}, {\tt SvRe}) and the basis for the CP-even, CP-odd Higgs scalars and the neutralinos has been extended by the $B-L$ fields. Finally, a mixing of left and right neutrinos to Majorana states is added.  The corresponding definitions in the model file of the \BLSSM read
\begin{lstlisting}[style=blssm,firstnumber=109]
(*--- Matter Sector ---- *)
DEFINITION[EWSB][MatterSector]= 
{    {{SdL, SdR}, {Sd, ZD}},
     {{SuL, SuR}, {Su, ZU}},
     {{SeL, SeR}, {Se, ZE}},
     {{phiL,phiR}, {SvRe, ZVR}}, 
     {{sigmaL,sigmaR}, {SvIm, ZVI}},
     {{phid, phiu,phi1, phi2}, {hh, ZH}}, 
     {{sigmad, sigmau,sigma1,sigma2}, {Ah, ZA}},
     {{SHdm,conj[SHup]},{Hpm,ZP}},
     {{fB, fW0, FHd0, FHu0,fBp,FC10,FC20}, {L0, ZN}}, 
     {{{fWm, FHdm}, {fWp, FHup}}, {{Lm,UM}, {Lp,UP}}},
     {{FvL,conj[FvR]},{Fvm,UV}},
     {{{FeL},{conj[FeR]}},{{FEL,ZEL},{FER,ZER}}},
     {{{FdL},{conj[FdR]}},{{FDL,ZDL},{FDR,ZDR}}},
     {{{FuL},{conj[FuR]}},{{FUL,ZUL},{FUR,ZUR}}}  
       }; 
\end{lstlisting} 

\paragraph{Phases for unrotated fields} For states which don't mix there is still a phase which is not fixed. This is in both models just the case for the gluino
\begin{lstlisting}[style=both,firstnumber=128]       
DEFINITION[EWSB][Phases]= 
{    {fG, PhaseGlu}
    }; 
\end{lstlisting}
With this definition, the physical gluino mass $M_{\tilde{g}}$ is related to the gaugino mass parameter $M_3$ by
\begin{equation}
M_{\tilde{g}} = e^{i \phi_{\tilde{g}}} M_3 
\end{equation}

\paragraph{Dirac Spinors} While \SARAH works internally often with Weyl spinors, Dirac spinors are commonly used by Monte-Carlo tools and also by \SPheno. Therefore, we have to define the relation between the two- and four-component fermions. Since there are no additional mass eigenstates in the \BLSSM compared to the MSSM but just some states come with more generations, the definition of Dirac spinors is the same in both models: 
\begin{lstlisting}[style=both,firstnumber=132]       
DEFINITION[EWSB][DiracSpinors]={
 Fd ->{  FDL, conj[FDR]},
 Fe ->{  FEL, conj[FER]},
 Fu ->{  FUL, conj[FUR]},
 Fv ->{  Fvm, conj[Fvm]},
 Chi ->{ L0, conj[L0]},
 Cha ->{ Lm, conj[Lp]},
 Glu ->{ fG, conj[fG]}
};	
\end{lstlisting}       
The first line defines a Dirac spinor for the down quarks
\begin{equation}
d = \left(\begin{array}{c} d_L \\ d_R^* \end{array}\right) 
\end{equation}
while the last line sets the Majorana gluino
\begin{equation}
\tilde{g} = \left(\begin{array}{c} \lambda_{\tilde{g}} \\ \lambda_{\tilde{g}}^* \end{array}\right)  
\end{equation}
In principle, one can also add the definition of Dirac spinors for the gauge eigenstates in exactly the same way. However, we are not interested in performing calculations for gauge eigenstates and I skip this part here. Interested readers can take a look at Appendix~\ref{app:BLSSMm} to the see the entire model file for the implementation of the \BLSSM in \SARAH.

\subsubsection{Parameter definitions}
\label{sec:parameters.m}
When the changes in {\tt B-L-SSM.m} are done, the model is already ready to be used with \SARAH. While in principle all calculations in \Mathematica can be performed, the different outputs need some more information. These are mostly formal points. All possible options which can be used to define the properties of a parameter in the file  {\tt parameters.m} are the following:
\begin{itemize}
\item \verb"Description": defines a string to describe the parameter.
\item \verb"OutputName": defines a string which is used for the parameter in the different outputs. No special characters should be used to be compatible with {\tt C++} and {\tt Fortran}. 
\item \verb"Real": defines if a parameter should be considered as real ({\tt True}) or complex ({\tt False}). Default is {\tt False}.
\item \verb"Form": can be used for matrices to define if it is diagonal ({\tt Diagonal}) or a scalar ({\tt Scalar}). By default no assumption is made.
\item \verb"LaTeX": defines the name of the parameter used in the \LaTeX\ output. Standard \LaTeX\ language
should be used \footnote{'{\tt \slash}' is interpreted by \Mathematica as escape sequence. Therefore, '{\tt \slash}' has to be replaced by '{\tt \slash\slash}' in the \LaTeX\ commands.}.
\item \verb"GUTnormalization": defines a GUT normalization for an Abelian gauge coupling. 
\item \verb"Dependence": defines a dependence on other parameters which should always be used.
\item \verb"DependenceOptional": defines a dependence which is optionally used in analytical calculations.
\item \verb"DependenceNum": defines a dependence which is used in numerical calculations.
\item \verb"DependenceSPheno": defines a dependence which is just used by \SPheno.
\item \verb"MatrixProduct": can be used to express a matrix as product of two other matrices. This can be used for instance to relate the CKM matrix to the quark rotation matrices.
\item \verb"LesHouches": defines the position of the parameter in a Les Houches spectrum file. For matrices just the name of the block is defined, for scalars the block and an entry has to be given: \{block, number\}.
\item \verb"Value": assigns a numerical value to the parameter.
\end{itemize}
Many of the above definitions are just optional  and are often not needed. I'll show some parts of {\tt parameters.m} to define the properties of new parameters in the \BLSSM, or to change properties of MSSM parameters. However, I won't show here all changes compared to the MSSM but pick just some important and interesting cases. The full list of changes is given in Appendix~\ref{app:paramerersfile}. \\

\paragraph*{Gauge sector} In the gauge sector we have three new gauge couplings: the one corresponding to the new $B-L$ gauge group and the two gauge couplings induced by gauge kinetic mixing. We define for all three couplings an output name, a \LaTeX\ name and the position in the Les Houches file which will become important later.  
\begin{lstlisting}[style=file,title=\hspace{13cm}parameters.m]
{g1BL,    {Description -> "Mixed Gauge Coupling 1",
           LesHouches -> {gauge, 10},
           LaTeX -> "g_{Y B}",
           OutputName -> gYB }},
{gBL1,    {Description -> "Mixed Gauge Coupling 2",
           LesHouches -> {gauge, 11},
           LaTeX -> "g_{B Y}",
           OutputName -> gBY}},
{gBL,     {Description -> "B-L-Coupling", 
           LaTeX -> "g_{B}",
           GUTnormalization -> Sqrt[3/2],
           LesHouches -> {gauge,4},
           OutputName -> gBL }},
\end{lstlisting}
Gauge couplings are by default taken to be real, i.e. it is not necessary to define this explicitly. We also used here $\sqrt{3/2}$ for the GUT normalization of the $B-L$ charge. The GUT normalization of the two off-diagonal charges are automatically set by \SARAH. Similarly, the properties of two new gaugino mass parameters are set, see  Appendix~\ref{app:paramerersfile}. \\

We have seen in eq.~(\ref{eq:rotZZp}) how the $\gamma-Z-Z'$ rotation matrix can be parametrized by two angles. To do this, we use the {\tt Dependence} statement for the parameter {\tt ZZ} used to label the rotation in the neutral gauge sector: 
\begin{lstlisting}[style=file,title=\hspace{13cm}parameters.m]
CW=Cos[ThetaW]; SW=Sin[ThetaW]; CWp=Cos[ThetaWp]; SWp=Sin[ThetaWp];
{ZZ,  {Description -> "Photon-Z-Z' Mixing Matrix",
       Dependence ->   {{CW,-SW CWp,  SW SWp },                             
                        {SW, CW CWp, -CW SW  },
                        {0 , SWp,        CWp }},
       Real ->True,
       LaTeX -> "Z^{\\gamma Z Z'}",
       LesHouches -> None,
       OutputName -> ZZ }},             
\end{lstlisting}
Here, we put this rotation matrix explicitly to real. The reason why we also add an output name is that this matrix shows up internally in \SPheno when it diagonalizes the gauge boson mass matrix. 

Since the rotation matrix is completely defined by the two rotation angles, it is not necessary to include it in a spectrum file. Therefore, {\tt  LesHouches -> None} is used. In eq.~(\ref{eq:ThetaWP}) an approximation for the new angle $\Theta'$ was given. It might be useful to use this approximation sometimes in numerical calculations in \Mathematica. Therefore, it is included. However, the better method  for high precision calculations with \SPheno is to use the numerical result for the rotation matrix calculated by \SPheno when diagonalizing the vector bosons mass matrix. To translate this matrix into the rotation angle, we use 
\begin{equation}
\Theta'=\arccos|Z^{\gamma Z Z'}_{33}| 
\end{equation}
\SPheno calculates this value, uses it internally to get the vertices, and writes it into the block {\tt ANGLES} as entry {\tt 10} of the spectrum file. The needed lines to define all that are:
\begin{lstlisting}[style=file,title=\hspace{13cm}parameters.m]              
{ThetaWp,  { LaTeX -> "{\\Theta'}_W",
             DependenceNum ->ArcTan[(2 g1BL Sqrt[g1^2+g2^2])
                      /(g1BL^2 + 16 (x1^2+x2^2)/(vd^2+vu^2) -g1^2-g2^2)]/2,
             Real ->True,
             DependenceSPheno -> ArcCos[Abs[ZZ[3,3]]],
             OutputName-> TWp,
             LesHouches -> {ANGLES,10}      }},
\end{lstlisting}

\paragraph*{Rotations in matter sector} In the matter sector we have to define new rotation matrices for the CP even and odd sneutrinos and for the neutrinos. The definitions are very short and just include the descriptions, the \LaTeX\ commands, the Les Houches and output names. For the neutrino rotation matrix the entries read
\begin{lstlisting}[style=file,title=\hspace{13cm}parameters.m] 
{UV,     {Description ->"Neutrino-Mixing-Matrix", 
          LaTeX -> "U^V",
          LesHouches -> UVMIX,
          OutputName-> UV      }},
\end{lstlisting}
Similarly, {\tt ZVR} and {\tt ZVI} are set. In addition, there are some rotation matrices which change compared to the MSSM. There is no need to modify the definition for the neutralino rotation matrix because no assumptions are made for these in the MSSM. However, the rotation matrices for scalar and pseudo-scalar Higgs are parametrized in the MSSM by two angles
\begin{equation}
 Z^A = \left(\begin{array}{cc} -\cos\beta & \sin\beta \\ \sin\beta & \cos\beta \end{array}\right) \hspace{1cm}
 Z^H = \left(\begin{array}{cc} -\sin\alpha & \cos\alpha \\ \cos\alpha & \sin\alpha \end{array}\right) \hspace{1cm}
\end{equation}
This parametrization is used to express  dependences optionally used in the MSSM. Since the rotation matrices have grown to $4\times 4$ matrices in the \BLSSM, we can no longer make use of that. Therefore, we can take the definition of the MSSM and overwrite the dependences by {\tt None}
\begin{lstlisting}[style=file,title=\hspace{13cm}parameters.m]
{ZH,   { Description->"Scalar-Mixing-Matrix",
         DependenceOptional->None}},
{ZA,   { Description->"Pseudo-Scalar-Mixing-Matrix",
         DependenceOptional->None}},
\end{lstlisting} 

We see here an feature which makes life often much simpler. It is actually not necessary to give the full definitions for particles and parameters in the model files for each model. One can make use of global definitions which are defined in the files 
\begin{verbatim}
 $PATH/Models/parameters.m
 $PATH/Models/particles.m
\end{verbatim}
The entries for the two Higgs rotation matrices in these files read
\begin{lstlisting}[style=file,title=\hspace{13cm}parameters.m]
{{ Description -> "Scalar-Mixing-Matrix", 
    LaTeX -> "Z^H",
    Real -> True, 
    DependenceOptional ->   {{-Sin[\[Alpha]],Cos[\[Alpha]]},
                              {Cos[\[Alpha]],Sin[\[Alpha]]}}, 
    LesHouches -> SCALARMIX,
    OutputName-> ZH     }},
             
{{ Description->"Pseudo-Scalar-Mixing-Matrix", 
    LaTeX -> "Z^A",
    Real -> True,
    DependenceOptional -> {{-Cos[\[Beta]],Sin[\[Beta]]},
                            {Sin[\[Beta]],Cos[\[Beta]]}}, 
    Value -> None, 
    LesHouches -> PSEUDOSCALARMIX,
    OutputName-> ZA      }}, 
\end{lstlisting}
We refereed to these in {\tt parameters.m} of the \BLSSM by using the same string as {\tt Description}, but we overwrote locally the dependences which changed.  

\subsubsection{Particles definitions}
We turn now to the particles. To define the properties of any particle present in the model, several options are available which can be put in {\tt particles.m}:
\begin{itemize}
\item \verb"Description": a string for defining the particle.
\item \verb"PDG": defines the PDG numbers of all generations.
\item \verb"PDG.IX": defines a nine-digit number of a particle supporting the proposal
Ref.~\cite{Brooijmans:2012yi,Basso:2012ew}. By default, the entries of {\tt PDG} are used in the output\footnote{ 
To switch to the new scheme,  either at the beginning of a \SARAH session or in the model files, one has to set {\tt UsePDGIX = True;}}.
\item \verb"ElectricCharge": defines the electric charge of a particle in units of $e$. This
information is exported to the \CalcHep/\CompHep and \UFO model files. 
\item \verb"Width": can be used to define a fixed numerical value for the width.
\item \verb"Mass": defines, how MC tools obtain a numerical value for the mass of the particle:
\begin{itemize}
\item a numerical value can be given
\item the keyword \verb"Automatic" assigns that \SARAH derives the tree level expression for the mass
from the Lagrangian. The mass is then calculated by using the values of the other
parameters.  
\item the keyword \verb"LesHouches" assigns that this mass is calculated numerically by a spectrum generator like \SPheno and provided via a Les Houches spectrum file. This is usually the preferred method because also loop corrections are included. 
\end{itemize}  
\item \verb"OutputName": defines the name used in the different outputs for other codes.
\item \verb"LaTeX": defines the name of the particle used in the \LaTeX\ output. 
\item \verb"FeynArtsNr": the number assigned to the particle used in the \FeynArts output
\item \verb"LHPC": defines the column and colour used for the particle in the steering file
for the  {\tt LHPC Spectrum Plotter} \footnote{{\tt http://lhpc.hepforge.org/}}. All colours available in {\tt gnuplot} can be used.
\item \verb"Goldstone": for each massive vector boson the name of corresponding Goldstone boson is given.
\end{itemize}

The properties of all particles appearing either as gauge or mass eigenstates, or just at intermediate steps can be defined in {\tt particles.m}. Usually, the user is only interested in the output for the mass eigenstates. Therefore, it is not really necessary to define the entire properties of all intermediate states and the gauge eigenstates. The only input which is usually helpful to have a nice looking \LaTeX\ output is to define the \LaTeX\ syntax for all particles which appear at any stage in the model. Again, I pick just some entries to demonstrate the procedure. The full changes compared to the MSSM are given in Appendix~\ref{app:particlesfile}.

\paragraph*{Gauge eigenstates and intermediate states} Intermediate states are those which don't show up in any Lagrangian. These are for instance, the superfields which get decomposed in components fields, or the real parts of complex scalars after symmetry breaking which directly mix to new mass eigenstates. Also Weyl spinors are considered as intermediate because the output is always in terms of Dirac spinors. For all of these particles it is usually sufficient to define a \LaTeX\ name to have a nice looking {\tt pdf} file at the end:
\begin{lstlisting}[style=file,title=\hspace{13cm}particles.m]
WeylFermionAndIndermediate = {
  ...
  (* Superfields *)
  {vR,   { Description -> "Right Neutrino Superfield" }},
  {C1,   { LaTeX  -> "\\hat{\\eta}" }},
  {C2,   { LaTeX  -> "\\hat{\\bar{\\eta}}" }},

  (* Intermediate Scalars *)
  {phi1,   { LaTeX  -> "\\phi_{\\eta}" }},
  {phi2,   { LaTeX  -> "\\phi_{\\bar{\\eta}}" }},

  {sigma1,   {   LaTeX -> "\\sigma_{\\eta}"}},
  {sigma2,   {   LaTeX -> "\\sigma_{\\bar{\\eta}}" }},
  
  ...
};
\end{lstlisting}
The properties of gauge eigenstates can be defined in the array {\tt ParticleDefinitions[GaugeES]}. However, since rarely calculations are done for these states, it is also often sufficient to give just the \LaTeX\ names. Of course, also all other information can be set as for the mass eigenstates if demanded.
\begin{lstlisting}[style=file,title=\hspace{13cm}particles.m]
ParticleDefinitions[GaugeES] = {
 ...
{SC10, {LaTeX -> "\\eta" }},
{SC20,  {LaTeX -> "\\bar{\\eta}"}},
...
\end{lstlisting}

\paragraph*{Mass eigenstates} More interesting are the mass eigenstates. The additional information given for those is used in the different output for \SPheno and the MC-tools. We begin with the new states which are not present in the MSSM: the $Z'$ and the corresponding ghost, as well as the real sneutrinos:
\begin{lstlisting}[style=file,title=\hspace{13cm}particles.m]
  {SvRe,  { Description -> "CP-even Sneutrino",
            LaTeX -> "\\nu^R",
            OutputName -> "nR",
            FeynArtsNr -> 41,
            LHPC -> {5, "blue"},
            PDG->{1000012,1000014,1000016,2000012,2000014,2000016},
            PDG.IX ->{200000001,200000002,200000003,
                      200000004,200000005,200000006} }},
  {SvIm, { Description -> "CP-odd Sneutrino",
           LaTeX -> "\\nu^I",
           OutputName -> "nI",
           FeynArtsNr -> 40,
           LHPC -> {5, "turquoise"},
           PDG->{4000012,4000014,4000016,5000012,5000014,5000016},
           PDG.IX ->{202000001,202000002,202000003,
                     202000004,202000005,202000006}}},  
           
  {VZp,  { Description -> "Z'-Boson",
           PDG -> {31},
           PDG.IX -> {122000002},
           Width -> Automatic, 
           Mass -> LesHouches,
           FeynArtsNr -> 10,
           LaTeX -> "{Z'}",
           Goldstone -> Ah[{2}],
           ElectricCharge -> 0,
           OutputName -> "Zp"}} 
  {gZp,   { Description -> "Z'-Ghost",  
            PDG -> 0,
            PDG.IX -> 0,
            Width -> 0, 
            Mass -> Automatic,
            FeynArtsNr -> 10,
            LaTeX -> "\\eta^{Z'}",
            ElectricCharge -> 0,
            OutputName -> "gZp"}},
\end{lstlisting}
Here, we defined that the second pseudo-scalar ({\tt Ah[\{2\}]}) is the Goldstone of the $Z'$. For the ghost we used the PDG {\tt 0} to make clear that this is not a physical state. \\
In addition, there are also particles where the number of generations has increased with respect to the MSSM. Since all MSSM particles are defined globally in {\tt \$PATH/SARAH/Models/particles.m} by the {\tt Description} statement, we can make use of that but just overwrite the lists of PDGs which become longer. The new CP even and odd Higgs states carry now four PDGs:
\begin{lstlisting}[style=file,title=\hspace{13cm}particles.m]
{hh ,  { Description -> "Higgs",
         PDG -> {25,35,9900025, 9900035},
         PDG.IX->{101000001,101000002,101000003,101000004}  }}, 
{Ah ,  { Description -> "Pseudo-Scalar Higgs",
         PDG -> {0,0,36,9900036},
         PDG.IX->{0,0,102000001,102000002} }},                
\end{lstlisting}
Here, we used for the first two entries of the pseudo-scalars {\tt 0}. This means that these two states are not physical, but the Goldstones of the massive, neutral gauge bosons. In the similar way, the additional PDGs for neutralinos {\tt Chi} and neutrinos {\tt Fv} are defined, see Appendix~\ref{app:particlesfile}.

\subsection{Running the model}
\label{sec:example_running}
When we are done with the model files, the model is initialized in \Mathematica via 
\begin{lstlisting}[style=mathematica]
<<[$PATH]/SARAH.m;
Start["B-L-SSM"];
\end{lstlisting}
After about 1 minute the message
\begin{lstlisting}[style=mathematica]
All Done. B-L-SSM is ready! 
\end{lstlisting}
should appear and no error messages or warning during the evaluation should show up. 
More detailed checks if the model implementation is self-consistent and if the model is working fine can be carried out by the command
\begin{lstlisting}[style=mathematica]
CheckModel[];
\end{lstlisting}
This function makes all checks listed in sec.~\ref{sec:checks}. With our implementation above there will be some messages like
\begin{lstlisting}[style=mathematica]
Lagrangian::PossibleMixing: Possible mixing between SvIm and SvRe induced by the term: 1/4 I (2 (sum[j2,1,3,conj[ZVR[gt2,j2]] sum[j1,1,3,conj[<<1>>] ml2[<<2>>]]]-sum[j2,1,3,conj[ZVI[<<2>>]] sum[j1,1,3,Times[<<2>>]]]+sum[j2,<<2>>,conj[ZVR[gt2,Plus[<<2>>]]] sum[j1,<<2>>,<<1>>]]-sum[j2,1,3,conj[ZVI[<<2>>]] sum[j1,1,3,Times[<<2>>]]])+<<6>>)

Lagrangian::PossibleMixing: Possible mixing between hh and Ah induced by the term: 1/2 I ((B[\[Mu]]-conj[B[\[Mu]]]) (ZA[gt2,2] ZH[gt1,1]+ZA[gt2,1] ZH[gt1,2])+(B[MuP]-conj[B[MuP]]) (ZA[gt2,4] ZH[gt1,3]+ZA[gt2,3] ZH[gt1,4]))
                                                                                                            
Lagrangian::PossibleMixing: Possible mixing between VP and VZ induced by the term: gBL1 (x1^2+x2^2) Cos[ThetaW] (gBL1 Cos[ThetaWp] Sin[ThetaW]-gBL Sin[ThetaWp])                                                                                                            
...
\end{lstlisting}
The last message is caused because it is not obvious from the Lagrangian that there is no $\gamma$--$Z$ mixing: the relation between the rotation angels $\Theta$, $\Theta'$ and all five involved gauge couplings ($g_{YY}$, $g_{BY}$, $g_{YB}$, $g_{BB}$, $g_2$) are not known at this stage. Hence, it is not obvious from the Lagrangian that these terms cancel. \\
The reason for the first two messages is that \SARAH found terms in the Lagrangian of the mass eigenstates which seem to cause a mixing between $\tilde \nu^R$  and $\tilde \nu^I$ as well as between $h$ and $A^h$. The origin of this is that we didn't define terms like  $B_\mu$, $B_\mu'$ or $Y_\nu$ explicitly as real. Therefore, \SARAH considers them to be complex.  In the complex case a mixing between CP even and odd scalars would occur unless specific relations among the phases are satisfied. This mixing would be missed in our implementation so far. We just have to keep that in mind that the model is supposed to be used for the CP conserving case, or just for parameters which satisfy conditions that the mixing between CP even and odd states vanishes. Of course, one can also use the entries in {\tt parameters.m} to define the parameters explicitly as real.\\
Even if a study of CP violation is clearly beyond the scope of this example, I want to show briefly what has to be changed to incorporate it. First, the definition of the VEVs has to be changed by introducing relative phases between the scalars:
\begin{lstlisting}[style=file]
(*--- VEVs ---- *)
DEFINITION[EWSB][VEVs]= 
{{SHd0, {vd,1/Sqrt[2]}, {sigmad,I/Sqrt[2]}, {phid,1/Sqrt[2]}},
 {SHu0, {vu,1/Sqrt[2]}, {sigmau,I/Sqrt[2]}, {phiu,1/Sqrt[2]},{etaU}},
 {SC10, {x1,1/Sqrt[2]}, {sigma1,I/Sqrt[2]}, {phi1,1/Sqrt[2]},{etaX1}},
 {SC20, {x2,1/Sqrt[2]}, {sigma2,I/Sqrt[2]}, {phi2,1/Sqrt[2]},{etaX2}},
 {SvL,  {0, 0},         {sigmaL,I/Sqrt[2]}, {phiL,1/Sqrt[2]}},
 {SvR,  {0, 0},         {sigmaR,I/Sqrt[2]}, {phiR,1/Sqrt[2]},{etaR}}     
};
\end{lstlisting} 
One could also work with complex VEVs
\begin{lstlisting}[style=file]
DEFINITION[EWSB][VEVs]= 
{{SHd0, {vdR,1/Sqrt[2]},{vdI,I/Sqrt[2]},
                   {sigmad, I/Sqrt[2]},{phid,1/Sqrt[2]}},
 {SHu0, {vuR,1/Sqrt[2]},{vuI,I/Sqrt[2]},
                   {sigmau, I/Sqrt[2]},{phiu,1/Sqrt[2]}},
 ...
 }
\end{lstlisting}
However, this is less common and mainly supposed to be used for generating \Vevacious model files. After introducing the phases also the definition of the rotations must be changed to respect the mixing between CP even and odd states:
\begin{lstlisting}[style=file]
(*--- Matter Sector ---- *)
DEFINITION[EWSB][MatterSector]= 
{  ...,
   {{phiL,phiR,sigmaL,sigmaR}, {Sv, ZV}}, 
   {{phid, phiu,phi1, phi2, sigmad, sigmau,sigma1,sigma2}, {hh, ZH}}, 
   ...    }
\end{lstlisting}
The rotation matrices and mass eigenstates in {\tt particles.m} and {\tt parameters.m} have to be adjusted accordingly: the matrices {\tt ZVR} and {\tt ZH} would no longer be real, and the states  {\tt Sv} need twelve PDGs, while {\tt hh} includes two Goldstones and six physical scalars. Also assignments of the Goldstones have to be adjusted in {\tt particles.m}
\begin{lstlisting}[style=file]
  {VZp,  { Description -> "Z'-Boson",
           Goldstone -> hh[{2}]}}  
  {VZ,  { Description -> "Z-Boson",
           Goldstone -> hh[{1}]}}  
\end{lstlisting}
However, as I said this is beyond the scope of the discussion here. We start now to study the CP conserving version of this model in detail.

\section{Example -- Part II: Masses, vertices, tadpoles and RGEs with \Mathematica}
\label{sec:example_mathematica}

\subsection{\LaTeX\ output}
\subsubsection{General information}
After we are done with the implementation of the model and all consistency checks are passed, we can generate a {\tt pdf} file  to get a first overview about the \BLSSM. The {\tt .tex} files generated by \SARAH include all information about the model, e.g. particle content and superpotential, and all information which \SARAH derives like RGEs as well as masses and vertices for specific eigenstates.
To get this output one has first to run the command {\tt ModelOutput} which tells \SARAH what it has to calculate and for which eigenstates. We are just interested in the eigenstates after EWSB:
\begin{lstlisting}[style=mathematica]
ModelOutput[EWSB]; 
\end{lstlisting}
This command calculates always all vertices for the considered eigenstates. Loop corrections and RGEs are not included in the calculations by default. To include them, the user can choose {\tt IncludeRGEs -> True} and {\tt IncludeLoopCorrections -> True}. This information will then also be included in the \LaTeX\ files. In principle, one can also use options for {\tt ModelOutput} to directly generate the \LaTeX\ output and all other outputs for the different codes by setting: {\tt WriteTeX  -> True}, {\tt WriteFeynArts -> True}, {\tt WriteCHep -> True}, {\tt WriteWHIZARD -> True}, {\tt WriteUFO -> True}. However, we won't make use of this here but discuss each output separately. \\

When \SARAH is done with all calculations, one can run 
\begin{lstlisting}[style=mathematica]
MakeTeX[WriteSARAH->True];
\end{lstlisting}
We used here the option that not only the information about the model and its physics is included, but also details about the implementation in \SARAH are attached to the pdf. Additional options which are available are
\begin{itemize}
\item \verb"FeynmanDiagrams": defines, if the Feynman diagrams should or should not be included in the output. By default they are included. To draw the Feynman diagrams, \SARAH makes use of the \LaTeX\ package {\tt feynmf} \cite{Ohl:FeynMF}
\item \verb"ShortForm": defines, if a shorter notation for the vertices should be used by not using a separate {\tt equation} environment for each vertex and skipping Feynman diagrams
\end{itemize}
When \SARAH is done with the output, the {\tt .tex} file are stored in
\begin{lstlisting}
$PATH/SARAH/Output/B-L-SSM/EWSB/TeX 
\end{lstlisting}
The main file which can be compiled with {\tt pdflatex} is {\tt B-L-SSM\_EWSB.tex}. In the case that the Feynman diagrams are included, the compilation is a bit more complicated because {\tt mpost} has to be used for each diagram after the first run of {\tt pdflatex}. Afterwards, a second run of {\tt pdflatex} is needed. \SARAH provides a shell script {\tt MakePDF.sh} which does take care of that. Thus, the easiest way to get the {\tt pdf} is
\begin{lstlisting}[style=terminal]
$ cd  $PATH/SARAH/Output/B-L-SSM/EWSB/TeX 
$ chmod 755 MakePDF.sh
$ ./MakePDF.sh
\end{lstlisting}
The second step is just necessary to make the script executable if it is not.

\subsubsection{Particles and parameters of the \BLSSM in \SARAH}
The reason why we have chosen the option {\tt WriteSARAH -> True} is that this includes helpful information about the implementation: the names for all particles and parameters in \SARAH are given, and it is shown how the pieces are called in \LaTeX\ and in the output for other codes. This output is given for all eigenstates. However, to follow the subsequent examples only the mass eigenstates after EWSB are necessary. So, I skip the output for the other eigenstates and show here only the corresponding tables for the mass eigenstates.

\paragraph{Particles} The whole list of fermions, scalars, vector bosons and ghosts are listed in Tabs.~\ref{tab:fermions}--\ref{tab:ghosts}. One sees that not only the names of each particle are given which are used at the different stage, but also what indices the particles carry. In the case of fermions, the Dirac spinors together with their Weyl components are listed. An alternative to get an overview about all particles during the \Mathematica session, is to use
\begin{lstlisting}[style=mathematica]
Particles[EWSB] 
\end{lstlisting}

\begin{table}[hbt]
\centering
\begin{tabular}{|ccc|} 
\hline 
\LaTeX & \SARAH & Output \\ 
\hline 
\(\tilde{\chi}^-_{{i}} = \left( \begin{array}{c} \lambda^-_{{i}}\\\lambda^{+,*}_{{i}}\end{array} \right) \) & \( \verb"Cha[{generation}]" = \left( \begin{array}{c} \verb"Lm[{generation}]" \\ \verb"conj[Lp[{generation}]]"\end{array} \right) \)  & \verb"C" \\ 
  \(\tilde{\chi}^0_{{i}} = \left( \begin{array}{c} \lambda^0_{{i}}\\\lambda^{0,*}_{{i}}\end{array} \right) \) & \( \verb"Chi[{generation}]" = \left( \begin{array}{c} \verb"L0[{generation}]" \\ \verb"conj[L0[{generation}]]"\end{array} \right) \)  & \verb"N" \\ 
  \(d_{{i \alpha}} = \left( \begin{array}{c} D_{L,{i \alpha}}\\D^*_{R,{i \alpha}}\end{array} \right) \) & \( \verb"Fd[{generation, color}]" = \left( \begin{array}{c} \verb"FDL[{generation, color}]" \\ \verb"conj[FDR[{generation, color}]]"\end{array} \right) \)  & \verb"d" \\ 
  \(e_{{i}} = \left( \begin{array}{c} E_{L,{i}}\\E^*_{R,{i}}\end{array} \right) \) & \( \verb"Fe[{generation}]" = \left( \begin{array}{c} \verb"FEL[{generation}]" \\ \verb"conj[FER[{generation}]]"\end{array} \right) \)  & \verb"e" \\ 
  \(u_{{i \alpha}} = \left( \begin{array}{c} U_{L,{i \alpha}}\\U^*_{R,{i \alpha}}\end{array} \right) \) & \( \verb"Fu[{generation, color}]" = \left( \begin{array}{c} \verb"FUL[{generation, color}]" \\ \verb"conj[FUR[{generation, color}]]"\end{array} \right) \)  & \verb"u" \\ 
  \(\nu_{{i}} = \left( \begin{array}{c} \lambda_{\nu,{i}}\\\lambda^*_{\nu,{i}}\end{array} \right) \) & \( \verb"Fv[{generation}]" = \left( \begin{array}{c} \verb"Fvm[{generation}]" \\ \verb"conj[Fvm[{generation}]]"\end{array} \right) \)  & \verb"nu" \\ 
  \(\tilde{g}_{{\alpha}} = \left( \begin{array}{c} \lambda_{{\tilde{g}},{\alpha}}\\\lambda^*_{{\tilde{g}},{\alpha}}\end{array} \right) \) & \( \verb"Glu[{color}]" = \left( \begin{array}{c} \verb"fG[{color}]" \\ \verb"conj[fG[{color}]]"\end{array} \right) \)  & \verb"go" \\ 
  \hline 
\end{tabular} 
\caption{Fermions in the \BLSSM after electroweak symmetry breaking.}
\label{tab:fermions}
\end{table}

\begin{table}[hbt]
\centering
\begin{tabular}{|ccc|ccc|} 
\hline 
\LaTeX & \SARAH & Output & \LaTeX & \SARAH & Output \\ 
\hline 
\(\tilde{d}_{{i \alpha}}\) & \verb"Sd[{generation, color}]" &\verb"sd"  & \(\tilde{u}_{{i \alpha}}\) & \verb"Su[{generation, color}]" &\verb"su" \\ 
\(\tilde{e}_{{i}}\) & \verb"Se[{generation}]" &\verb"se"  & \(\nu^i_{{i}}\) & \verb"SvIm[{generation}]" &\verb"nI" \\ 
\(\nu^R_{{i}}\) & \verb"SvRe[{generation}]" &\verb"nR"  & \(h_{{i}}\) & \verb"hh[{generation}]" &\verb"h" \\ 
\(A^0_{{i}}\) & \verb"Ah[{generation}]" &\verb"Ah"  & \(H^-_{{i}}\) & \verb"Hpm[{generation}]" &\verb"{Hm, Hp}" \\ 
\hline 
\end{tabular} 
\caption{Scalars in the \BLSSM after electroweak symmetry breaking.}
\label{tab:scalars}
\end{table}

\begin{table}[hbt]
\centering
\begin{tabular}{|ccc|ccc|} 
\hline 
\LaTeX & \SARAH & Output & \LaTeX & \SARAH & Output \\ 
\hline 
\(g_{{\alpha \rho}}\) & \verb"VG[{color, lorentz}]" &\verb"g"  & \(\gamma_{{\rho}}\) & \verb"VP[{lorentz}]" &\verb"A" \\ 
\(Z_{{\rho}}\) & \verb"VZ[{lorentz}]" &\verb"Z"  & \({Z'}_{{\rho}}\) & \verb"VZp[{lorentz}]" &\verb"Zp" \\ 
\(W^-_{{\rho}}\) & \verb"VWm[{lorentz}]" &\verb"{Wm, Wp}" & & & \\ 
 \hline 
\end{tabular} 
\caption{Vector bosons in the \BLSSM after electroweak symmetry breaking.}
\label{tab:vector}
\end{table}

\begin{table}[hbt]
\centering
\begin{tabular}{|ccc|ccc|} 
\hline 
\LaTeX & \SARAH & Output & \LaTeX & \SARAH & Output \\ 
\hline 
\(\eta^G_{{\alpha}}\) & \verb"gG[{color}]" &\verb"gG"  & \(\eta^{\gamma}\) & \verb"gP" &\verb"gA" \\ 
\(\eta^Z\) & \verb"gZ" &\verb"gZ"  & \(\eta^{Z'}\) & \verb"gZp" &\verb"gZp" \\ 
\(\eta^-\) & \verb"gWm" &\verb"gWm"  & \(\eta^+\) & \verb"gWmC" &\verb"gWpC" \\ 
\hline 
\end{tabular} 
\caption{Ghost particles in the \BLSSM after electroweak symmetry breaking.}
\label{tab:ghosts}
\end{table}

\paragraph*{Parameters} All parameters which are present at some stage in the \BLSSM are listed in Tab.~\ref{tab:parameters}. This includes not only the fundamental parameters like gauge couplings, superpotential couplings and soft-breaking terms, but also rotation matrices and angles, VEVs as well as auxiliary parameters which just show up via dependences defined in {\tt parameters.m}. One can get the entire list of parameters also during the \Mathematica session by using 
\begin{lstlisting}[style=mathematica]
parameters
\end{lstlisting}

\begin{table}[hbt]
\begin{tabular}{|ccc|ccc|ccc|} 
\hline 
\LaTeX & \SARAH & Output & \LaTeX & \SARAH & Output & \LaTeX & \SARAH & Output \\ 
\hline 
\hline 
\(g_1\) & \verb"g1" & \verb"g1"  & \(g_2\) & \verb"g2" & \verb"g2"  & \(g_3\) & \verb"g3" & \verb"g3" \\ 
\(g_{B}\) & \verb"gBL" & \verb"gBL"  & \(g_{Y B}\) & \verb"g1BL" & \verb"gYB"  & \(g_{B Y}\) & \verb"gBL1" & \verb"gBY" \\ 
\({\mu_{\eta}}\) & \verb"MuP" & \verb"MuP"  & \(B_{\eta}\) & \verb"B[MuP]" & \verb"BMuP"  & \(\mu\) & \verb"\[Mu]" & \verb"Mu" \\ 
\(B_{\mu}\) & \verb"B[\[Mu]]" & \verb"Bmu"  & \(Y_d\) & \verb"Yd" & \verb"Yd"  & \(T_d\) & \verb"T[Yd]" & \verb"Td" \\ 
\(Y_e\) & \verb"Ye" & \verb"Ye"  & \(T_e\) & \verb"T[Ye]" & \verb"Te"  & \(Y_u\) & \verb"Yu" & \verb"Yu" \\ 
\(T_u\) & \verb"T[Yu]" & \verb"Tu"  & \(Y_x\) & \verb"Yn" & \verb"Yx"  & \(T_x\) & \verb"T[Yn]" & \verb"Tx" \\ 
\(Y_\nu\) & \verb"Yv" & \verb"Yv"  & \(T_\nu\) & \verb"T[Yv]" & \verb"Tv"  & \(m_q^2\) & \verb"mq2" & \verb"mq2" \\ 
\(m_l^2\) & \verb"ml2" & \verb"ml2"  & \(m_{H_d}^2\) & \verb"mHd2" & \verb"mHd2"  & \(m_{H_u}^2\) & \verb"mHu2" & \verb"mHu2" \\ 
\(m_d^2\) & \verb"md2" & \verb"md2"  & \(m_u^2\) & \verb"mu2" & \verb"mu2"  & \(m_e^2\) & \verb"me2" & \verb"me2" \\ 
\(m_{\nu}^2\) & \verb"mvR2" & \verb"mv2"  & \(m_{\eta}^2\) & \verb"mC12" & \verb"mC12"  & \(m_{\bar{\eta}}^2\) & \verb"mC22" & \verb"mC22" \\ 
\(M_1\) & \verb"MassB" & \verb"M1"  & \(M_2\) & \verb"MassWB" & \verb"M2"  & \(M_3\) & \verb"MassG" & \verb"M3" \\ 
\({M}_{BL}\) & \verb"MassBp" & \verb"MBp"  & \({M}_{B B'}\) & \verb"MassBBp" & \verb"MBBp"  & \(v_d\) & \verb"vd" & \verb"vd" \\ 
\(v_u\) & \verb"vu" & \verb"vu"  & \(v_{\eta}\) & \verb"x1" & \verb"x1"  & \(v_{\bar{\eta}}\) & \verb"x2" & \verb"x2" \\ 
\(Z^{\gamma Z Z'}\) & \verb"ZZ" & \verb"ZZ"  & \(Z^{W}\) & \verb"ZW" & \verb"ZW"  & \(Z^{\tilde{W}}\) & \verb"ZfW" & \verb"ZfW" \\ 
\(\phi_{\tilde{g}}\) & \verb"PhaseGlu" & \verb"pG"  & \(Z^D\) & \verb"ZD" & \verb"ZD"  & \(Z^U\) & \verb"ZU" & \verb"ZU" \\ 
\(Z^E\) & \verb"ZE" & \verb"ZE"  & \(Z^i\) & \verb"ZVI" & \verb"ZVI"  & \(Z^R\) & \verb"ZVR" & \verb"ZVR" \\ 
\(Z^H\) & \verb"ZH" & \verb"ZH"  & \(Z^A\) & \verb"ZA" & \verb"ZA"  & \(Z^+\) & \verb"ZP" & \verb"ZP" \\ 
\(N\) & \verb"ZN" & \verb"ZN"  & \(U\) & \verb"UM" & \verb"UM"  & \(V\) & \verb"UP" & \verb"UP" \\ 
\(U^V\) & \verb"UV" & \verb"UV"  & \(U^e_L\) & \verb"ZEL" & \verb"ZEL"  & \(U^e_R\) & \verb"ZER" & \verb"ZER" \\ 
\(U^d_L\) & \verb"ZDL" & \verb"ZDL"  & \(U^d_R\) & \verb"ZDR" & \verb"ZDR"  & \(U^u_L\) & \verb"ZUL" & \verb"ZUL" \\ 
\(U^u_R\) & \verb"ZUR" & \verb"ZUR"  & \(e\) & \verb"e" & \verb"el"  & \(\Theta_W\) & \verb"ThetaW" & \verb"TW" \\ 
\(\beta\) & \verb"\[Beta]" & \verb"betaH"  & \({\Theta'}_W\) & \verb"ThetaWp" & \verb"TWp"  & \(\alpha_S\) & \verb"AlphaS" & \verb"aS" \\ 
\(\alpha^{-1}\) & \verb"aEWinv" & \verb"aEWinv"  & \(v\) & \verb"v" & \verb"v"  & \({\beta'}\) & \verb"BetaP" & \verb"Bp" \\ 
\(x\) & \verb"vX" & \verb"vX"  & \(\tan\Big(\beta'\Big)\) & \verb"TBetaP" & \verb"TBp"  & \(Mass[VWm]^*\) & \verb"Mass[VWm]" & \verb"Mass[VWm]" \\ 
\(G_f\) & \verb"Gf" & \verb"Gf" & & && & & \\ 
 \hline 
\end{tabular} 
\caption{All parameters in the \BLSSM with their names used internally by \SARAH as well as the names for the \LaTeX\ and other outputs.}
\label{tab:parameters}
\end{table}

\subsection{Extracting mass matrices, tadpole equations and vertices}
We can finally do some physics. At first, we want to study the analytical properties of the \BLSSM within \Mathematica. For this purpose I'll give some example how to extract mass matrices, vertices or tadpole equations and how to deal with them. To improve the readability I'll give the input in \Mathematica format but the output of \Mathematica will be translated into \LaTeX. If the user just wants to see the expressions without modifying them it is also possible to generate the entire \LaTeX\ output for the model and just read the {\tt pdf} as just shown in the previous section. 

\subsubsection{Tadpole equations}  We start with the tadpole equations. All four minimum conditions are returned via
\begin{lstlisting}[style=mathematica]
TadEquations={TadpoleEquation[vd], TadpoleEquation[vu], 
                        TadpoleEquation[x1], TadpoleEquation[x2]}
\end{lstlisting}
and read
\begin{align} 
0 &= - v_u \Re B_{\mu} + \frac{1}{8} v_d \Big(2 \Big(g_1 g_{B Y}  + g_{Y B} g_{B} \Big)\Big(v^2_{\eta}- v^2_{\bar{\eta}} \Big)+ \Big(g_{1}^{2} + g_{Y B}^{2} + g_{2}^{2}\Big)\Big( v^2_d- v^2_u \Big)+ v_d \Big(m_{H_d}^2 + |\mu|^2\Big)\\ 
0 &= \frac{1}{8} v_u \Big(2 \Big(g_1 g_{B Y}  + g_{Y B} g_{B} \Big)\Big(v_{\bar{\eta}}^{2}- v_{\eta}^{2} \Big) + \Big(g_{1}^{2} + g_{Y B}^{2} + g_{2}^{2}\Big)\Big(v_{u}^{2}- v_{d}^{2} \Big)\Big) - v_d {\Re\Big(B_{\mu}\Big)}  + v_u \Big(m_{H_u}^2 + |\mu|^2\Big)\\ 
0 &= - v_{\bar{\eta}} \Re B_{\eta}  + \frac{1}{4} v_{\eta} \Big(2 \Big(g_{B}^{2} + g_{B Y}^{2}\Big)\Big(v^2_{\eta}- v^2_{\bar{\eta}}\Big) + \Big(g_1 g_{B Y}  + g_{Y B} g_{B} \Big)\Big(v^2_d- v^2_u \Big)\Big) + v_{\eta} \Big(m_{\eta}^2 + |{\mu_{\eta}}|^2\Big)\\ 
0 &= \frac{1}{4} v_{\bar{\eta}} \Big(-2 \Big(g_{B}^{2} + g_{B Y}^{2}\Big)\Big(v^2_{\eta}- v^2_{\bar{\eta}} \Big) + \Big(g_1 g_{B Y}  + g_{Y B} g_{B} \Big)\Big(v_{u}^{2}- v_{d}^{2}  \Big)\Big) + v_{\bar{\eta}} \Big(m_{\bar{\eta}}^2 + |{\mu_{\eta}}|^2\Big) - v_{\eta} {\Re\Big(B_{\eta}\Big)} 
\end{align} 
I saved them in a new list {\tt TadEquations} which we will use in the following. Alternatively, one can also use the content of {\tt TadpleEquations[EWSB]} which contains all tadpole equations after EWSB separated by commas. To get the same {\tt TadEquations} as above, we make an equation out of any entry in {\tt TadpoleEquations[EWSB]} by
\begin{lstlisting}[style=mathematica]
TadEquations = Map[# == 0 &, TadpoleEquations[EWSB]];
\end{lstlisting}
Let's gain some understanding of these equations. A convenient choice is solve them for $\mu$, $\mu'$ $B_\mu$ and $B_\mu'$. For simplicity we do this by restricting ourself to the real case ({\tt conj[x\_]->x}) and work in the triangle basis where $g_{YB}$ disappears:
\begin{lstlisting}[style=mathematica]
solutionTad=Solve[TadEquations 
     /. \[Mu] conj[\[Mu]] -> AbsMu2 /. MuP conj[MuP] -> AbsMuP2 
     /. conj[x_] -> x /. gBL1 -> 0,   
 {B[\[Mu]], B[MuP], AbsMu2, AbsMuP2}] 
\end{lstlisting}
In addition, we introduced new parameters for $|\mu|^2$ and $|\mu'|^2$. There are mainly two reasons for this: (i) we can solve the equations in that way for $|\mu|^2$;  (ii) if {\tt X} and {\tt B[X]} appear in the equations, \Mathematica interprets {\tt B[X]} as function of {\tt X} instead as an independent parameter. The consequence is that it can't solve the equations. To circumvent this, a replacement like {\tt B[X]->BX} would be necessary.\\

\begin{figure}[hbt]
\centering 
\includegraphics[width=0.6\linewidth]{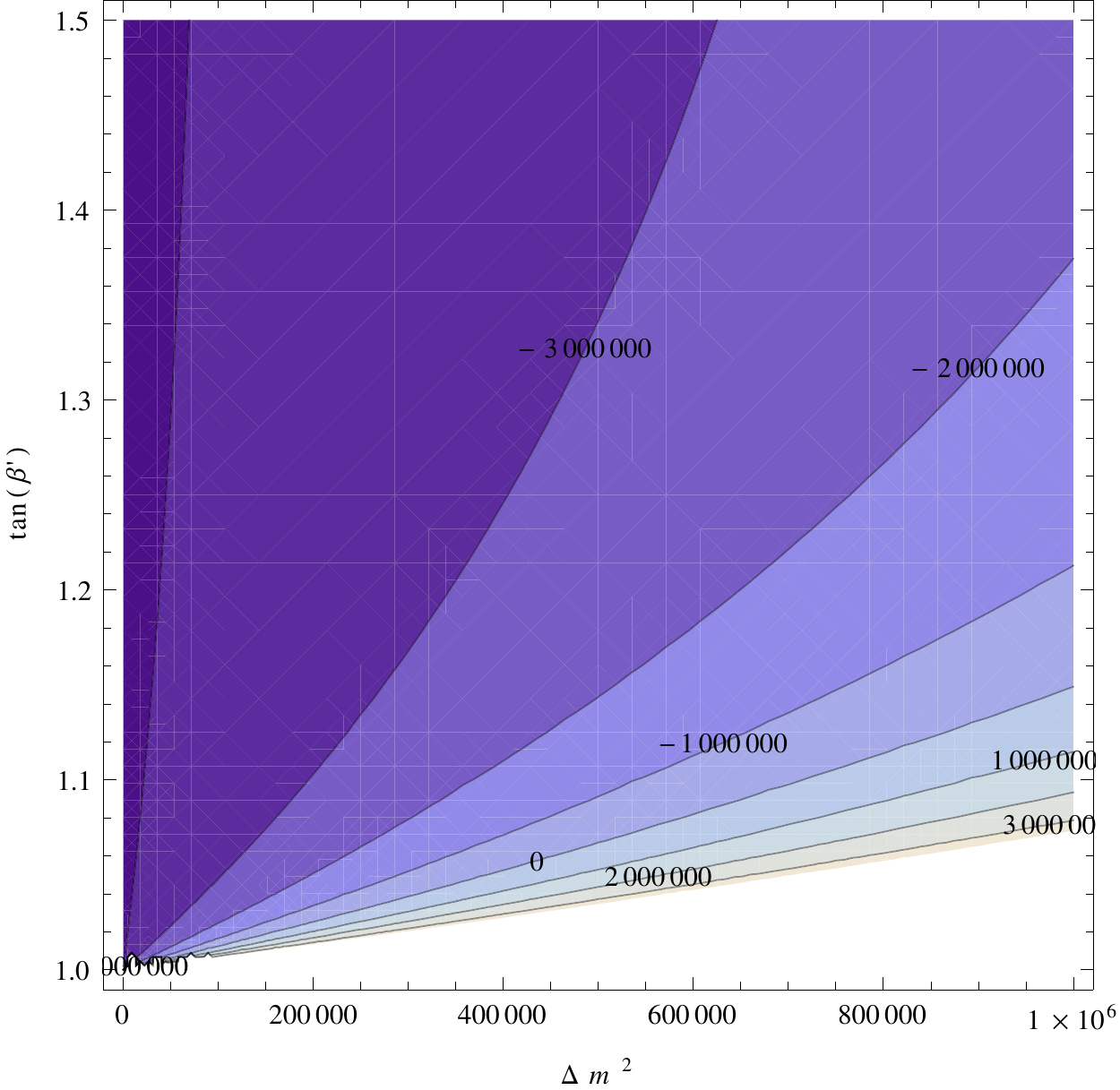}
\caption{$|\mu'|^2$ in the $(\Delta m^2,\tan\beta')$ plane using eq.~(\ref{eq:absmup}) with $\Delta m^2 = m_{\eta}^2-m_{\bar \eta}^2$. I have chosen $M_{Z'} = 2.5$~TeV and $m_{\eta}^2=1~\text{TeV}^2$. }
\label{fig:abmup2}
\end{figure}

We are just interested for the moment in the solutions for $|\mu|^2$ and $|\mu'|^2$. They read
\begin{align}
|\mu|^2 =& \frac{1}{8} (- v_{u}^{2}  + v_{d}^{2})^{-1} (+8 m_{H_d}^2 v_{d}^{2} +g_{1}^{2} v_{d}^{4} +g_{Y B}^{2} v_{d}^{4} +g_{2}^{2} v_{d}^{4} -8m_{H_u}^2 v_{u}^{2} - g_{1}^{2} v_{u}^{4} +- g_{Y B}^{2} v_{u}^{4} \nonumber \\ 
 &- g_{2}^{2} v_{u}^{4} +2 g_{Y B} g_{B} v_{d}^{2} v_{\eta}^{2} +2 g_{Y B} g_{B} v_{u}^{2} v_{\eta}^{2} +-2 g_{Y B} g_{B} v_{d}^{2} v_{\bar{\eta}}^{2} +-2 g_{Y B} g_{B} v_{u}^{2} v_{\bar{\eta}}^{2} ) \\
|\mu'|^2 =& -\frac{1}{4} (- v_{\bar{\eta}}^{2}  + v_{\eta}^{2})^{-1} (+4 m_{\eta}^2 v_{\eta}^{2} +g_{Y B} g_{B} v_{d}^{2} v_{\eta}^{2} +- g_{Y B} g_{B} v_{u}^{2} v_{\eta}^{2} +2 g_{B}^{2} v_{\eta}^{4} +-4 m_{\bar{\eta}}^2 v_{\bar{\eta}}^{2} +g_{Y B} g_{B} v_{d}^{2} v_{\bar{\eta}}^{2} \nonumber \\ 
&+- g_{Y B} g_{B} v_{u}^{2} v_{\bar{\eta}}^{2} +-2 g_{B}^{2} v_{\bar{\eta}}^{4} ) 
\end{align}
For a better understanding, we can make the approximation of vanishing kinetic mixing ($g_{YB}=0$): in this limit, as we will see below, the vector boson masses are given by $M_Z = \frac{1}{4}(g_1^2+g_2^2) v^2$ and $M_Z' = g_{B}^2 x^2$, with $v=\sqrt{v_d^2+v_u^2}$ and $x_=\sqrt{ v_{\eta}^{2} +  v_{\bar{\eta}}^{2}}$. In addition, we make the replacements $v_d \to v \sin\beta$, $v_u \to v \cos\beta$, $x_1 \to x \sin\beta'$, and $x_2 \to x \cos\beta'$ and express $\beta$, $\beta'$ by $\tan\beta$ ({\tt TB}) and $\tan\beta'$ ({\tt TBp}):
\begin{lstlisting}[style=mathematica]
SimplySolution=Simplify[solutionTad 
   /. {x1 -> x Sin[BetaP], x2 -> x Cos[BetaP], 
                     vd -> v Sin[Beta], vu -> v Cos[Beta]}  
   /. { Beta -> ArcTan[TB], BetaP -> ArcTan[TBp]}
   /. {g1BL->0, v -> 2/Sqrt[(g1^2 + g2^2)] MZ, -> MZp/gBL}  ]
\end{lstlisting}
We find quite simple expressions:
\begin{align}
\label{eq:absmu}
|\mu|^2 =&  \frac{1}{2 (-1 + \tan^{2}\beta)}  \Big(- (2 m_{H_d}^2  + M_Z^{2}) \tan^2\beta^{2}  + 2 m_{H_u}^2  + M_Z^{2}\Big) \\
\label{eq:absmup}
|\mu'|^2= & \frac{1}{2 (-1 + \tan^{2}\beta')}  \Big(- \Big(2 m_{\eta}^2  + M_{Z'}^{2}\Big)\tan^2\beta'  + 2 m_{\bar \eta}^2  + M_{Z'}^{2}\Big)
\end{align}
The first expression is just the one of the MSSM. Thus, any new contribution to $\mu$ comes only from gauge kinetic mixing. The equation for $|\mu'|$ looks very similar. However, the large ratio between $M_Z$ and $M_{Z'}$ gives a much larger constraint: for radiative symmetry breaking $\tan\beta'$ is usually restricted to be close to $1$ to minimize the negative contributions. We can check this by assuming  $\Delta m^2 = m_{\eta}^2-m_{\bar \eta}^2$
together with $M_{Z'} = 2.5$~TeV and $m_{\eta}^2=1~\text{TeV}^2$. $|\mu'|^2$ will then just be a function of $\tan\beta'$ and $\Delta m^2$. Using the {\tt ContourPlot} function of \Mathematica 
\begin{lstlisting}[style=mathematica]
ContourPlot[
  AbsMuP2 /. SimplySolution[[1]] 
    /. {mC12 -> mC22 - deltaM, MZp -> 2500} /. mC22 -> 10^6 
  {deltaM, 0, 10^6}, {TBp, 1, 1.5}, 
  ContourLabels -> True, 
  FrameLabel->{\[CapitalDelta] Superscript[m, 2],"tan(\[Beta]')"}] 
\end{lstlisting}
we get the plot shown in Fig.~\ref{fig:abmup2}. 
We see that the numbers for $|\mu'|$ quickly drop and the entire area with $\tan\beta' > 1.1$ is ruled out.

\subsubsection{Mass Matrices}  We turn now to the mass matrices.
First, I have to provide the proof that our approximation for the vector boson masses in the limit of vanishing gauge kinetic mixing to obtain eqs.~(\ref{eq:absmu})--(\ref{eq:absmup}) is correct. To do that we check the vector boson mass matrix. That's done via
\begin{lstlisting}[style=mathematica]
FullSimplify[ MassMatrix[VectorBoson] 
        /. {x1 -> x Sin[BetaP], x2 -> x Cos[BetaP], 
            vd -> v Sin[Beta], vu -> v Cos[Beta]]}]
\end{lstlisting}
where all three possibilities ({\tt VectorBoson=VP}, {\tt VectorBoson=VZ}, {\tt VectorBoson=VZP}) return the same mass matrix:
\begin{equation}
\left(
\begin{array}{ccc}
\frac{1}{4} g_{1}^{2} v^{2}  + g_{B Y}^{2} x^{2} & -\frac{1}{4} g_1 g_2 v^{2} & \frac{1}{4} g_1 g_{Y B} v^{2}  + g_{B} g_{B Y} x^{2} \\
-\frac{1}{4} g_1 g_2 v^{2} & \frac{1}{4} g_{2}^{2} v^{2} & -\frac{1}{4} g_{Y B} g_2 v^{2} \\
\frac{1}{4} g_1 g_{Y B} v^{2}  + g_{B} g_{B Y} x^{2} & -\frac{1}{4} g_{Y B} g_2 v^{2} & \frac{1}{4}  g_{Y B}^{2} v^{2}  + g_{B}^{2} x^{2} 
\end{array}\right)
\end{equation}
This matrix is block diagonal for $g_{BY}=g_{YB}=0$ with a upper $2\times 2$ matrix which is identical to the SM. Thus, we can see that gauge-kinetic mixing in this model is an import effect because it leads to $Z$--$Z'$ mixing already at tree-level. 

In the scalar sector the mass matrix for the CP odd scalars is printed   via 
\begin{lstlisting}[style=mathematica]
MassMatrix[Ah] 
\end{lstlisting}
and reads 
\begin{equation} 
m^2_{A^0} = \left( 
\begin{array}{cccc}
m_{\sigma_{d}\sigma_{d}} &{\Re\Big(B_{\mu}\Big)} &0 &0\\ 
{\Re\Big(B_{\mu}\Big)} &m_{\sigma_{u}\sigma_{u}} &0 &0\\ 
0 &0 &m_{\sigma_{\eta}\sigma_{\eta}} &{\Re\Big(B_{\eta}\Big)}\\ 
0 &0 &{\Re\Big(B_{\eta}\Big)} &m_{\sigma_{\bar{\eta}}\sigma_{\bar{\eta}}}\end{array} 
\right) +  \xi_{Z}m^2(Z) +  \xi_{{Z'}}m^2({Z'}) 
 \end{equation} 
with
\begin{align} 
m_{\sigma_{d}\sigma_{d}} &= \frac{1}{8} \Big(2 \Big(g_1 g_{B Y}  + g_{Y B} g_{B} \Big)\Big(- v_{\bar{\eta}}^{2}  + v_{\eta}^{2}\Big) + \Big(g_{1}^{2} + g_{Y B}^{2} + g_{2}^{2}\Big)\Big(- v_{u}^{2}  + v_{d}^{2}\Big)\Big) + m_{H_d}^2 + |\mu|^2\\ 
m_{\sigma_{u}\sigma_{u}} &= \frac{1}{8} \Big(2 \Big(g_1 g_{B Y}  + g_{Y B} g_{B} \Big)\Big(- v_{\eta}^{2}  + v_{\bar{\eta}}^{2}\Big) + \Big(g_{1}^{2} + g_{Y B}^{2} + g_{2}^{2}\Big)\Big(- v_{d}^{2}  + v_{u}^{2}\Big)\Big) + m_{H_u}^2 + |\mu|^2\\ 
m_{\sigma_{\eta}\sigma_{\eta}} &= \frac{1}{4} \Big(2 \Big(g_{B}^{2} + g_{B Y}^{2}\Big)\Big(- v_{\bar{\eta}}^{2}  + v_{\eta}^{2}\Big) + \Big(g_1 g_{B Y}  + g_{Y B} g_{B} \Big)\Big(- v_{u}^{2}  + v_{d}^{2}\Big)\Big) + m_{\eta}^2 + |{\mu_{\eta}}|^2\\ 
m_{\sigma_{\bar{\eta}}\sigma_{\bar{\eta}}} &= \frac{1}{4} \Big(-2 \Big(g_{B}^{2} + g_{B Y}^{2}\Big)\Big(- v_{\bar{\eta}}^{2}  + v_{\eta}^{2}\Big) + \Big(g_1 g_{B Y}  + g_{Y B} g_{B} \Big)\Big(- v_{d}^{2}  + v_{u}^{2}\Big)\Big) + m_{\bar{\eta}}^2 + |{\mu_{\eta}}|^2
\end{align} 
and gauge fixing contributions $m^2(Z) \xi_{Z}$ and  $m^2(Z') \xi_{Z'}$.  We see that the matrix is block-diagonal. That means that there is no mixing between the CP odd components of the Higgs doublets and bileptons. However, this is a statement which is only strictly true at tree-level. Therefore, we kept the mixing of both states in the model definition. One self-consistency check to see that the Goldstone degrees of freedom appear correctly is quickly done: this matrix should have two zero eigenvalues in Landau gauge. For this purpose and for simplicity we solve the tadpoles with respect to the soft-breaking scalar masses and plug the solution into the pseudo-scalar mass matrix. In addition, we are going to Landau gauge ({\tt  RXi[\_] -> 0})
\begin{lstlisting}[style=mathematica]
Solve[TadEquations, {mHd2, mHu2, mC12,mC22}]; 
MassMatrix[Ah] /. % /. RXi[_] -> 0;
Eigenvalues[%] //. a_ conj[x_] + a_ x_ -> 2 a  Re[x]
\end{lstlisting}
In the last line, we replaced $a x^* + a x \to 2 a \Re(x)$. The outcome is this handy list of eigenvalues:
\begin{equation}
\left\{0,\hspace{0.5cm} 0, \hspace{0.5cm}
\frac{1}{2 v_{\eta} v_{\bar{\eta}}}  \Big(2 v_{\bar{\eta}}^{2} {\Re\Big(B_{\eta}\Big)}  + 2 v_{\eta}^{2} {\Re\Big(B_{\eta}\Big)} \Big),\hspace{0.5cm}
\frac{1}{2  v_{d} v_{u}} \Big(2 v_{d}^{2} {\Re\Big(B_{\mu}\Big)}  + 2 v_{u}^{2} {\Re\Big(B_{\mu}\Big)} \Big)\right\}
\end{equation}
So, we find the expected two massless modes. The physical pseudo-scalars have at tree-level the same expressions as in the MSSM by replacing the corresponding VEVs and $B$-terms for the $B-L$ sector.  \\

\begin{figure}[hbt]
\includegraphics[width=0.3\linewidth]{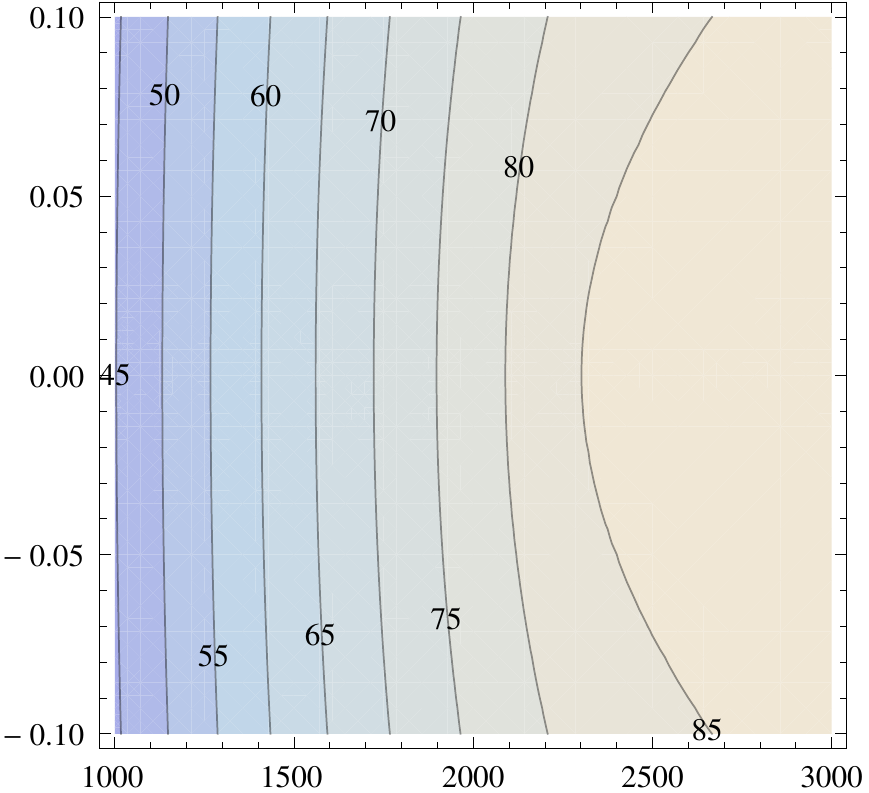} \hfill
\includegraphics[width=0.3\linewidth]{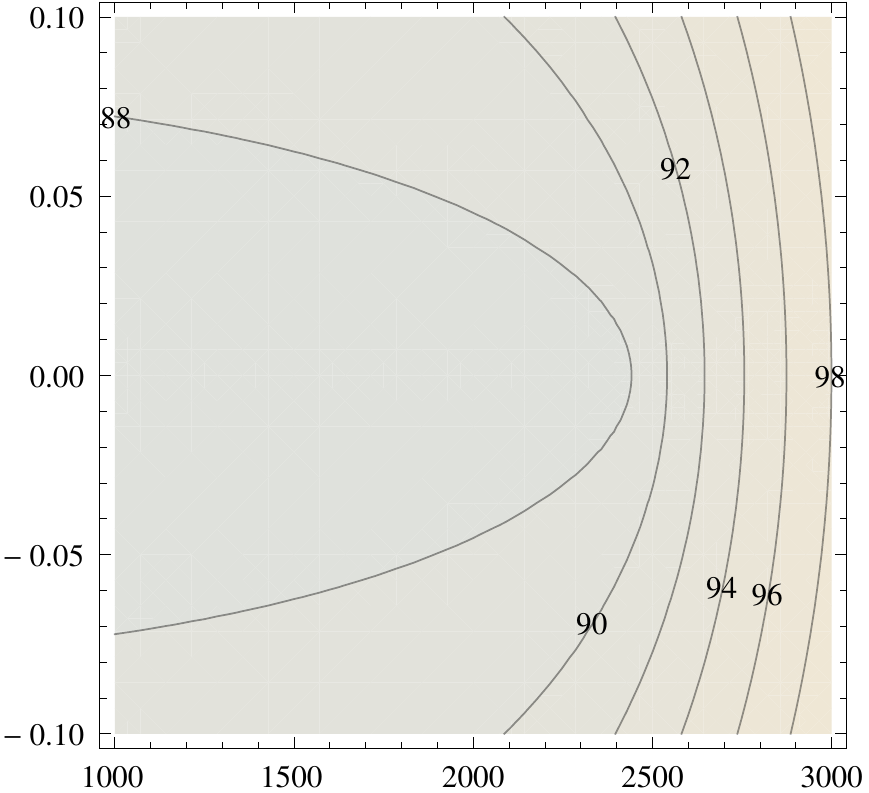} \hfill
\includegraphics[width=0.3\linewidth]{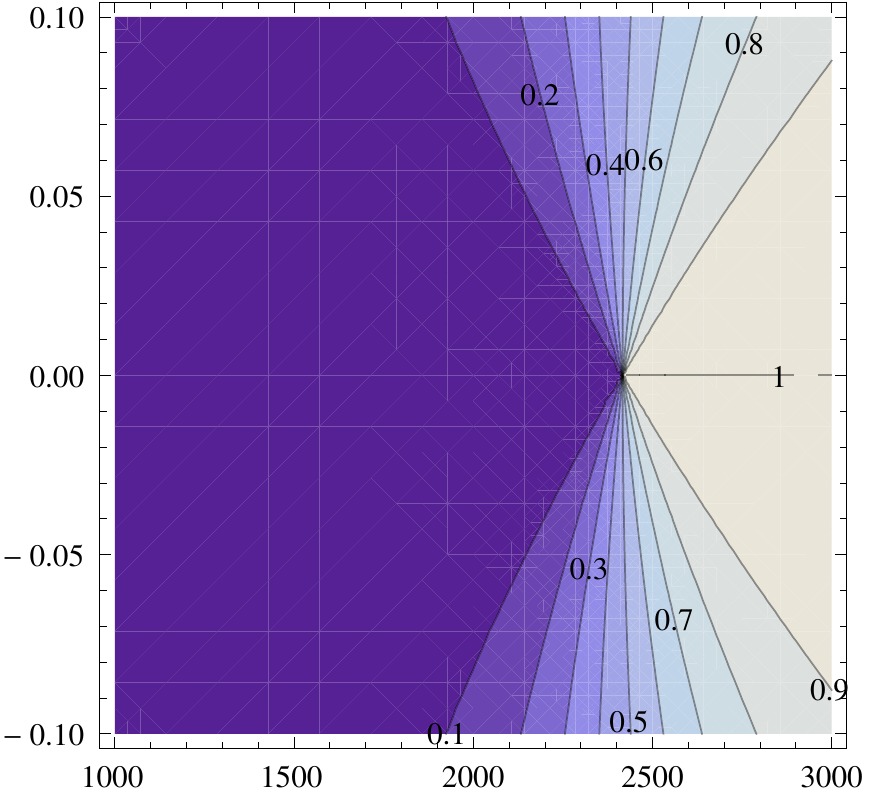} 
\caption{The two lightest eigenvalues of the scalar mass matrix at tree-level (left and middle) and the bilepton fraction of the lightest eigenstates (right) in the $(x,g_{YB})$ plane. }
\label{fig:mHHkm}
\end{figure}

The scalar mass matrix is given by
\begin{lstlisting}[style=mathematica]
MassMatrix[hh] 
\end{lstlisting}
and is a bit more complicated. We parametrize it by 
\begin{equation}
 \left( 
\begin{array}{cccc}
m_{\phi_{d}\phi_{d}} &m_{\phi_{u}\phi_{d}} &m_{\phi_{\eta}\phi_{d}} &m_{\phi_{\bar{\eta}}\phi_{d}}\\ 
m_{\phi_{d}\phi_{u}} &m_{\phi_{u}\phi_{u}} &m_{\phi_{\eta}\phi_{u}} &m_{\phi_{\bar{\eta}}\phi_{u}}\\ 
m_{\phi_{d}\phi_{\eta}} &m_{\phi_{u}\phi_{\eta}} &m_{\phi_{\eta}\phi_{\eta}} &m_{\phi_{\bar{\eta}}\phi_{\eta}}\\ 
m_{\phi_{d}\phi_{\bar{\eta}}} &m_{\phi_{u}\phi_{\bar{\eta}}} &m_{\phi_{\eta}\phi_{\bar{\eta}}} &m_{\phi_{\bar{\eta}}\phi_{\bar{\eta}}}\end{array} 
\right) 
 \end{equation} 
with
\begin{align} 
m_{\phi_{d}\phi_{d}} &= \frac{1}{8} \Big(2 \Big(g_1 g_{B Y}  + g_{Y B} g_{B} \Big)\Big(- v_{\bar{\eta}}^{2}  + v_{\eta}^{2}\Big) + \Big(g_{1}^{2} + g_{Y B}^{2} + g_{2}^{2}\Big)\Big(3 v_{d}^{2}  - v_{u}^{2} \Big)\Big) + m_{H_d}^2 + |\mu|^2\\ 
m_{\phi_{d}\phi_{u}} &= -\frac{1}{4} \Big(g_{1}^{2} + g_{Y B}^{2} + g_{2}^{2}\Big)v_d v_u  - {\Re\Big(B_{\mu}\Big)} \\ 
m_{\phi_{u}\phi_{u}} &= \frac{1}{8} \Big(2 g_1 g_{B Y} \Big(v_{\bar{\eta}}^{2}- v_{\eta}^{2} \Big) + 2 g_{Y B} g_{B} \Big(v_{\bar{\eta}}^{2}- v_{\eta}^{2}  \Big) + \Big(- g_{1}^{2}  - g_{2}^{2}  - g_{Y B}^{2} \Big)\Big(v_{d}^{2}-3 v_{u}^{2} \Big)\Big) + m_{H_u}^2 + |\mu|^2\\ 
m_{\phi_{d}\phi_{\eta}} &= \frac{1}{2} \Big(g_1 g_{B Y}  + g_{Y B} g_{B} \Big)v_d v_{\eta} \\ 
m_{\phi_{u}\phi_{\eta}} &= -\frac{1}{2} \Big(g_1 g_{B Y}  + g_{Y B} g_{B} \Big)v_u v_{\eta} \\ 
m_{\phi_{\eta}\phi_{\eta}} &= \frac{1}{4} \Big(2 \Big(g_{B}^{2} + g_{B Y}^{2}\Big)\Big(3 v_{\eta}^{2}  - v_{\bar{\eta}}^{2} \Big) + \Big(g_1 g_{B Y}  + g_{Y B} g_{B} \Big)\Big(- v_{u}^{2}  + v_{d}^{2}\Big)\Big) + m_{\eta}^2 + |{\mu_{\eta}}|^2\\ 
m_{\phi_{d}\phi_{\bar{\eta}}} &= -\frac{1}{2} \Big(g_1 g_{B Y}  + g_{Y B} g_{B} \Big)v_d v_{\bar{\eta}} \\ 
m_{\phi_{u}\phi_{\bar{\eta}}} &= \frac{1}{2} \Big(g_1 g_{B Y}  + g_{Y B} g_{B} \Big)v_u v_{\bar{\eta}} \\ 
m_{\phi_{\eta}\phi_{\bar{\eta}}} &= - \Big(g_{B}^{2} + g_{B Y}^{2}\Big)v_{\eta} v_{\bar{\eta}}  - {\Re\Big(B_{\eta}\Big)} \\ 
m_{\phi_{\bar{\eta}}\phi_{\bar{\eta}}} &= \frac{1}{4} \Big(-2 \Big(g_{B}^{2} + g_{B Y}^{2}\Big)\Big(-3 v_{\bar{\eta}}^{2}  + v_{\eta}^{2}\Big) + \Big(g_1 g_{B Y}  + g_{Y B} g_{B} \Big)\Big(- v_{d}^{2}  + v_{u}^{2}\Big)\Big) + m_{\bar{\eta}}^2 + |{\mu_{\eta}}|^2
\end{align}  
One can see that this matrix is in general not block diagonal, i.e. there is already a mixing between the MSSM doublets and the bileptons at tree-level. 
However, all terms $m_{\phi_i \phi_j}$ with $i=d,u$ and $j=\eta,\bar \eta$ are proportional to $\Big(g_1 g_{B Y}  + g_{Y B} g_{B} \Big)$, i.e. this mixing is only visible if gauge kinetic mixing is taken into account. That's another reason why gauge-kinetic mixing is in general a very important effect in this model. We can also try to get an estimate of the size of this mixing. For this purpose, we plug the solution of the tadpole equations in the Higgs mass matrix and fix some numerical values: $g_1=0.36$, $g_2=0.63$, $g_{BL}=0.5$, $\mu=\mu'=1$~TeV, $B_\mu=B_\mu'=1~\text{TeV}^2$ \footnote{While there is not much space to play with the gauge couplings, the other parameters could be chosen differently, of course.}.
\begin{lstlisting}[style=mathematica]
numMhh = MassMatrix[hh] /. solutionTad2 /. gBL1->0 /. conj[x_]->x
   /.{x1 -> x Sin[BetaP], x2 -> x Cos[BetaP], vd -> v Sin[Beta], vu -> v Cos[Beta]}
   /. {BetaP -> ArcTan[1.1],  Beta -> ArcTan[10.], v -> 246}
   /. {g1 -> 0.36, g2 -> 0.63, gBL -> 0.5}
   /. { B[\[Mu]] -> 10^6, \[Mu] -> 10^3,B[MuP] -> 10^6, MuP -> 10^3} 
\end{lstlisting}
{\tt numMhh} is now the mass matrix which just depends on the bilepton VEV $x$ and the off-diagonal gauge coupling. We can write a simple function which diagonalizes this mass matrix for given values of these two parameters. Furthermore, this function extracts the two lightest masses as well as the bilepton admixture of the lightest eigenstates and returns these three values:
\begin{lstlisting}[style=mathematica]
FunctionHiggEigenvalues[xInput_, g1BLinput_] := Block[{eig},
   eig = Eigensystem[simpHH /. {x -> xInput, g1BL -> g1BLinput}];
   bileptonFrac = Drop[eig[[2, -1]], {3, 4}];
   bileptonFrac=bileptonFrac.bileptonFrac;
   Return[{Sqrt[eig[[1, -1]]], Sqrt[eig[[1, -2]]], bileptonFrac}];
   ]; 
\end{lstlisting}
We can use this new function with the {\tt ContourPlot} command of \Mathematica
\begin{lstlisting}[style=mathematica]
ContourPlot[FunctionHiggEigenvalues[x, g1BL][[NUMBER]], 
   {x, 1000, 3000}, {g1BL, -0.1,0.1}, 
   ContourLabels -> True, ImageSize -> 250]
\end{lstlisting}
where {\tt NUMBER}={\tt 1},{\tt 2},{\tt 3} should be used to get all three plots shown in Fig.~\ref{fig:mHHkm}. We see at these plots that the mixing is especially large when both states are close in mass and can be easily $O(10\%)$ and more. 

We can now turn the sfermion sector. The matrices there are usually quite lengthy. I just want to pick out one important effect which we see in the diagonal entries of the charged sleptons for instance. The entry corresponding to $\tilde \tau_L$ in the $6\times 6$ mass matrix of the sfermions ({\tt Se}) is shown via
\begin{lstlisting}[style=mathematica]
MassMatrix[Se][[3, 3]] 
\end{lstlisting}
We can re-write the terms a bit by first assuming that only third generation Yukawas a non-zero. For this purpose, we expand the sum and put all entries of $Y_e$ to zero but the (3,3) one. In addition, we make the assumption that gauge-kinetic mixing vanishes for simplicity and we can use the  usual replacements for the VEVs
\begin{lstlisting}[style=mathematica]
  % /. sum[a_, b_, c_, d_] :> Sum[d, {a, b, c}] 
    /. Ye[3, 3] -> Ytau /. Ye[a__] -> 0 
    /. {x1 -> x Sin[BetaP], x2 -> x Cos[BetaP], vd -> v Sin[Beta], vu -> v Cos[Beta]}
    /. {v -> 2/Sqrt[(g1^2 + g2^2)] MZ, x -> MZp/gBL, gBL1 -> 0, g1BL -> 0}
    /. conj[x_] -> x
\end{lstlisting}
The entries read then 
\begin{equation}
-\frac{1}{4 (g_{1}^{2} + g_{2}^{2})} (2 (g_1^2- g^2_2)M_Z^{2} \cos2 \beta    + -8 M_Z^{2} Y_\tau^{2} (\sin\beta)^{2}  + (g_{1}^{2} + g_{2}^{2})(-4 m_{l,{3 3}}^{2}  + M_{Z'}^{2} \cos2 {\beta'}   )) 
\end{equation}
The important point is the appearance of $M_Z'$ which gives large negative contributions because the lower limit on this mass is about 2.5~TeV. Thus, $\tan\beta'$ must be close to 1 to minimize this term and to prevent tachyons.\\

The last mass matrix we want to check is the one of the neutralinos. This mass matrix is returned by
\begin{lstlisting}[style=mathematica]
MassMatrix[Chi]
\end{lstlisting}
and reads
\begin{equation} 
 \left( 
\begin{array}{ccccccc}
M_1 &0 &-\frac{1}{2} g_1 v_d  &\frac{1}{2} g_1 v_u  &{M}_{B B'} &- g_{B Y} v_{\eta}  &g_{B Y} v_{\bar{\eta}} \\ 
0 &M_2 &\frac{1}{2} g_2 v_d  &-\frac{1}{2} g_2 v_u  &0 &0 &0\\ 
-\frac{1}{2} g_1 v_d  &\frac{1}{2} g_2 v_d  &0 &- \mu  &-\frac{1}{2} g_{Y B} v_d  &0 &0\\ 
\frac{1}{2} g_1 v_u  &-\frac{1}{2} g_2 v_u  &- \mu  &0 &\frac{1}{2} g_{Y B} v_u  &0 &0\\ 
{M}_{B B'} &0 &-\frac{1}{2} g_{Y B} v_d  &\frac{1}{2} g_{Y B} v_u  &{M}_{BL} &- g_{B} v_{\eta}  &g_{B} v_{\bar{\eta}} \\ 
- g_{B Y} v_{\eta}  &0 &0 &0 &- g_{B} v_{\eta}  &0 &- {\mu_{\eta}} \\ 
g_{B Y} v_{\bar{\eta}}  &0 &0 &0 &g_{B} v_{\bar{\eta}}  &- {\mu_{\eta}}  &0\end{array} 
\right) 
 \end{equation}   
The upper $4 \times 4$ block is the one known from the MSSM. The lower $3 \times 3$ block is the counterpart in $B-L$ sector of this model. Here, we make a similar observation as for the scalar Higgs mass matrix: both blocks are only coupled if gauge-kinetic mixing is taken into account. \\

In the same way all other mass matrices of the model can be checked with \SARAH and interesting observations can be made like the CP even and odd mass-splitting for the sneutrinos.

\subsubsection{Vertices} 
\paragraph*{Single vertices} We continue with vertices and show how they can be handled in \SARAH. Let's assume one is interested in the coupling between two up-quarks and the neutral CP even scalars:
\begin{lstlisting}[style=mathematica]
Vertex[{bar[Fd], Fd, hh}] 
\end{lstlisting}
gives
\begin{align} 
 &-i \frac{1}{\sqrt{2}} \delta_{\alpha \beta} \sum_{b=1}^{3}U^{u,*}_{L,{j b}} \sum_{a=1}^{3}U^{u,*}_{R,{i a}} Y_{u,{a b}}   Z_{{k 2}}^{H} \Big(\frac{1-\gamma_5}{2}\Big)\\ 
 & + \,-i \frac{1}{\sqrt{2}} \delta_{\alpha \beta} \sum_{b=1}^{3}\sum_{a=1}^{3}Y^*_{u,{a b}} U_{R,{j a}}^{u}  U_{L,{i b}}^{u}  Z_{{k 2}}^{H} \Big(\frac{1+\gamma_5}{2}\Big)
\end{align} 
$U_L$, $U_R$ and $Z^H$ are rotation matrices as shown in Tab.~\ref{tab:parameters}, $i,j,k$ are the generation indices of the external states and $\alpha$, $\beta$ are the colour indices of the quarks. One sees that only the projection on the second gauge eigenstate ($Z_{{k2}}$) contributes which corresponds to the up-Higgs. Thus, this is the same vertex as in the MSSM and the new $B-L$ sector does not contribute here. That's different if one considers for instance the neutrino-scalar vertex:
\begin{lstlisting}[style=mathematica]
Vertex[{Fv,Fv, hh}] 
\end{lstlisting}
returns
\begin{align} 
 &-i \frac{1}{\sqrt{2}} \Big(\sum_{b=1}^{3}U^{V,*}_{j b} \sum_{a=1}^{3}U^{V,*}_{i 3 + a} Y_{\nu,{a b}}   Z_{{k 2}}^{H} +\sum_{b=1}^{3}U^{V,*}_{i b} \sum_{a=1}^{3}U^{V,*}_{j 3 + a} Y_{\nu,{a b}}   Z_{{k 2}}^{H} \nonumber \\ 
 &+\Big(\sum_{b=1}^{3}U^{V,*}_{j 3 + b} \sum_{a=1}^{3}U^{V,*}_{i 3 + a} Y_{x,{a b}}   + \sum_{b=1}^{3}U^{V,*}_{i 3 + b} \sum_{a=1}^{3}U^{V,*}_{j 3 + a} Y_{x,{a b}}  \Big)Z_{{k 3}}^{H} \Big)\Big(\frac{1-\gamma_5}{2}\Big)\\ 
  & + \,-i \frac{1}{\sqrt{2}} \Big(\sum_{b=1}^{3}\sum_{a=1}^{3}Y^*_{\nu,{a b}} U_{{j 3 + a}}^{V}  U_{{i b}}^{V}  Z_{{k 2}}^{H} +\sum_{b=1}^{3}\sum_{a=1}^{3}Y^*_{\nu,{a b}} U_{{i 3 + a}}^{V}  U_{{j b}}^{V}  Z_{{k 2}}^{H} \nonumber \\ 
 &+\Big(\sum_{b=1}^{3}\sum_{a=1}^{3}Y^*_{x,{a b}} U_{{j 3 + a}}^{V}  U_{{i 3 + b}}^{V}  + \sum_{b=1}^{3}\sum_{a=1}^{3}Y^*_{x,{a b}} U_{{i 3 + a}}^{V}  U_{{j 3 + b}}^{V} \Big)Z_{{k 3}}^{H} \Big)\Big(\frac{1+\gamma_5}{2}\Big)
 \end{align} 
 We find here also projections on the third gauge eigenstate which comes from the $\hat \nu \hat \eta \hat \nu$-term in the superpotential. In general, many vertices get modified with respect to the MSSM and a discussion of all effects is far beyond the scope of this manuscript. I just want to pick out one more vertex: the electron--$Z$ interaction:
\begin{lstlisting}[style=mathematica]
Vertex[{bar[Fe],Fe, VZ}] 
\end{lstlisting}
We find that this vertex receives important modification due to the $Z$--$Z'$ mixing:
\begin{align} 
 &\frac{i}{2} \delta_{i j} \Big(- \Big(g_1 + g_{B Y}\Big)\cos{\Theta'}_W  \sin\Theta_W   + g_2 \cos\Theta_W  \cos{\Theta'}_W   + \Big(g_{Y B} + g_{B}\Big)\sin{\Theta'}_W  \Big)\Big(\gamma_{\mu}\cdot\frac{1-\gamma_5}{2}\Big)\\ 
  & + \,-\frac{i}{2} \delta_{i j} \Big(\Big(2 g_1  + g_{B Y}\Big)\cos{\Theta'}_W  \sin\Theta_W   - \Big(2 g_{Y B}  + g_{B}\Big)\sin{\Theta'}_W  \Big)\Big(\gamma_{\mu}\cdot\frac{1+\gamma_5}{2}\Big)
\end{align} 
Working in the triangle basis, the contributions from $g_{BY}$ vanish. However, the coupling compared to SM expectation gets still modified by the presence of $\sin\Theta'_W$. This gives strong constraints on the angle $\Theta'_W$.  That's of course the case for any $Z$-interaction in this model. \\

\paragraph*{All vertices}  It has been already be mentioned that it is also possible to calculate all vertices at once. The command to do this is
\begin{lstlisting}[style=mathematica]
MakeVertexList[EWSB]
\end{lstlisting}
This creates lists
\begin{lstlisting}[style=mathematica]
SA`VertexList[$TYPE] 
\end{lstlisting}
for the different generic types of vertices ({\tt \$TYPE}={\tt SSS}, {\tt SSSS}, {\tt SSV}, {\tt SSVV}, {\tt SVV}, {\tt FFS}, {\tt FFV}, {\tt VVV}, {\tt VVVV}, {\tt GGS}, {\tt GGV}). One can also play a bit with these lists. For instance, to get all fermion interactions with the $Z'$ one can use the {\tt Select} command of \Mathematica 
\begin{lstlisting}[style=mathematica]
Select[SA`VertexList[FFV], FreeQ[#, VZp] == False &] 
\end{lstlisting}
Similarly, all scalar interactions can be extracted where the off-diagonal gauge couplings show up: 
\begin{lstlisting}[style=mathematica]
Select[SA`VertexList[SSSS], 
      (FreeQ[#, g1BL] == False) &]; 
\end{lstlisting}
If one compares the length of this list with the length of all four scalar interactions in general
\begin{lstlisting}[style=mathematica]
Length[%] - Length[SA`VertexList[SSSS]]
\end{lstlisting}
it turns out that this is the case for any vertex. That's of course not surprising because of the $D$-term contributions, but it underlines the importance of this effect again.

\subsection{Understanding the RGEs}
\subsubsection{Analytical results}
The full two-loop RGEs of the \BLSSM are calculated just via
\begin{lstlisting}[style=mathematica]
CalcRGEs[];
\end{lstlisting}
The options for {\tt CalcRGEs} are
\begin{itemize}
\item \verb"TwoLoop": defines, if two-loop RGEs should be calculated. This is done by default.
\item \verb"ReadLists": defines, if the results from previous calculations should be read instead of calculating the RGEs again. 
\item \verb"VariableGenerations": defines, if the generations of some particles should be treated as free parameters. The RGEs are then expressed in terms of \verb"NumberGenertions[X]", where \verb"X" is the name of the superfield.
\item \verb"NoMatrixMultiplication": can be set if the RGEs should not be expressed in terms of matrix multiplication but by using sums over indices. 
\item \verb"IgnoreAt2Loop": can be used to define parameters which should be put to zero in the two-loop calculation.
\item \verb"WriteFunctionsToRun": defines, if a file should be written to evaluate the RGEs numerically in \Mathematica. This is done by default and we are going to make use of it. 
\end{itemize}
When the calculation is finished, the full two-loop RGEs are saved in different arrays:
\begin{itemize}
 \item  \verb"Gij": Anomalous dimensions of all chiral superfields
 \item \verb"BetaYijk": Trilinear superpotential parameters ($Y_d$, $Y_e$, $Y_u$, $Y_x$, $Y_\nu$)
 \item \verb"BetaMuij": Bilinear superpotential parameters ($\mu$, $\mu'$)
 \item \verb"BetaTijk": Trilinear soft-breaking parameters ($T_d$, $T_e$, $T_u$, $T_x$, $T_\nu$)
 \item \verb"BetaBij": Bilinear soft-breaking parameters ($B_\mu$, $B_\mu'$)
 \item \verb"Betam2ij": Scalar squared masses ($m_{\tilde{q}}^2$, $m_{\tilde{d}}^2$, $m_{\tilde{u}}^2$, $m_{\tilde{e}}^2$, $m_{\tilde{l}}^2$, $m_{H_d}^2$, $m_{H_u}^2$, $m_{\eta}^2$, $m_{\bar{\eta}}^2$)
 \item \verb"BetaMi": Gaugino masses ($M_1$, $M_2$, $M_3$, $M_B$, $M_{BB'}$)
 \item \verb"BetaGauge": Gauge couplings ($g_1$, $g_2$, $g_3$, $g_{B}$, $g_{BY}$, $g_{YB}$)
 \item \verb"BetaVEVs": VEVs ($v_d$, $v_u$, $v_\eta$, $v_{\bar \eta}$)
\end{itemize}
\verb"BetaQijkl", \verb"BetaWijkl", \verb"BetaLi", \verb"BetaSLi", \verb"BetaDGi" and \verb"BetaFIi" are empty in this model. All lists are three dimensional arrays where each entry gives (i) the parameter, (ii) the one-loop $\beta$-function (up to a factor $1/16\pi^2$), and (iii) the two-loop $\beta$-function  (up to a factor $1/(16\pi^2)^2$). To check the order in which the RGEs for the trilinear superpotential parameters are given, one can use
\begin{lstlisting}[style=mathematica]
Transpose[BetaYijk][[1]] 
\end{lstlisting}
what returns
\begin{lstlisting}[style=mathematica]
{Yd[i1, i2], Ye[i1, i2], Yu[i1, i2], Yn[i1, i2], Yv[i1, i2]} 
\end{lstlisting}
Thus, if we want to see the one-loop RGE for the electron Yukawa coupling we have to check 
\begin{lstlisting}[style=mathematica]
BetaYijk[[2,2]] 
\end{lstlisting}
and get 
\begin{align}
&+3 {Y_e  Y_{e}^{\dagger}  Y_e} +{Y_e  Y_{\nu}^{\dagger}  Y_\nu}-\frac{1}{10} Y_e \Big(3 \Big(3 \sqrt{10} g_1 g_{B Y}  + 3 \sqrt{10} g_{Y B} g_{B} \nonumber \\ 
 & + 5 \Big(2 g_{2}^{2}  + g_{B}^{2} + g_{B Y}^{2}\Big) + 6 g_{1}^{2}  + 6 g_{Y B}^{2} \Big)-30 \mbox{Tr}\Big({Y_d  Y_{d}^{\dagger}}\Big)-10 \mbox{Tr}\Big({Y_e  Y_{e}^{\dagger}}\Big) \Big) 
\end{align}
One sees that gauge kinetic mixing is also taken here into account. In the limit $g_{B}=g_{BY}=g_{YB}=0$ this reproduces the well known MSSM result. Maybe, more interesting are new features in the gauge sector. To get just all one-loop RGEs at once, we can execute 
\begin{lstlisting}[style=mathematica]
BetaGauge /. {a_, b_, c_} -> {a, b}
\end{lstlisting}
That returns
\begin{align} 
\beta_{g_1}^{(1)} & =  
\frac{3}{5} \Big(11 g_{1}^{3}  + 4 \sqrt{10} g_{1}^{2} g_{B Y}  + g_1 \Big(11 g_{Y B}^{2}  + 15 g_{B Y}^{2}  + 2 \sqrt{10} g_{Y B} g_{B} \Big) + g_{Y B} \Big(15 g_{B}  + 2 \sqrt{10} g_{Y B} \Big)g_{B Y} \Big)\\ 
\beta_{g_2}^{(1)} & =  
g_{2}^{3}\\ 
\beta_{g_3}^{(1)} & =  
-3 g_{3}^{3} \\ 
\beta_{g_{B}}^{(1)} & =  
\frac{3}{5} \Big(11 g_{Y B}^{2} g_{B} +4 \sqrt{10} g_{Y B} g_{B}^{2} +15 g_{B}^{3} +11 g_1 g_{Y B} g_{B Y} +2 \sqrt{10} g_1 g_{B} g_{B Y} +2 \sqrt{10} g_{Y B} g_{B Y}^{2} \nonumber \\ 
 &+15 g_{B} g_{B Y}^{2} \Big)\\ 
\beta_{g_{Y B}}^{(1)} & =  
\frac{3}{5} \Big(g_1 \Big(15 g_{B}  + 2 \sqrt{10} g_{Y B} \Big)g_{B Y}  + g_{1}^{2} \Big(11 g_{Y B}  + 2 \sqrt{10} g_{B} \Big) + g_{Y B} \Big(11 g_{Y B}^{2}  + 15 g_{B}^{2}  + 4 \sqrt{10} g_{Y B} g_{B} \Big)\Big)\\ 
\beta_{g_{B Y}}^{(1)} & =  
\frac{3}{5} \Big(11 g_{1}^{2} g_{B Y}  + g_1 \Big(11 g_{Y B} g_{B}  + 2 \sqrt{10} \Big(2 g_{B Y}^{2}  + g_{B}^{2}\Big)\Big) + g_{B Y} \Big(15 \Big(g_{B}^{2} + g_{B Y}^{2}\Big) + 2 \sqrt{10} g_{Y B} g_{B} \Big)\Big)
\end{align} 
The expressions for $g_2$ and $g_3$ are just the MSSM results but $g_1$ gets modified by gauge-kinetic mixing. Solving these equations analytically is no longer possible but we are going to study them numerically. \\

Finally, we can also check analytical expressions for soft-breaking terms. We will do this at the example of the bilepton masses which are given in the last two entries of {\tt Betam2ij}. Even more interesting is the difference between both $\beta$-functions: we see from eq.~(\ref{eq:absmup}) that a large mass splitting between both masses is needed to get radiative symmetry breaking. Thus, starting with the same values at the GUT scale, the differences in the $\beta$-functions are crucial in order to break $B-L$ or not. To see the difference, we can use 
\begin{lstlisting}[style=mathematica]
Betam2ij[[-1, 2]] - Betam2ij[[-2, 2]] 
\end{lstlisting}
and find
\begin{equation}
2 \sqrt{6} g_{B} \sigma_{1,4}  + 2 \sqrt{6} g_{B Y} \sigma_{1,1}  -4 \mbox{Tr}\Big({T_x^*  T_x}\Big)  -4 m_{\eta}^2 \mbox{Tr}\Big({Y_x  Y_x^*}\Big)  -8 \mbox{Tr}\Big({m_{\nu}^2  Y_x  Y_x^*}\Big)
\end{equation}
$\sigma_{1,4}$ and $\sigma_{1,1}$ are abbreviations for often appearing traces over scalar masses. These can be found in {\tt TraceAbbr}:
\begin{align} 
\sigma_{1,1} & = \frac{1}{20} \Big(4 \sqrt{15} g_1 \Big(-2 \mbox{Tr}\Big({m_u^2}\Big)  - \mbox{Tr}\Big({m_l^2}\Big)  - m_{H_d}^2  + m_{H_u}^2 + \
\mbox{Tr}\Big({m_d^2}\Big) + \mbox{Tr}\Big({m_e^2}\Big) + \mbox{Tr}\Big({m_q^2}\Big)\Big)\nonumber \\ 
 &-5 \sqrt{6} g_{B Y} \Big(-2 m_{\bar{\eta}}^2  + 2 \mbox{Tr}\Big({m_l^2}\Big)  -2 \mbox{Tr}\Big({m_q^2}\Big)  + 2 m_{\eta}^2  - \mbox{Tr}\Big({m_e^2}\Big)  - \mbox{Tr}\Big({m_{\nu}^2}\Big)  + \mbox{Tr}\Big({m_d^2}\Big) + \mbox{Tr}\Big({m_u^2}\Big)\Big)\Big)\\
\sigma_{1,4} & = \frac{1}{20} \Big(4 \sqrt{15} g_{Y B} \Big(-2 \mbox{Tr}\Big({m_u^2}\Big)  - \mbox{Tr}\Big({m_l^2}\Big)  - m_{H_d}^2  + m_{H_u}^2 + \mbox{Tr}\Big({m_d^2}\Big) + \mbox{Tr}\Big({m_e^2}\Big) + \mbox{Tr}\Big({m_q^2}\Big)\Big)\nonumber \\ 
 &-5 \sqrt{6} g_{B} \Big(-2 m_{\bar{\eta}}^2  + 2 \mbox{Tr}\Big({m_l^2}\Big)  -2 \mbox{Tr}\Big({m_q^2}\Big)  + 2 m_{\eta}^2  - \mbox{Tr}\Big({m_e^2}\Big)  - \mbox{Tr}\Big({m_{\nu}^2}\Big)  + \mbox{Tr}\Big({m_d^2}\Big) + \mbox{Tr}\Big({m_u^2}\Big)\Big)\Big)
\end{align}
If one starts with a model in which all scalars unify as we have in mind according to sec.~\ref{sec:CBLSSM}, one gets $\sigma_{1,1}=\sigma_{1,4}=0$. This is a RGE invariant and does always hold. Thus, the only possibility to get a large mass splitting are large values for $Y_x$, $T_x$ and soft-masses $m_{\nu}^2$, $m_\eta^2$. We will come back to this when we study the model with \SPheno.

\subsubsection{Numerical results}
\label{sec:SolveRGEs}
The RGEs for the gauge coupling demand a closer look. However, the analytical expressions at one-loop are already a bit complicated and it is hard to learn something from them. Before turning to the full numerical analysis of the entire set of RGEs with \SPheno there is the possibility to study the RGEs in \Mathematica first: {\tt CalcRGEs} creates a file which contains all $\beta$-functions in a format which can be used with {\tt NDSolve} in \Mathematica to solve the RGEs. This file is loaded via
\begin{lstlisting}[style=mathematica]
<< "$PATH/SARAH/Output/B-L-SSM/RGEs/RunRGEs.m" 
\end{lstlisting}
In addition, also a function is provided to perform the RGE running:
\begin{lstlisting}[style=mathematica]
RunRGEs[values, start, finish, Options] 
\end{lstlisting}
The input is:
\begin{itemize}
 \item {\tt values}: all non-zero values for parameters at the scale where the running starts
 \item {\tt start}: logarithm of the scale where the running starts
 \item {\tt finish}: logarithm of the scale where the running should end
 \item {\tt Options}: optionally two-loop contributions can be turned off ({\tt TwoLoop->False})
\end{itemize}
Since the running at one-loop for $g_1$, $g_2$ and $g_3$ is the same as in the MSSM in the limit of vanishing gauge kinetic mixing, we should find the same unification at about $2\cdot 10^{16}$~GeV. This can be tested via 
\begin{lstlisting}[style=mathematica]
solutionMSSM = 
  RunRGEs[{g1->0.46, g2->0.64, g3->1.09}, 3, 17, TwoLoop->False];
Plot[{g1[t], g2[t], g3[t]} /. solutionMSSM, {t, 3, 17}, 
   Frame->True, Axes->False, 
   FrameLabel->{"log[Q/GeV]",Subscript["g", "i"]}] 
\end{lstlisting}
Here, I used for the SM gauge couplings as input the $\overline{\text{DR}}$ values which are found when including thresholds discussed in sec.~\ref{sec:thresholds} with SUSY states of about 1~TeV. The plot which is created via these two commands is shown in Fig.~\ref{fig:RGE} on the left.  We find a value of about 0.7 of the gauge couplings at the unification scale:
\begin{lstlisting}[style=mathematica,mathescape]
g2[Log[10, 2 10^16]] /. solutionMSSM

$\hookrightarrow$ {0.697832}
\end{lstlisting}

\begin{figure}[!tb]
\includegraphics[width=0.45\linewidth]{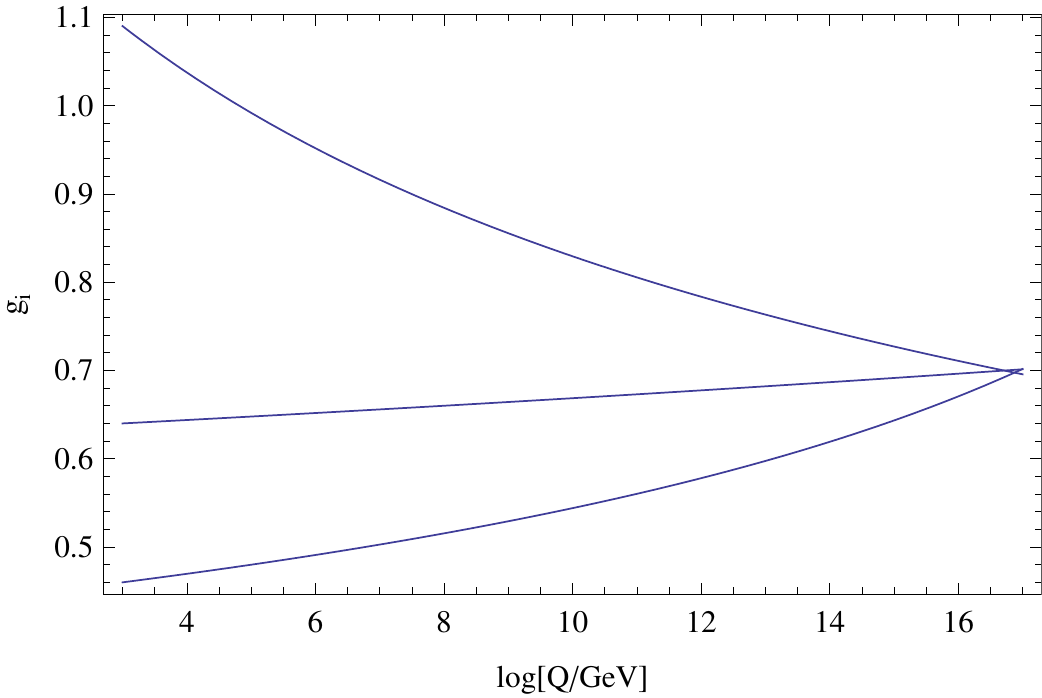}  \hfill
\includegraphics[width=0.45\linewidth]{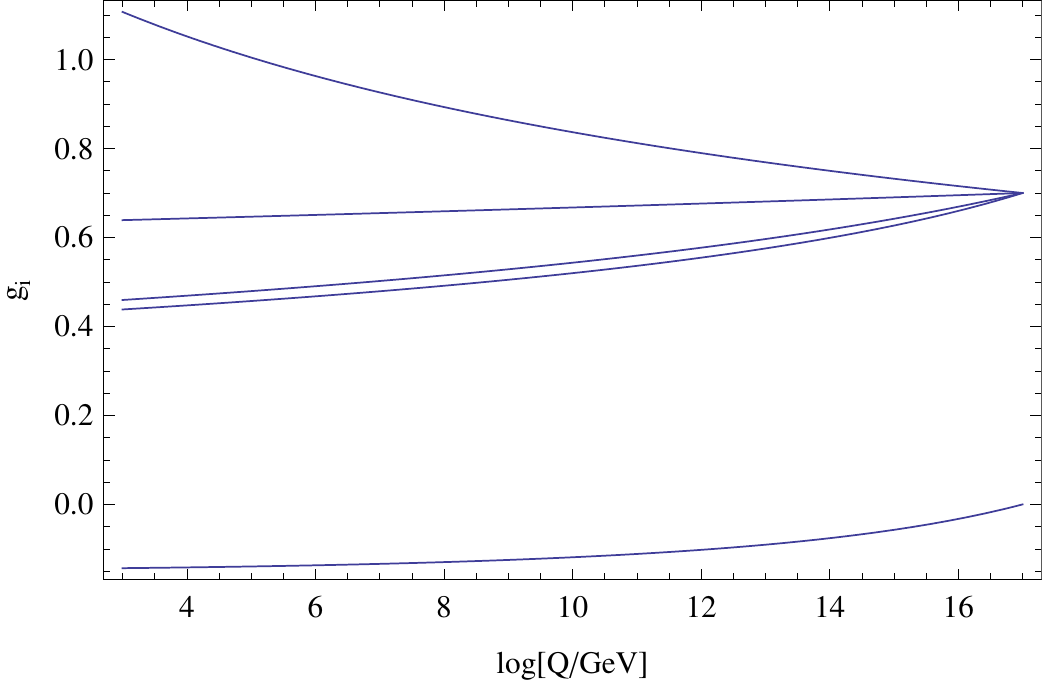}
\caption{On the left: running of the three gauge couplings in the MSSM limit; on the right: running in the $B-L$-SSM including gauge-kinetic mixing.}
\label{fig:RGE}
\end{figure}

We can now use this value and demand a strict unification ($g_1=g_2=g_3=g_B$) and run down the RGEs
\begin{lstlisting}[style=mathematica]
solutionBLSSM = 
  RunRGEs[{g1->0.7, g2->0.7, g3->0.7, gBL->0.7}, 16, 3, TwoLoop->False];
\end{lstlisting}
We can check what we get for $g_1$ at 1~TeV. And actually the naive try
\begin{lstlisting}[style=mathematica]
g1[3] /. solutionBLSSM 
\end{lstlisting}
returns 0.476. That's quite a bit away from the input value of 0.46 we started with. The reason is that we missed to perform the rotations to go to the correct basis, see eq.~(\ref{eq:Triangle1}). What we have to associate with the physical hypercharge coupling is
\begin{lstlisting}[style=mathematica]
(g1[3]*gBL[3] - g1BL[3]*gBL1[3])/
  Sqrt[gBL[3]^2 + gBL1[3]^2] /. solutionBLSSM 
\end{lstlisting}
This indeed returns exactly the input value of 0.46. We can now plot the proper couplings via 
\begin{lstlisting}[style=mathematica]
Plot[{(g1[t]*gBL[t] - g1BL[t]*gBL1[t])/Sqrt[gBL[t]^2 + gBL1[t]^2], 
   g2[t], g3[t],  Sqrt[gBL[t]^2 + gBL1[t]^2], 
   (g1BL[t]*gBL[t] + gBL1[t]*g1[t])/Sqrt[gBL[t]^2 + gBL1[t]^2]} 
   /. solutionBLSSM, {t, 3, 17}, 
  Frame -> True, Axes -> False, 
  FrameLabel -> {"log[Q/GeV]", Subscript["g", "i"]}] 
\end{lstlisting}
The plot is shown on the right in Fig.~\ref{fig:RGE}. One sees here that the off-diagonal coupling gets negative. The values of the five physical, running couplings at 3~TeV are 
\begin{lstlisting}[style=mathematica,mathescape]
{(g1[t]*gBL[t] - g1BL[t]*gBL1[t])/Sqrt[gBL[t]^2 + gBL1[t]^2], 
  g2[t], g3[t],  Sqrt[gBL[t]^2 + gBL1[t]^2], 
 (g1BL[t]*gBL[t] + gBL1[t]*g1[t])/Sqrt[gBL[t]^2 + gBL1[t]^2]} 
   /. t->Log[10,3000] /. solutionBLSSM
   
$\hookrightarrow$ {0.464058, 0.640818, 1.07975, 0.442469, -0.142595}
\end{lstlisting}
Here, the GUT normalization is still included for the $U(1)$ couplings. The not normalized values are 
\begin{eqnarray}
\label{eq:runningGauge}
& g_1 = 0.359458 \,,\hspace{1cm} g_{BL} = 0.541912 \,,\hspace{1cm} \tilde{g} = -0.110454 &
\end{eqnarray}

With the same procedure we could now also start to analyse the running of the superpotential terms and the soft-masses. An estimate of the running gaugino masses based on an universal GUT values is obtained by
\begin{lstlisting}[style=mathematica]
solGauginos = 
  RunRGEs[{g1 -> 0.7, g2 -> 0.7, g3 -> 0.7, gBL -> 0.7, 
           MassB -> 1000, MassWB -> 1000, MassG -> 1000, 
           MassBp -> 1000, MassBBp -> 0}, 
  17, 3, TwoLoop -> False]; 
\end{lstlisting}
and we find 
\begin{lstlisting}[style=mathematica,mathescape]
{MassB[t], MassWB[t], MassG[t], MassBp[t],MassBBp[t]} 
  /. t->Log[10,3000]  /. solGauginos 
  
$\hookrightarrow$ {480.986, 838.056, 2379.32, 399.549, -128.763}
\end{lstlisting}
The hierarchy is similar to the CMSSM, but the $B'$ soft mass is smaller than all other gaugino masses. So, it might be that this particle would be a new dark matter candidate. I'll come back to that in the next section. We also see that the off-diagonal terms runs negative similar to the off-diagonal coupling. In the limit of vanishing kinetic mixing it is easy to explain the hierarchy of the gaugino masses: combining eq.~(\ref{eq:BetaG}) and eq.~(\ref{eq:BetaM}) we see that the value $M_i/g_i^2$ is a constant at one-loop, i.e. 
\begin{equation}
\label{eq:MiGi}
M_i = \frac{g^2_i}{g_{GUT}^2}M_{1/2}  
\end{equation}
Plugging in the numbers from eq.~(\ref{eq:runningGauge}) we get for the gaugino mass terms
\begin{equation}
M_1 \simeq 439.5~\text{GeV} \,,\hspace{1cm} M_2 \simeq 838.0~\text{GeV}\,,\hspace{1cm} M_3 \simeq 2379.3~\text{GeV}\,,\hspace{1cm} M_{B'} \simeq 399.5~\text{GeV}
\end{equation}
The values for $M_1$ and $M_{B'}$ are a bit different because eq.~(\ref{eq:MiGi}) doesn't include the effect of gauge kinetic-mixing. Nevertheless, there is a nice agreement with the numerical results.

\section{Example -- Part III: Mass spectrum, decays, flavour observables and  fine-tuning with \SPheno}
\label{sec:example_spheno}

\subsection{Calculating the mass spectrum with \SPheno}
We start now to make use of the different outputs \SARAH provides to use the derived information about a model with other tools. 
Maybe, the most important interface it the one to \SPheno which gives a very flexible, fully functional and highly precise spectrum generator for the model under consideration. A similar functionality to get a tailor made spectrum generator based on \SARAH became available with \FlexibleSUSY \footnote{I'm going to concentrate here on \SPheno. Readers interested in \FlexibleSUSY might check the homepage: {\tt http://flexiblesusy.hepforge.org/}}.\\
Before we can use \SPheno we have to provide an additional input file for \SARAH. I'll start with a description what this file is supposed to do. 

\subsubsection{Defining the boundary conditions}
In general, there are two different kinds of \SPheno versions the user can create which need a different amount of input
\begin{itemize}
 \item {\bf 'GUT Version'}: in a GUT version of \SPheno a RGE running between the electroweak, SUSY and GUT scale is supported. The user can define appropriate boundary conditions at each of these three scales. Furthermore, also threshold effects by including additional scales where heavy particles are integrated out can optionally be included. Finally, the user can define a condition which has to be satisfied to identify the GUT scale. The most common choice is the unification scale of gauge couplings, but also other choices like Yukawa unification are possible. In addition, these version include also the possible to define the entire input at the SUSY scale and skip the RGE running to the GUT scale. 
 \item {\bf 'Low Scale' Version}: in a low scale version no RGE running is included, but the \SPheno version expects all free parameters to be given at the SUSY scale. 
\end{itemize}
I concentrate on the first option because we are interested in a GUT model. Actually, the input for the second version is much shorter and can be easily derived form the information given here. The file to define the properties should be called {\tt SPheno.m} and must be located in
\begin{lstlisting}
$PATH/SARAH/Models/B-L-SSM/
\end{lstlisting}
as well. The entire file is shown in Appendix~\ref{app:sphenofile} and I'll discuss here the main parts of it.

\paragraph*{Input parameters} We have collected in eq.~(\ref{eq:BLinput}) a list of all input parameters we want to use. These parameters should be given to \SPheno  in our numerical studies via a Les Houches input file. We make the choice that all input parameters which are not a matrix are included in the block {\tt MINPAR} of the Les Houches input file. This can be done via
\begin{lstlisting}[style=file,firstnumber=1,title=\hspace{14cm}SPheno.m]
MINPAR={{1, m0},
        {2, m12},
        {3, TanBeta},
        {4, SignumMu},
        {5, Azero},
        {6, SignumMuP}, 
        {7, TBetaP},
        {8, MZp}};
\end{lstlisting}
The exact meaning of this definition will become clear when we discuss the Les Houches input in sec.~\ref{sec:spheno_LH}. In addition, it would also be possible to use the array {\tt EXTPAR} to define input which will be given via the block {\tt EXTPAR} in the Les Houches file. This might be useful to remove parameters not present in the MSSM from {\tt MINPAR}. In that case the corresponding lines in {\tt SPheno.m} would read
\begin{lstlisting}[style=file]
MINPAR={{1, m0},
        {2, m12},
        {3, TanBeta},
        {4, SignumMu},
        {5, Azero},
EXTPAR={{101, SignumMuP}, 
        {102, TBetaP},
        {103, MZp}};
\end{lstlisting}
\SARAH is completely agnostic concerning official SLHA conventions for the {\tt MINPAR} and {\tt EXTPAR} blocks \cite{Skands:2003cj,Allanach:2008qq}. It just refers to the definition as given by the user. However, it might be helpful to stick for models which are covered by SLHA to the corresponding conventions. \\

In any case, three of the input parameters are always real. This information is set via
\begin{lstlisting}[style=file,firstnumber=10,title=\hspace{14cm}SPheno.m]
RealParameters = {TanBeta, TBetaP,m0};
\end{lstlisting}
This definition is especially for $\tan\beta$ and $\tan\beta'$ important because we will use trigonometric functions with these parameters as argument. If not defined as real, \Fortran might return {\tt NaNs}. \\

\paragraph*{Tadpole equations} We choose to solve the tadpole equations with respect to $\mu$, $B_\mu$, $\mu'$ and $B_\mu'$:
\begin{lstlisting}[style=file,firstnumber=11,title=\hspace{14cm}SPheno.m]
ParametersToSolveTadpoles = {B[\[Mu]],B[MuP],\[Mu],MuP};
\end{lstlisting}
In that way \SARAH uses the {\tt Solve} command of \Mathematica to get the analytical solutions and exports them to \Fortran code. For other choices of parameters, no analytical solution might exist. In these cases it is possible to use solve the equations numerically during the run with \SPheno. I give some more information about this in Appendix~\ref{app:tadpoles_num}. One could give also the solutions based on simplified assumptions or define some assumptions which should be used for solving the equations. This is briefly discussed in Appendix~\ref{app:tadpoles_more}.

\paragraph*{Renormalization scale} \SPheno can be used either with a fixed renormalization scale or it calculates the scale dynamically. A convenient choice is often $M_{SUSY} = \sqrt{m_{\tilde{t}_1} m_{\tilde{t}_2}}$. In our case the stops are part of the six generations of up-squarks. Since we have always large $A$-terms the stops will always correspond to the lightest and heaviest of the mixed states. So we can put as renormalization scale $M_{SUSY} = \sqrt{m_{\tilde{u}_1} m_{\tilde{u}_6}}$. If that the hierarchy of the up-squark masses is not clear, another possibility would be to give an analytical expression of the determinant of the stop mass matrix as renormalization scale. This is for instance done in the MSSM by setting
\begin{lstlisting}[style=file]
RenormalizationScale = Sqrt[(mq2[3, 3] + (vu^2*conj[Yu[3, 3]]*Yu[3, 3])/2)*(mu2[3, 3] + (vu^2*conj[Yu[3, 3]]*Yu[3, 3])/2)-((vd*\[Mu]*conj[Yu[3, 3]] - vu*conj[T[Yu][3, 3]])*(vd*conj[\[Mu]]*Yu[3, 3] - vu*T[Yu][3, 3]))/2]; 
\end{lstlisting}
Here, we skipped $D$-terms which are negligible in the MSSM. However, this does no longer hold in the \BLSSM because of the new contributions from the bilepton VEVs. Hence, the definition of the renormalization scale using that approach would even be a bit more lengthy. We keep here for simplicity and since this is a save choice in this model $M_{SUSY} = \sqrt{m_{\tilde{u}_1} m_{\tilde{u}_6}}$. \\
In the very first iteration when the stop masses are unknown, \SPheno needs a crude first guess of the renormalization scale. We choose $\sqrt{m_0^2 + 4 M_{1/2}^2}$. However, also a constant like 1~TeV could be used. The two lines to define both scales for the \BLSSM are
\begin{lstlisting}[style=file,firstnumber=13,title=\hspace{14cm}SPheno.m]
RenormalizationScale = MSu[1]*MSu[6];
RenormalizationScaleFirstGuess = m0^2 + 4 m12^2;
\end{lstlisting}

\paragraph*{GUT condition} The condition for the GUT scale in our model is $g_1 = g_2$. However, because of gauge kinetic mixing one has to be careful by choosing the correct $g_1$: in the running the general $2\times 2$ gauge coupling matrix is used, i.e. for the check of the GUT scale it is necessary to rotate it to the triangle form:
\begin{lstlisting}[style=file,firstnumber=16,title=\hspace{14cm}SPheno.m]
ConditionGUTscale = (g1*gBL-g1BL*gBL1)/Sqrt[gBL^2+gBL1^2] == g2;
\end{lstlisting}
We demand that gauge-kinetic mixing vanishes at the GUT scale and this condition will simplify to {\tt g1 == g2}. Nevertheless, one should keep the full form to stabilize numerics in the first few iterations. \\

\paragraph*{Boundary conditions} Now, we define the boundary conditions at the GUT scale. First, we make sure again that the $U(1)$ coupling matrix is in the triangle form. In order to further stabilize numerics in the first iterations, where the unification might not be too good, we average $g_1$ and $g_2$ before we set the couplings in the $B-L$ sector and the off-diagonal ones. All other entries just parametrize in an obvious form the boundary conditions from eqs.~(\ref{eq:boundary1})--(\ref{eq:boundary4}):
\begin{lstlisting}[style=file,firstnumber=19,title=\hspace{14cm}SPheno.m]
BoundaryHighScale={
{g1,(g1*gBL-g1BL*gBL1)/Sqrt[gBL^2+gBL1^2]},
{g1,Sqrt[(g1^2+g2^2)/2]},
{g2,g1},
{gBL, g1},
{g1BL,0},
{gBL1,0},
{T[Ye], Azero*Ye},
{T[Yd], Azero*Yd},
{T[Yu], Azero*Yu},
{T[Yv], Azero*Yv},
{T[Yn], Azero*Yn},
{mq2, DIAGONAL m0^2},
{ml2, DIAGONAL m0^2},
{md2, DIAGONAL m0^2},
{mu2, DIAGONAL m0^2},
{me2, DIAGONAL m0^2},
{mvR2, DIAGONAL m0^2},
{mHd2, m0^2},
{mHu2, m0^2},
{mC12, m0^2},
{mC22, m0^2},
{MassB, m12},
{MassWB,m12},
{MassG,m12},
{MassBp,m12},
{MassBBp,0}
};
\end{lstlisting}
Note, the keyword {\tt DIAGONAL} and the usage of the parameters defined via {\tt MINPAR}. \\

At the SUSY scale we rotate again the gauge couplings to the triangle basis because the $(2,1)$ entry won't be zero any more due RGE effects. In addition, we calculate $v_{\eta}$ and $v_{\bar{\eta}}$ from the input values of $\tan\beta'$ and $M_Z'$. Finally, the input parameters for $Y_x$ and $Y_\nu$ are used here. Since $Y_x$ and $Y_\nu$ are matrices, it is not possible to define them via {\tt MINPAR} or {\tt EXTPAR}. Therefore, {\tt LHInput[x]} is used. With this command, \SPheno expects the parameters are given via blocks {\tt YXIN} and {\tt YNUIN} in the Les Houches file \footnote{The convention is that the block name for the output is used together wit the suffix {\tt IN}}. 
\begin{lstlisting}[style=file,firstnumber=48,title=\hspace{14cm}SPheno.m]
BoundarySUSYScale = {
 {g1T,(g1*gBL-g1BL*gBL1)/Sqrt[gBL^2+gBL1^2]},
 {gBLT, Sqrt[gBL^2+gBL1^2]},
 {g1BLT,(g1BL*gBL+gBL1*g1)/Sqrt[gBL^2+gBL1^2]},
 {g1, g1T},
 {gBL, gBLT},
 {g1BL, g1BLT},
 {gBL1,0},
 {vevP, MZp/gBL},
 {betaP,ArcTan[TBetaP]},
 {x2,vevP*Cos[betaP]},
 {x1,vevP*Sin[betaP]},
 {Yv,  LHInput[Yv]},
 {Yn,  LHInput[Yn]}
};
\end{lstlisting}
The boundary conditions at the EWSB scale are similar to the ones at the SUSY scale and skipped here. 
%
To further stabilize numerics in the first run up to the GUT scale, it is useful to give approximate values for the initialization of the gauge couplings which are not present in the SM :
\begin{lstlisting}[style=file,firstnumber=78,title=\hspace{14cm}SPheno.m]
InitializationValues = {
 {gBL, 0.5},
 {g1BL, -0.06},
 {gBL1, -0.06}
 }
\end{lstlisting}
\paragraph*{Decays} Finally, we tell \SARAH that it should make use of the default conventions to write code to calculate two- and three-body decays with \SPheno:
\begin{lstlisting}[style=file,firstnumber=97,title=\hspace{14cm}SPheno.m]
ListDecayParticles = Automatic;
ListDecayParticles3B =Automatic;
\end{lstlisting}
By convention, the following decays are included that way: (i) all two-body decays of SUSY particles, Higgs states, the top quark and additional vector bosons; (ii) three-body decays of SUSY fermions in three other fermions and decays of SUSY scalars in another scalar and two fermions. \\

\paragraph*{Precision} We are mostly going to neglect neutrino masses in the following. However, if a study of the neutrino phenomenology should be included, it might be necessary to calculate the masses of the neutrino eigenstates with a higher precision as this is usually done in \SPheno. The reason is the potential large hierarchy between left and right neutrinos. Details about this are given in Appendix~\ref{app:quadruple}.

\subsubsection{Obtaining and running the \SPheno code}
\label{sec:getSPheno}
To obtain the \SPheno output  we run in \Mathematica after \SARAH is loaded and the \BLSSM is initialized the command
\begin{lstlisting}[style=mathematica]
MakeSPheno[];
\end{lstlisting}
The different options are:
\begin{itemize}
\item \verb"ReadLists->True": can be used if all vertices and RGEs have already
been calculated for the model and the former results should be  used.
\item \verb"InputFile": defines the name of the \SPheno input file. By default \verb"SPheno.m" is used.
\item  \verb"StandardCompiler->Compile": defines the compiler which should be set as standard in the {\tt Makefile}. 
      Default is {\tt gfortran}.
\item \verb"IncludeFlavorKit": can be used to disable the output of flavour observables based on \FlavorKit. 
\end{itemize}
When executing {\tt MakeSPheno}, \SARAH calculates first all information which it needs, i.e. it is not necessary that the user has done the calculation of vertices or RGEs before. When \SARAH is done, the source code for \SPheno is stored in {\tt \$SPATH/SARAH/Output/B-L-SSM/EWSB/SPheno/}. 
The compilation of this codes is done as follows: one has to enter the directory of the \SPheno installation, 
a new sub-directory has to be created and the code must copied into this directory:
\begin{lstlisting}[style=terminal]
$ cd $PATH/SPHENO
$ mkdir BLSSM
$ cp $PATH/SARAH/Output/B-L-SSM/EWSB/SPheno/* BLSSM/
\end{lstlisting}
Afterwards, the code is compiled via 
\begin{lstlisting}[style=terminal]
$ make Model=BLSSM 
\end{lstlisting}
and a new executable {\tt SPhenoBLSSM} is created which we will use in the following.

\subsubsection{Setting the input for {\tt SPhenoBLSSM}}
\label{sec:spheno_LH}
An input file, by default called {\tt LesHouches.in.BLSSM}, is needed to run {\tt SPhenoBLSSM}. \SARAH writes a template 
for that file which has been copied to the {\tt BLSSM} subdirectory of \SPheno together with the \Fortran code. We move it to the root
directory of \SPheno 
\begin{lstlisting}[style=terminal]
$ cp BLSSM/LesHouches.in.BLSSM . 
\end{lstlisting}
By doing this we can work now from the \SPheno main directory and we don't have to give the file as argument when 
running \SPheno. Thus, \SPheno can be called via 
\begin{lstlisting}[style=terminal]
$ ./bin/SPhenoBLSSM
\end{lstlisting}
Alternatively, one can keep the Les Houches file in the {\tt BLSSM} and work with it via
\begin{lstlisting}[style=terminal]
$ ./bin/SPhenoBLSSM BLSSM/LesHouches.in.BLSSM
\end{lstlisting}

However, before we can run \SPheno we first have to define the input parameters: for that purpose we have to fill the template written by \SARAH with numbers. I'm going to discuss the different blocks appearing in the Les Houches file briefly. More details especially about the block {\tt SMINPUTS} can also be found in the official references for SLHA \cite{Skands:2003cj,Allanach:2008qq}.\\
At the very beginning, the block {\tt MODSEL} is given: 
\begin{lstlisting}[style=file,firstnumber=1,title=\hspace{11cm}LesHouches.in.BLSSM]
Block MODSEL      #  
1 1               #  1/0: High/low scale input 
2 1               # Boundary Condition  
6 1               # Generation Mixing
\end{lstlisting}
This block fixes the general setup the user wants to use:
\begin{itemize}
 \item[{\tt 1}]: defines, if a GUT scale input is used ({\tt 1}) or all parameters should be given at the SUSY scale ({\tt 0})
 \item[{\tt 2}]: in principle, it is possible to define in {\tt SPheno.m} different boundary conditions for the GUT input, for instance if different SUSY breaking mechanism should be studied. This flag can be used to choose one. Since we haven't made use of that, this flag has no effect here.
 \item[{\tt 5}]: if put to {\tt 1}, CP violation is allowed and the phase of the CKM matrix is included. 
 \item[{\tt 6}]: if put to {\tt 1}, flavour violation is allowed and all off-diagonal entries in soft- or superpotential-parameters can receive non-zero values. If put to {\tt 0}, the CKM matrix is taken to be the identity matrix.
 \item[{\tt 12}]: this flag can be given optionally to fix the SUSY scale to a constant value.
\end{itemize}
The block {\tt SMINPUTS} contains all important values for the SM parameters like Fermi constant $G_F$, strong coupling constant $\alpha_S(M_Z)$, pole mass of the $Z$-boson, third generation fermion masses $m_{t}$, $m_{b}$, $m_{\tau}$. Also other parameters can be set as explained in the SLHA write ups but this is usually not necessary.
\begin{lstlisting}[style=file,firstnumber=5,title=\hspace{11cm}LesHouches.in.BLSSM]
Block SMINPUTS    # Standard Model inputs 
2 1.166370E-05    # G_F,Fermi constant 
3 1.187000E-01    # alpha_s(MZ) SM MSbar 
4 9.118870E+01    # Z-boson pole mass 
5 4.180000E+00    # m_b(mb) SM MSbar 
6 1.735000E+02    # m_top(pole) 
7 1.776690E+00    # m_tau(pole) 
\end{lstlisting}
\begin{table}[t]
\centering
\begin{tabular}{|c|ccccccc|}
 \hline 
  & $m_0$ [GeV] & $M_{1/2}$  [GeV] & $\tan\beta$ & $A_0$  [GeV] & $\tan\beta'$ & $M_{Z'}$  [GeV] & diag($Y_x$) \\
\hline  
 EP1 & 1700 & 1500 & 7.0 & -1400 & 1.20 & 2500 & (0.42,\,0.42,\,0.38) \\
 EP2 & 1700 & 1500 & 7.0 & -1400 & 1.06 & 4500 & (0.42,\,0.42,\,0.05) \\
 \hline
\end{tabular}
\caption{Input parameters for the example points EP1 and EP2. In addition, we use $\text{sign}(\mu)=\text{sign}(\mu')=1$ and $Y_\nu=0$. }
\label{tab:EP}
\end{table}
We turn now to the input to fix the parameter point we want to study. I have chosen two points as examples, {\bf EP1} and {\bf EP2}, with slightly different input. All necessary input parameters are given in Tab.~\ref{tab:EP}.
The neutrino Yukawa coupling is highly constrained by neutrino data and we can ignore it here for our purposes. It will become important if the user wants to study lepton-flavour violation for instance. The input is highly motivated by the observations we made at the analytical level: we need small $\tan\beta'$ and large $m_0$, $A_0$, $Y_x$ to break $B-L$ radiatively. One can test that the values of $Y_x$ in Tab.~\ref{tab:EP} are close to the Landau pole. When using even larger values \SPheno will stop with an error message. The values for EP1 are set in the Les Houches file via
\begin{lstlisting}[style=file,firstnumber=12,title=\hspace{11cm}LesHouches.in.BLSSM]
Block MINPAR      # Input parameters 
1   1.7000000E+03    # m0
2   1.5000000E+03    # m12
3   7.0000000E+00    # TanBeta
4   1.0000000E+00    # SignumMu
5  -1.4000000E+03    # Azero
6   1.0000000E+00    # SignumMuP
7   1.2000000E+00    # TBetaP
8   2.5000000E+03    # MZp
Block YXIN    #  
1 1   4.2000000E-01  # Yx(1,1)
2 2   4.2000000E-01  # Yx(2,2)
3 3   3.8000000E-01  # Yx(3,3)
Block YVIN    #  
1 1   0.0000000E+00  # Yv(1,1)
2 2   0.0000000E+00  # Yv(2,2)
3 3   0.0000000E+00  # Yv(3,3)
\end{lstlisting}
Finally, we have some more switches in the block {\tt SPhenoInput}:
\begin{lstlisting}[style=file,firstnumber=31,title=\hspace{11cm}LesHouches.in.BLSSM]
Block SPhenoInput   # SPheno specific input 
  1 -1              # error level 
  2  0              # SPA conventions 
  7  0              # Skip 2-loop Higgs corrections 
  8  3              # Method used for two-loop calculation 
  9  1              # Gaugeless limit used at two-loop 
 10  0              # safe-mode used at two-loop 
 11 1               # calculate branching ratios 
 13 1               # 3-Body decays: none (0), fermion (1), scalar (2), both (3) 
 12 1.000E-04       # write only branching ratios larger than this  
 15 1.000E-30       # write only decay if width larger than this  
 31 -1              # fixed GUT scale (-1: dynamical GUT scale) 
 32 0               # Strict unification 
 34 1.000E-04       # Precision of mass calculation 
 35 40              # Maximal number of iterations
 37 1               # Set Yukawa scheme  
 38 2               # 1- or 2-Loop RGEs 
 50 0               # Majorana phases: use only positive masses 
 51 0               # Write Output in CKM basis 
 52 0               # Write spectrum in case of tachyonic states 
 55 1               # Calculate one loop masses 
 57 1               # Calculate low energy constraints 
 60 1               # Include possible, kinetic mixing 
 65 1               # Solution tadpole equation 
 75 1               # Write WHIZARD files 
 76 1               # Write HiggsBounds file 
 86 0.              # Maximal width to be counted as invisible
510 0.              # Write tree level values for tadpole solutions 
515 0               # Write parameter values at GUT scale 
520 1.              # Write effective Higgs couplings (HB blocks) 
525 0.              # Write contributions to diphoton decay of Higgs 
530 1.              # Write Blocks for Vevacious 
\end{lstlisting}
These flags can be used to adjust the calculations done by \SPheno and the output. All possible flags are listed in Appendix~\ref{app:SPhenoFlags}. I’m explaining here just the most important ones:
\begin{itemize}
 \item {\bf Loop-level}: to turn off all loop corrections to the masses, flag {\tt 55} is put to  {\tt 0}. The two-loop corrections in the Higgs sector are turned on/off by {\tt flag 7}. {\tt flag 8} chooses the method to calculate the two-loop contributions: {\tt 1} purely numerical effective potential calculation, {\tt 2} semi-analytical effective potential calculation, {\tt 3} diagrammatic calculation, {\tt 8}/{\tt 9} results from literature (if available)
 \item {\bf Decays and branching ratios}: all decays are turned on/off via flag {\tt 11}, while three-body decays can be adjusted via flag {\tt flag 13}. 
 \item {\bf Precision observables}: the calculation of precision and flavour observables is turned on and off using flag {\tt 57}
 \item {\bf Gauge-kinetic mixing}: \SPheno can be forced to ignore gauge kinetic mixing by setting flag {\tt 60} to {\tt 0}
 \item {\bf Output for other codes}: the output of \HB and \HS input files are switched on/off via flag {\tt 76}, while {\tt flag 75} is responsible for the output of the parameter file for \WHIZARD. The \Vevacious specific blocks in the spectrum file are included or excluded with flag {\tt 530}
 \item {\bf Fine-tuning}: the calculation of the fine-tuning is skipped when setting flag {\tt 550} to {\tt 0}
\end{itemize}

\subsubsection{Calculating the spectrum and Higgs couplings with \SPheno}
When the input file is filled with numbers we can run the point as explained above. 
The entire output including all parameters, masses, branching ratios and low-energy 
observables is saved by default in {\tt SPheno.spc.BLSSM}. This file is rather lengthy and contains a lot of information. 
I will pick some parts of it and discuss them.\\

Sometimes, it is convenient to work with in- and output files with other names. In that case, the names can be used as argument for {\tt SPhenoBLSSM}. For instance, we can make two input files for the points EP1 and EP2 and run them via 
\begin{lstlisting}[style=terminal]
$ ./bin/SPhenoBLSSM LesHouches.in.BLSSM_EP1 SPheno.spc.BLSSM_EP1
$ ./bin/SPhenoBLSSM LesHouches.in.BLSSM_EP2 SPheno.spc.BLSSM_EP2
\end{lstlisting}
The entire output is written to the files given as second argument, i.e. {\tt SPheno.spc.BLSSM\_EP1} and {\tt SPheno.spc.BLSSM\_EP2}.

\paragraph*{The running parameters and mass spectrum for EP1} First, we check if gauge coupling unification at about $10^{16}$~GeV remains. From the block {\tt gaugeGUT} we see that the unification scale is a bit higher than in the MSSM and that we have no strict unification here because there is a small offset of $g_3$. This behaviour is also known from the MSSM and one assumes that higher order corrections as well as threshold corrections from super heavy particles are responsible for an exact unification once taken into account. 
\begin{lstlisting}[style=file,firstnumber=38,title=\hspace{11cm}SPheno.spc.BLSSM (EP1)]
Block gaugeGUT Q=  5.55462899E+16  # (GUT scale)
   1    7.14501610E-01  # g1(Q)^DRbar
   2    7.14501610E-01  # g2(Q)^DRbar
   3    6.87174622E-01  # g3(Q)^DRbar
   4    7.14501610E-01  # gBL(Q)^DRbar 
\end{lstlisting}

The running gauge couplings at the SUSY scale are given in the block {\tt Gauge}. The SUSY scale shown as $Q$ in the head of the block is about 3~TeV, i.e. the stop masses used to fix the scale are quite heavy. That's a consequence of the large  $m_0$ and $M_{1/2}$ we used to get radiative $B-L$ breaking. Note, $g_1$ in this block is the value without GUT normalization. Thus, it is related to $g_1$ used in sec.~\ref{sec:SolveRGEs} by a factor of $\sqrt{5/3}$. We find also an off-diagonal coupling of -0.11 and $g_B$ close to $0.55$ as expected from our estimates with \Mathematica, see eq.~(\ref{eq:runningGauge}). 
\begin{lstlisting}[style=file,firstnumber=51,title=\hspace{11cm}SPheno.spc.BLSSM (EP1)]
Block GAUGE Q=  3.04424240E+03  # (SUSY Scale)
   1    3.63178061E-01  # g1
   2    6.44147228E-01  # g2
   3    1.00570244E+00  # g3
   4    5.48624610E-01  # gBL
  10   -1.13408283E-01  # gYB
  11    0.00000000E+00  # gBY 
\end{lstlisting}

Many more blocks appear after the one for the gauge couplings and contain all other running parameters  at the SUSY scale. We just want to take a look at two other blocks: {\tt BL} and {\tt MSOFT}. The first one contains the soft-terms, VEVs and $\mu'$ in the $B-L$ sector, the second one the soft-terms Higgs and gaugino masses from the MSSM part.
\begin{lstlisting}[style=file,firstnumber=58,title=\hspace{11cm}SPheno.spc.BLSSM (EP1)]
Block BL Q=  3.04424240E+03  # (SUSY Scale)
   1    1.59624321E+03  # MuP
   2    4.75415053E+06  # BMuP
  11    8.49277085E+05  # mC12
  12    3.74696940E+06  # mC22
  32   -2.21262550E+02  # MBBp
  31    6.59015204E+02  # MBp
  41    3.50066906E+03  # x1
  42    2.91722422E+03  # x2
  43    4.55684990E+03  # vX
\end{lstlisting}
\begin{lstlisting}[style=file,firstnumber=75,title=\hspace{11cm}SPheno.spc.BLSSM (EP1)]  
Block MSOFT Q=  3.04424240E+03  # (SUSY Scale)
  21    4.00188260E+06  # mHd2
  22   -4.80480045E+06  # mHu2
   1    7.87335844E+02  # M1
   2    1.32773211E+03  # M2
   3    3.57502541E+03  # M3  
\end{lstlisting}
We see here that $m_{H_u}^2$ is negatives as expected to break the electroweak symmetry. However, both soft-terms for the bileptons are actually positive and one might wonder if $B-L$ is really broken radiatively. It is broken because the full condition is not that a soft-term has to be negative, but that the determinant of the mass matrix has a negative eigenvalue in the limit of vanishing VEVs. For the lower $2\times 2$ block of the scalar mass matrix this conditions reads
\begin{equation}
(m_{\eta} + |\mu'|^2)(m_{\bar\eta} + |\mu'|^2) - |B_{\mu'}|^2  < 0
\end{equation}
Because of the large value of $B_{\mu'}$ the condition is fulfilled for this point and $B-L$ is broken. 
Another observation is that the blino soft mass $M_{B'}$ is lighter than the bino one as we already expected from sec.~\ref{sec:SolveRGEs}. Actually, $M_{B'}$ is also smaller than the other gaugino masses and the $\mu$-term not shown here. So, one would expect that the lightest neutralino is a blino. However, we will see below that this is not the case.\\

After all blocks with the running parameters, all masses are printed in the block {\tt MASS}. 
\begin{lstlisting}[style=file,firstnumber=264,title=\hspace{11cm}SPheno.spc.BLSSM (EP1)]
Block MASS  # Mass spectrum
#   PDG code      mass          particle
   1000001     3.30871123E+03  # Sd_1
   1000003     3.60959661E+03  # Sd_2
   1000005     3.62130996E+03  # Sd_3
   2000001     3.62131463E+03  # Sd_4
   2000003     3.73450325E+03  # Sd_5
   2000005     3.73452286E+03  # Sd_6
   1000002     2.67104525E+03  # Su_1
   1000004     3.32204106E+03  # Su_2
   1000006     3.61199579E+03  # Su_3
   2000002     3.61201405E+03  # Su_4
   2000004     3.73368042E+03  # Su_5
   2000006     3.73369930E+03  # Su_6
   1000011     1.76023835E+03  # Se_1
   1000013     1.77259986E+03  # Se_2
   1000015     1.77264366E+03  # Se_3
   2000011     2.11331234E+03  # Se_4
   2000013     2.11823100E+03  # Se_5
   2000015     2.11824845E+03  # Se_6
   1000012     4.96686479E+02  # SvRe_1
   1000014     8.25895521E+02  # SvRe_2
   1000016     8.25895521E+02  # SvRe_3
   2000012     2.11130838E+03  # SvRe_4
   2000014     2.11640817E+03  # SvRe_5
   2000016     2.11642628E+03  # SvRe_6
   4000012     2.11130838E+03  # SvIm_1
   4000014     2.11640817E+03  # SvIm_2
   4000016     2.11642628E+03  # SvIm_3
   5000012     2.69896818E+03  # SvIm_4
   5000014     2.99406918E+03  # SvIm_5
   5000016     2.99406918E+03  # SvIm_6
        25     1.24182913E+02  # hh_1
        35     3.55295631E+02  # hh_2
   9900025     3.06408049E+03  # hh_3
   9900035     3.96149850E+03  # hh_4
        36     3.06214679E+03  # Ah_3
   9900036     3.09582809E+03  # Ah_4
        37     3.06360579E+03  # Hpm_2
        23     9.11887000E+01  # VZ
        24     8.06613172E+01  # VWm
         1     5.00000000E-03  # Fd_1
         3     9.50000000E-02  # Fd_2
         5     4.18000000E+00  # Fd_3
         2     2.50000000E-03  # Fu_1
         4     1.27000000E+00  # Fu_2
         6     1.73500000E+02  # Fu_3
        11     5.10998930E-04  # Fe_1
        13     1.05658372E-01  # Fe_2
        15     1.77669000E+00  # Fe_3
        12     1.31876738E-11  # Fv_1
        14     1.36424206E-11  # Fv_2
        16     1.83674189E-11  # Fv_3
       112     1.48043893E+03  # Fv_4
       114     2.06846561E+03  # Fv_5
       116     2.06846561E+03  # Fv_6
   1000021     3.75512792E+03  # Glu
   1000024     1.37525092E+03  # Cha_1
   1000037     2.27493791E+03  # Cha_2
        31     2.50003732E+03  # VZp
   1000022     7.85565922E+02  # Chi_1
   1000023     1.29973769E+03  # Chi_2
   1000025     1.37521122E+03  # Chi_3
   1000035     1.70198093E+03  # Chi_4
   1000032     2.27046891E+03  # Chi_5
   1000036     2.27465029E+03  # Chi_6
   1000039     3.64727768E+03  # Chi_7 
\end{lstlisting}
The squarks ({\tt Su} and {\tt Sd}) are very heavy, and also the charged sleptons ({\tt Se}) have masses well above 1~TeV. However, there are some light CP even sneutrinos ({\tt SvRe}) while the CP odd ones are much heavier ({\tt SvIm}). Thus, the mass splitting between the CP eigenstates of the sneutrinos is another interesting and import aspect of the \BLSSM. Actually, scrolling down, we see that all other SUSY states like neutralinos ({\tt Chi}) and charginos ({\tt Cha}) are actually heavier than $\tilde \nu_1^R$. Thus, the LSP and therefore the DM candidate is the lightest CP even sneutrino for this point. The hierarchy of the heavy neutrino ({\tt Fv}) reflects the input values of $Y_x$.  We see also that the $Z'$ mass is not exactly identical to the input. The reason is that the input is taken to be the tree-level mass in the limit of vanishing gauge-kinetic mixing, while here the full one-loop mass including all mixing effects is shown. However, the differences are very small and one has usually not to worry about them.\\
In the Higgs sector, we have the lightest scalar ({\tt hh\_1}) in the mass range preferred by measurements. Also the second scalar $h_2$ is not much heavier. In contrast, the two heavier scalars as well as the two physical pseudo-scalars ({\tt Ah}) and the charged Higgs ({\tt Hpm}) are all about 3~TeV. \\
This is maybe a good time to briefly comment on the importance of the loop corrections in this model. We can turn off the two-loop corrections via
\begin{lstlisting}[style=file,numbers=none]
Block SPhenoInput   # SPheno specific input
...
 7  1              # Skip 2-loop Higgs corrections
\end{lstlisting}
and we get $m^{(1L)}_{h_1}=115.9~GeV$ and $m^{(1L)}_{h_1}=356.0~GeV$. Turning even off one-loop corrections via
\begin{lstlisting}[style=file,numbers=none]
  55 0               # Calculate one loop masses
\end{lstlisting}
we find $m^{(T)}_{h_1}=87.2$~GeV and $m^{(T)}_{h_2}=353.9$~GeV. The corrections for $m_{h_1}$ are similar as in the MSSM and one can guess already from these numbers that this state is mostly a MSSM-like doublet. To confirm this guess, we can check the mixing matrix in the scalar sector which is given in the block {\tt SCALARMIX}:
\begin{lstlisting}[style=file,firstnumber=516,title=\hspace{11cm}SPheno.spc.BLSSM (EP1)]
Block SCALARMIX Q=  3.04424240E+03  # ()
  1  1     1.48960470E-01   # ZH(1,1)
  1  2     9.88343642E-01   # ZH(1,2)
  1  3    -2.39894518E-02   # ZH(1,3)
  1  4    -2.03009771E-02   # ZH(1,4)
  2  1     4.79675110E-03   # ZH(2,1)
  2  2     3.09937950E-02   # ZH(2,2)
  2  3     7.20236031E-01   # ZH(2,3)
  2  4     6.93019794E-01   # ZH(2,4)
  ...
\end{lstlisting}
These are the entries of the $Z^H$ rotation matrix and we concentrate here one only on the first two eigenstates. 
We see that bilepton admixture of the first eigenstate is just $O(10^{-2})\%$ \footnote{The exact value is $(2.3 \cdot 10^{-2})^2+(1.1 \cdot 10^{-2})^2$}, while the second one is nearly a pure bilepton. So, the mixing between both sectors for this point is moderately small because the masses are not too close.\\

The mixing matrix for the neutralinos is shown in the block {\tt NMIX}:
\begin{lstlisting}[style=file,firstnumber=555,title=\hspace{11cm}SPheno.spc.BLSSM (EP1)]
Block NMIX Q=  3.04424240E+03  # ()
  1  1     9.95429472E-01   # Real(ZN(1,1),dp)
  1  2    -1.92612245E-03   # Real(ZN(1,2),dp)
  1  3     2.32548331E-02   # Real(ZN(1,3),dp)
  1  4    -1.10389021E-02   # Real(ZN(1,4),dp)
  1  5    -2.68824741E-02   # Real(ZN(1,5),dp)
  1  6    -6.00307896E-02   # Real(ZN(1,6),dp)
  1  7     6.42452196E-02   # Real(ZN(1,7),dp)
  ...
\end{lstlisting}
So, we see that the lightest state has a bino fraction of about 99\% \footnote{Given by $(0.995)^2$}, a wino fraction of $O(10^{-4}\%)$ \footnote{Given by  $(1.9\cdot10^{-3})^2$}, a Higgsino fraction of roughly 0.07\% \footnote{Given by $(2.3\cdot10^{-2})^2 + (1.1\cdot10^{-2})^2$} and a similar blino fraction \footnote{Given by $(2.6\cdot10^{-2})^2$}, and finally a bileptino fraction of a bit less than 1\% \footnote{Given by  $(6.0\cdot10^{-2})^2+(6.4\cdot10^{-2})^2$}. This is surprising because we have seen above that the blino soft-term is smaller than the bino one. Why isn't then the LSP a blino? To understand this, we can check the mixing of the second lightest neutralino. Since the default convention by \SPheno is that all Majorana masses are negative, some entries of the rotation matrix are imaginary. Hence, we find the composition of $\tilde \chi^0_2$ in {\tt IMNMIX}:
\begin{lstlisting}[style=file,firstnumber=605,title=\hspace{11cm}SPheno.spc.BLSSM (EP1)]
Block IMNMIX Q=  3.04424240E+03  # ()
  ...
  2  1    -6.79846118E-02   # Aimag(ZN(2,1))
  2  2     3.37464672E-05   # Aimag(ZN(2,2))
  2  3     3.48780244E-03   # Aimag(ZN(2,3))
  2  4     1.61010850E-03   # Aimag(ZN(2,4))
  2  5    -6.48176532E-01   # Aimag(ZN(2,5))
  2  6    -2.47517778E-02   # Aimag(ZN(2,6))
  2  7     7.58035531E-01   # Aimag(ZN(2,7)) 
\end{lstlisting}
We see that in the $B-L$ sector there is a large mixing between blino and bileptino. The reason is that the mixing entries are proportional to $M_{Z'}$ and not to $M_Z$ as for the MSSM part. Therefore, one doesn't have a pure blino what would be lighter than the bino. One can check that in the limit of very heavy $\mu'$, the bileptinos decouple and the blino is indeed the LSP \cite{O'Leary:2011yq}. All other rotation matrices are given as well in the spectrum file but not further discussed here. \\

\paragraph*{Higgs Couplings} Other important information about the properties of all scalars and pseudoscalars for the given parameter point is shown in the two blocks {\tt HiggsBoundsInputHiggsCouplingsFermions} and {\tt HiggsBoundsInputHiggsCouplingsBosons}. These blocks give the coupling of all Higgs states in the model normalized to the SM expectation as defined in eq.~(\ref{eq:HiggsCouplingRatios}).

Thus, the values in these blocks for the lightest Higgs, which we associated with the SM-like state, should be close to one, while those for the second lightest Higgs are expected to be much smaller. This is exactly what we observe:
\begin{lstlisting}[style=file,firstnumber=805,title=\hspace{11cm}SPheno.spc.BLSSM (EP1)]
Block HiggsBoundsInputHiggsCouplingsFermions # 
    1.21544807E+00    ...   3  25   5    5 # h_1 b b coupling 
    1.21544807E+00    ...   3  25   3    3 # h_1 s s coupling 
    9.94987651E-01    ...   3  25   6    6 # h_1 t t coupling  
    9.94987651E-01    ...   3  25   4    4 # h_1 c c coupling 
    1.21544807E+00    ...   3  25  15   15 # h_1 tau tau coupling 
    1.21544807E+00    ...   3  25  13   13 # h_1 mu mu coupling  
    1.22177746E-03    ...   3  35   5    5 # h_2 b b coupling 
    1.22177746E-03    ...   3  35   3    3 # h_2 s s coupling 
    9.79053104E-04    ...   3  35   6    6 # h_2 t t coupling  
    9.79053104E-04    ...   3  35   4    4 # h_2 c c coupling 
    1.22177746E-03    ...   3  35  15   15 # h_2 tau tau coupling 
    1.22177746E-03    ...   3  35  13   13 # h_2 mu mu coupling  
 ...   
Block HiggsBoundsInputHiggsCouplingsBosons # 
    9.98814843E-01   3   25  24   24 # h_1 W W coupling 
    1.00445622E+00   3   25  23   23 # h_1 Z Z coupling  
    0.00000000E+00   3   25  23   22 # h_1 Z gamma coupling 
    1.00085782E+00   3   25  22   22 # h_1 gamma gamma coupling 
    9.89551529E-01   3   25  21   21 # h_1 g g coupling 
    0.00000000E+00   4   25  21   21        23 # h_1 g g Z coupling 
    9.83376076E-04   3   35  24   24 # h_2 W W coupling 
    1.10657726E-03   3   35  23   23 # h_2 Z Z coupling  
    0.00000000E+00   3   35  23   22 # h_2 Z gamma coupling 
    9.83659658E-04   3   35  22   22 # h_2 gamma gamma coupling 
    9.78781315E-04   3   35  21   21 # h_2 g g coupling 
    0.00000000E+00   4   35  21   21        23 # h_2 g g Z coupling 

...
\end{lstlisting}
Since the couplings squared to down-type quarks and leptons for $h_1$ are about 20\% larger than in the SM we can expect that this point doesn't explain the measurements too well. To quantize that we will use \HS later, see sec.~\ref{sec:HS}. These coupling ratios can also be used to get the production cross section of the Higgs states at the LHC in the different channels by rescaling the SM results. This is done in the blocks {\tt HiggsLHC7} and {\tt HiggsLHC8}. The SM cross section for a given Higgs mass have been obtained by fitting the data from 
\begin{verbatim}
https://twiki.cern.ch/twiki/bin/view/LHCPhysics/CERNYellowReportPageAt7TeV
https://twiki.cern.ch/twiki/bin/view/LHCPhysics/CERNYellowReportPageAt8TeV
\end{verbatim}
Since this fit is only valid in a finite range for the Higgs mass\footnote{Gluon fusion and vector boson fusion from 50--1000~GeV, all other channels from 50--300~GeV. For 13 and 14~TeV only cross sections are provided for $m_h = 125.5 \pm 0.5$~GeV. That's the reason why no {\tt HiggsLHC13} and {\tt HiggsLHC14} blocks are given.} also the results are only given for Higgs masses which lie in the fit range of each channel. 
\begin{lstlisting}[style=file,firstnumber=789,title=\hspace{11cm}SPheno.spc.BLSSM (EP1)]
Block HiggsLHC7 # Higgs production cross section at LHC7 [pb] 
  1        25        1.54190769E+01 # Gluon fusion 
  2        25        1.22391804E+00 # Vector boson fusion 
  3        25        5.54211509E-01 # W-H production 
  4        25        3.24580015E-01 # Z-H production 
  5        25        8.74181618E-02 # t-t-H production 
  1        35        2.29725104E-03 # Gluon fusion 
  2        35        2.30558730E-04 # Vector boson fusion 
Block HiggsLHC8 # Higgs production cross section at LHC8 [pb] 
  1        25        1.95249057E+01 # Gluon fusion 
  2        25        1.57753091E+00 # Vector boson fusion 
  3        25        7.08633453E-01 # W-H production 
  4        25        4.05291805E-01 # Z-H production 
  5        25        1.32262917E-01 # t-t-H production 
  1        35        3.22819525E-03 # Gluon fusion 
  2        35        3.49847940E-04 # Vector boson fusion  
\end{lstlisting}

\paragraph*{A point with a light scalar}
We want briefly discuss the particularities of the point EP2 given in Tab.~\ref{tab:EP}. To run this point, we have only to change three lines in the Les Houches input file compared to EP1:
\begin{lstlisting}[style=file,title=\hspace{11cm}LesHouches.in.BLSSM]
Block MINPAR      # Input parameters 
...
7   1.060E+00    # TBetaP
8   4.500000E+03    # MZp 
Block YXIN    #  
...
3 3   5.00000000E-02         # Yx(3,3)
\end{lstlisting}
The interesting aspect of this point is the Higgs sector which now contains a light scalar:
\begin{lstlisting}[style=file,title=\hspace{11cm} SPheno.spc.BLSSM (EP2)]
Block MASS  # Mass spectrum
...
        25     7.78960599E+01  # hh_1
        35     1.25607177E+02  # hh_2
\end{lstlisting}
This point is mostly a bilepton, of course. However, it has a doublet fraction of about 1\% doublet as we can see from the scalar mixing matrix
\begin{lstlisting}[style=file,title=\hspace{11cm}SPheno.spc.BLSSM (EP2)]
Block SCALARMIX Q=  3.04382955E+03  # ()
  1  1     2.41678467E-02   # ZH(1,1)
  1  2     1.60726196E-01   # ZH(1,2)
  1  3    -6.98258293E-01   # ZH(1,3)
  1  4    -6.97150171E-01   # ZH(1,4)
  2  1    -1.47075505E-01   # ZH(2,1)
  2  2    -9.75679078E-01   # ZH(2,2)
  2  3    -1.13971461E-01   # ZH(2,3)
  2  4    -1.15886318E-01   # ZH(2,4)
...
\end{lstlisting}
Thus, it will be interesting to see if it passes all tests from Higgs searches. We are going to check that with \HB in sec.~\ref{sec:HBHS}. Another important aspect of this point is the Higgs mass at the different  loop levels:
\begin{align}
& m_{h_1}^{(T)} = 87.1~\text{GeV}\,,\hspace{1cm} m_{h_2}^{(T)} = 184.3~\text{GeV} \\
& m_{h_1}^{(1L)} = 68.3~\text{GeV}\,,\hspace{1cm} m_{h_2}^{(2L)} = 117.5~\text{GeV} \\
& m_{h_1}^{(2L)} = 77.9~\text{GeV}\,,\hspace{1cm} m_{h_2}^{(2L)} = 125.6~\text{GeV}
\end{align}
At tree-level the lighter mass is the doublet, and the heavier one is the bilepton. Thus, there is a level crossing at one-loop compared to tree-level. The bilepton mass nearly changes by a factor of 3 when including radiative corrections. Therefore, to have some trust in the mass prediction, also two-loop corrections for the bileptons are crucial for this point. These corrections are even a bit larger than for the MSSM-like particle and give a push of about 10~GeV.

\subsection{Decay widths and branching}
\SPheno modules by \SARAH do not only calculate the mass spectrum and effective couplings for the Higgs scalars, but they provide also functions 
for calculating decays. To adjust the output of the decays, the important flags in the Les Houches input are 
\begin{lstlisting}[style=file,title=\hspace{11cm}LesHouches.in.BLSSM]
Block SPhenoInput   # SPheno specific input
 ...
 11 1               # calculate branching ratios (BR)
 13 1               # 3-Body decays: none (0), fermion (1), scalar (2), both (3)
 12 1.000E-04       # write only BR larger than this value 
 15 1.000E-30       # write only decay if width larger than this value  
\end{lstlisting}
With flag {\tt 11} the entire calculation of all decays can be turn on and off. Note, that for the \HB and \HS output discussed in sec.~\ref{sec:HBHS} the calculation of at least two-body decays is essential. {\tt flag 13} can be used to turn on/off the three-body decays separately: either only three-body decays of fermions, or only of scalars, or of both can be calculated. Usually, the scalar three-body decays are quite time consuming because of the many decay channels. 
This can sometimes be helpful because the calculation of the three-body decays can be time consuming and is not always needed. With flag {\tt 12} a lower limit of the branching ratios which should be shown in the SLHA file is given, while {\tt flag 15} gives a lower limit on the width to be listed in the SLHA file. \\

For our discussion, we pick out some decays of specific states which give some impression of the main differences compared to the MSSM. For instance, we have seen that for EP1 the neutralino is actually not the LSP. So, it should decay. This is exactly what we find:
\begin{lstlisting}[style=file,firstnumber=1495,title=\hspace{11cm}SPheno.spc.BLSSM (EP1)]
DECAY   1000022     3.45203467E-17   # Chi_1
#    BR          NDA  ID1   ID2
     1.00E+00    2    16    1000012   # BR(Chi_1 -> Fv_3 SvRe_1 ) 
\end{lstlisting}
The width is very small. The reason is that it is suppressed twice: by the small blino admixture to the neutralino and the small right fraction of the neutrino. \\

The second lightest neutralino consists mostly of a $B-L$ states, i.e the coupling to the second lightest Higgs (bilepton) is larger than the lightest one. This explains why $\tilde{\chi}_2$ mostly decays into the second lightest scalar but not the lightest one despite the larger phase space. 
\begin{lstlisting}[style=file,firstnumber=1498,title=\hspace{11cm}SPheno.spc.BLSSM (EP1)]
DECAY   1000023     1.51576185E-03   # Chi_2
#    BR         NDA    ID1       ID2
     1.74E-03   2      1000022   25   # BR(Chi_2 -> Chi_1 hh_1 )
     9.81E-01   2      1000022   35   # BR(Chi_2 -> Chi_1 hh_2 )
     1.01E-02   2      1000022   23   # BR(Chi_2 -> Chi_1 VZ ) 
\end{lstlisting}
\vspace{0.5cm}

Also the heavy neutrinos are expected to decay. Since these are mostly right-handed states the coupling to the lightest neutralino and sneutrino is much larger than for the light neutrino. Thus, the width of these states is much larger than the width of the lightest neutralino even if the involved particles seem to be similar. 
\begin{lstlisting}[style=file,firstnumber=1598,title=\hspace{11cm}SPheno.spc.BLSSM (EP1)]
DECAY       112     3.53565312E-02   # Fv_4
#    BR          NDA  ID1      ID2
     9.99E-01    2    1000022  1000012   # BR(Fv_4 -> Chi_1 SvRe_1 ) 
\end{lstlisting}
\vspace{0.5cm}

Another interesting topic in these models are the decays of the $Z'$: new decay channels can alter significantly the width of the $Z'$ and have an impact on the limits of $Z'$ masses  from collider searches \cite{Krauss:2012ku,Hirsch:2012kv}. However, for our point EP1 we see that 98\% of the final states are SM fermions. The reason is that the important channels in right-neutrinos are kinematically forbidden. 
\begin{lstlisting}[style=file,firstnumber=1433,title=\hspace{11cm}SPheno.spc.BLSSM (EP1)]
DECAY        31     2.05629381E+01   # VZp
#    BR        NDA  ID1        ID2
     1.50E-03  2    1000022    1000023   # BR(VZp -> Chi_1 Chi_2 )
     6.39E-04  2    1000022    1000035   # BR(VZp -> Chi_1 Chi_4 )
     1.05E-01  2         -1          1   # BR(VZp -> Fd_1^* Fd_1 )
     1.05E-01  2         -3          3   # BR(VZp -> Fd_2^* Fd_2 )
     1.05E-01  2         -5          5   # BR(VZp -> Fd_3^* Fd_3 )
     1.17E-01  2        -11         11   # BR(VZp -> Fe_1^* Fe_1 )
     1.17E-01  2        -13         13   # BR(VZp -> Fe_2^* Fe_2 )
     1.17E-01  2        -15         15   # BR(VZp -> Fe_3^* Fe_3 )
     2.64E-02  2         -2          2   # BR(VZp -> Fu_1^* Fu_1 )
     2.64E-02  2         -4          4   # BR(VZp -> Fu_2^* Fu_2 )
     2.62E-02  2         -6          6   # BR(VZp -> Fu_3^* Fu_3 )
     7.61E-02  2         12         12   # BR(VZp -> Fv_1 Fv_1 )
     7.61E-02  2         14         14   # BR(VZp -> Fv_2 Fv_2 )
     7.61E-02  2         16         16   # BR(VZp -> Fv_3 Fv_3 )
     1.19E-03  2         25         23   # BR(VZp -> hh_1 VZ )
     1.84E-02  2        -24         24   # BR(VZp -> VWm VWm^* ) 
\end{lstlisting}
\vspace{0.5cm}

This changes in EP2 where the $Z'$ is much heavier than the right neutrinos. Here, the BR in right neutrinos is about 10\% and much more important than the channels in sleptons which are also kinematically allowed for $M_{Z'} = 4.5$~TeV. 
\begin{lstlisting}[style=file,firstnumber=1432,title=\hspace{11cm}SPheno.spc.BLSSM (EP2)]
DECAY        31     4.18911354E+01   # VZp
#    BR          NDA ID1       ID2
     ...
     6.74E-02    2        12         12   # BR(VZp -> Fv_1 Fv_1 )
     6.74E-02    2        14         14   # BR(VZp -> Fv_2 Fv_2 )
     6.74E-02    2        16         16   # BR(VZp -> Fv_3 Fv_3 )
     1.01E-01    2       112        112   # BR(VZp -> Fv_4 Fv_4 )
     ...
     4.47E-03    2  -1000011    1000011   # BR(VZp -> Se_1^* Se_1 )
     4.32E-03    2  -1000013    1000013   # BR(VZp -> Se_2^* Se_2 )
     4.32E-03    2  -1000015    1000015   # BR(VZp -> Se_3^* Se_3 )
     1.33E-03    2  -2000011    2000011   # BR(VZp -> Se_4^* Se_4 )
     1.26E-03    2  -2000013    2000013   # BR(VZp -> Se_5^* Se_5 )
     1.26E-03    2  -2000015    2000015   # BR(VZp -> Se_6^* Se_6 )
     7.94E-03    2   1000012    5000012   # BR(VZp -> SvRe_1 SvIm_4 )
     6.82E-04    2   2000012    4000012   # BR(VZp -> SvRe_4 SvIm_1 )
     6.45E-04    2   2000014    4000014   # BR(VZp -> SvRe_5 SvIm_2 )
     6.45E-04    2   2000016    4000016   # BR(VZp -> SvRe_6 SvIm_3 )
     1.71E-03    2       -24         24   # BR(VZp -> VWm VWm^* ) 
\end{lstlisting}
\vspace{0.5cm}

We go back to EP1 and change topics a bit: we are no longer interested in new effects compared to the MSSM, but want to know how good the BR of the light Higgs reproduce the SM expectations. The width and BRs for this point calculated by \SPheno are
\begin{lstlisting}[style=file,firstnumber=1343,title=\hspace{11cm}SPheno.spc.BLSSM (EP1)]
DECAY        25     4.36597968E-03   # hh_1
#    BR          NDA   ID1   ID2
     2.12E-03    2     22    22   # BR(hh_1 -> VP VP )
     8.89E-02    2     21    21   # BR(hh_1 -> VG VG )
     1.84E-02    2     23    23   # BR(hh_1 -> VZ VZ )
     1.65E-01    2     24   -24   # BR(hh_1 -> VWm^* VWm_virt )
     2.34E-04    2     -3     3   # BR(hh_1 -> Fd_2^* Fd_2 )
     6.27E-01    2     -5     5   # BR(hh_1 -> Fd_3^* Fd_3 )
     2.52E-04    2    -13    13   # BR(hh_1 -> Fe_2^* Fe_2 )
     7.27E-02    2    -15    15   # BR(hh_1 -> Fe_3^* Fe_3 )
     2.42E-02    2     -4     4   # BR(hh_1 -> Fu_2^* Fu_2 ) 
\end{lstlisting}
These values have to be compared with those of the SM for a Higgs mass of 124.2~GeV. The numbers for the SM can be found online at
\begin{verbatim}
https://twiki.cern.ch/twiki/bin/view/LHCPhysics/CERNYellowReportPageBR3 
\end{verbatim}
and the different BRs and the total width are:
\begin{eqnarray*}
&\text{BR}(h\to b\bar b) = 5.89\cdot 10^{-1} \,,\hspace{1cm}
\text{BR}(h\to \tau\bar \tau) = 6.45\cdot 10^{-1} &\\
&\text{BR}(h\to \mu\bar \mu) = 2.24\cdot 10^{-4} \,,\hspace{1cm}
\text{BR}(h\to c\bar c) = 2.97\cdot 10^{-2} &\\
&\text{BR}(h\to s\bar s) = 2.51\cdot 10^{-1} \,,\hspace{1cm}
\text{BR}(h\to gg) = 8.62\cdot 10^{-2} &\\
&\text{BR}(h\to \gamma\gamma) = 2.28\cdot 10^{-3} \,,\hspace{1cm}
\text{BR}(h\to Z\gamma) = 1.47\cdot 10^{-3} &\\
&\text{BR}(h\to WW) = 2.02\cdot 10^{-1} \,,\hspace{1cm}
\text{BR}(h\to ZZ) = 2.45\cdot 10^{-2} &\\
\\
&\Gamma_h = 3.96~\text{MeV}&
\end{eqnarray*}
These BR are similar to the ones we got with \SPheno but they don't agree exactly. Also the expected width is smaller by about 10\% than the one calculated by \SPheno. The reason for this is the enhanced coupling of the Higgs to bottom quarks for this point which was already visible from the \HB blocks as discussed in the last subsection. 

\subsection{Flavour and precision observables}
The \SPheno modules written by \SARAH contain already out of the box the routines to calculate many quark and lepton flavour violating observables. In addition, also other observables like $(g-2)_l$ and $\delta\rho$ are calculated. We are going to start with a short discussion of the results which can be obtained just by running \SARAH and \SPheno out of the box. In a second step, I show how the \FlavorKit functionality  \cite{Porod:2014xia} can be used to implemented Wilson coefficients for new operators, and how to use these coefficients to calculate new observables with \SPheno.

\subsubsection{Observables out of the box}
To turn the calculation of low-energy observables on, the  Les Houches input file must contain:
\begin{lstlisting}[style=file,title=\hspace{11cm}LesHouches.in.BLSSM]
Block SPhenoInput   # SPheno specific input 
...
57 1               # Calculate low energy constraints  
\end{lstlisting}
In that case the QFV and LFV observables are given in the blocks {\tt FlavorKitQFV} and {\tt FlavorKitLFV}. $\delta\rho$, $(g-2)_l$ and EDMs are shown in {\tt SPhenoLowEnergy}. This block reads for EP1:
\begin{lstlisting}[style=file,firstnumber=964,title=\hspace{11cm}SPheno.spc.BLSSM (EP1)]
Block SPhenoLowEnergy # low energy observables 
      20    7.44947152E-16  # (g-2)_e
      21    3.18490399E-11  # (g-2)_mu
      22    9.02332214E-09  # (g-2)_tau
      23    0.00000000E+00  # EDM(e)
      24    0.00000000E+00  # EDM(mu)
      25    0.00000000E+00  # EDM(tau)
      39   -1.62703955E-04  # delta(rho) 
\end{lstlisting}
While the EDMs vanish because we didn't include CP violation, the other observables don't receive a large contribution in this model because of the heavy SUSY spectrum. So, let's turn to the flavour observables.\\

All LFV rates are smaller than $10^{-35}$ and can be interpreted as numerical zeros. For QFV there are, of course, the non-vanishing SM contributions which have to be taken into account as well. For QFV observables, \SPheno does not only give the absolute size of the observable like the corresponding branching ratio or mass splitting, but also the observable normalized to the SM expectation. For this purpose \SPheno actually calculates each observable internally twice. In the second calculation all non-SM contributions are dropped. It is more convenient to confront the value normalized to the SM prediction of \SPheno with exclusion limits from experiment. The reason is that by taking this ratio uncertainties, e.g. in the hadronic parameters, drop out. Also a constant shift just caused by missing higher order corrections in \SPheno does not lead to the false impression of a deviation from the SM as long as the ratio is close to 1. We see for our point that SUSY and $B-L$ contributions change the prediction of at most 1.5\% compared to SM expectation. Thus, this point is in total agreement with all limits. The reason for this are, of course, again the heavy sfermions in general and the weak coupling of the few light states. Even if these results are not very exciting I show the entire output of \SPheno for QFV observables to give an overview what is calculated:
\begin{lstlisting}[style=file,firstnumber=972,title=\hspace{11cm}SPheno.spc.BLSSM (EP1)]
Block FlavorKitQFV # quark flavor violating observables 
     200    3.17688828E-04  # BR(B->X_s gamma)
     201    1.00853596E+00  # BR(B->X_s gamma)/BR(B->X_s gamma)_SM
     300    6.58819362E-04  # BR(D->mu nu)
     301    9.99994713E-01  # BR(D->mu nu)/BR(D->mu nu)_SM
     400    6.23567236E-03  # BR(Ds->mu nu)
     401    9.99994126E-01  # BR(Ds->mu nu)/BR(Ds->mu nu)_SM
     402    6.08680525E-02  # BR(Ds->tau nu)
     403    9.98785723E-01  # BR(Ds->tau nu)/BR(Ds->tau nu)_SM
     500    5.76751780E-07  # BR(B->mu nu)
     501    9.99956660E-01  # BR(B->mu nu)/BR(B->mu nu)_SM
     502    1.27192820E-04  # BR(B->tau nu)
     503    9.91058613E-01  # BR(B->tau nu)/BR(B->tau nu)_SM
     600    7.06814633E-01  # BR(K->mu nu)
     601    9.99999621E-01  # BR(K->mu nu)/BR(K->mu nu)_SM
     602    2.43654009E-05  # R_K = BR(K->e nu)/(K->mu nu)
     603    2.39724507E-05  # R_K^SM = BR(K->e nu)_SM/(K->mu nu)_SM
    1900    1.79264760E+01  # Delta(M_Bs)
    1901    1.00256419E+00  # Delta(M_Bs)/Delta(M_Bs)_SM
    1902    4.00744431E-01  # Delta(M_Bd)
    1903    1.00272001E+00  # Delta(M_Bd)/Delta(M_Bd)_SM
    4000    2.52888015E-15  # BR(B^0_d->e e)
    4001    1.01492301E+00  # BR(B^0_d->e e)/BR(B^0_d->e e)_SM
    4002    7.84581655E-14  # BR(B^0_s->e e)
    4003    1.01404869E+00  # BR(B^0_s->e e)/BR(B^0_s->e e)_SM
    4004    1.08030916E-10  # BR(B^0_d->mu mu)
    4005    1.01492301E+00  # BR(B^0_d->mu mu)/BR(B^0_d->mu mu)_SM
    4006    3.35173125E-09  # BR(B^0_s->mu mu)
    4007    1.01404869E+00  # BR(B^0_s->mu mu)/BR(B^0_s->mu mu)_SM
    4008    2.26121864E-08  # BR(B^0_d->tau tau)
    4009    1.01492426E+00  # BR(B^0_d->tau tau)/BR(B^0_d->tau tau)_SM
    4010    7.10840026E-07  # BR(B^0_s->tau tau)
    4011    1.01404999E+00  # BR(B^0_s->tau tau)/BR(B^0_s->tau tau)_SM
    5000    1.64101883E-06  # BR(B-> s e e)
    5001    9.91378447E-01  # BR(B-> s e e)/BR(B-> s e e)_SM
    5002    1.59074205E-06  # BR(B-> s mu mu)
    5003    9.91246240E-01  # BR(B-> s mu mu)/BR(B-> s mu mu)_SM
    6000    1.10869943E-07  # BR(B -> K mu mu)
    6001    9.98828312E-01  # BR(B -> K mu mu)/BR(B -> K mu mu)_SM
    7000    4.14273034E-05  # BR(B->s nu nu)
    7001    9.99749124E-01  # BR(B->s nu nu)/BR(B->s nu nu)_SM
    7002    1.91819884E-06  # BR(B->D nu nu)
    7003    9.99750026E-01  # BR(B->D nu nu)/BR(B->D nu nu)_SM
    8000    1.30902890E-10  # BR(K^+ -> pi^+ nu nu)
    8001    9.99859854E-01  # BR(K^+ -> pi^+ nu nu)/BR(K^+ -> pi^+ nu nu)_SM
    8002    3.06205871E-11  # BR(K_L -> pi^0 nu nu)
    8003    9.99751687E-01  # BR(K_L -> pi^0 nu nu)/BR(K_L -> pi^0 nu nu)_SM
    9100    2.08773176E-15  # Delta(M_K)
    9102    1.00002301E+00  # Delta(M_K)/Delta(M_K)_SM
    9103    3.30952503E-03  # epsilon_K
    9104    1.00227602E+00  # epsilon_K/epsilon_K^SM 
\end{lstlisting}
\vspace{0.5cm}

I have claimed above that we can neglect the neutrino Yukawa couplings for most studies because they have hardly an impact on the calculation since they are highly constrained. To show this, we can try what happens if we turn on this coupling by using arbitrary values of $O(0.01)$ for some entries of $Y_\nu$: 
\begin{lstlisting}[style=file,firstnumber=21,title=\hspace{11cm}LesHouches.in.BLSSM]
Block YVIN    #  
1 1   0.010000E+00         # Yv(1,1)
1 2   0.001000E+00         # Yv(1,2)
1 3   0.001000E+00         # Yv(1,3)
2 1   0.010000E+00         # Yv(2,1)
2 2   0.010000E+00         # Yv(2,2)
2 3   0.001000E+00         # Yv(2,3)
3 1   0.000000E+00         # Yv(3,1)
3 2   0.001000E+00         # Yv(3,2)
3 3   0.010000E+00         # Yv(3,3)
\end{lstlisting}
We find only a small impact on most masses and the QFV observables. However, the light neutrino masses are much too large
\begin{lstlisting}[style=file,firstnumber=315,title=\hspace{13cm}SPheno.spc.BLSSM]
Block MASS  # Mass spectrum
  ...
        12     5.40862949E-04  # Fv_1
        14     2.46024560E-03  # Fv_2
        16     5.03020429E-03  # Fv_3
  ...
\end{lstlisting}
Thus, it would be necessary to make the right-neutrino much heavier to get a kind of seesaw suppression. However, also some 
flavour observables, in particular $\mu$--$e$ conversion, are already in conflict with the experimental values shown in Tab.~\ref{tab:LFVbounds}. 
\begin{lstlisting}[style=file,firstnumber=1027,title=\hspace{13cm}SPheno.spc.BLSSM]
Block FlavorKitLFV # lepton flavor violating observables 
     701    2.12125239E-14  # BR(mu->e gamma)
     702    8.56147942E-18  # BR(tau->e gamma)
     703    1.01032494E-17  # BR(tau->mu gamma)
     800    1.20856362E-11  # CR(mu-e, Al)
     801    2.17524735E-11  # CR(mu-e, Ti)
     802    2.94406896E-11  # CR(mu-e, Sr)
     803    3.31181138E-11  # CR(mu-e, Sb)
     804    1.78405656E-11  # CR(mu-e, Au)
     805    1.67814615E-11  # CR(mu-e, Pb)
     901    1.28640195E-12  # BR(mu->3e)
     902    6.31256061E-16  # BR(tau->3e)
     903    9.44251544E-15  # BR(tau->3mu)
     904    4.22162097E-16  # BR(tau- -> e- mu+ mu-)
     905    1.23482830E-14  # BR(tau- -> mu- e+ e-)
     906    3.37021529E-24  # BR(tau- -> e+ mu- mu-)
     907    2.29773463E-23  # BR(tau- -> mu+ e- e-)
    1001    1.94197824E-17  # BR(Z->e mu)
    1002    6.42222866E-19  # BR(Z->e tau)
    1003    1.23585715E-18  # BR(Z->mu tau)
    1101    1.94824828E-16  # BR(h->e mu)
    1102    1.83418137E-15  # BR(h->e tau)
    1103    1.40346574E-15  # BR(h->mu tau)
    2001    9.16839984E-19  # BR(tau->e pi)
    2002    1.35797684E-19  # BR(tau->e eta)
    2003    1.49779387E-19  # BR(tau->e eta')
    2004    1.78707828E-18  # BR(tau->mu pi)
    2005    2.64136021E-19  # BR(tau->mu eta)
    2006    2.89107562E-19  # BR(tau->mu eta')
\end{lstlisting}
\begin{table}[tb!]
\centering
\begin{tabular}{|c|c|}
\hline
LFV Process & Bound   \\
\hline
    $\mu \rightarrow  e \gamma$ & $5.7\times 10^{-13}$~\cite{Adam:2013mnn} \\
    $\tau \to e \gamma$ & $3.3 \times 10^{-8}$~\cite{Aubert:2009ag}\\
    $\tau \to \mu \gamma$ & $4.4 \times 10^{-8}$~\cite{Aubert:2009ag}\\
    $\mu \rightarrow e e e$ &  $1.0 \times 10^{-12}$~\cite{Bellgardt:1987du}\\
    $\tau \rightarrow \mu \mu \mu$ & $2.1\times10^{-8}$~\cite{Hayasaka:2010np}\\
    $\tau^- \rightarrow e^- \mu^+ \mu^-$ &  $2.7\times10^{-8}$~\cite{Hayasaka:2010np}\\
    $\tau^- \rightarrow \mu^- e^+ e^-$ &  $1.8\times10^{-8}$~\cite{Hayasaka:2010np} \\
    $\tau \rightarrow e e e$ & $2.7\times10^{-8}$~\cite{Hayasaka:2010np} \\
    $\mu^-, \mathrm{Ti} \rightarrow e^-, \mathrm{Ti}$ &  $4.3\times 10^{-12}$~\cite{Dohmen:1993mp}\\
    $\mu^-, \mathrm{Au} \rightarrow e^-, \mathrm{Au}$ & $7\times 10^{-13}$~\cite{Bertl:2006up}  \\
\hline
\end{tabular}
\caption{Current experimental bounds for some low-energy LFV observables.}
\label{tab:LFVbounds}
\end{table}
This brings us already to our end of the short excursion to flavour observables which are included in \SARAH and \SPheno. 
I discuss now what can be done if your favourite observable is not yet calculated by \SPheno.

\subsubsection{Adding new observables}
I show now how the \FlavorKit functionality can be used to implement new observables in \SPheno. To make it even more interesting we choose a process for which \SPheno doesn't even now the Wilson coefficients. Namely, we decide to study the flavour violating, radiative decays of the top quark:
\begin{equation}
 \text{BR}(t \to q \gamma) \hspace{1cm} \text{with} \hspace{0.2cm} q=u,c
\end{equation}
The process is interesting, because it is highly suppressed in the SM by GIM but can receive large contributions in SUSY models \cite{Liu:2004qw}. 
The transition amplitude can be expressed by \cite{AguilarSaavedra:2002ns} 
\begin{equation}
\mathscr{M} = \bar{u}(p_q) [i \sigma^{\mu\nu} q_\nu (A_\gamma + B_\gamma \gamma_5)] u(p_t) \epsilon_\mu^*(q)
\end{equation}
with the quark momenta $p_q$ and $p_t$. The partial width can be expressed using $A_\gamma$ and $B_\gamma$
as
\begin{equation}
\label{eq:partialWidth}
\Gamma(t \to q \gamma) = \frac{1}{\pi}\left(\frac{m_t^2-m_q^2}{m_t}\right)^3 \left(|A_\gamma|^2 + |B_\gamma|^2 \right) 
\end{equation}
and the BRs we are interested in are
\begin{equation}
\text{BR}(t \to q \gamma) = \Gamma(t \to q \gamma)/\Gamma_{tot}
\end{equation}
with the total width $\Gamma_{tot}$ of the top quark. \\
One can also work in the chiral basis using the effective Lagrangian 
\begin{equation}
\La = A_R \mathscr{O}^\gamma_R + A_L \mathscr{O}^\gamma_L 
\end{equation}
with the operators
\begin{align}
 \mathscr{O}^\gamma_L =& \epsilon_\mu \bar{u}(p_q)[ i \sigma^{\mu\nu} p_\nu P_R ]u(p_t)\\
 \mathscr{O}^\gamma_R =& \epsilon_\mu \bar{u}(p_q)[ i \sigma^{\mu\nu} p_\nu P_L ]u(p_t)
\end{align}
The relations between the coefficients are just
\begin{align}
A_\gamma =& A_L + A_R \\
B_\gamma =& A_L - A_R
\end{align}
We are going to make use of this relation in the following by first calculating $A_L$, $A_R$ and translating them later into $A_\gamma$, $B_\gamma$ to calculate the partial width according to eq.~(\ref{eq:partialWidth}). \\

Our to-do list is the following:
\begin{enumerate}
 \item Get the generic expressions for $A_L$, $A_R$
 \item Implement those in \SARAH 
 \item Implement the formula for the BRs in \SARAH
 \item Make sure that all information is found by \SARAH and included in the \SPheno output
\end{enumerate}
Even of this sounds like a lot of work involving several loop  calculations and hacking some code, this is not the case at all. Each step is fully automatized. The user just has to create three small input files. The first input file is needed for \PreSARAH. \PreSARAH is a \Mathematica package which calculates one-loop amplitudes in a generic way using \FeynArts and \FormCalc and extracts the Wilson coefficients for the operators the user needs. The results are then translated into input files which can be used by \SARAH. So, we create the file {\tt TopPhotonQ.m} with the following content:
\begin{lstlisting}[style=file,title=\hspace{13cm} TopPhotonQ.m]
NameProcess="TopPhotonQ";

ConsideredProcess = "2Fermion1Vector";
FermionOrderExternal={1,2};
NeglectMasses={3};  


ExternalFields= {bar[TopQuark], TopQuark, Photon};
CombinationGenerations = {{3,2}, {3,1}};


AllOperators={
   {OTgQSL,Op[7] Pair[ec[3],k[1]]}, 
   {OTgQSR,Op[6] Pair[ec[3],k[1]]}
};

OutputFile = "TopPhotonQ.m";

Filters = {};
\end{lstlisting}
This defines a new process called {\tt TopPhotonQ} which involves two fermions and one vector boson ({\tt ConsideredProcess}). The Fierz ordering of the external states is defined via {\tt FermionOrderExternal}. \PreSARAH is absolutely agnostic concerning particle physics and it tries always to calculate the most general case. For us, this would mean that the results are a function of three masses: those of the fermions and the one of the vector boson. 
To make sure that in the generic expression the photon mass doesn't show  up, we put {\tt NeglectMasses=\{3\}}. The reason is that the photon is the third particle defined in the list of all external states. The information in {\tt ExternalFields} is not used at all by \PreSARAH. \PreSARAH just includes the information in the output used for \SARAH. \SARAH knows then what a 'top quark' and a 'photon' is. Also 
{\tt CombinationGenerations} is not used by \PreSARAH but just passed to \SARAH and \SPheno. This list contains all combinations of external generation indices for which the coefficients are later calculated by \SPheno. We need here (3,2) for top-charm and (3,1) for top-up operators. The two operators from above are called {\tt OTgQSL} ($A_L$) and {\tt OTgQSR} ($A_R$) and their expressions are given in \FeynArts syntax using {\tt ec[3]} for the helicity of the third particle and {\tt k[1]} as momentum of the first particle. The meaning of these symbols and how to use for instance Dirac matrices in the definition of operators is explained in the \FeynArts manual and briefly summarized in the \FlavorKit reference as well. Note, in the case of Wilson coefficients for QFV observables, \SARAH will automatically generate for each Wilson coefficient {\tt X} another coefficient {\tt XSM} which just includes SM contributions.
Finally, we tell \PreSARAH that the output should be written into the file {\tt TopPhotonQ.m} and we don't want to filter out any diagrams ({\tt Filters=\{\}}). 

We run now this file in \Mathematica with \PreSARAH similar as we run models with \SARAH:
\begin{lstlisting}[style=mathematica]
<<$PATH/PreSARAH/PreSARAH.m
Start["TopPhotonQ.m"];
\end{lstlisting}
The output file is located in the \PreSARAH output directory and has to be copied to the \FlavorKit directory of \SARAH. Because of obvious reasons we choose the {\tt QFV/Operators} subdirectory. 
\begin{lstlisting}[style=terminal]
$ cp $PATH/PreSARAH/Output/TopPhotonQ.m $PATH/SARAH/FlavorKit/QFV/Operators/
\end{lstlisting}
If we would have put the file in {\tt LFV/Operators}, the coefficients would have been calculated at $Q=M_Z$ instead of $Q=160$~GeV. We are already done with steps (1) and (2) of the to-do list. Now we have to teach \SPheno have to calculate the branching ratios. For this purpose we need two files which we have to put into
\begin{lstlisting}
$PATH/SARAH/FlavorKit/QFV/Observables/
\end{lstlisting}
The first file is a steering file in \Mathematica syntax. It defines the name for the process used by \SARAH internally, what operators are needed to calculate the process and what observables should show up in the spectrum file later. 
\begin{lstlisting}[style=file,title=\hspace{13cm} TqGamma.m]
NameProcess = "TqGamma";
NameObservables = {{BrTuGamma, 210, "BR(t->u gamma)"},
    {ratioTuGamma, 211, "BR(t->u gamma)/BR(t->c gamma)_SM"},
    {BrTcGamma, 212, "BR(t->c gamma)"},
    {ratioTcGamma, 213, "BR(t->c gamma)/BR(t->c gamma)_SM"},
};

NeededOperators = {OTgQSL, OTgQSR,
                   OTgQSLSM, OTgQSRSM};

Body = "TqGamma.f90"; 
\end{lstlisting}
{\tt NameObservables} is an array containing all observables which should show up in the spectrum file. The first part of each entry gives the name of a variable, the second one the number used in the Les Houches block {\tt FlavorKitQFV} and the third one the comment which is used in the Les Houches file to make clear to  which variable the shown number belongs. All operators, which we need to calculate the observables, are given in {\tt NeededOperators}. As mentioned above, \SARAH creates not only routines to calculate the Wilson coefficients including all new physics, but also coefficients in the SM limit. For our purpose, we need both sets of coefficients, because we want not only to calculate the BR but also normalize it to the SM expectation calculated under the same assumptions. The name of another file is  given at the end of the steering file. This file contains the 'body' of the \Fortran routine to calculate the observable {\tt TqGamma.f90}. With 'body' I mean that the head of the routine is automatically generated by \SARAH. The user can start at the stage of initializing variables needed for the calculation of the observable. The entire file  {\tt TqGamma.f90} looks as follows:
\begin{lstlisting}[style=file,title=\hspace{13cm} TqGamma.f90]
Real(dp) :: width, widthSM, norm
Complex(dp) :: Agamma, Bgamma, AgammaSM, BgammaSM
Integer :: i1, gt1, gt2

Do i1=1,2

If (i1.eq.1) Then         ! t -> u gamma
 gt1 = 3
 gt2 = 1
Elseif (i1.eq.2) Then     ! t -> c gamma
 gt1 = 3
 gt2 = 2
End if

Agamma=OTgQSL(gt1,gt2)+OTgQSR(gt1,gt2)
Bgamma=OTgQSL(gt1,gt2)-OTgQSR(gt1,gt2)
AgammaSM=OTgQSLSM(gt1,gt2)+OTgQSRSM(gt1,gt2)
BgammaSM=OTgQSLSM(gt1,gt2)-OTgQSRSM(gt1,gt2)

norm=1/Pi*((mf_u(gt1)**2-mf_u(gt2)**2)/(2._dp*mf_u(gt1)))**3

width=norm*(Abs(Agamma)**2+Abs(Bgamma)**2)
widthSM=norm*(Abs(AgammaSM)**2+Abs(BgammaSM)**2)



If (i1.eq.1) Then
 BrTuGamma = width/gTFu(3)
 ratioTuGamma=width/widthSM
Elseif (i1.eq.2) Then 
 BrTcGamma = width/gTFu(3)
 ratioTcGamma=width/widthSM
End if

End do
\end{lstlisting}
In the first three lines we initialize a few local variables we need. 
In the entire routine we make a loop over two iterations. In the first iteration we pick up the generation indices (3,1) for the top-u decay, in the second one the indices (3,2). These indices are saved in the variables {\tt gt1} and {\tt gt2} which are then used as argument for the  coefficients. We first express $A_\gamma$ and $B_\gamma$ by $A_L$ and $A_R$ and do the same with the coefficients for the SM part only. In line 20, we calculate the overall normalization factor $((m_t^2-m_q^2)/(2 m_t))^3/\pi$ and plug everything into eq.~(\ref{eq:partialWidth}). The variable {\tt width} in line 22 is the partial width including all contributions, {\tt widthSM} in line 23 is the partial width only with SM contributions. Our final observables are then calculated using the total width of the top ({\tt gTFu(3)}) or taking the ratio of both widths. We are now done with step (3) of the to-do list. Step (4) happens automatically if all files have been put into the correct directories. We can now generate the \SPheno code again.
\begin{lstlisting}[style=mathematica]
<<$PATH/SARAH.m
Start["B-L-SSM"];
MakeSPheno[ReadLists->True];
\end{lstlisting}
To save time, I used the option {\tt ReadLists->True}, i.e. \SARAH reads the list with all analytical results from the previous run. The output has to be copied again in the {\tt BLSSM} subdirectory of \SPheno and can be compiled as usual. After running \SPheno with the input for EP1 we get the new entries in the block {\tt FlavorKitQFV}
\begin{lstlisting}[style=file,firstnumber=972,title=\hspace{11cm}SPheno.spc.BLSSM (EP1)]
Block FlavorKitQFV # quark flavor violating observables 
     200    3.17688828E-04  # BR(B->X_s gamma)
     210    2.98458324E-16  # BR(t->u gamma)
     211    7.63379583E-01  # BR(t->u gamma)/BR(t->c gamma)_SM
     212    4.32829318E-14  # BR(t->c gamma)
     213    7.63840319E-01  # BR(t->c gamma)/BR(t->c gamma)_SM
 ...
\end{lstlisting}
Obviously, the numbers and comments show up as expected. We see some deviations from the SM prediction. However, this is still far away from the experimental limits which still allow branching ratios in the percent range \cite{Abe:1997fz}. To get large effects of this size one might search for points with light stops for instance. 

One can also compare these numbers with the SM prediction in literature. The partial with is strongly suppressed by GIM mechanism and therefore rather sensitive on the values of the CKM matrix as well as on the running quark masses in the loop. Therefore, the predicted rates in the SM come with a sizable, theoretical error.  In Ref.~\cite{AguilarSaavedra:2002ns} SM prediction for  $\text{BR}(t \to q \gamma)$  and other flavour violating top decays were given. The number for the radiative decay into a charm quark reads
\begin{equation}
 \text{BR}(t \to c \gamma) = (4.6 \substack{+1.2 \\-1.0} \pm 0.4 \substack{+ 1.6 \\ -0.5 } ) \times 10^{-14}
\end{equation}
and agrees with our calculation within errors. $\text{BR}(t \to u \gamma)$ is supposed to be suppressed by a factor $|V_{ub}/V_{cb}|^2 \simeq 7.9 \cdot 10^{-3}$ where $V_{qb}$ are the entries of the CKM matrix. That's also similar to what we find. 

\subsection{Getting the fine-tuning}
One of the main motivation for SUSY was naturalness: it solves the hierarchy problem of the SM by stabilizing the unprotected Higgs mass. However, with the more and more severe limits on the SUSY masses, the question about the fine-tuning rises again. \SARAH and \SPheno provide functions to calculate the fine-tuning according to eq.~(\ref{eq:measure}). The user can choose the list of parameters which should be included in the fine-tuning calculation. For this purpose, {\tt SPheno.m} of the model has to be extended by:
\begin{lstlisting}[style=file,firstnumber=99,title=\hspace{13cm}SPheno.m]
IncludeFineTuning = True;
FineTuningParameters={
  {m0,1/2}, {m12,1}, {Azero,1}, 
  {\[Mu],1}, {B[\[Mu]],1}, {MuP,1}, {B[MuP],1}
}; 
\end{lstlisting}
The list {\tt FineTuningParameters} contains the parameters which are varied at the GUT and a numerical coefficient to 'normalize' the fine-tuning. The factor $1/2$ for $m_0$ is there because at the GUT scale the boundary conditions are in terms of $m_0^2$. We see that the fine-tuning can not only be calculated with respect to the input parameters defined in {\tt MINPAR}. Also other parameters which for instance are fixed by the tadpoles equations can be included. In principle, one can also calculate the fine-tuning with respect to SM parameters like the top Yukawa coupling ({\tt Yu[3,3]}) or the strong interaction ({\tt g3}) by including those in the list above. \\

After editing {\tt SPheno.m}, it is necessary to reproduce the \SPheno code with \SARAH and to compile the new version. Exactly the same steps are in sec.~\ref{sec:getSPheno} are used for that. When \SPheno is compiled, the fine-tuning is calculated and included in spectrum file if the corresponding flag is set in the Les Houches input file:
\begin{lstlisting}[style=file,title=\hspace{12cm}LesHouches.in.BLSSM]
Block SPhenoInput   # SPheno specific input 
...
550 1.              # Calculate Fine-Tuning  
\end{lstlisting}
Running our point EP1 with the new version of {\tt SPhenoBLSSM}, we find before the output of the decays start the block {\tt FineTuning}. This block contains the following entries:
\begin{lstlisting}[style=file,firstnumber=1112,title=\hspace{12cm}SPheno.spc.BLSSM\_EP1]
Block FineTuning #  
       0    1.30058900E+03  #  Overall FT 
       1    1.28554566E+02  # m0
       2    1.30058900E+03  # m12
       3    1.30781970E+02  # Azero
       4    1.25153715E+03  # \[Mu]
       5    1.04705641E+01  # B[\[Mu]]
       6    5.17087156E+02  # MuP
       7    4.87134742E+02  # B[MuP]       
\end{lstlisting}
The overall fine-tuning is given n the first entry coming with number {\tt 0}. All other entries list the fine-tuning with respect to the different parameters. This makes is obvious what parameters contribute mostly to the fine-tuning. In our example these are mainly $M_{1/2}$ and $\mu$ which have a similar fine-tuning. Moreover, one sees that even the additional parameters from the $B-L$ sector can have quite some impact on the fine-tuning. The reason is that the tadpole equations are coupled because of gauge-kinetic mixing. We can check this assumption by using the flag 
\begin{lstlisting}[style=file,title=\hspace{12cm}LesHouches.in.BLSSM]
Block SPhenoInput   # SPheno specific input 
..
60 0               # Include possible, kinetic mixing 
\end{lstlisting}
to turn off gauge-kinetic mixing. The impact on the overall fine-tuning is moderately small. However, the contributions of $\mu'$ and $B_\mu'$ do vanish in this limit as expected. Also the fine-tuning with respect to $M_{1/2}$ becomes smaller because the off-diagonal gauge couplings and gaugino do not further contribute to the running of the gauginos.
\begin{lstlisting}[style=file,firstnumber=1112]
Block FineTuning #  
       0    1.27402534E+03  #  Overall FT 
       1    1.60632359E+01  # m0
       2    1.07256471E+03  # m12
       3    2.11712917E+02  # Azero
       4    1.27402534E+03  # \[Mu]
       5    1.14380486E+01  # B[\[Mu]]
       6    1.05503076E-08  # MuP
       7    3.38297907E-09  # B[MuP]
\end{lstlisting}

\section{Example -- Part IV: Higgs constraints, vacuum stability, dark matter and collider studies}
\label{sec:example_tools}
\subsection{Checking Higgs constraints with \HB and \HS}
\label{sec:HBHS}
\HB and \HS are dedicated tools to study the Higgs properties of a given parameter point in a particular model. While \HB checks the parameter point against exclusion limits from Higgs searches, \HS gives a $\chi^2$ value to express how good the point reproduces the Higgs measurements. In general, \HS and \HB can handle different inputs: either the cross sections for all necessary processes can be given at the parton or hadron level, or the effective couplings of the Higgs states to all SM particles are taken as input. In addition, the masses and widths of all CP even and odd as well as charged Higgs states are always needed. 
\SPheno provides all input for the effective coupling approach. The information is given in the SLHA spectrum file and in addition in separated files (called {\tt MH\_GammaTot.dat}, {\tt  MHplus\_GammaTot.dat}, {\tt BR\_H\_NP.dat}, {\tt BR\_Hplus.dat}, {\tt BR\_t.dat}, {\tt effC.dat}). While SLHA files can be used with \HB for models with up to five neutral scalars, the  separated files can be used with even up to nine neutral and nine charged scalars. Since the second input works in more cases I'm going to concentrate on it. First, I discuss how exclusion limits are checked with \HB, afterwards I show the usage of \HS. 

\subsubsection{\HB}
In the same directory in which the \SPheno spectrum file is located, also all other input files for \HB and \HS are saved by \SPheno. The (relative) path to this directory has to be given as last argument to \HB when executing it. Thus, working from the directory {\tt \$PATH}, \HB is started via:
\begin{lstlisting}[style=terminal]
$ ./HiggsBounds/HiggsBounds LandH effC 6 1 SPHENO/
\end{lstlisting}
From other directories, one can use absolute paths:
\begin{lstlisting}[style=terminal]
$ $PATH/HiggsBounds/HiggsBounds LandH effC 6 1 $PATH/SPHENO/
\end{lstlisting}
The other arguments are the data which should be used. Here, we have chosen {\tt LandH} which incorporated data from LEP and hadron colliders. Other options would be {\tt onlyL} for only LEP data, or {\tt onlyH} for only Tevatron and LHC data, or {\tt onlyP} for only data which has been published. Then, we turn on the effective coupling input {\tt effC} via separated files. The other possible option with the data provided by \SPheno would be {\tt SLHA} which uses also the effective coupling approach. The number of neutral scalar \footnote{Sum of CP even and physical CP odd scalars, i.e. we have 4+2=6 in the \BLSSM. In the case of CP violation, the number of physical mass eigenstates has to be used.} and charged scalars are given as integer. \HB checks all files for consistency and if no problem appears, it writes the results to a file called {\tt HiggsBounds\_results.txt} in the same directory where the input is located. For our standard point EP1 the results look as follows
\begin{lstlisting}[mathescape,style=file,title=\hspace{12cm}HiggsBounds\_results.txt]
 # generated with HiggsBounds version 4.1.3 on 22.01.2015 at 08:41
 # settings: LandH, effC
 #
 # column abbreviations
 #   n          : line id of input
 #   Mh(i)      : Neutral Higgs boson masses in GeV
 #   Mhplus(i)  : Charged Higgs boson masses in GeV
 #   HBresult   : scenario allowed flag    (1: allowed,                      0: excluded, -1: unphysical)
 #   chan       : most sensitive channel (see below).                  chan=0 if no channel applies
 #   obsratio   : ratio [sig x BR]_model/[sig x BR]_limit             (<1: allowed, >1: excluded)
 #   ncomb      : number of Higgs bosons combined in                    most sensitive channel
 #
 # channel numbers used in this file
 #         682 : (p p)->h1->Z Z-> l l l l (low mass)                    where h1 is SM-like (CMS-PAS-HIG-13-002)
 # (for full list of processes, see Key.dat)
 #
 #cols:  n           Mh(1)           Mh(2)           Mh(3)           
         1         124.183         355.296         3064.08         
|\Suppressnumber|
 $\hookrightarrow$      Mh(4)           Mh(5)           Mh(6)           Mhplus(1)  
 $\hookrightarrow$     3961.50         3062.15         3095.83         3063.61         
                   
                   
 $\hookrightarrow$       HBresult  chan    obsratio     ncomb 
 $\hookrightarrow$       1          682    0.525609         1  |\Reactivatenumber|
 \end{lstlisting}
All information to understand the output is already given in the file: \HB finds that the strongest constraints come from $p p \to h_1 \to Z Z \to 4l$ at CMS ({\tt chan=682}) but the rate normalized to the exclusion limit ({\tt obsratio}) is smaller than 1, i.e. the point is allowed ({\tt HBresult=1}). A list with all  processes which are implemented and which were checked is also written to {\tt Key.dat} in the same directory. \\
We are far away from any exclusion limit, i.e. we don't have to worry about small uncertainties in the Higgs masses because they won't change the overall result. For point which are closer to the border, one has to think more about this. In these cases it might be helpful to provide an input file {\tt MHall\_uncertainties.dat} which includes the uncertainties in the Higgs mass calculation. I'll give more details about this in the \HS part. If this file is provided, \HB runs several times varying all Higgs masses in the range of their uncertainty and checks for the strongest constraints. 
 
\paragraph*{Checking light singlet}
We can also run the point EP2 which has a light singlet of just 79 GeV and find that also this point is allowed by all Higgs searches because of the highly reduced coupling of this scalar to SM particles:  
\begin{lstlisting}[style=file,title=\hspace{12cm}HiggsBounds\_results.txt]
# channel numbers used in this file
 #           1 : (e e)->(h1)Z->(b b-bar)Z                               (hep-ex/0602042, table 14b (LEP))
 # (for full list of processes, see Key.dat)
 #
 #cols:  n          ...  HBresult  chan    obsratio     ncomb
 #
         1          ...        1     1    0.505675         1
\end{lstlisting}
The most dominant search channel comes from LEP, but the rate is just about half of the one needed to rule this point out.

\subsubsection{\HS}
\label{sec:HS}
\HS is the complement to \HB and checks how good a point reproduces the Higgs mass and rate measurements. The syntax is very similar to \HB and to run it with the data for our standard point we have to call from the directory {\tt \$PATH}
\begin{lstlisting}[style=terminal]
$ ./HiggsSignals/HiggsSignals latestresults peak 2 effC 6 1 SPHENO/ 
\end{lstlisting}
It would be possible to use also here absolute paths. The first three arguments are different compared to \HS and have the following meaning: (i) what experimental data should be used (refers to the corresponding sub-directory in {\tt \$PATH/HIGGSSIGNALS/Expt\_tables/}), (ii) the $\chi^2$ method ({\tt peak} for peak centred, {\tt mass} for mass centred, or {\tt both}) \footnote{It is recommended by the \HS authors at the moment to use {\tt peak} because for the mass centred method not much data is provided by ATLAS and CMS.}   (iii) parametrization of the Higgs mass uncertainty ({\tt 1}: box, {\tt 2}: Gaussian; {\tt 3}: box and Gaussian). The results are written into the file {\tt HiggsBounds\_results.txt} which reads for EP1:
\begin{lstlisting}[mathescape,style=file,title=\hspace{12cm}HiggsSignals\_results.txt]
 # generated with HiggsSignals version 1.2.0 on 22.01.2015 at 11:07
 # settings: latestresults, effC, peak, gaussian
 #
 # column abbreviations
 #   n          : line id of input
 #   Mh(i)      : Neutral Higgs boson masses in GeV
 #   Mhplus(i)  : Charged Higgs boson masses in GeV
 #   csq(mu)    : Chi^2 from the signal strengths observables
 #   csq(mh)    : Chi^2 from the Higgs mass observables
 #   csq(tot)   : total Chi^2
 #   nobs(mu)   : number of signal strength observables
 #   nobs(mh)   : number of Higgs mass observables
 #   nobs(tot)  : total number of observables
#   Pvalue     : Probability, given csq(tot) and ndf=nobs(tot)-  0
 #
 #cols:  n           Mh(1)           Mh(2)           Mh(3)       
              1     124.183         355.296         3064.08         
|\Suppressnumber|             
 $\hookrightarrow$     Mh(4)           Mh(5)           Mh(6)       Mhplus(1)         
 $\hookrightarrow$     3961.50         3062.15         3095.83         3063.61         
    
 $\hookrightarrow$    csq(mu)         csq(mh)         csq(tot)      nobs(mu) 
 $\hookrightarrow$    124.810         5.34802         130.158        80     
 
 
 $\hookrightarrow$     nobs(mh)    nobs(tot)          Pvalue
 $\hookrightarrow$     4           84                 0.933103E-03 |\Reactivatenumber|
\end{lstlisting}
Also, the \HS output is rather self-explaining. The important numbers are the $\chi_\mu^2$ ({\tt csq(mu)}) for the Higgs rates, the $\chi_m^2$ ({\tt csq(mh)}) for the Higgs mass  and the combined $\chi_{tot}^2$ ({\tt csq(tot)}). The combined one is also translated into a $p$-value ({\tt Pvalue})\footnote{The $p$-value in this context is the ratio of $\chi_{tot}^2$ divided by the numbers of degrees of freedom (ndf). For ndf \HS takes the numbers of observables. To change this and to define the number of free parameters in the model, {\tt Nparam} has to be set in the file {\tt usefulbits\_HS.f90} in the \HS source code.}.
However, one warning appears in the terminal when running \HS in that way:
\begin{lstlisting}[style=terminal]
Optional datafile SPHENO/MHall_uncertainties.dat not found. Using default values. 
\end{lstlisting}
That means that the file {\tt MHall\_uncertainties.dat} was not found. That's not surprising because it wasn't created by \SPheno. This file contains the theoretical uncertainty of the Higgs mass prediction. Thus, to produce this file some estimate of the size of missing higher order corrections to the Higgs masses is needed. This is something what \SPheno can't do automatically at the moment. However, \HS assumes no theoretical uncertainty if the file is missing. That's, of course, unrealistic. The theoretical uncertainty for the corrections included in the \SARAH-\SPheno interface is expected to be similar to the one in the MSSM using standard two-loop corrections. Thus, we put 3.0~GeV for the two light scalars and 1.0 GeV for all others\footnote{An estimate for this uncertainty could be obtained by varying the renormalization scale in \SPheno in the range $[Q/2,2 Q]$ and checking the impact on the masses.}. We create {\tt MHall\_uncertainties.dat} and put it in the \SPheno directory where also all other input files for \HB/\HS are stored. The content of  {\tt MHall\_uncertainties.dat} reads
\begin{lstlisting}[style=file]
1  3.0 3.0 1.0 1.0 1.0 1.0 1.0 
\end{lstlisting}
Now, running \HS again we find that the $\chi^2$ values become slightly smaller.
\begin{lstlisting}[mathescape,firstnumber=16,style=file,title=\hspace{12cm}HiggsSignals\_results.txt]
csq(mu)         csq(mh)         csq(tot)   nobs(mu) 
80.8857         3.46415         84.3498        80     
|\Suppressnumber|     
 $\hookrightarrow$  nobs(mh)    nobs(tot)    Pvalue
 $\hookrightarrow$    4           84         0.468755      |\Reactivatenumber|
\end{lstlisting}

\paragraph*{Light singlet}
We want to run also the second example point which has a light singlet. We keep our estimate of the theoretical uncertainty and find for this point: 
\begin{lstlisting}[mathescape,firstnumber=16,style=file,title=\hspace{12cm}HiggsSignals\_results.txt]
#cols:  n           Mh(1)           Mh(2)       ...  
        1          77.8961         125.607      ...    
        
 |\Suppressnumber|            
 $\hookrightarrow$     csq(mu)         csq(mh)         csq(tot)  
 $\hookrightarrow$     81.0244         1.35131         82.3757       
 
 $\hookrightarrow$     nobs(mu) nobs(mh)   nobs(tot)            Pvalue        
 $\hookrightarrow$      80     4            84                 0.529729     |\Reactivatenumber|
\end{lstlisting}
The couplings of the SM-like Higgs to the SM-fermions don't change significantly between EP1 and EP2, i.e. also $\chi^2_{\mu}$ stayed the same. However, the mass of the SM-like state is a bit closer to the best-fit of the measurements, hence  $\chi^2_m$ has slightly decreased.

\subsection{Checking the vacuum stability with \Vevacious}
The parameter points EP1 and EP2 have passed the first checks. The mass spectrum looks promising and they are consistent with all bounds from flavour and Higgs physics.
As next step we want to check if the points have really a stable vacuum: since \SPheno found a solution for the tadpole equations, it is sure that the given parameters are at least at a local minimum with respect to the scalar potential where the set $\{v_d,v_u,x_1,x_2\}$ of VEVs is non-zero. However, this doesn't ensure that this is also the global minimum. First, there might be a deeper minimum for other values of $\{v_d,v_u,x_1,x_2\}$. Those minima are in general are ruled out because they would predict another mass for the $Z$-boson. Another possibility is that other particles could receive VEVs as well. These can either be points with spontaneous $R$-parity violation where the sneutrino get a VEV ($\{v_d,v_u,x_1,x_2,v^i_{\tilde{\nu}_L}, v^i_{\tilde{\nu}_R}\}$), points where charge is broken by slepton VEVs ($\{v_d,v_u,x_1,x_2,v^i_{\tilde{e}_L}, v^i_{\tilde{e}_R}\}$), or points where charge and colour are broken by squark VEVs ($\{v_d,v_u,x_1,x_2,v^i_{\tilde{d}_R}, v^i_{\tilde{d}_R},v^i_{\tilde{u}_L},v^i_{\tilde{u}_R}\}$). The last two possibilities are completely forbidden and points would always be ruled out by that. However, the dangerous regions for charge or colour breaking are those where the trilinear soft-terms are large compared to the soft-masses in the stop or stau sector \cite{Nilles:1982dy, AlvarezGaume:1983gj, Derendinger:1983bz, Claudson:1983et, Kounnas:1983td, Drees:1985ie, Gunion:1987qv, Komatsu:1988mt, Langacker:1994bc, Casas:1995pd, Casas:1996de,Camargo-Molina:2013sta}. This is not the case for the points EP1 and EP2 and we don't have to worry about that. Spontaneous $R$-parity violation is not completely forbidden and could lead to a different phenomenology. However, in our approach it is also very likely that  the electroweak VEV changes at the global minimum where the sneutrinos gain non-zero VEVs. Hence, such a scenario is ruled out as well by the $Z$ mass. We are going to check the stability of the vacuum with neglecting and with including the possibility of sneutrino VEVs. For this purpose we use the package \Vevacious \cite{Camargo-Molina:2013qva}. \\

\Vevacious  is a tool to check for the global minimum of the one-loop effective potential for a given model allowing for a particular set of non-zero VEVs. For this purpose \Vevacious finds first all tree-level minima by using {\tt HOM4PS2} \cite{lee2008hom4ps}. Afterwards, it minimizes the one-loop effective potential starting from these minima using {\tt minuit}  \cite{James:1975dr}. If the input minimum turns out not to be the global one, life-time of meta-stable vacua can be calculated using {\tt Cosmotransitions} \cite{Wainwright:2011kj}.\\
\Vevacious takes the tadpole equations, the polynomial part of the  scalar potential and all mass 
matrices as input. All of this information has to be expressed including all VEVs which should be tested. That means, that to check for charge and colour breaking minima the stop and stau have to show up in the potential and mass matrices and the entire mixing triggered by these VEVs should be included. To take care of all that, the corresponding input files can be generated by \SARAH as explained below.

\subsubsection{Finding the global minimum without sneutrino VEVs}
If one is just interested in the global minimum for the case that no other VEVs are allowed, it is straightforward to get the model files for \Vevacious: the standard implementation of the model can be used together with the command {\tt MakeVevacious}:
\begin{lstlisting}[style=mathematica]
<<$PATH/SARAH/SARAH.m;
Start["B-L-SSM"];
MakeVevacious[];
\end{lstlisting}
{\tt MakeVevacious} comes with some options which I list for completeness. However, we stick to the default settings. 
\begin{itemize}
\item {\tt ComplexParameters}: defines, if specific parameters should be treated as complex. By default, all parameters are assumed to be real in the \Vevacious output. 
 \item {\tt IgnoreParameters}: defines, if a given set of parameters should be set to zero when
 writing the \Vevacious model files.
 \item {\tt OutputFile}: defines the name for the model files. By default {\tt BLSSM.vin} is used.
 \item {\tt Scheme}, defines, the renormalization scheme. For SUSY models \SARAH uses  $\overline{\text{DR}}'$ and for non-SUSY $\overline{\text{MS}}$ by default
  \end{itemize}
One sees from the first option that the parameters are handled less general in the \Vevacious output as this is usually done by \SARAH. The reason is that the evaluation of a parameter point with \Vevacious  can be very time consuming. Thus, doing reasonable approximations might be an option to speed this up. \\
 
As soon as the model file is created, it is convenient to copy them to the model directory of the local \Vevacious installation. In addition, one can also generate a new subdirectory which contains the  \SPheno spectrum files for the \BLSSM used as input  for \Vevacious, as well as the output written by \Vevacious
\begin{lstlisting}[style=terminal]
$ cd $PATH/VEVACIOUS
$ mkdir BLSSM/
$ cp $PATH/SARAH/Output/B-L-SSM/Vevacious/BLSSM.vin models/
$ cp $PATH/SPHENO/SPheno.spc.BLSSM BLSSM/
\end{lstlisting}
These steps are just optional: the user can give in the initialization file used by \Vevacious, which is  discussed in a second, also paths to other locations of the model and spectrum file. Independent of the location of the files, one has to write this initialization file for a new study. The easiest way is to start with the file included in the \Vevacious package in the subdirectory {\tt bin} and edit it
\begin{lstlisting}[style=terminal]
$ cd $PATH/VEVACIOUS/bin
$ cp VevaciousInitialization.xml VevaciousInitialization_BLSSM.xml 
\end{lstlisting}
The only changes we apply here is to give the paths for {\tt HOM4PS2} \footnote{\tt http://www.math.nsysu.edu.tw/~leetsung/works/HOM4PS\_soft\_files/} and {\tt CosmoTransitions} \footnote{\tt http://chasm.uchicago.edu/cosmotransition} which are used by \Vevacious. I'm going to assume here that these are installed in the same directory {\tt \$PATH} as all other tools are. The other information needed is the location of the model and spectrum files as mentioned above. 
\begin{lstlisting}[style=file,numbers=none,title=\hspace{10cm}VevaciousInitialization\_BLSSM.xml ]
...
  
  <!-- path to Hom4PS2 -->
  <hom4ps2_dir>
  $PATH/HOM4PS2/
  </hom4ps2_dir>

...

  <!-- path to CosmoTransitions -->
  <ct_path>
  $PATH/CosmoTransitions_package_v1.0.2
  </ct_path> 

...

  <!-- path to model file -->
  <model_file>
  $PATH/Vevacious-1.1.01/models/BLSSM.vin
  </model_file>

...

  <!-- path to spectrum file -->  
  <slha_file>
  $PATH/Vevacious-1.1.01/BLSSM/SPheno.spc.BLSSM
  </slha_file>
\end{lstlisting}
For all other settings like what homotopy method should be used, what's the tolerance to consider extrema as identical, what's the necessary survival probability to label a meta-stable point 'long-lived', how should \Vevacious try to get away from saddle points we keep the default values. Interested reader might take a look at the \Vevacious manual for more details about these options.\\

When all adjustments of the initialization file are done, we can run \Vevacious on EP1 by calling from the directory {\tt \$PATH/Vevacious/BLSSM}
\begin{lstlisting}[style=terminal]
$ ./../bin/Vevacious.exe --input=./../bin/VevaciousInitialization_BLSSM.xml 
\end{lstlisting}
 After about 30s \Vevacious is done with checking for the global minimum of the one-loop effective potential. Since it hasn't started {\tt CosmoTransitions} to calculate the tunnelling time, the point is stable. This can also be seen from the file {\tt SPheno.spc.BLSSM} where a new block has been appended:
\begin{lstlisting}[style=file,firstnumber=1542,title=\hspace{11cm}SPheno.spc.BLSSM (EP1)]
BLOCK VEVACIOUSRESULTS # results from Vevacious 
    0    0    1.00000000E+000    stable    # stability of input
    0    1    -1.00000000E+000    unnecessary    # tunneling time                 in Universe ages / calculation type
    0    2    0.00000000E+000    1.0    # estimated best tunneling                 temperature / survival probability at this                 temperature
    1    0    -5.25790342E+011    relative_depth    # DSB vacuum                  potential energy
    1    1    3.58009590E+001    vd    # DSB vacuum VEV
    1    2    2.37944201E+002    vu    # DSB vacuum VEV
    1    3    3.50031616E+003    x1    # DSB vacuum VEV
    1    4    2.91749649E+003    x2    # DSB vacuum VEV
    2    0    -5.25790342E+011    relative_depth    # panic vacuum                  potential
    2    1    3.58009590E+001    vd    # panic vacuum VEV
    2    2    2.37944201E+002    vu    # panic vacuum VEV
    2    3    3.50031616E+003    x1    # panic vacuum VEV
    2    4    2.91749649E+003    x2    # panic vacuum VEV 
\end{lstlisting}
This block contains a flag to assign the stability ({\tt [0,0]}={\tt 1}: stable, {\tt [0,0]}={\tt 0} long-lived, {\tt [0,0]}={\tt -1}: short-lived, {\tt [0,0]}={\tt -2}: thermally excluded), the tunnelling time if calculated (entry {\tt [0,1]}) and the temperature at which tunnelling is likely to happen (entry {\tt [0,2]}). Afterwards the input VEVs are repeated (entries {\tt [1,1]}--{\tt [1,4]}) and the depth of the potential at the input minimum is given (entry {\tt [1,0]}). Finally, the depth of the global minimum together with the VEVs at that minimum are shown (entries {\tt [2,0]}--{\tt [2,4]}). Obviously, the entries {\tt [1,X]} and {\tt [2,X]} are identical. \\

If the user is interested in some more information about all possible minima found at tree-level and one-loop, he/she can check the file  {\tt Vevacious\_tree-level\_extrema.txt}. This file contains all VEV combinations which are actually a minimum of the tree-level potential. Also the depth of the potential at each minimum is given at tree-level and one-loop level:
\begin{lstlisting}[style=file,title=\hspace{10cm}Vevacious\_tree-level\_extrema.txt]
{
{ { vd->(-143.325), vu->(-953.137), x1->(0),x2 ->(0) }, 
   TreeLevelPotentialValue -> -13789506454.7, 
   EffectivePotentialValue -> -2.12169006e+13 }
{ { vd->(-35.8005), vu->(-238.079), x1->(-3500.67), x2->(-2917.22) }, 
   TreeLevelPotentialValue -> -5.29194675525e+11, 
   EffectivePotentialValue -> -2.17385124968e+13 }
{ { vd->(-35.8005), vu->(-238.079), x1->(3500.67), x2->(2917.22) }, 
   TreeLevelPotentialValue -> -5.29194675525e+11, 
   EffectivePotentialValue -> -2.17385124968e+13 }
{ { vd->(0), vu->(0), x1->(-3503.34), x2->(-2919.45) }, 
   TreeLevelPotentialValue -> -5.29142228921e+11, 
   EffectivePotentialValue -> -2.17384087462e+13 }
{ { vd->(0), vu->(0), x1 ->(0), x2->(0) }, 
   TreeLevelPotentialValue -> 0.0, 
   EffectivePotentialValue -> -2.12127237174e+13 }
{ { vd->(0), vu->(0), x1->(3503.34), x2->(2919.45) }, 
   TreeLevelPotentialValue -> -5.29142228921e+11, 
   EffectivePotentialValue -> -2.17384087462e+13 }
{ { vd->(35.8005), vu->(238.079), x1->(-3500.67), x2->(-2917.22) }, 
   TreeLevelPotentialValue -> -5.29194675525e+11, 
   EffectivePotentialValue -> -2.17385124968e+13 }
{ { vd->(35.8005), vu->(238.079), x1->(3500.67), x2->(2917.22) }, 
   TreeLevelPotentialValue -> -5.29194675525e+11, 
   EffectivePotentialValue -> -2.17385124968e+13 }
{ { vd->(143.325), vu->(953.137), x1->(0), x2->(0) }, 
   TreeLevelPotentialValue -> -13789506454.7, 
   EffectivePotentialValue -> -2.12169006e+13 }
} 
\end{lstlisting}
We can see from that file that there are actually four minima which are not related by a phase transformation of the VEVs. The minimum without symmetry  breaking (all VEVs are zero) has a depth of 0 at tree-level as expected but receives large loop corrections. Nevertheless, the depth is still much less than for all other combinations where at least one symmetry (electroweak or $B-L$ is broken). There is just one minimum where both symmetries are broken and this corresponds to our input minimum. The full list of minima found at one-loop is given in the file {\tt Vevacious\_loop-corrected\_minima.txt}. All additional minima listed there are small variations of the tree-level ones. \\

We can do the same check for EP2 and find that also this point is stable. 

\subsubsection{Checking for spontaneous $R$-parity violation}
As mentioned above, one can't be completely sure that the point is stable if \Vevacious doesn't find a deeper minimum if the first check is passed. There is still the possibility that additional particles might receive VEVs. We are checking here the possibility of spontaneous $R$-parity violation. For this purpose, it is necessary to create a new \SARAH model file. We call it {\tt B-L-SSM\_RpV.m}. The simplest way is to take our {\tt B-L-SSM.m} file as basis and apply the following changes:
\begin{lstlisting}[style=file,numbers=none,title=\hspace{13cm}BLSSM\_RpV.m]
Model`Name = "BLSSMRpV";

DEFINITION[EWSB][VEVs]= 
{ ...,
   {SvL, {vL[3], 1/Sqrt[2]}, {sigmaL, \[ImaginaryI]/Sqrt[2]},{phiL,1/Sqrt[2]}},
   {SvR, {vR[3], 1/Sqrt[2]}, {sigmaR, \[ImaginaryI]/Sqrt[2]},{phiR,1/Sqrt[2]}},
};
 
(*--- Matter Sector ---- *)
DEFINITION[EWSB][MatterSector]= 
{
...
  {{phid, phiu,phi1, phi2,phiL,phiR}, {hh, ZH}}, 
  {{sigmad, sigmau,sigma1,sigma2,sigmaL,sigmaR}, {Ah, ZA}},
  {{SHdm,conj[SHup],SeL, SeR},{Hpm,ZP}},
  {{fB, fW0, FHd0, FHu0,fBp,FC10,FC20,FvL,conj[FvR]}, {L0, ZN}}, 
  {{{fWm, FHdm,FeL}, {fWp, FHup,conj[FeR]}}, {{Lm,UM}, {Lp,UP}}}
...
       }; 
\end{lstlisting}
First, we change the name of the model to make sure that no files of the other implementation are overwritten. The main modification is to give VEVs to the left and right sneutrinos as done by the changes in {\tt DEFINITION[EWSB][VEVs]}. However, we didn't consider the most general case where all three generations get VEVs, but restrict VEVs to the third generation only. Even in this case we have to deal with a 6-dimensional parameter space. In the general case, we would even have 10 VEVs and running \Vevacious would take significant longer. Therefore, it's always good to check what degrees of freedom can be rotated away. The other lines are a consequence of $R$-parity violation: a mixing between the CP even and odd sneutrinos and the Higgs scalars happens, the charged Higgs scalar mix with the charged sleptons. In the fermionic sector the charginos mix similarly with the charged leptons, and the neutralinos with the neutrinos. We have to include this mixing because \Vevacious checks not only the tree-level potential but also the one-loop effective potential. This mixing will give additional contributions to the one-loop corrections. If we are really just interested in the \Vevacious output, we can skip the modifications of {\tt particles.m} and {\tt parameters.m} and come back directly to your study with \Vevacious: the remaining steps are the same as for the $R$-parity conserving case: (i) running the {\tt B-L-SSM\_RpV} with \SARAH, (ii) running {\tt MakeVevacious[]}, (iii) copying the file to the \Vevacious installation, (iv) creating a new initialization file {\tt VevaciousInitialization\_BLSSM\_RpV.xml}  with the location of the model file, (v) running \Vevacious. \\

We find that both parameter points pass also this check. For example, for EP1 the \Vevacious output reads: 
\begin{lstlisting}[style=file,firstnumber=1542,title=\hspace{11cm}SPheno.spc.BLSSM (EP1)]
BLOCK VEVACIOUSRESULTS # results from Vevacious version 1.1.01, documented in arXiv:1307.1477, arXiv:1405.7376 (hep-ph)
    0    0    1.00000000E+000    stable    # stability of input
    0    1    -1.00000000E+000    unnecessary    # tunneling time ...
    0    2    0.00000000E+000    1.0  # ... tuneling temperature
    1    0    -5.24262779E+011    relative_depth    # DSB vacuum  ...
    1    1    0.00000000E+000    vL3    # DSB vacuum VEV
    1    2    0.00000000E+000    vR3    # DSB vacuum VEV
    1    3    3.58548559E+001    vd    # DSB vacuum VEV
    1    4    2.26723606E+002    vu    # DSB vacuum VEV
    1    5    3.49979972E+003    x1    # DSB vacuum VEV
    1    6    2.91786084E+003    x2    # DSB vacuum VEV
    2    0    -5.24262779E+011    relative_depth    # panic vacuum ...
    2    1    0.00000000E+000    vL3    # panic vacuum VEV
    2    2    0.00000000E+000    vR3    # panic vacuum VEV
    2    3    3.58548559E+001    vd    # panic vacuum VEV
    2    4    2.26723606E+002    vu    # panic vacuum VEV
    2    5    3.49979972E+003    x1    # panic vacuum VEV
    2    6    2.91786084E+003    x2    # panic vacuum VEV
\end{lstlisting}

\vspace{0.5cm}
I want to show that things are not always so boring and unexpected things can happen in such complicated potentials. For this purpose, I modify the input parameters for EP1 a bit:
\begin{eqnarray}
& A_0 = -1600~\text{GeV}, \hspace{1cm} Y_x = \text{diag}(0.39,0.40,0.41)& 
\end{eqnarray}
Even with these modifications, all right-sneutrino soft-terms are still positive:
\begin{lstlisting}[style=file,numbers=none,title=\hspace{12cm}SPheno.spc.BLSSM]
Block mv2 Q=  3.04177370E+03  # (SUSY Scale)
  1  1     1.03252784E+06   # Real(mv2(1,1),dp)
  2  2     9.42973968E+05   # Real(mv2(2,2),dp)
  3  3     8.52847167E+05   # Real(mv2(3,3),dp) 
\end{lstlisting}
Nevertheless, we find that at the global minimum $R$-parity is broken by sneutrino VEVs: 
\begin{lstlisting}[style=file,firstnumber=1542,title=\hspace{11cm}SPheno.spc.BLSSM]
BLOCK VEVACIOUSRESULTS # results from Vevacious version 1.1.02, documented in arXiv:1307.1477, arXiv:1405.7376 (hep-ph)
    0    0   -2.00000000E+000    long-lived_but_thermally_excluded  #
    0    1    4.89995748E+031    direct_path    # tunneling time in                          Universe ages / calculation type
    0    2    1.11583595E+003    0.0    # estimated best tunneling                     temperature / survival probability                     at this temperature
    1    0    -6.30122774E+011    relative_depth    # DSB vacuum ...
    1    1    0.00000000E+000    vL3    # DSB vacuum VEV
    1    2    0.00000000E+000    vR3    # DSB vacuum VEV
    1    3    3.57990259E+001    vd    # DSB vacuum VEV
    1    4    2.39981235E+002    vu    # DSB vacuum VEV
    1    5    3.66612055E+003    x1    # DSB vacuum VEV
    1    6    3.05552200E+003    x2    # DSB vacuum VEV
    2    0    -6.99210151E+011    relative_depth    # panic vacuum ...
    2    1    0.00000000E+000    vL3    # panic vacuum VEV
    2    2    1.57052814E+003    vR3    # panic vacuum VEV
    2    3    4.56069965E-001    vd    # panic vacuum VEV
    2    4    3.14986273E+000    vu    # panic vacuum VEV
    2    5    2.41594641E+003    x1    # panic vacuum VEV
    2    6    2.05701579E+003    x2    # panic vacuum VEV
\end{lstlisting}
We see that the stability is labelled as 'long lived but thermally excluded' ({\tt [0,0]}={\tt -2}). This means that the point is long-lived at zero temperature but quickly decays if temperature effects are taken into account. In that case the entry {\tt [0,2]} shows at which temperature the tunnelling is likely to happen. Thus, this point is actually ruled out. \\

One sees at this example that the condition
\begin{equation}
 m^2_{\tilde \nu^C} < 0
\end{equation}
sometimes used in literature for distinguishing $R$-parity violation and conservation is not necessary. On the other hand, it is also possible to find points where this condition is fulfilled, but $R$-parity is still unbroken at the global minimum \cite{CamargoMolina:2012hv}, i.e. it is also not sufficient. Therefore, one shouldn't rely on such simple minded conditions but perform always a numerical check to test the vacuum stability. The same statement holds for charge and colour breaking minima: analytical thumb rules like \cite{Nilles:1982dy, AlvarezGaume:1983gj, Derendinger:1983bz, Claudson:1983et, Kounnas:1983td}
\begin{align}
A_{\tau}^{2} <& 3 ( m_{H_{d}}^{2} + |\mu|^{2} + m_{\tau_L}^{2} + m_{\tau_R}^{2} ) \\
A_{t}^{2} <& 3 ( m_{H_{u}}^{2} + |\mu|^{2} + m_{t_L}^{2} + m_{t_R}^{2} )  
\end{align}
which are, unfortunately, still widely used in literature don't bear up against numerical checks and turn out to be pretty useless \cite{Camargo-Molina:2014pwa,Chattopadhyay:2014gfa}. These conditions miss the large majority of points which actually suffer from an unstable ew vacuum.

\subsection{Calculating the dark matter properties with \MO}
As next step we want to study the dark matter (DM) properties of the model by using \MO. \MO is a tool which not only calculates the relic density for one or more dark matter candidates, but it also gives cross sections for direct and indirect DM searches. To enable these calculations, \MO needs in general three inputs
\begin{enumerate}
 \item The model files to implement a new model
 \item A steering file to coordinate the different calculations
 \item Numerical values for all parameters
\end{enumerate}
I'll show step by step how these three points are addressed. 

\subsubsection{Implementing new models in \MO}
The calculation of the cross section and all necessary decay widths are done by \CalcHep which comes together with \MO. Thus, a new model in \MO is implemented by providing the corresponding \CalcHep model files. That means, one can use the \SARAH output for \CalcHep to work with \MO:
\begin{lstlisting}[style=mathematica]
<<$PATH/SARAH/SARAH.m;
Start["B-L-SSM"];
MakeCHep[];
\end{lstlisting}
By just running {\tt MakeCHep[]}, the default options are used. 
However, there are several options to adjust the output:
\begin{itemize}
\item \verb"FeynmanGauge": defines, if Feynman gauge should be supported beside Landau gauge. This is done by default. 
\item \verb"CPViolation": defines, if parameters should be handled as complex. By default, all parameters are treated as real because \CalcHep is not really optimized for the usage of complex parameters and this option should be used carefully.  
\item \verb"ModelNr": numbers the model files. \SARAH starts by default with {\tt 1}.
\item \verb"CompHep": can be used to write model files in \CompHep instead of \CalcHep format.
\item \verb"NoSplittingWith": \SARAH does not decompose four-scalar interactions in pairs of 
two scalar interactions with auxiliary fields if particular fields are involved. Such a decomposition is usually done because of the implicit colour structure in \CalcHep which doesn't allow four-point interactions of coloured states. To keep the model files shorter, \SARAH makes the same decomposition also for non-coloured states. 
\item \verb"NoSplittingOnly": One can define particles, for which \SARAH does not decompose four-scalar interactions in pairs of 
two scalar interactions with auxiliary fields if  {\it only} the given fields are involved the interaction. 
\item \verb"UseRunningCoupling": defines, if $\alpha_S$ should run in the model files.
\item \verb"SLHAinput": defines, if parameter values should be read from a spectrum file.
\item \verb"CalculateMasses": defines, if tree-level masses should be calculated internally by \CalcHep.
\item \verb"RunSPhenoViaCalcHep": writes C code to run \SPheno from the graphical interface of \CalcHep 
to calculate the spectrum  on the fly.
\item \verb"IncludeEffectiveHiggsVertices": defines, if effective Higgs vertices $h\gamma\gamma$ and $hgg$ should be included.
\item \verb"DMcandidate1": sets the first DM candidate.
\item \verb"DMcandidate2": sets optionally a second DM candidate.
\end{itemize}

For our example we can stick to the default options. I'll just comment on two important switches which demand a further explanation:

\paragraph*{Mass spectrum} By using {\tt SLHAinput -> True} the model files are written in a way that \CalcHep respectively \MO expect all input parameters to be provided in a spectrum file which is called {\tt SPheno.spc.BLSSM}. \CalcHep and \MO are going to read this file and extract all important information using the {\tt SLHA+} functionality \cite{Belanger:2010st} from it. With the other options \MO/\CalcHep expect either all masses and rotation matrices given in the file {\tt vars.mdl} ({\tt SLHAinput -> False, CalculateMasses -> False}), or it expects all fundamental parameters (soft-terms, couplings and VEVs) as input and diagonalizes the mass matrices internally ({\tt SLHAinput -> False, CalculateMasses -> True}).

\paragraph*{Dark Matter candidates} One can work either with one or two dark matter candidates in \MO. The first DM candidate is the lightest particle of all states having a particular charge under a discrete symmetry To define the symmetry and the charge, the option {\tt DMcandidate1->Value} is used. There are two possibilities for {\tt Value}: (i) when set to {\tt Default}, the DM candidate is the lightest odd particle odd under the first $Z_2$ defined as global symmetry; (ii) for any other choice, one can give first the name of the global symmetry and then the quantum number with respect to that symmetry {\tt GlobalSymmetry == Charge}.\\

\vspace{0.5cm}
When \SARAH is finished with {\tt MakeCHep}, the \CalcHep model files are located in the directory
\begin{verbatim}
$PATH/SARAH/Output/B-L-SSM/EWSB/CHep/ 
\end{verbatim}
To implement the model in \MO, a new project has to be created and the files have to be copied in the working directory of this project:
\begin{lstlisting}[style=terminal]
$ cd $PATH/MICROMEGAS
$ ./newProject BLSSM
$ cd BLSSM
$ cp $PATH/SARAH/Output/B-L-SSM/EWSB/CHep/*  work/models
\end{lstlisting}

\subsubsection{Setting up the DM calculations}
To use the model with \MO a steering or 'main' file has to be provided either in \Fortran or {\tt C} language,  and must be compiled. Examples for these files are delivered with \MO and called {\tt main.F} and {\tt main.c}.  \SARAH writes also two examples which can be used for the following calculations:
\begin{itemize}
 \item {\tt CalcOmega.cpp}: this file calculates only the DM relic density $\Omega h^2$ and prints the result at the screen and into a file called {\tt omg.out}
 \item {\tt CalcOmega\_with\_DDetection.cpp}: this file calculates the DM relic density $\Omega h^2$ and in addition some direct detection rates: (i) spin independent cross section with proton and neutron in pb, (ii) spin dependent cross section with proton and neutron in pb, (iii) recoil events in the 10 -- 50~keV region at \({}^{73}\)Ge, \({}^{131}\)Xe, \({}^{23}\)Na and \({}^{127}\)I nuclei. The output is also written into a file called {\tt omg.out}. Note, the syntax for the direct detection calculations has been changed in \MO compared to earlier versions. \SARAH includes also a file {\tt CalcOmega\_with\_DDetection\_old.cpp} which is compatible with versions {\tt 2.X} of \MO. 
\end{itemize}
We are going to choose the second file which includes the calculation of direct detection rates. There are even more calculations \MO can do like  indirect detection rates. Those can be added as well to the main file provided by \SARAH or the user can write an own file. For this purpose, it might be helpful to take a look at {\tt main.F} or {\tt main.c} which show the different options to turn on specific calculations and outputs.\\
The steering files written by \SARAH were copied together with all model files into the working directory of the current project. We can move it to the main project directory and compile it
\begin{lstlisting}[style=terminal]
$ mv work/models/CalcOmega_with_DDetection.cpp .
$ make main=CalcOmega_with_DDetection.cpp
\end{lstlisting}
A new binary {\tt CalcOmega\_with\_DDetection} is now available. The only missing piece are the input parameters.

\subsubsection{Running \MO with \SPheno spectrum files}
Providing the numerical parameters is pretty easy because \MO/\CalcHep can read the \SPheno spectrum file. However, the user must make sure that no complex rotation matrices show up in the spectrum file: in the case of Majorana matrices and no CP violation, there are two equivalent outputs: (i) all Majorana masses are positive, but some entries of the corresponding rotation matrices are complex; (ii) all mixing matrices are real, but some masses are negative. \CalcHep can just handle the second case with real matrices. Hence, one has to use the flag 
\begin{lstlisting}[style=file,numbers=none,title=\hspace{12cm}LesHouches.in.BLSSM]
Block SPhenoInput   # SPheno specific input 
 ...
 50 0               # Majorana phases: use only positive masses  
\end{lstlisting}
to get the spectrum according to that convention. Afterwards, the spectrum file just has to be moved to the same directory as {\tt CalcOmega\_with\_DDetection}. We copy it there and start the calculation:
\begin{lstlisting}[style=terminal]
$ cp $PATH/SPHENO/SPheno.spc.BLSSM .
$ ./CalcOmega_with_DDetection
\end{lstlisting}
The first run can take some time, even up to several hours depending on the computer power: \MO has to compile all necessary annihilation channels of the DM candidate for that particular parameter point. All further evaluations of similar points are done in a second or less. However, as soon as new channels are needed, \MO has to compile new amplitudes and the computation slows down extremely again. This can happen for instance, if the DM candidate changes or if the second lightest state becomes close in mass and co-annihilation has to be included. As soon as the run is done, we see on the screen:
\begin{lstlisting}[style=terminal]
...
Xf=2.43e+01 Omega h^2=1.22e-01

# Channels which contribute to 1/(omega) more than 1%.
# Relative contributions in % are displayed
   98% ~nR1 ~nR1 ->h2 h2 

==== Calculation of CDM-nucleons amplitudes  =====
         TREE LEVEL
CDM-nucleon micrOMEGAs amplitudes:
proton:  SI  1.217E-10  SD  0.000E+00
neutron: SI  1.225E-10  SD  0.000E+00
         BOX DIAGRAMS
CDM-nucleon micrOMEGAs amplitudes:
proton:  SI  1.217E-10  SD  0.000E+00
neutron: SI  1.225E-10  SD  0.000E+00
CDM-nucleon cross sections[pb]:
 proton  SI 6.453E-12  SD 0.000E+00
 neutron SI 6.539E-12  SD 0.000E+00

======== Direct Detection ========
73Ge: Total number of events=6.71E-07 /day/kg
Number of events in 10 - 50 KeV region=3.56E-07 /day/kg
131Xe: Total number of events=1.07E-06 /day/kg
Number of events in 10 - 50 KeV region=5.49E-07 /day/kg
23Na: Total number of events=6.66E-08 /day/kg
Number of events in 10 - 50 KeV region=3.57E-08 /day/kg
I127: Total number of events=1.05E-06 /day/kg
Number of events in 10 - 50 KeV region=5.48E-07 /day/kg
\end{lstlisting}
In the first line, the freeze out temperature and the relic density is given. We find that this point falls into the preferred 2$\sigma$ region
\begin{equation}
 0.1153<\Omega_{CDM}h^2<0.1221
\end{equation}
combining Planck, WMAP polarization, high-resolution CMB data, and baryon acoustic oscillation results \cite{Ade:2013zuv}.

The important channels contributing to the annihilation follow in the next lines. This point is a bit boring, because the annihilation in two bileptons makes 98\% of the entire annihilation. All other individual channels are not printed because they are below 1\%. This threshold can be changed in  {\tt CalcOmega\_with\_DDetection.cpp} by changing the cut to lower values:
\begin{lstlisting}[style=file,firstnumber=18,title=\hspace{10cm}CalcOmega\_with\_DDetection.cpp]
double cut = 0.01;               // cut-off for channel output
\end{lstlisting}
\vspace{0.5cm}
The same information is also written in the file {\tt omg.out}. The style of this file is inspired by the SLHA format: 
\begin{lstlisting}[style=file,title=\hspace{14cm}omg.out]
1 0.122031 # relic density 
100 0.980976 # ~nR1 ~nR1 -> h2 h2
201 0.000000000006453 #
202 0.000000000000000 #
203 0.000000000006539 #
204 0.000000000000000 # 
301 0.000001 #
302 0.000001 #
303 0.000000 #
304 0.000001 #
\end{lstlisting}
Because of this format, one can append this file to the spectrum file to save the dark matter results together with the other information, and read it later with a standard SLHA parser. 
\begin{lstlisting}[style=terminal]
$ cat SPheno.spc.BLSSM > SPheno.spc.BLSSM_with_MO
$ echo 'Block DARKMATTER #' >> SPheno.spc.BLSSM_with_MO
$ cat omg.out >> SPheno.spc.BLSSM_with_MO
\end{lstlisting}
This is for instance done automatically when running scans with \SSP and including \MO. \\

The values shown for the direct detection rates can be compared with limits from experiments. For this purpose it is helpful to multiply these values by a factor of $10^{-36}$ to get the rates in $\text{cm}^2$ which is usually used to present the direct detection limits in the $(m_{DM},\sigma)$ plane.\\

We can do the same for the EP2. This point has a neutralino LSP, i.e. \MO has to compile again many channels and we have to wait again some time for the results. The output on the screen looks less promising:
\begin{lstlisting}[style=terminal]
Xf=2.15e+01 Omega h^2=2.66e+01

# Channels which contribute to 1/(omega) more than 1%.
# Relative contributions in % are displayed
   28% ~N1 ~N1 ->e3 E3 
   27% ~N1 ~N1 ->e2 E2 
   27% ~N1 ~N1 ->e1 E1 
    5% ~N1 ~N1 ->u3 U3 
    3% ~N1 ~N1 ->Wm Wp 
    2% ~N1 ~N1 ->u2 U2 
    2% ~N1 ~N1 ->u1 U1 
    1% ~N1 ~N1 ->Z Z 
    1% ~N1 ~N1 ->h2 h2 
...    
\end{lstlisting}
Despite the many different channels which contribute to the annihilation, the relic density is much too high. This is not surprising because it is well known that for a neutralino LSP often particular conditions are needed to fulfil the relic density bounds. Either a charged particle close in mass, resonances, or a large Higgsino fraction are needed. This holds not only for a bino LSP in the CMSSM but also for a blino LSP in the constrained \BLSSM as we have it here \cite{Basso:2012gz}.

\subsection{Monojet events with \WHIZARD}
We change topics again and enter the wide field of collider studies with Monte Carlo (MC) tools. That's nothing what can be addressed in detail in this manuscript. Tools like \CalcHep, \WHIZARD, \MG, \Herwig or \Sherpa are very powerful and offer a rather unlimited number of possibilities what can be done. Therefore, I'm just going to show at two examples how the output of \SARAH can be used together with \WHIZARD and \MG to perform simple studies. As soon as a model is implemented in these tools and is working fine for one study, it can be used in the same way as all models delivered with the different tools. Thus, to become more familiar with these tools, one can check for the many examples and tutorials which can be found online. \\

Actually, there is one big advantage when working with model files produced by \SARAH: the chosen MC tool needs not only the model files containing all vertices but also numerical values for all parameters have to be provided. This can be a delicate task especially in supersymmetric models coming with a lots of parameters and rotation matrices. When using numerical values for all these parameters obtained with another code, one has to make always sure that the conventions which are used in the model file and these of the spectrum generator are identical. This problem is absent when working with model files produced by \SARAH and spectrum files generates with a \SPheno version also produced by \SARAH. In that case, the implementation of models in the MC tool and in \SPheno are based on single model file in \SARAH. Thus, the same conventions are used for sure in both parts.

\subsubsection{Introduction}
\WHIZARD \cite{Kilian:2007gr} is a fast tree-level MC generator for events at parton level. \WHIZARD makes use of \OMEGA \cite{Moretti:2001zz} to generate the matrix elements, i.e. strictly speaking a model implementation in \WHIZARD means that model files for \WHIZARD and \OMEGA have to be generated and included in both codes. \SARAH is going to take care of both. 
I'll first show how the \WHIZARD and \OMEGA model files are generated with \SARAH and how they are compiled. In the second step, I'll show how the parameters are passed from \SPheno to \WHIZARD. Finally, events for the process
\begin{equation}
 p p \to j \tilde{\chi}_1 \tilde{\chi}_1 \hspace{1cm} \sqrt{s}=14~\text{TeV} 
\end{equation}
are generated. The jet $p_T$ and rapidity distributions are plotted using intrinsic \WHIZARD functions.

\subsubsection{Generating the model files for \WHIZARD/\OMEGA}
For the process we are interested in, we just need vertices which involve fermions. Thus, we can neglect all vertices which only come with scalars and vectors. This is sometimes helpful because the compilation of the model files with \WHIZARD/\OMEGA can be quite time and memory consuming for complicated models. So, we run
\begin{lstlisting}[style=mathematica]
MakeWHIZARD[Exclude->{SSSS,SSVV,SSV,SVV,GGV,GGS,VVV,VVVV}]
\end{lstlisting}
to include only {\tt FFV} and {\tt FFS} vertices. This shortcut is very helpful for our purposes here to get quickly some results. However, it has to be used carefully in order to make sure that no relevant vertices are dropped. In the case that all vertices should be kept, there is another possibility to speed up compilation a bit: usually, \SARAH splits the entire list of vertices in pieces containing 150 vertices and writes for each part a separate file. Especially for {\tt SSSS} and {\tt SSS} interactions even 150 vertices can cause a large file which needs some time to be compiled. Thus, for complicated models where the expressions for the vertices are lengthy, it might be helpful to go even for less couplings per file. That's done by the option {\tt MaximalCouplingsPerFile -> X} with some integer {\tt X}. A good choice for the full model files for the \BLSSM is 50 or less. There are some more flags which can be used to adjust the \WHIZARD output. The full list of options is:
\begin{itemize}
\item \verb"MaximalCouplingsPerFile": defines the maximal number of vertices per file.
\item \verb"WriteOmega": defines, if the model files for \OMEGA should be written.
\item \verb"WriteWHIZARD": defines, if the model files for \WHIZARD should be written.
\item \verb"WOModelName": defines the name for the model in the output.
\item \verb"Version": defines, for which version of \WHIZARD the files are generated. By default \verb"2.2.0" is used.
\item \verb"ReadLists": defines, if the information from a former evaluation should be used.
\end{itemize}

\subsubsection{Compiling the model files}
After the interface has completed, the generated files are stored in the directory
\begin{verbatim}
$PATH/SARAH/Output/B-L-SSM/EWSB/WHIZARD_Omega/ 
\end{verbatim}
In order to use the model with \WHIZARD and \OMEGA, the generated code must be compiled and installed.
In most cases this is done by
\begin{lstlisting}[style=terminal] 
$ cd $PATH/SARAH/Output/B-L-SSM/EWSB/WHIZARD_Omega
$ ./configure --prefix=$PATH/WHIZARD/ WO_CONFIG=$PATH/WHIZARD/bin/
$ make
$ make install
\end{lstlisting}
If \WHIZARD has not been installed globally in the home directory of the current user, \WHIZARD won't be able to find the binaries. Thus, the {\tt WO\_CONFIG} environment variable was used to point explicitly to the binaries.  By default, the {\tt configure} script would install the compiled model into \verb".whizard" in the home directory of the user. If the user wants to have several \WHIZARD installations or install \WHIZARD locally, it might be better to provide a model just for one installation. For these cases the installation path has been defined via the \verb"--prefix" option of the {\tt configure} script. More information on the available options is shown with the command
\begin{lstlisting}[style=terminal]
./configure  --help
\end{lstlisting}
The configure script prints also another import information, namely the name of the model which is used to load it in \WHIZARD:
\begin{lstlisting}[style=terminal]
...
###########################################################
configure: collecting models
###########################################################

found: blssm_sarah

configure: writing whizard/Makefile.src
configure: writing omega/Makefile.src

...
\end{lstlisting}
Thus, the model is called {\tt blssm\_sarah}. This name could be changed by using the option {\tt WOModelName} of {\tt MakeWHIZARD}. \\

The model files produced by \SARAH are supposed to be used with \WHIZARD {\tt 2.x}. The possibility to patch these files for a use with \WHIZARD {\tt 1.x} does exit in principle. However, I won't go into detail here and highly recommend to use version 2.

\subsubsection{Parameter values from \SPheno}
\WHIZARD is able to read Les Houches files for the MSSM generated by public spectrum generators using SLHA conventions. However, \WHIZARD does not provide a possibility to read spectrum files which go beyond that. Therefore, to link \WHIZARD and \SPheno, all \SPheno modules created by \SARAH write the information about the parameters and masses into an additional file. This file is written in the \WHIZARD specific format and can be directly read by \WHIZARD.  In our example the file is called {\tt WHIZARD.par.BLSSM} and it is written to the same directory where \SPheno writes the standard spectrum file.  One just has to make sure that the corresponding flag is turned on the Les Houches input for \SPheno to get this output:
\begin{lstlisting}[style=file,numbers=none,title=\hspace{12cm}LesHouches.in.BLSSM] 
Block SPhenoInput   # SPheno specific input 
...
75 1                # Write WHIZARD files 
\end{lstlisting}
The parameter file can then be included in the {\tt Sindarin} input file for \WHIZARD  via
\begin{lstlisting}[style=file,numbers=none] 
include("$PATH/SPHENO/WHIZARD.par.BLSSM") 
\end{lstlisting}

\subsubsection{{\tt Sindarin} input and running \WHIZARD}
\WHIZARD comes with its own steering language called {\tt Sindarin}. With {\tt Sindarin} all settings to define a process in a specific model, to apply cuts and even to make plots can be put in one single input file. The input file {\tt BLSSM\_monojet.sin} for our example of monojets at the LHC  might look as follows: 
\begin{lstlisting}[style=file,title=\hspace{12cm}BLSSM\_monojet.sin] 
model = blssm_sarah

include("$PATH/SPHENO/WHIZARD.par.BLSSM")

Mu1 = 0.
Md1 = 0.
Mu2 = 0.
Md2 = 0.

alias parton = u1:u1bar:u2:u2bar:d2:d2bar:d1:d1bar:G
alias jet = parton

process monojet = parton, parton => jet, N1, N1

compile

sqrts = 14 TeV

beams = p, p => pdf_builtin
cuts = all Pt >= 50 GeV [jet]
integrate (monojet) { iterations = 5:20000 }

$description = "Monojets"
$y_label = "$N_{\textrm{events}}$"
$title = "Jet-$p_T$ in $pp\to j\tilde\chi^0\tilde\chi^0$"
$x_label = "$p_T(j)$/GeV"
histogram pt_jet (0 GeV, 1000 GeV, 10 GeV)
$title = "Jet rapidity in $pp\to j\tilde\chi^0\tilde\chi^0$"
$x_label = "$\eta(j)$"
histogram eta_jet (-5, 5, 0.1)
analysis = record pt_jet (eval Pt [extract index 1 [jet]]);
           record eta_jet (eval Eta [extract index 1 [jet]])

simulate (monojet) { n_events = 100000 }
compile_analysis { $out_file = "monojet.dat" }
\end{lstlisting}

First, we set the model and tell \WHIZARD where it finds the spectrum file written by \SPheno. In general, the \SPheno file contains non-zero and different masses for all SM fermions. However, to group fermions together into one object, those have to have the same masses. Therefore, we put all first and second generation quark masses explicitly to zero in lines 5--8. Afterwards, we can combine these quarks, their antiparticles and the gluon into one object called {\tt parton}. For the final state we define another object {\tt jet} which consists of the same particles. When we now define a process involving {\tt parton} and {\tt jet}, \WHIZARD will generate all non-vanishing subprocesses on parton level. A name for the process ({\tt monojet}) is given. This name is used in the following to refer to this process. Thus, one can also define several processes in one file, treat them separately, and run one after the other. \\
The next steps are to compile the process (line 15), set the beam energy (line 17) and define the pdf set which should be used (line 19). We apply a $p_t$ cut on the jet of 50~GeV. The process is now fully set and can be integrated (line 20). To improve numerics, we use 5 iterations\footnote{For more complicated processes it is often useful to make two runs via {\tt integrate (process)  \{iterations= 15:20000, 25:20000\}} to further improve numerics}. \\
Lines 22--32 are used to generate figures directly while running \WHIZARD. The figures will show a histogram of the jet $p_T$ from 0 to 1000~GeV in bins of 10~GeV and the rapidity of the jet from -5 to -5 in bins of 0.1. Note, {\tt pt\_jet} and {\tt eta\_jet} are undefined at this stage but just variables. The {\tt analysis} command is used to tell \WHIZARD what is meant by both, {\tt pt\_jet} and {\tt eta\_jet}. In the last two lines, the number of events and the name for the output file are given. \\

\begin{figure}[h]
    \centering
        \includegraphics[trim=1.5cm 8cm 4cm 3cm, clip=true, width=.55\textwidth]{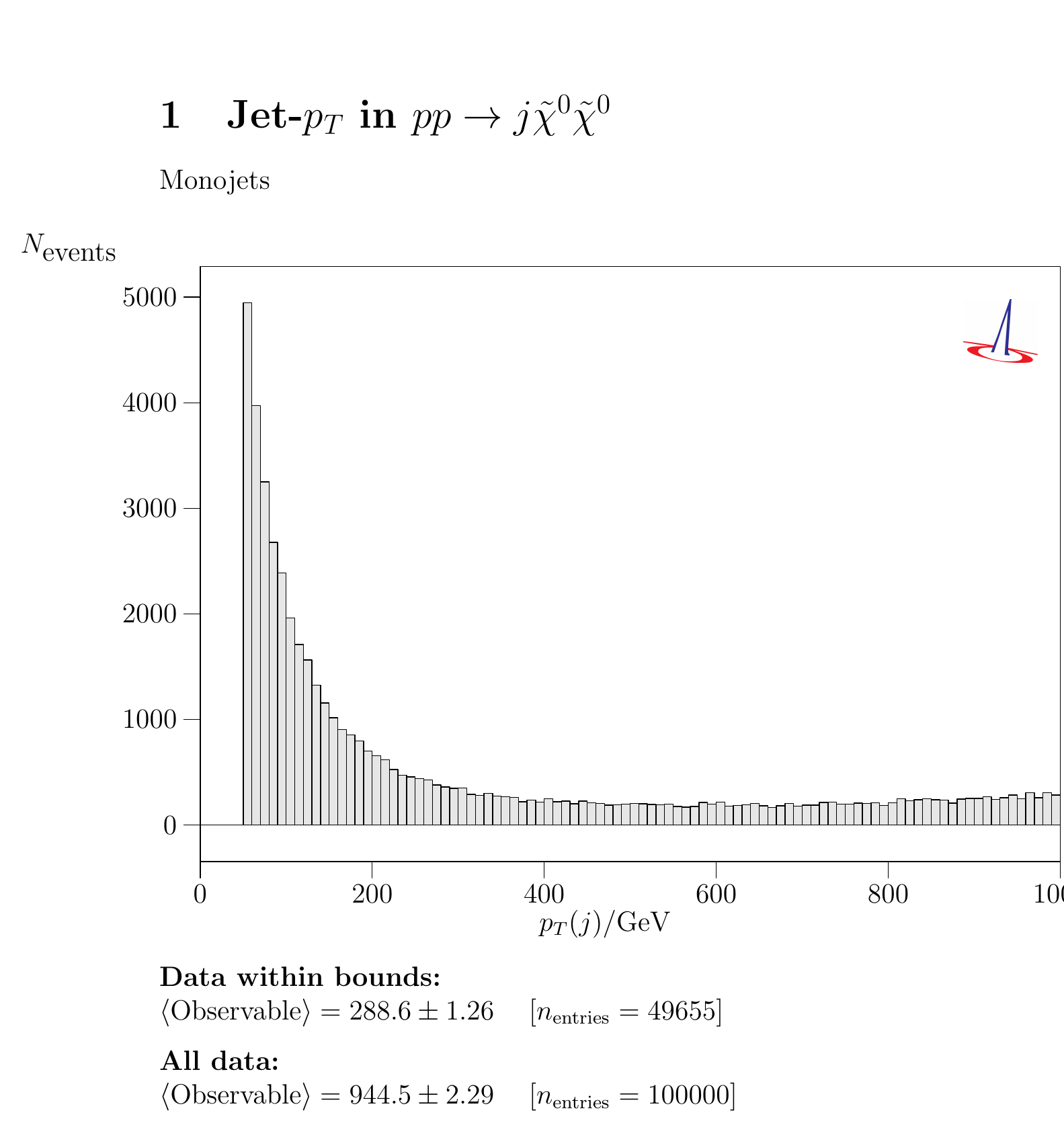} \\
        \includegraphics[trim=1.5cm 8cm 4cm 3cm, clip=true, width=.55\textwidth]{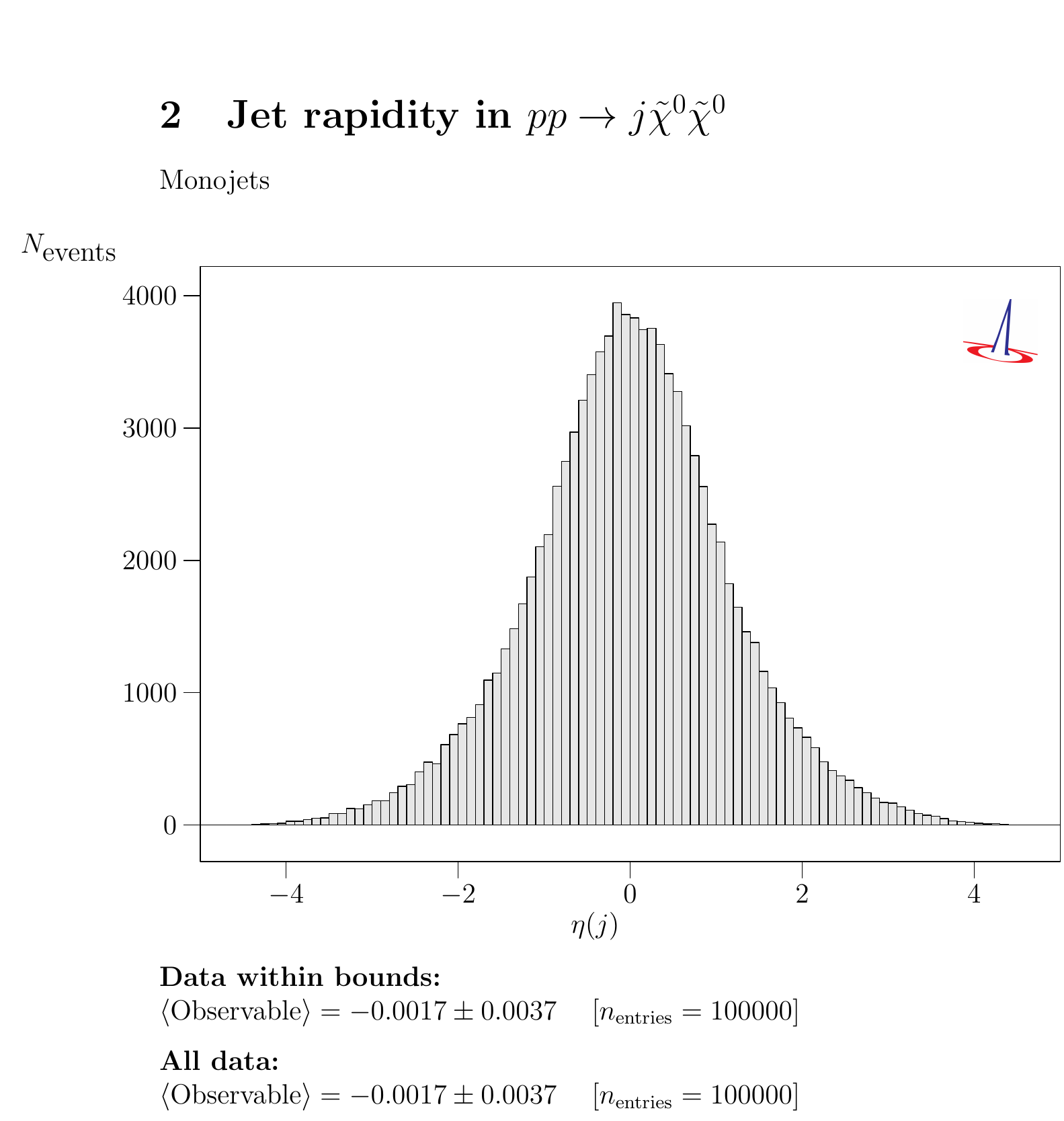}
    \caption{Plots produces by \WHIZARD for the monojet event $p p \to j \tilde{\chi}^0_1 \tilde{\chi}^0_1$. On the left: transversal momentum distribution $p_T(j)$; on the right: rapidity distributions $\eta(j)$.}
    \label{fig:WHIZARD}
\end{figure}

We save this file in the root directory of \WHIZARD ({\tt \$PATH/WHIZARD}). However, running it in the same directory would give some mess because \WHIZARD produces several output files. Therefore, we generate a new sub-directory which contains at the end the entire \WHIZARD output:
\begin{lstlisting}[style=terminal]
$ cd $PATH/WHIZARD
$ mkdir run_BLSSM_monojet
$ cd run_BLSSM_monojet
$ ./../bin/whizard BLSSM_monojet.sin
\end{lstlisting}
The last line runs the executable {\tt whizard} in the binary directory on our {\tt Sindarin} input file. Note, we didn't move {\tt BLSSM\_monojet.sin} to the sub-directory {\tt run\_BLSSM\_monojet}. The reason is that we might want to clean this directory by {\tt rm *} in order to make a new run with other settings. \\
After some time, \WHIZARD is done and has created a {\tt pdf} including both plots shown in Fig.~\ref{fig:WHIZARD}. The output directory includes also all events in the \WHIZARD native format called {\tt evx}. To turn on the output of other formats, it's possible to add the flags to the {\tt Sindarin} input file:
\begin{lstlisting}[style=file]
sample_format = <format> 
\end{lstlisting}
where {\tt <format>} can be for instance {\tt lhef} to get files in the Les Houches accord event format. For a complete list of all supported formats, I refer to the \WHIZARD manual. 

\subsection{Dilepton analysis with \MG}
As second example for doing a collider study with \SARAH model files, I'll show the usage of \UFO model files with \MG \cite{Degrande:2011ua}. The \UFO format is also supported by other tools like \Herwig or \Sherpa and the user can pick his/her favourite MC program. The command to generate the \UFO files is
\begin{lstlisting}[style=mathematica]
MakeUFO[]
\end{lstlisting}
As option one can give a list of generic vertices which shouldn't be included in the output similar to {\tt MakeWHIZARD}: {\tt Exclude -> \$LIST}. By default, four scalar vertices are excluded and this is sufficient for us to have a speedy output. The \UFO model files for the \BLSSM is written to 
\begin{verbatim}
$PATH/SARAH/Output/B-L-SSM/EWSB/UFO
\end{verbatim}
All files included in this directory have to be copied to a new sub-directory in \MG's model directory:
\begin{lstlisting}[style=terminal]
cd $PATH/MADGRAPH/
mkdir models/BLSSM
cp $PATH/SARAH/Output/B-L-SSM/EWSB/UFO/* models/BLSSM
\end{lstlisting}
Now, we can import this model in \MG and work with it. For this purpose one can either start the interactive mode by running 
\begin{lstlisting}[style=terminal]
./bin/mg5_aMC
\end{lstlisting}
or one can make a short input file including all necessary commands and give it as argument:
\begin{lstlisting}[style=terminal]
./bin/mg5_aMC Input_pp-MuMu.txt
\end{lstlisting}
Here, I used a file {\tt Input\_pp-MuMu.txt} which contains the following lines:
\begin{lstlisting}[style=file,title=\hspace{12cm}Input\_pp-MuMu.txt]
import model BLSSM -modelname
define p d1 d1bar d2 d2bar u1 u1bar u2 u2bar
generate p p > e2 e2bar
output ppMuMu
exit
\end{lstlisting}
In the first line we import the model in \MG. The option {\tt -modelname} is used to keep the names of the particles as given in the model files. By default, \MG will try to use the default naming conventions. However, this would fail for this model, because there are more than two CP even scalars and {\tt h3} can be used as name for the CP odd one as \MG wants to do\footnote{One has also to be careful that \MG uses the PDGs of sfermions according to their flavour. However, in the case of flavour violation this is longer possible. Moreover, \SPheno versions by \SARAH always sort them by their mass. Hence, it is always safer to use the option {\tt -modelname}. One can use the command {\tt display particles} in the interactive \MG session to see the names of all particles present in the loaded model.}. We define a multi-particle called {\tt p} which consists of all light quarks. We can skip the gluon because it won't contribute to our process. The muon is the second lepton which is called {\tt e2} and the anti-muon is accordingly {\tt e2bar}. Thus, in the third line we generate the process $pp\to \mu\bar \mu$. The output for \ME is written to a new subdirectory {\tt ppMuMu} and we close \MG when it is done via {\tt exit}. \\

After \MG has created the output for \ME and finished, we can enter the new subdirectory {\tt ppMuMu}. The important settings to generate events are done via the files in the {\tt Cards}-directory: the file {\tt param\_card.dat} is used to give the input for all parameters, {\tt run\_card.dat} controls the event generations. In the last file, the user can for instance set the beam type and energy, define the renormalization scale,  apply cuts, and fix the number of events. \\

We want to use, of course, the spectrum file as written by \SPheno. However, there is one caveat:  \ME has problems with reading the \HB specific blocks in the \SPheno spectrum file ({\tt HiggsBoundsInputHiggsCouplingsFermions} and 
{\tt HiggsBoundsInputHiggsCouplingsBosons}). If these blocks are included, \ME won't accept the file.  Therefore, we either modify the output by hand and delete these blocks or we re-generate the file by changing the options in the Les Houches input file. The \HB blocks are  disabled by the flag
\lstset{frame=shadowbox}
\begin{lstlisting}[style=file,numbers=none,title=\hspace{12cm}LesHouches.in.BLSSM]
Block SPhenoInput   # SPheno specific input 
...
 11 1               # calculate branching ratios 
...
520 0.              # Write effective Higgs couplings
\end{lstlisting}
In addition, we turned on the decays just in case that this was not done before: \ME is going to read the decay blocks from \SPheno to know the widths of all particles. If those widths are not provided via the SLHA file it is necessary to calculate them first with \ME before generating events. \\
When we have the spectrum file in the correct form, we can copy this file to the {\tt Cards} directory as {\tt param\_card.dat}. 
\begin{lstlisting}[style=terminal]
$ cd $PATH/MADGRAPH/ppMuMu/
$ cp $PATH/SPHENO/SPheno.spc.BLSSM Cards/param_card.dat
\end{lstlisting}
The other settings we have to do demand small modifications on the run-card: we want to generate one million events and we want to apply a cut on the invariant mass of leptons to get rid of the $Z$-peak. The number of events is set here:
\begin{lstlisting}[style=file,firstnumber=27,title=\hspace{13cm}run\_card.dat]
#*******************************************************************
# Number of events and rnd seed                                    *
# Warning: Do not generate more than 1M events in a single run     *
# If you want to run Pythia, avoid more than 50k events in a run.  *
#*******************************************************************
  1000000 = nevents ! Number of unweighted events requested
\end{lstlisting}
\vspace{0.5cm}
And the cuts are applied here:
\begin{lstlisting}[style=file,firstnumber=281,title=\hspace{13cm}run\_card.dat]
#*******************************************************************
# Minimum and maximum invariant mass for pairs                     *
# WARNING: for four lepton final state mmll cut require to have    *
#          different lepton masses for each flavor!                *           
#*******************************************************************
 ...
 ...
 ...
 200   = mmll ! min invariant mass of l+l- (same flavour) lepton pair
\end{lstlisting}
\vspace{0.5cm}
We are now ready to generate the events. This can either be done again in the interactive mode by starting
\begin{lstlisting}[style=terminal]
./bin/madevent
\end{lstlisting}
or we can directly start the event generation with
\begin{lstlisting}[style=terminal]
./bin/generate_events 0 0 
\end{lstlisting}
The two {\tt 0}'s are used as argument because we don't want to make any further modifications on the {\tt param}- or {\tt run}- card, and we also don't want to run {\tt pythia} or any detector simulation. When starting \ME in that way a long list of warnings appears on the screen:
\begin{lstlisting}[style=terminal]
...
WARNING: information about "imuvmix [6, 2]" is missing (full block missing) using default value: 0.0. 
...
\end{lstlisting}
The reason is that the \UFO model files by \SARAH in general can handle complex parameters. However, \SPheno does only print the real parts if we don't turn on CP violation. The zeros for all imaginary parts are not given explicitly in the spectrum file. Thus, \ME doesn't find an input for the imaginary parts and takes them as zero as it should. In addition, \ME prints a warning for each parameter where this happens. Thus, we don't have to worry about these many warnings. \\
\ME will give a status update in a new browser window. When it is done, the events are saved in the Les Houches event format and can be processed further. \\

\begin{figure}[hbt]
\centering
\includegraphics[width=0.6\linewidth]{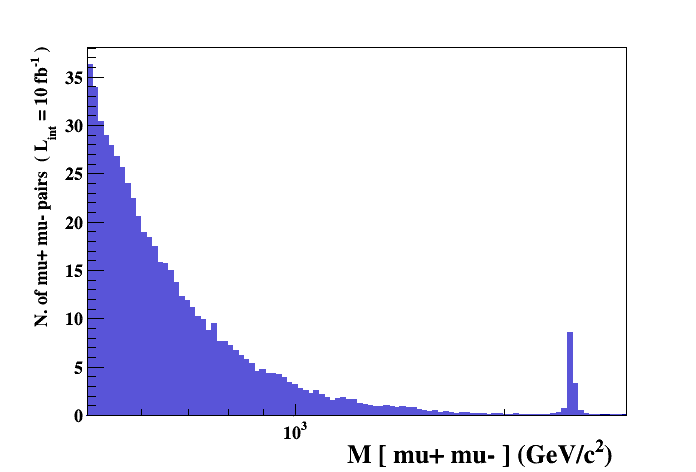}
\caption{Invariant mass of the $\mu$-pair.}
\label{fig:MGmumu}
\end{figure}

We are just going to make a plot to check if the $Z'$ peak shows up. This can for instance be done with \MA \cite{Conte:2012fm} which I assume here to be installed as well in {\$PATH}. We make another short input file called {\tt plotMuMu.txt} and save it in {\tt \$PATH/MADANALYSIS}. The content of the file is the following
\begin{lstlisting}[style=file,title=\hspace{13cm}plotMuMu.txt]
import $PATH/MADGRAPH/ppMuMu/Events/run_01/unweighted_events.lhe.gz
plot M(mu+ mu-) 100 500 3000  [logX]
submit ppMuMu
\end{lstlisting}
In the first line we import the unweighted events which are generated by \MG and which are saved by default in the {\tt LHE} format. In the second line, we make a histogram of the invariant mass of the muon pair in the mass range of 500 to 3000~GeV using 100 bins. For the $x$-axis we use a log scale ({\tt logX}). Finally, everything is submitted to be evaluated by \MA and the output directory should be called {\tt ppMuMu}. We run \MA on that file:
\begin{lstlisting}[style=terminal]
./bin/ma5  plotMuMu.txt
\end{lstlisting}
The output of \MA is stored in {\tt \$PATH/MADANALYSIS/ppMuMU} and contains also the plot shown in Fig.~\ref{fig:MGmumu} with the expected peak at 2.5~TeV.

\section{Example -- Part V: Making scans}
\label{sec:example_scans}

We have learned in the last sections how \SARAH can be used together with other tools to study all aspects of a model. Of course, it is often not sufficient to consider just one single parameter point. SUSY models like the \BLSSM have even in their constrained version a large parameter space which wants to be explored. Thus, at some point one has to start making scans to check many different points. I'll discuss two possibilities how to perform scans: the first one is only using functions the {\tt Linux} bash provides together with simple scripts\footnote{Unfortunately, I don't have experience with Mac. However, I expect that MacOS provides similar functions}. That might be sufficient to check quickly the dependence of a few observables on a single parameter. Afterwards, I'll introduce the \Mathematica package \SSP which is a dedicated tool for more sophisticated scans. 

\subsection{Using shell scripts}
\begin{figure}[hbt]
\centering
\includegraphics[width=0.6\linewidth]{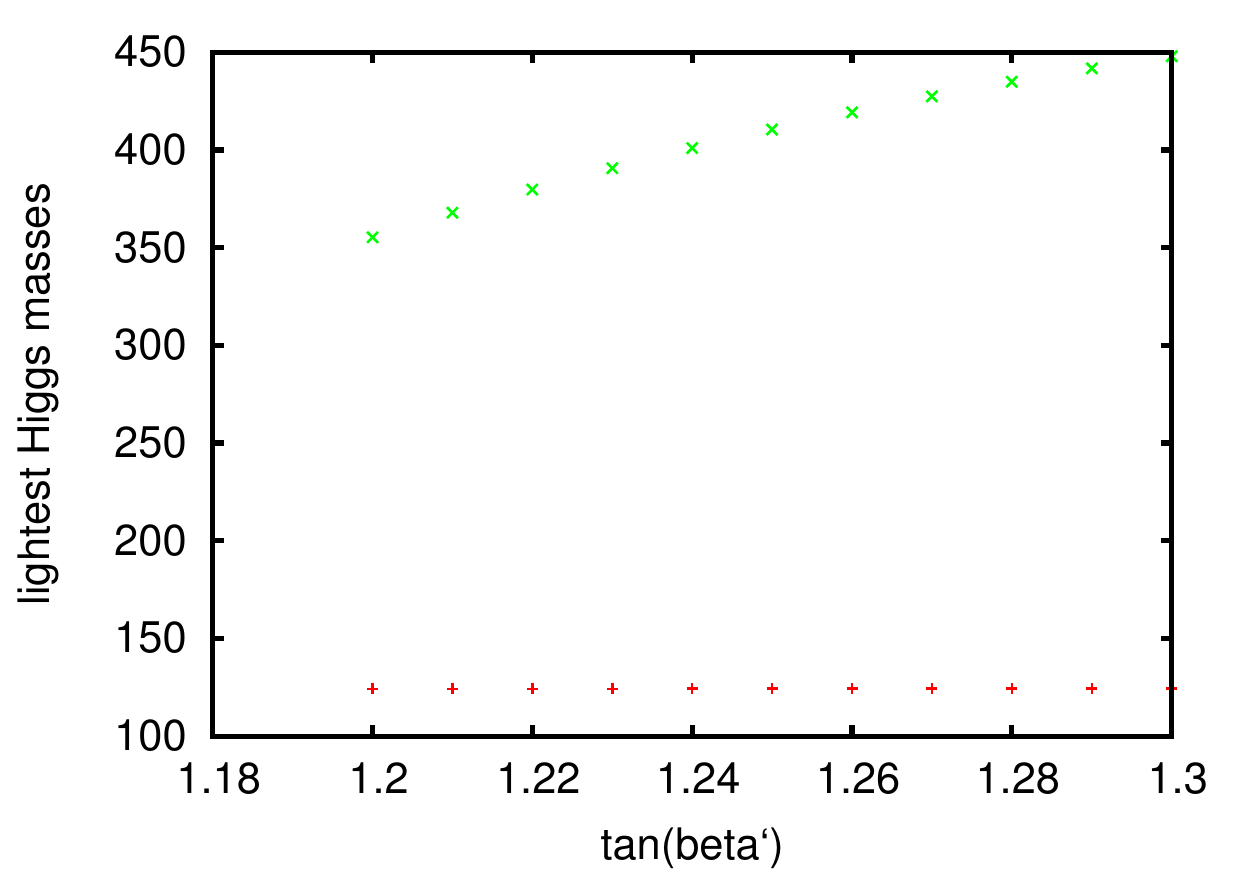}
\caption{Simple plot created with a shell script and {\tt gnuplot} showing the two lightest scalar masses as function of $\tan\beta'$.}
\label{fig:gnuplot}
\end{figure}

Let's assume that one is just interested in the dependence of the two lightest Higgs masses on $\tan\beta'$ in a small range starting from our parameter point EP1. In principle, one doesn't need any additional software to do a small scan but {\tt Linux} provides everything what's needed. For this purpose we create a file called {\tt LesHouches.in.BLSSM\_Template} which is the input file for EP1 with just one change: we replace the input value for $\tan\beta'$ by an unique string
\begin{lstlisting}[style=file,numbers=none, title=\hspace{11cm}LesHouches.in.BLSSM\_Template]
Block MINPAR      # Input parameters 
...
6   1.0000000E+00    # SignumMuP
7   TBPINPUT         # TBetaP
8   2.5000000E+03    # MZp 
\end{lstlisting}
Now, we can write a short bash script which makes a loop over all numbers from 1.2 to 1.3 in steps of 0.01 using the {\tt seq} command. For each value we use the {\tt sed} command to replace the string {\tt TBPINPUT} in {\tt LesHouches.in.BLSSM\_Template} by the value of the loop variable and to generate a complete input file {\tt LesHouches.in.BLSSM} in that way. We run \SPheno with that file and use {\tt grep} and {\tt sed} to extract the masses of the two lightest Higgs states. These numbers are 'piped' into two files called {\tt results\_hX.dat} with {\tt X}={\tt 1},{\tt 2} using \verb">>". The full script called {\tt RunSPheno.sh} reads:
\lstset{escapeinside={}}
\begin{lstlisting}[style=file,title=\hspace{13cm}RunSPheno.sh]
#!/bin/bash
 rm results_h1.dat
 rm results_h2.dat
 for i in $(seq 1.20 0.01 1.30)
 do
 sed -e "s#TBPINPUT#$i#" LesHouches.in.BLSSM_Template > LesHouches.in.BLSSM
 rm SPheno.spc.BLSSM
 ./bin/SPhenoBLSSM
 mh1=`cat SPheno.spc.BLSSM | grep "# hh_1" | grep -v DECAY | sed 's/\(.*\) 25 \(.*\)\# hh_1/\2/g'`
 mh2=`cat SPheno.spc.BLSSM | grep "# hh_2" | grep -v DECAY | sed 's/\(.*\) 35 \(.*\)\# hh_2/\2/g'`
 echo "$i $mh1" >> results_h1.dat
 echo "$i $mh2" >> results_h2.dat
 done 
\end{lstlisting}
We see here that a very handy method to extract single lines from the \SPheno spectrum file is to use {\tt grep} with the comments appearing in the spc file ({\tt \# ...}). The {\tt sed} commands after {\tt grep} are used to cut the PDG and the comment appearing in the same line in the spectrum file, i.e. the variable {\tt mh1} just contains a real number at the end. \\
The script has to be saved in the \SPheno root directory where also {\tt LesHouches.in.BLSSM\_Template} is located. Otherwise, the paths must be adjusted accordingly. We run the script via
\begin{lstlisting}[style=terminal]
$ ./RunSPheno.sh
\end{lstlisting}
When the script is finished, the file {\tt results\_h1.dat} just contain in each line a pair of the  $\tan\beta'$ value and of the corresponding Higgs mass 
\begin{lstlisting}[style=file,title=\hspace{13.5cm}results\_h1.dat]
1.20     1.24183203E+02  
1.21     1.24194534E+02  
1.22     1.24203648E+02  
...
\end{lstlisting}
We can plot both masses as function of $\tan\beta'$ for instance with {\tt gnuplot} which is also included in many Linux distributions. For this purpose we write a short input file ({\tt gnuplot\_mh.txt}) like 
\begin{lstlisting}[style=file,title=\hspace{13cm}gnuplot\_mh.txt]
set terminal postscript eps 25 color solid linewidth 3 enhanced;
set output 'TBpMh.eps';
set xlabel 'tan(beta`)';
set ylabel 'lightest Higgs masses';
set key off;
plot "results_h1.dat", "results_h2.dat";
exit 
\end{lstlisting}
I used here basic {\tt gnuplot} commands to adjust the output format (line 1), the name of the output file ({\tt TBpMH.eps}), the labels for the axes (lines 3 \& 4), disabling the legend  (line 5) and plotting the content of the two files with our data (line 6). 
We run {\tt gnuplot} on that file
\begin{lstlisting}[style=terminal]
$ gnuplot gnuplot_mh.txt
\end{lstlisting}
and get the plot shown in Fig.~\ref{fig:gnuplot}.\\

There are now many possibilities to improve this ansatz. One can include easily in the script to run \SPheno also other codes; the scans can be varied by playing with {\tt seq}, more observables can be stored; the appearance of the plot can improved by using the full power {\tt gnuplot} provides to polish the layout; and so on.  However, I think there is no need to invent the wheel again and again. There are  public tools which can be used for scanning and plotting. I'll discuss briefly \SSP now which is one of these tools.

\subsection{Making scans with \SSP}
A tool which is optimized for parameter scans using \SPheno and the other tools discussed so far, is the \Mathematica package {\tt SSP} ({\tt SARAH Scan and Plot}). \SSP provides functions for simple random or grid scans, but can also make use of intrinsic \Mathematica functions to sample the parameter space or to include constraints directly during the scan. I want to discuss here two simple examples. First, a linear scan in $M_{Z'}$
\begin{equation}
M_{Z'} \in [2.5,4.0]~\text{TeV} 
\end{equation}
Second, a grid scan in the range
\begin{equation}
\tan\beta' \in [1.20,1.25] \hspace{1cm} M_{Z'} \in [2.5,3.0]~\text{TeV} 
\end{equation}
All other parameters are set to the values of EP1. For more complicated scans one can also study the examples which are delivered with \SSP.

\subsubsection{General setup}
First, we need a file which contains information about the location and usage of all the different tools. For this purpose ,we rename the file {\tt DefaultSettings.in} included in the \SSP package to {\tt DefaultSettings.in.BLSSM}. The content should look like follows: 
\begin{lstlisting}[style=file,title=\hspace{11cm}DefaultSettings.in.BLSSM]
DEFAULT[SPheno] = "$PATH/SPHENO/bin/SPhenoBLSSM"; 
DEFAULT[SPhenoInputFile] = "LesHouches.in.BLSSM";
DEFAULT[SPhenoSpectrumFile] = "SPheno.spc.BLSSM";

DEFAULT[MicroOmegas] = "$PATH/MICROMEGAS/BLSSM/CalcOmega_with_DDetection_MO4"; 
DEFAULT[MicroOmegasInputFile] = "SPheno.spc.BLSSM"; 
DEFAULT[MicroOmegasOutputFile] = "omg.out";    
DEFAULT[DarkMatterCandidate] = ALL; 

DEFAULT[HiggsBounds] = "$PATH/HIGGSBOUNDS/HiggsBounds LandH effC 6 1"; 
DEFAULT[HiggsSignals] = "$PATH/HIGGSBOUNDS/HiggsSignals latestresults peak 2 effC 6 1"; 

DEFAULT[VevaciousBin] = "$PATH/VEVACIOUS/bin/Vevacious.exe";
DEFAULT[VevaciousInit] = "$PATH/VEVACIOUS1/bin/VevaciousInitialization_BLSSM.xml";
\end{lstlisting}
Of course, {\tt \$PATH} has to be replaced everywhere by the installation directory of the different tools. The absolute path to the executable has to be defined for \SPheno and the name for the in- and output has to be given. Also the path for the executable for \MO is set. The names of the spectrum file used as input, and the output file written by \MO is the other information necessary to include \MO in the scan. In addition, one can define if \MO should only calculate the relic density if a specific particle is the LSP. In that case the PDG has to be given, i.e. either {\tt 1000022} for a neutralino LSP or {\tt 1000012} for a CP even sneutrino LSP. We use here {\tt ALL} to calculate the relic for any particle. One could also use 
\begin{lstlisting}[style=file]
DEFAULT[DarkMatterCandidate] = 1000022 | 1000012;  
\end{lstlisting}
to just consider a subset of particles. The lines below give the commands to run \HB and \HS as explained in sec.~\ref{sec:HBHS}. Finally, to run \Vevacious the path to the executable as well as the desired initialization file have to be given as done in the last two lines. 

\subsubsection{Defining a scan}
A second input file defines the scan we want to make. \SARAH also writes templates for this file during the \SPheno output which could be used as starting point. The file names of these templates start with {\tt  SSP\_Template}. We call the file for our examples here {\tt BLSSM\_TBpMZp.m}. The different parts are:\\
At the very beginning, the file which contains the information about the installation of the codes involved in the scans is loaded. This is the file we have set up in the first step. Then, identifiers for all scans which we want to make are defined using the list {\tt RunScans}. We just perform two scans here as said above called {\tt MZpLinear} and {\tt MZpTBpGrid}, but there is in principle no limit how many scans are done within a single file. 
By default, \SSP always runs \SPheno. We also want to include here \HB and \HS and put therefore the flags to {\tt True}. To include \MO as well, it would just be necessary to put also that flag to {\tt True}. However, this would slow down the scan significantly because different LSPs show up in the range we have chosen and \MO would need a long time to compile all amplitudes. Hence, I skip it for the example here but for practical applications it an easily be included. For the same reason, I have also not included \Vevacious in the scan.  
\begin{lstlisting}[style=file,title=\hspace{12cm}BLSSM\_TBpMZp.m]
(* SETUP *)
LoadSettings="DefaultSettings.m.BLSSM";  

RunScans = {MZpLinear,MZpTBpGrid}; 
 
DEFINITION[a_][IncludeMicrOmegas]=False; 
DEFINITION[a_][IncludeVevacious]=False;
DEFINITION[a_][IncludeHiggsBounds]=True;
DEFINITION[a_][IncludeHiggsSignals]=True;    
\end{lstlisting}Note, we applied all definitions to any scan defined in this file because we used {\tt DEFINITION[a\_]}. To use different options for the different scans, {\tt DEFINITION[\$NAMEofSCAN]} can be used. \\

Now, the main part which defines all input parameters and ranges follows. \SSP is very agnostic concerning the underlying model. Therefore, it is first necessary to tell it what blocks are actually needed for a scan before the values can be defined ({\tt DEFINITION[a\_][Blocks]=\{...\}}). These are the blocks which we discussed in sec.~\ref{sec:spheno_LH}. The main part of the input is setting the numbers for these blocks and their different entries. In this context, the blocks are defined as arrays: first the block number appears, then we can give the numerical value. Fixed values are assigned by the flag {\tt Value}. Thus, the blocks {\tt MODSEL}, {\tt SMINPUTS} and {\tt SPhenoInput}  which just come with fixed values read:
\begin{lstlisting}[style=file,firstnumber=10,title=\hspace{12cm}BLSSM\_TBpMZp.m]
DEFINITION[a_][Blocks]={MODSEL,SMINPUTS,SPhenoInput,MINPAR,YVIN,YXIN}; 
 
DEFINITION[a_][MODSEL]={ 
{{1},{Value->1}}, 
{{2},{Value->1}}, 
{{6},{Value->0}} 
}; 
DEFINITION[a_][SMINPUTS]={ 
{{2},{Value->1.166390*10^-5}}, 
{{3},{Value->0.1172}}, 
{{4},{Value->91.18760}}, 
{{5},{Value->4.2}}, 
{{6},{Value->172.9}}, 
{{7},{Value->1.777}} 
}; 
DEFINITION[a_][SPhenoInput]={ 
{{1},{Value->-1}}, (* error level *)
{{2},{Value->0}},  (* SPA conventions *) 
{{7},{Value->0}},
{{8},{Value->3}},
{{11},{Value->1}}, (* Calculate widhts and BRs *)
{{12},{Value->0.0001}}, (* minimal BR to write out *) 
{{13},{Value->0}}, (* Enable 3-body decays *) 
{{34},{Value->0.0001}}, (* precision of masses *) 
{{38},{Value->2}}, (* 1/2 - Loop RGEs *) 
{{50},{Value->0}},
{{51},{Value->0}}, (* Switch to CKM matrix *) 
{{55},{Value->1}}, (* 1 - Loop masses *)
{{57},{Value->0}}, (* low energy constraints *)
{{60},{Value->1}}, (* Include possible, kinetic mixing *)
{{65},{Value->1}}, (* Solution tadpole equation *)
{{75},{Value->1}}, (* Write WHIZARD files *) 
{{76},{Value->1}},  (* Write HiggsBounds files *) 
{{86},{Value->0.}},  (* Maximal width to be counted as invisible in Higgs decays; -1: only LSP *) 
{{550},{Value->1}},  (* Calculate Fine-Tuning *) 
{{530},{Value->1.}}  (* Write Blocks for Vevacious *) 
}; 
\end{lstlisting}
The other blocks showing up in the Les Houches file are those to set the parameters for the scans. Those are defined in a similar way:
\begin{lstlisting}[style=file,firstnumber=48,title=\hspace{12cm}BLSSM\_TBpMZp.m]
DEFINITION[MZpLinear][MINPAR]={ 
{{1},{Value->1700}} (*m0*), 
{{2},{Value->1700}} (*m12*), 
{{3},{Value->7}} (*TanBeta*), 
{{4},{Value->1}} (*SignumMu*), 
{{5},{Value->-1400}} (*Azero*), 
{{6},{Value->1}} (*SignumMuP*), 
{{7},{Value->1.20}} (*TBetaP*), 
{{8},{Min->2500, Max->4000, Steps->40, Distribution->LINEAR}} (*MZp*)}; 

DEFINITION[MZpTBpGrid][MINPAR]={ 
{{1},{Value->1700}} (*m0*), 
{{2},{Value->1700}} (*m12*), 
{{3},{Value->7}} (*TanBeta*), 
{{4},{Value->1}} (*SignumMu*), 
{{5},{Value->-1400}} (*Azero*), 
{{6},{Value->1}} (*SignumMuP*), 
{{7},{Min->1.2, Max->1.25, Steps->15, Distribution->LINEAR}} (*TBetaP*), 
{{8},{Min->2500, Max->3000, Steps->15, Distribution->LINEAR}} (*MZp*)}; 
 
DEFINITION[a_][YVIN]={ 
{{1,1},{Value->0}} ,
{{3,3},{Value->0}} }; 
 
DEFINITION[a_][YXIN]={ 
{{1,1},{Value->0.42}} , 
{{2,2},{Value->0.42}} , 
{{3,3},{Value->0.30}} }; 
\end{lstlisting}
We see that we can use for both scans exactly the same blocks but for {\tt MINPAR}. That means the majority of blocks had just to be defined once using again {\tt DEFINITION[a\_]}. For the two versions of {\tt MINPAR} we gave the name of the scans as arguments. Also the scan ranges are defined easily: \\
The linear scan is set up by varying $M_{Z'}$ ({\tt MINPAR[8]}) in the range between 2500 and 4000 using 40 steps with a linear distribution. The grid scan is set by varying $\tan\beta'$ and $M_{Z'}$ ({\tt MINPAR[7]}, {\tt MINPAR[8]}) within the given limits and assuming linear distributions in both directions. There are also other options possible, e.g. a logarithmic distribution ({\tt Distribution->LOG}) or random distribution ({\tt Distribution->RANDOM}). Also relations to other parameters can be given. For instance, to scale in a scan $m_0$ the same way as $M_{1/2}$, one could use
\begin{lstlisting}[style=file,numbers=none]
DEFINITION[MZpTBpGridZoom][MINPAR]={ 
{{1},{Min->1500, Max->3000, Steps->25, Distribution->LINEAR}} (*m0*), 
{{2},{Value->MINPAR[1]}} (*m12*)
...
\end{lstlisting}
For the possibility to perform basic Marcov-Chain Monte-Carlo runs, to apply fits during the scan, or to make a sampling of the parameter space, I refer to the \SSP manual and the examples which come with \SSP. \\

Finally, we want to get some figures automatically when the scan is finished. For the linear scan we want to plot (i) the two lightest CP even sneutrinos, (ii) the lightest CP even and odd sneutrino, (iii) the two lightest neutralino masses, and (iv) the composition of the lightest neutralino. To make the plots a bit more appealing, we first generate a basic style ({\tt BasicStyle}) which is applied to all figures. Since our figures are a composition of two respectively four plots, we have to define also styles with two ({\tt StyleDefault2}) and four ({\tt StyleDefault4}) colours. That's done by mapping our colours on the basic style. These definitions so far a purely \Mathematica commands. 
When this is done, the plots themselves are defined quickly: we use the option {\tt P2D} of \SSP for a 2-dimensional plots, and set for each figure what parameters and observables should be shown, what style should be used, what's the label of the $y$ axis and what's the name for the output file. 
\begin{lstlisting}[style=file,firstnumber=78,title=\hspace{12cm}BLSSM\_TBpMZp.m]
BasicStyle={Frame->True, Axes->False,
         FrameTicksStyle -> Directive[Black, 10],
         ImageSize -> 200, ImageMargins->10, 
         Joined->True,FrameLabel->{UseLaTeX["$M_{Z'}$~[GeV]"],yAxis}};
StyleDefault2 = Map[Join[BasicStyle,{PlotStyle->#}]&,{Red,Green}];
StyleDefault4 = Map[Join[BasicStyle,{PlotStyle->#}]&,{Red,Green,Blue,Black}];

DEFINITION[MZpLinear][Plots]={ 
   {P2D, {MINPAR[8],{MASS[1000012],MASS[1000014]}},
      StyleDefault2  /. yAxis->UseLaTeX["$m_{\\tilde \\nu^R_{1,2}}$~[GeV]"],
      "MZp_MSvRe.pdf"},
   {P2D, {MINPAR[8],{MASS[1000012],MASS[4000012]}},
      StyleDefault2  /. yAxis->UseLaTeX["$m_{\\tilde \\nu^R_1},m_{\\tilde \\nu^I_1}$~[GeV]"],
      "MZp_MSvRe1_SvIm1.pdf"},
   {P2D, {MINPAR[8],{Abs[MASS[1000022]],Abs[MASS[1000023]]}},
      StyleDefault2/. yAxis->UseLaTeX["$m_{\\tilde \\chi^0_{1,2}}$~[GeV]"],
     "MZp_MChi.pdf"},
   {P2D, {MINPAR[8],{Log[10,NMIX[1,1]^2+NMIX[1,2]^2], Log[10,NMIX[1,3]^2+NMIX[1,4]^2],
    Log[10,NMIX[1,5]^2 ], Log[10,NMIX[1,6]^2 +NMIX[1,7]^2]}}, 
      StyleDefault4/. yAxis->UseLaTeX["$\\log(Z^N_{i1})$~[GeV]"],
     "MZp_ZN.pdf"}
};  
\end{lstlisting}
Note, I used here the keyword {\tt UseLaTeX} together with \LaTeX\ syntax for the labels. By doing this, \SSP calls the script {\tt fragmaster}\footnote{\tt http://www.ctan.org/tex-archive/support/fragmaster} which makes use of {\tt psfrag} to get nice looking labels. \\

The plots for the grid scan are actually even simpler to define because we can use the same style for each plot. The flag {\tt P3D} performs 3-dimensional plots (contour plots) based on the \Mathematica function {\tt ListContourPlot}. Thus, one can set the options for these plots by using the {\tt SetOptions} command of \Mathematica. 
We are going to make four plots again: (i,ii) the two lightest Higgs masses in the $(\tan\beta', M_{Z'})$ plane, and the results of (iii) \HB and (iv) \HS in the same plane: 
\begin{lstlisting}[style=file,firstnumber=105,title=\hspace{12cm}BLSSM\_TBpMZp.m] 
StyleDefault={Frame->True, Axes->False,
              FrameLabel->{tan\[Beta],Subscript["M","Z'"]},
              FrameTicksStyle -> Directive[Black, 10],
              ContourLabels->True, ImageSize -> 200};
              
DEFINITION[MZpTBpGrid][Plots]={ 
   {P3D, {MINPAR[7],MINPAR[8], MASS[25]},StyleDefault,"tbp_MZp_Mass25.pdf"},
   {P3D, {MINPAR[7],MINPAR[8], MASS[35]},StyleDefault,"tbp_MZp_Mass35.pdf"},
   {P3D, {MINPAR[7],MINPAR[8], HIGGSBOUNDS[10]},StyleDefault,"tbp_MZp_HB10.pdf"},
   {P3D, {MINPAR[7],MINPAR[8], HIGGSSIGNALS[10]},StyleDefault,"tbp_MZp_HS10.pdf"}
}; 
\end{lstlisting}
When we are done with setting up the entire scan, it is started by running \Mathematica with
\begin{lstlisting}[style=mathematica]
<<$PATH/SSP/SSP.m;
Start["BLSSM_TBpMZp.m"];
\end{lstlisting}

\begin{figure}[hbt]
\centering
\includegraphics[width=0.45\linewidth]{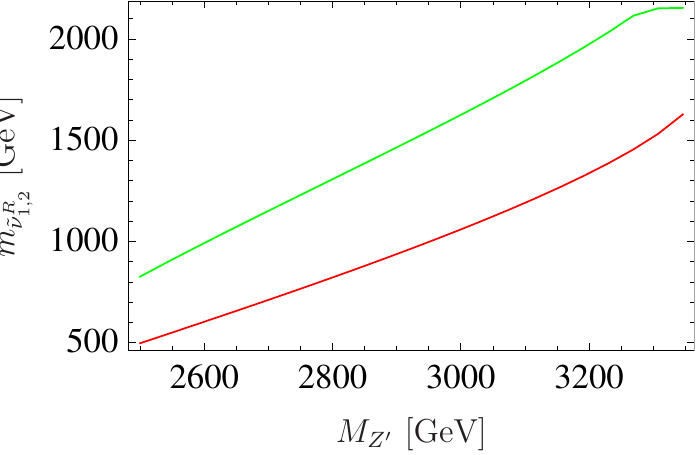} \hfill 
\includegraphics[width=0.45\linewidth]{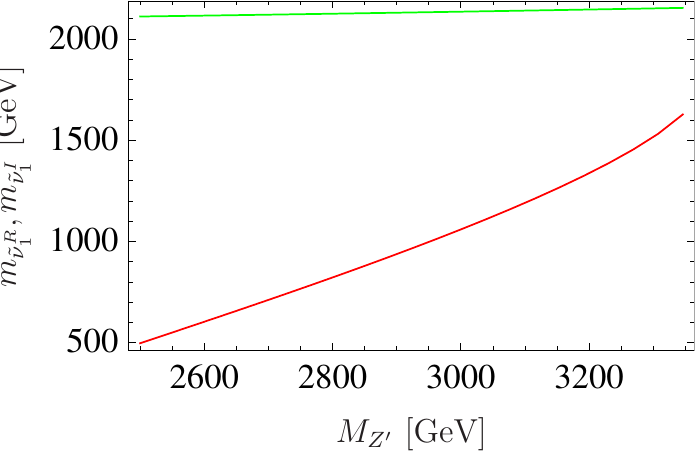} \\
\includegraphics[width=0.45\linewidth]{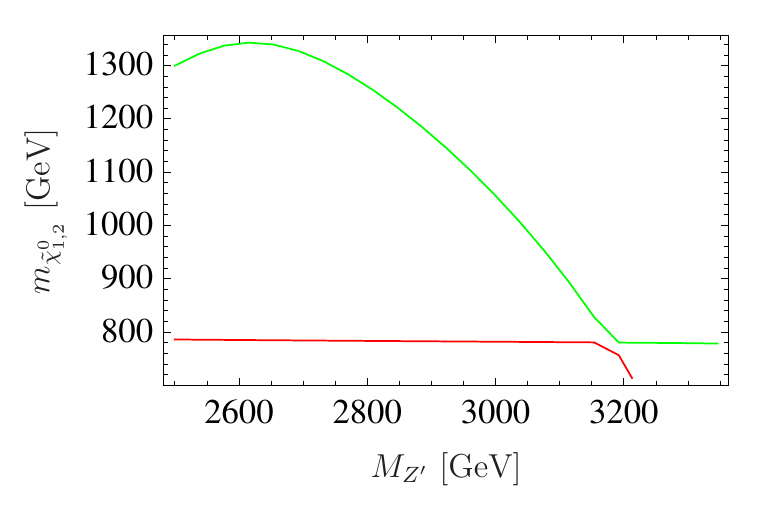} \hfill 
\includegraphics[width=0.45\linewidth]{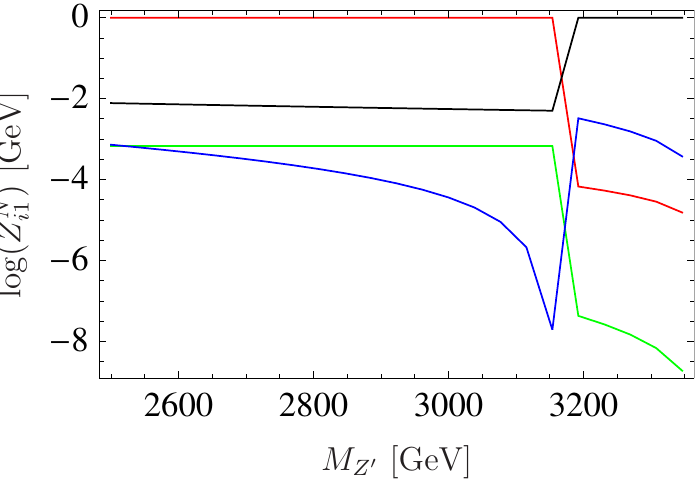}
\caption{The two lightest CP even sneutrino masses (top left), the lightest CP even and odd sneutrino mass (top right), the lightest CP even sneutrino and the lightest neutralino mass (bottom left), and the decomposition of the lightest neutralino (bottom right). Plots are produced with \SSP using EP1 and a variation of $\tan\beta'$.}
\label{fig:SSPlinear}
\end{figure}

\begin{figure}[hbt]
\centering
\includegraphics[width=0.45\linewidth]{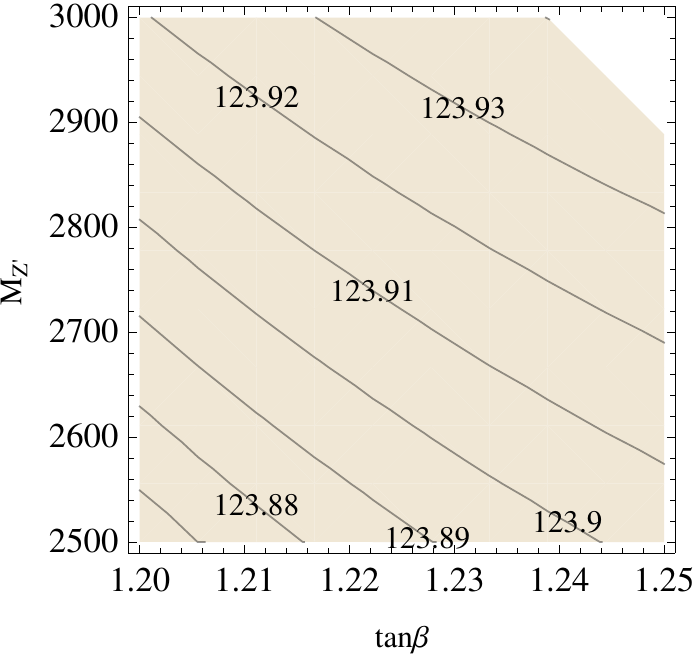} \hfill 
\includegraphics[width=0.45\linewidth]{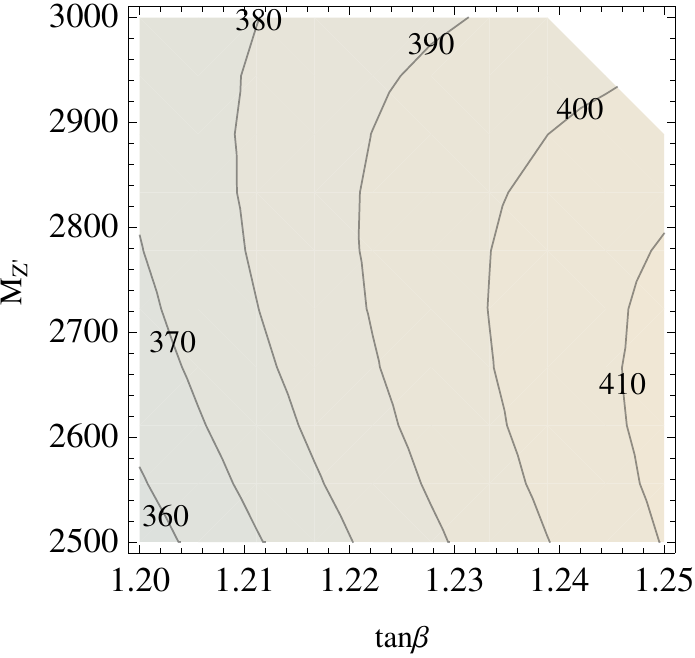} \\
\includegraphics[width=0.45\linewidth]{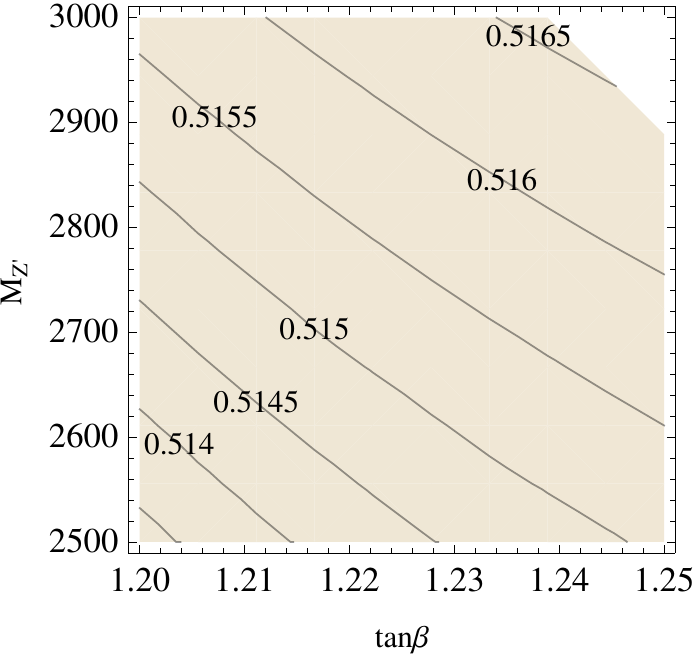} \hfill 
\includegraphics[width=0.45\linewidth]{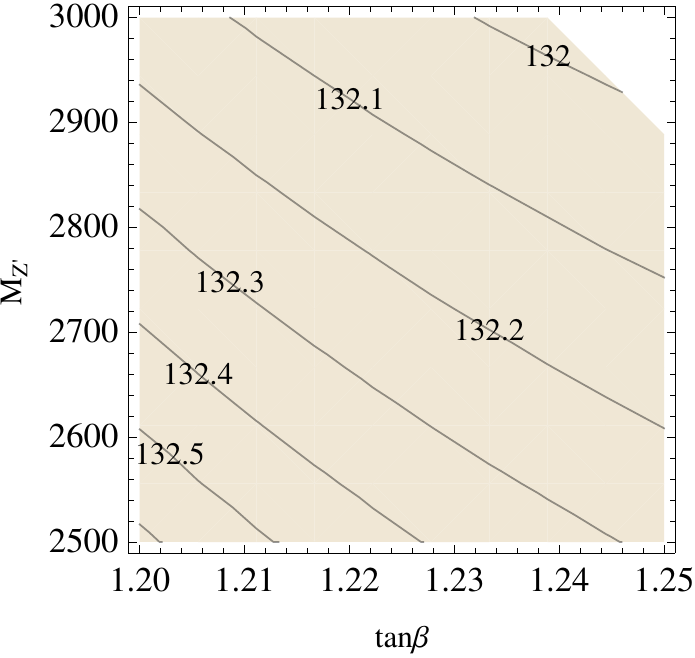}
\caption{The two lightest scalar Higgs masses (top row), the '{\tt obsratio}' as calculated by \HB (bottom left), and the total $\chi^2$ as calculated by \HS. Plots are produced with \SSP using EP1 and making a grid scan in the $(\tan\beta', M_{Z'})$ plane.}
\label{fig:SSPgrid}
\end{figure}

The output is stored in the sub-directories 
\begin{verbatim}
$PATH/SSP/Output/MZpLinear
$PATH/SSP/Output/MZpTBpGrid
\end{verbatim}
These directories contain not only the scan data saved in the Les Houches format ({\tt SpectrumFiles.spc}) and \Mathematica format ({\tt Data.m}) but also the plots we wanted to see. I'll show these in Fig.~\ref{fig:SSPlinear} for the linear scan and in Fig.~\ref{fig:SSPgrid} for the grid scan.

\subsubsection{Using data of a scan in \Mathematica}
\begin{figure}[hbt]
\centering
\includegraphics[width=0.55\linewidth]{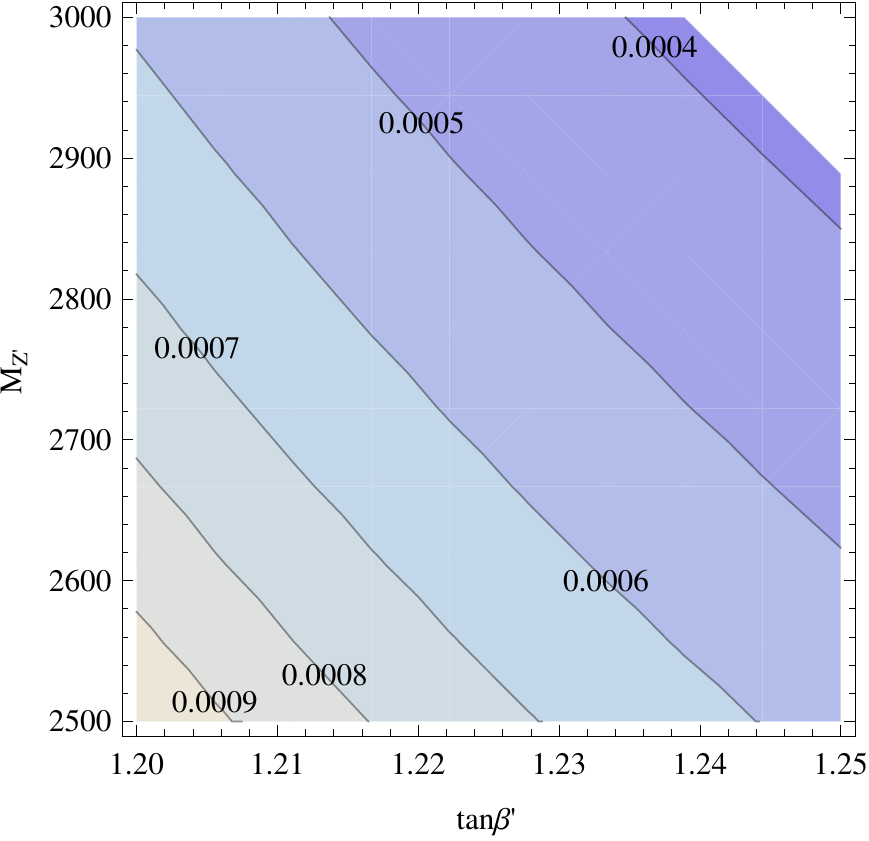} 
\caption{Example for contour plots with \SSP and \Mathematica: the doublet fraction of the second lightest scalar. We used here EP1 and varied $\tan\beta'$ and $M_{Z'}$.}
\label{fig:SSPmath}
\end{figure}

Of course, it is also possible to use the results of scans performed by \SSP later in \Mathematica. For this purpose \SSP provides a function {\tt MakeSubNum} to translate the data saved in the SLHA or \Mathematica format into a list of \Mathematica substitutions. These substitutions can then be used to either to apply cuts or to extract points, or to make more plots.\\ 
To load and format the data of the grid scan, we can either use
\begin{lstlisting}[style=mathematica]
<<$PATH/SSP/SSP.m
data=Get["$PATH/SSP/Output/MZpTBpGrid/Data.m"];
SubData = MakeSubNum/@ data;
\end{lstlisting}
or 
\begin{lstlisting}[style=mathematica]
<<$PATH/SSP/SSP.m
ReadSpectrumFile["$PATH/SSP/Output/MZpTBpGrid/SpectrumFiles.spc", "ENDOFPARAMETERFILE"];
SubData = MakeSubNum/@ AllLesHouchesInput;
\end{lstlisting}
With both options we get an array of substitutions called {\tt SubData} which we can use. However, there is one caveat: the data files for large scans are huge. These file include any information as calculated by \SPheno and the other tools. Often, not all information is really needed, but one is only interested in the behaviour of a subset of parameters or observables. Therefore, it often saves a lot of time and memory to extract the information from the big files which is actually needed, and store that information in smaller files. This can be done for instance under  {\tt Linux} with a small shell script using again {\tt grep} and the comments appearing in each line of the \SPheno output: 
\begin{lstlisting}[style=file,title=\hspace{14cm}extract.sh]
#!/bin/bash

cat *$ | grep --regexp="Block MINPAR" \
--regexp="# TBp" \
--regexp="# MZp" \
--regexp="Block MASS" \
--regexp="# hh_1" \
--regexp="# hh_2" \
--regexp="# SvRe_1" \
--regexp="# Chi_1" \
--regexp="Block SCALARMIX" \
--regexp="ZH(1,1)" \
--regexp="ZH(1,2)" \
--regexp="ZH(2,1)" \
--regexp="ZH(2,2)" \
--regexp="# FlavorKitQFV" \
--regexp="# BR(B->X_s gamma)/BR(B->X_s gamma)_SM" \
--regexp="ENDOFPARAMETERFILE" | grep -v DECAY > SmallSpectrum.out
\end{lstlisting}
This script takes as argument the name of the file containing all spectra, extracts the data and writes the necessary lines in another file called {\tt SmallSpectrum.out}. 
We call this script {\tt extract.sh} and save it in {\tt \$PATH}. It can be used then via
\begin{lstlisting}[style=terminal]
./extract  SSP/Output/TBpMZpGrid/SpectrumFiles.spc
\end{lstlisting}
The small spectrum file is now loaded much faster  
\begin{lstlisting}[style=mathematica]
<<$PATH/SSP/SSP.m
ReadSpectrumFile["$PATH/SSP/Output/MZpTBpGrid/SmallSpectrum.out", "ENDOFPARAMETERFILE"];
SubData = MakeSubNum/@ AllLesHouchesInput;
\end{lstlisting}
and we can use it for instance to make some more plots. For instance, to make a contour plot of the doublet fraction of the second lightest Higgs in the $(\tan\beta',M_{Z'})$ plane, we can use
\begin{lstlisting}[style=mathematica]
ListContourPlot[
 Table[{ MINPAR[7], MINPAR[8],
            SCALARMIX[2, 1]^2 + SCALARMIX[2, 2]^2} /. SubData[[k]], 
            {k, 1, Length[SubData]}], 
   Frame -> True, Axes -> False,ContourLabels -> True, 
   FrameLabel -> {"tan\[Beta]'", Subscript["M", "Z'"]}]
\end{lstlisting}
The obtained plot is shown in Fig.~\ref{fig:SSPmath}. 

We can also apply some cuts and collect points with a neutralino mass below 500~GeV 
\begin{lstlisting}[style=mathematica]
MChi500=Select[SubData, (Abs[MASS[1000022]] < 500 ) /. # &];
\end{lstlisting}
and check for which values of $M_{Z'}$ this occurs:
\begin{lstlisting}[style=mathematica,mathescape]
MINPAR[8] /. Chi500 
$\hookrightarrow$ {3000., 2944.44, 3000., 2888.89, 2944.44, 2888.89}
\end{lstlisting}
Thus, this is only the case for heavy $Z'$ masses. Similarly, one can also extract all points with a neutralino LSP by demanding that the lightest neutralino is lighter than the lightest CP even sneutrino:
\begin{lstlisting}[style=mathematica]
ChiLSP = Select[
   SubData, (Abs[MASS[1000022]] < Abs[MASS[1000012]] ) /. # &];
\end{lstlisting}
or check what is the lightest mass appearing for the second scalar:
\begin{lstlisting}[style=mathematica]
LightestHiggs = Select[SubData, 
   (Abs[MASS[35]] ==  Min[Table[MASS[35] /. SubData[[i]], {i, 1, Length[SubData]}]]) /. # &];
\end{lstlisting}
The relevant information about this point is shown via 
\begin{lstlisting}[style=mathematica,mathescape]
{MINPAR[7], MINPAR[8], MASS[25], MASS[35], 
       SCALARMIX[1, 1]^2 + SCALARMIX[1, 2]^2} /. LightestHiggs 
$\hookrightarrow$ {1.2, 2500., 123.863, 355.269, 0.999013}
\end{lstlisting}

We see that there is now an infinite numbers of  possibilities to study data within \Mathematica using the {\tt Select} or also other \Mathematica commands.

\section{Summary}
\label{sec:summary}
In the first part of this paper I have given an overview what models \SARAH can handle and what calculations it can do for these models. In addition, I have discussed to what other HEP tools the information derived by \SARAH can be linked. In the second part I have discussed in great detail how all aspects of a SUSY model can be studied with \SARAH and the related tools. For this purpose I choose the \BLSSM as example. The implementation of the \BLSSM in \SARAH was explained, and it was shown what can be done within \Mathematica to gain some understanding about the model. Afterwards, I have explained how \SARAH in combination with \SPheno is used to calculate the mass spectrum, decays, flavour and precision observables, and the fine-tuning. The next step was to check parameter points with \HB and \HS for their Higgs properties, with \Vevacious for their vacuum stability, and with \MO for their dark matter relic density. I have given two short examples for collider studies using \SARAH model files. First, monojet events with \WHIZARD were generated. Second, a simple dilepton analysis with \MG was done. Finally, I discussed possibilities how to perform parameter scans using either shell scripts or \SSP.\\
This manuscripts hopefully shows how helpful \SARAH can be to study models beyond the SM or MSSM. Because of a very high level of automatization the user can get quickly results with a precision which is otherwise just available for the MSSM. Of course, also the possibility to make mistakes is highly reduces compared to a home-brewed calculation. I hope that the detailed explanation of a specific example simplifies the first contact of interested users with the many different tools which are available today.

\section*{Acknowledgements}
I'm very grateful to Mark D. Goodsell and Kilian Nickel for providing routines for the two-loop calculation via the \SARAH--\SPheno interface. In particular I thank Werner Porod who raised my interest in supersymmetry. This was the starting point of the entire development of \SARAH. I'm in debt to Martin Hirsch, Avelino Vicente, Daniel Busbridge, James Scoville, Alexander Voigt, Peter Athron, Roberto Ruiz de Austri Bazan, Moritz McGarrie, Lorenzo Basso  and many others for testing \SARAH, helpful suggestions, and also their bug reports. Finally, it has been a pleasure to work with Jose Eliel Camargo-Molina, Ben O'Leary and again Werner Porod and Avelino on \Vevacious and \FlavorKit. 
I thank Manuel Krauss, Lukas Mitzka, Tim Stefaniak and Avelino Vicente for their remarks on the manuscript. 

\begin{appendix}
\section{Some more details}
I couldn't address all interesting topics in the main part. Therefore, I give in this appendix to selected topics a few more details. 
\subsection{Flags in \SARAH model files}
\label{app:sarahflags}
There are a different flags to enable or disable distinctive features which might be present in some models:
\begin{itemize}
 \item \verb"AddDiracGauginos = True/False;", default: \verb"False", includes/excludes Dirac Gaugino mass terms
 \item \verb"AddFIU1 = True/False;", default: \verb"False", includes/excludes Fayet-Iliopoulos $D$-terms
 \item \verb"NoU1Mixing = True/False;", default: \verb"False", disables effects from gauge-kinetic mixing
 \item \verb"IgnoreGaugeFixing=True/False;", default: \verb"False", disables the calculation of the gauge fixing and ghost terms. Note, that this is just possible at tree-level. For loop calculations the ghosts are needed. 
\end{itemize}
Specific parts of the Lagrangian are turned off via
\begin{itemize}
 \item \verb"AddTterms = True/False;" includes/excludes trilinear soft breaking couplings
 \item \verb"AddBterms = True/False;" includes/excludes bilinear soft breaking couplings
 \item \verb"AddLterms = True/False;" includes/excludes linear soft breaking couplings
 \item \verb"AddSoftScalarMasses = True/False;" includes/excludes soft-breaking scalar masses
 \item \verb"AddSoftGauginoMasses = True/False;" includes/excludes Majorana masses for gauginos
 \item \verb"AddSoftTerms = True/False;" includes/excludes all soft-breaking terms
 \item \verb"AddDterms = True/False;" includes/excludes all $D$-terms
 \item \verb"AddFterms = True/False;" includes/excludes all $F$-terms
\end{itemize}
By default all terms are included. In particular the last two flags have to be used very carefully.

\subsection{Parts of the Lagrangian in \SARAH}
\SARAH saves the different parts of the Lagrangian which it has derived in different variables. This happens for all eigenstates ({\tt \$EIGENSTATES}) and the user has access to this information:
\begin{itemize}
\item \verb"LagSV[$EIGENSTATES]": parts with scalars and vector bosons (i.e. kinetic terms for scalars)
\item \verb"LagFFV[$EIGENSTATES]": parts with fermions and vector bosons (i.e. kinetic terms of fermions)
\item \verb"LagSSSS[$EIGENSTATES]": parts with only scalars (i.e. scalar potential)
\item \verb"LagFFS[$EIGENSTATES]": parts with fermions and scalars
\item \verb"LagVVV[$EIGENSTATES]": parts with three vector bosons
\item \verb"LagVVVV[$EIGENSTATES]": parts with four vector bosons
\item \verb"LagGGS[$EIGENSTATES]": parts with ghosts and scalars  
\item \verb"LagGGV[$EIGENSTATES]": parts with ghosts and vector bosons
\item \verb"LagSSA[$EIGENSTATES]": parts with scalars and auxiliary fields
\end{itemize}
In addition, the different steps to derive the Lagrangian of the gauge eigenstates  are also saved in different variables:
\begin{itemize}
 \item Superpotential: {\tt Superpotential}
 \item Fermion - scalar interactions coming from the superpotential: {\tt Wij}
 \item F-Terms: {\tt FTerms}
 \item Scalar soft-breaking masses: {\tt SoftScalarMass }
 \item Gaugino masses: {\tt SoftGauginoMass }
 \item Soft-breaking couplings: {\tt SoftW}
 \item Kinetic terms for scalars: {\tt KinScalar}
 \item Kinetic terms for fermions: {\tt KinFermion}
 \item D-Terms: {\tt DTerms}
 \item Interactions between gauginos and a scalar and a fermion: {\tt FSGaugino}
 \item Trilinear self-interactions of gauge bosons: {\tt GaugeTri}
 \item Quartic self-interactions of gauge bosons: {\tt GaugeQuad}
 \item Interactions between vector bosons and gauginos: {\tt BosonGaugino}
\end{itemize}

\subsection{A more precise mass calculation}
\label{app:quadruple}
In some cases a numerical more precise calculation is needed to diagonalize mass matrices in \SPheno. This is the case if the hierarchy in the mass matrix very large. In that case double precision with about 15 digits precision might not be sufficient. The best example are models with $R$-parity violation where neutrinos and neutralinos mix. Another example are seesaw type-I like models where TeV-scale right-neutrinos mix with the left-neutrinos. In this case one has to go for quadruple precision which gives a precision of about 32 digits. To enable quadruple precision for specific masses, two small changes are necessary:\\
\begin{enumerate}
 \item In {\tt SPheno.m} used to set up the \SPheno output, one has to define for which particles the higher precision is needed. This is done with the variable {\tt QuadruplePrecision} which accepts a  list of mass eigenstates. If we just want to have the masses of the neutrinos, which are called {\tt Fv} in the considered model, with higher precision, the corresponding line reads 
 \begin{lstlisting}[style=file,numbers=none,title=\hspace{12cm}SPheno.m]
  QuadruplePrecision = {Fv};
 \end{lstlisting}
\item We must change the {\tt Makefile} of \SPheno located in the {\tt src} directory and remove the compiler flag {\tt -DONLYDOUBLE}. This flags forces all calculations just to be done with double precision. 
\begin{lstlisting}[style=file,title=\hspace{12cm}Makefile]
  PreDef = -DGENERATIONMIXING  
\end{lstlisting}
By doing that, the routines necessary for a higher precision get compiled. To make sure that everything is consistent, it might a good idea to re-compile the entire code after changing the {\tt Makefile}:
\begin{lstlisting}[style=terminal]
 cd $PATH/SPHENO
 make cleanall
 make MODEL=$NAME
\end{lstlisting}
\end{enumerate}

\subsection{More about tadpole equations and \SPheno}
\subsubsection{Numerical solutions}
\label{app:tadpoles_num}
In the main part of this paper we solved the tadpole equations for the \SPheno output with respect to $\mu$, $\mu'$, $B_\mu$ and $B_\mu'$ for which an analytical solution exists. This must not always be the case. For instance, if one wants not to use $\tan\beta'$ and $M_{Z'}$ as input, but obtain $x_1$ and $x_2$ from the minimum conditions, an analytical solution does not exist. To solve the equations numerically and to define the initialization used by the {\tt Broydn} routines used for that, {\tt SPheno.m} has to contain the following lines:
\begin{lstlisting}[style=file]
ParametersToSolveTadpoles = {\[Mu],B[\[Mu]], x1, x2};
NumericalSolutionTadpoleEquations = True;
InitializationTadpoleParameters = { \[Mu] -> m0, B[\[Mu]]-> m0^2, x1->m0, x2->m0};
\end{lstlisting}
The first line is the same as for the analytical approach and defines that the tadpole equations have to be solved with respect to
$\mu$, $B_\mu$, $x_1$, and $x_2$ this time. Without the other two lines, {\tt Mathematica's} function {\tt Solve}  would try to find an analytical solution but fail. \SARAH would then stop the output with an error message. However, due to the second line the attempts to solve the tadpole equations 
inside \Mathematica are skipped. The third line assumes that $\mu$, $x_1$ and $x_2$ are $O(m_0)$ and $B_\mu$ is $O(m_0^2)$ 
These values are used in the numerical routines for initializing the calculation. 
Of course, other possible and reasonable choices would have been to relate $\mu, B_\mu$ with the running 
soft-breaking terms of the Higgs, and $x_1$ and $x_2$ to $\mu'$ which is now used as input: 
\begin{lstlisting}[style=file]
InitializationTadpoleParameters = { \[Mu] -> Sqrt[mHd2], B[\[Mu]]-> mHd2, x1->MuP, x2->MuP};
\end{lstlisting}
Also constant values can be used
\begin{lstlisting}[style=file]
InitializationTadpoleParameters = { \[Mu] -> 10^3, B[\[Mu]]-> 10^6, x1->10^3, x2->10^3};
\end{lstlisting}
Usually, the time needed to find the solution changes only slightly with the chosen initialization values
as long as they are not completely off. Note, all choices above would only find the solution which is the closest one to the initialization values. However, the equations are cubic in the VEVs and there will be in general many solutions. Thus, it would be necessary to check if the found vacuum is the global one or at least long-lived. This could be done for instance with \Vevacious.

\subsubsection{Assumptions and fixed solutions}
\label{app:tadpoles_more}
\paragraph*{Assumptions} It is possible to define a list with replacements which are done by \SARAH when it tries to solve
the tadpole equations. For instance, to approximate some matrices as diagonal and to assume that all parameters are real, one could use 
\begin{lstlisting}[style=file,numbers=none]
AssumptionsTadpoleEquations = {Yx[a__]->Delta[a] Yx[a], 
    T[Yx][a__]->Delta[a] T[Yx][a], conj[x_]->x}; 
\end{lstlisting}
That has, of course, no impact on our example because these matrices don't show up in the tadpole equations. However, in the $R$-parity violating case with sneutrinos VEVs this might help to find analytical solutions which don't exists in the most general case. 
\paragraph*{Fixed solutions} There might be cases in which an analytical solution exists when some approximations are made, but \Mathematica doesn't find this solution. Then, it might be useful to give the solutions as input in {\tt SPheno.m}. This can be done via 
\begin{lstlisting}[style=file,numbers=none]
UseGivenTapdoleSolution=True;

SubSolutionsTadpolesTree = {x1 -> sol1Tree, x2 -> sol2Tree,...};
SubSolutionsTadpolesLoop = {x1 -> sol1Loop, x2 -> sol2Loop, ....};
\end{lstlisting}
Note, the solutions have to be given for the tree-level and loop corrected tadpole equations. In the loop-corrected tadpole equations the one-loop contributions are parametrized by {\tt Tad1Loop[i]} where $i$ is an integer counting the equations. 

\subsection{Thresholds in SUSY models}
\label{app:thresholds}
\paragraph*{Assumptions} It is possible to include threshold effects in the RGE evaluation with \SPheno. I concentrate here on the simpler case where the gauge symmetry doesn't change. In that case  \SARAH can derive the RGEs for all scales from the RGEs of the highest scale above all thresholds as follows:
\begin{itemize}
\item The number of generations of the fields which are supposed to be integrated out during the RGE evaluation are parametrized by new variables $n_{gen}(\Phi_i)$. All gauge group constants like the Dynkin index $S(R)$ are expressed as function of \(n_{gen}(\Phi_i)\). $n_{gen}(\Phi_i)$ is dynamically adjusted during the \SPheno run when the RGEs cross the different thresholds.
\item The superpotential and soft-couplings which involve the heavy are set to zero when a threshold is crossed. For instance, we take the Yukawa-like coupling $Y_\Phi^{ij} \Phi_i \phi_j H$ which involves three generations of the heavy field $\Phi$. At the threshold of $\Phi_k$, the $k$-th row of
$Y_\Phi$ is set to zero.  
\end{itemize}
In addition, two assumptions have to be satisfied: (i) the difference between the masses of the scalar and fermionic component of the heavy superfield is negligible, i.e. the masses coming from superpotential interactions are much larger than the soft-breaking term; (ii) these masses are a consequence of bilinear terms in the superpotential. Both assumptions are fulfilled for instance for very heavy vector-like particles or for singlets which have a large Majorana mass. 

\paragraph*{Procedure} There are two steps necessary to implement thresholds according to the above assumptions. First, small changes in the model file of the considered model are necessary: the heavy states have to be 'deleted' at the SUSY scale. This is done by adding the superfields to the array {\tt DeleteFields}. 
\begin{lstlisting}[style=file,numbers=none]
 DeleteFields = {...};
\end{lstlisting}
This ensures that the heavy particle are not take into account in the calculation of mass matrices, vertices or loop corrections at the SUSY scale. \\

The second step is to add the thresholds to {\tt SPheno.m}. For this purpose, the threshold scales have to be defined via the array {\tt Thresholds}. In this array, the numerical value of each threshold scale has to be fixed by some parameter which will be known by \SPheno. In addition, the fields must stated which should be integrated out at that scale. Afterwards, the boundary conditions for all threshold scales can be define via the arrays {\tt BoundaryConditionsUp} and {\tt BoundaryConditionsDown}. The conditions in the first array are applied during the evaluation from $M_Z$ to $M_{GUT}$, the conditions of the second array when running down from $M_{GUT}$.
\begin{lstlisting}[style=file,numbers=none,title=\hspace{13cm}SPheno.m]
...
 Thresholds = {{Scale1, {HeavyFields1}},
               {Scale2, ... }};
...               
BoundaryConditionsUp[[x]] = { ...};
BoundaryConditionsDown[[x]] = { ...};
\end{lstlisting}

\paragraph*{Seesaw Type--I} To exemplify these steps, I'll discuss briefly the simplest model with a threshold scale: Seesaw type--I. In this model, the MSSM particle content is extended by three generations of right handed neutrino superfields. In addition, a neutrino Yukawa coupling $Y_\nu$ between the left- and right neutrinos is present. The mass of the right neutrino is fixed by a Majorana mass term $M_\nu$. We assume that $M_\nu  \gg M_{SUSY}$, i.e. the right neutrinos should be integrated out and shouldn't play any role at $M_{SUSY}$.  This will generate the Weinberg operator $W_\nu$ which couples in its supersymmetric version two left-lepton superfields with two up-Higgs superfields. 
\begin{lstlisting}[style=file,numbers=none,title=\hspace{12cm}Seesaw1.m]
...
SuperFields[[8]] = {v, 3, vR, 0, 1,  1, RpM};

...
SuperPotential = Yu u.q.Hu - Yd d.q.Hd -Ye e.l.Hd + \[Mu] Hu.Hd 
                  + Yv v.l.Hu + Mv/2 v.v +WOp/2 l.Hu.l.Hu;
...
DeleteParticles={v}; 
\end{lstlisting}
It might look a bit odd that $M_\nu$ and $W_\nu$ show up in the same superpotential. However, we will make sure that during the numerical analysis the Weinberg operator just gets initialized when the right neutrinos are integrated out. The following lines in {\tt SPheno.m} of the Seesaw I take care of that:

\begin{lstlisting}[style=file,numbers=none,title=\hspace{12cm}SPheno.m]
Thresholds={
{Abs[MvIN[1,1]],{v[1]}},
{Abs[MvIN[2,2]],{v[2]}},
{Abs[MvIN[3,3]],{v[3]}}
};

BoundaryConditionsUp=Table[{},{Length[Thresholds]}];
BoundaryConditionsDown=Table[{},{Length[Thresholds]}];

BoundaryConditionsDown[[1]]={
{WOp[index1,index2],
    WOp[index1,index2] - Yv[1,index1] Yv[1,index2]/MassOfv[1]} };
    
BoundaryConditionsDown[[2]]={
{WOp[index1,index2],
    WOp[index1,index2] - Yv[2,index1] Yv[2,index2]/MassOfv[2]} };
    
BoundaryConditionsDown[[3]]={
{WOp[index1,index2], - Yv[3,index1] Yv[3,index2]/MassOfv[3]}   }; 
\end{lstlisting}
Here, we defined three threshold scales which are given by the diagonal entries of the input value of $M_\nu$ ({\tt MvIN[X,X]}). At each scale the corresponding generation of the right neutrino superfields is integrated out. Then, we initialize the three boundary conditions for each threshold scale when running up and down. While we need not define any boundary condition when running up, we initialize the Weinberg operator when running down. The shift of the coefficients of the Weinberg operator at each threshold scale $I$ are given by
\begin{equation}
\delta W_\nu^{ij}  = -\frac{Y_\nu^{Ii} Y_\nu^{Ij}}{m_{\nu}^{I}}
\end{equation}
where $m_{\nu}^{I}$ is the $I$-th eigenvalue of the running matrix $M_\nu$ which in general is not diagonal. 

\section{Flags in \SPheno input file}
\label{app:SPhenoFlags}
There are many options which can be used in the block {\tt SPhenoInput} in the Les Houches input file to set up the calculations and the output done by \SPheno: 
\begin{itemize}
\item[\tt 1] sets the error level; default is {\tt 0}
\item[\tt 2] if {\tt 1}, the SPA conventions are used; default is {\tt 0}
\item[\tt 7] if {\tt 1}, skips two loop Higgs masses; default is {\tt 0} 
\item[\tt 8] Method to calculate two-loop corrections; default is {\tt 3}
  \begin{itemize}
    \item[\tt 1]: fully numerical method
    \item[\tt 2]: semi-analytical method
    \item[\tt 3]: diagrammatic calculation
    \item[\tt 8/9]: using results from literature if available; {\tt 8} includes only $\alpha_S$ corrections
  \end{itemize} 
\item[\tt 9] if {\tt 1}, two-loop corrections are calculated in gauge-less limit; default is {\tt 1} 
\item[\tt 10] if {\tt 1}, safe mode is used for the numerical derivative in the two-loop Higgs calculations; default is {\tt 0}
\item[\tt 11] if {\tt 1}, the branching ratios of the SUSY and Higgs particles are calculated; default is {\tt 1}
\item[\tt 12] defines minimum value for a branching ratios to be included in output; default is $10^{-4}$
\item[\tt 13] adjusts the three-body decays: {\tt 0}: no three-body decays are calculated; {\tt 1} only three-body decays of fermions are calculated; {\tt 2} only three-body decays of scalars are calculated; {\tt 3} three-body decays of fermions and scalars are calculated; default is {\tt 1}
\item[\tt 14] if {\tt 1}, the running parameters at the mass scale of the decaying particle are calculated. Otherwise, the parameters at the standard renormalization scale are used; default is {\tt 1}
\item[\tt 15] defines minimum value for a width to be included in output; default is $10^{-30}$
\item[\tt 31] positive values are uses as GUT scale; otherwise a dynamical GUT scale fulfilling the given condition is used; default is {\tt -1}
\item[\tt 32] if {\tt 1}, forces strict unification, i.e. $g_1=g_2=g_3$; default is {\tt 0}
\item[\tt 33] if set, a fixed renormalization scale is used
\item[\tt 34] sets the relative precision of the mass calculation; default is $10^{-4}$
\item[\tt 35] sets the maximal number of iterations in the calculation  of the masses; default is 40
\item[\tt 36] sets the minimal number of iterations before \SPheno stops because of tachyon in the spectrum; default is 5
\item[\tt 37] defines if CKM matrix is taken to be in the up- ({\tt 1}) or down- ({\tt 2}) quark sector; default is {\tt 1}
\item[\tt 38] sets the loop order of the RGEs: {\tt 1} or {\tt 2} can be used; default is {\tt 2}
\item[\tt 39] if {\tt 1}, writes output using SLHA1 format; default is {\tt 0}
\item[\tt 41] sets the width of the Z-boson $\Gamma_Z$, default is 2.49~GeV
\item[\tt 42] sets the width of the W-boson $\Gamma_W$, default is 2.06 GeV
\item[\tt 50] if {\tt 1}, negative fermion masses are rotated to real ones by multiplying the rotation matrix with $i$; default is {\tt 1}
\item[\tt 51] if {\tt 0}, the parameters \(Y_u,Y_d,T_u,T_d,m_q^2,m_d^2,m_u^2\) are not rotated into the SCKM basis in the spectrum file; default is {\tt 0}
\item[\tt 52] if {\tt 1}, a negative mass squared is always ignored and set 0; default is {\tt 0}
\item[\tt 53] if {\tt 1}, a negative mass squared at $M_Z$ is always ignored and set 0; default is {\tt 0}
\item[\tt 54] if {\tt 1}, the output is written even if there has been a problem during the run; default is {\tt 0}
\item[\tt 55] if {\tt 0}, the loop corrections to all masses are skipped; default is {\tt 1}
\item[\tt 57] if {\tt 0}, the calculation of the low energy observables is skipped; default is {\tt 1}
\item[\tt 58] if {\tt 0}, the calculation of $\delta_{VB}$ in the boundary conditions at the SUSY scale is skipped; default is {\tt 1}
\item[\tt 60] if {\tt 0}, possible effects from kinetic mixing are neglected; default is {\tt 1}
\item[\tt 61] if {\tt 0}, the RGE running of SM parameters is skipped in a low scale input; default is {\tt 1}
\item[\tt 62] if {\tt 0}, the RGE running of SUSY parameters to the low scale is skipped for the calculation of the flavour and precision observables; default is {\tt 1}
\item[\tt 63] if {\tt 0}, the RGE running of SM parameters to the low scale is skipped for the calculation of the flavour and precision observables; default is {\tt 1}
\item[\tt 64] if {\tt 1}, the running parameters at the scale $Q=160$ are written in the spectrum file; default is {\tt 0}
\item[\tt 65] can be used if several, independent solution to the tadpole equations exists; default is {\tt 1}. An integer is used to pick one solution
\item[\tt 75] if {\tt 1}, a file containing all parameters in \WHIZARD format is created; default is {\tt 1}
\item[\tt 76] if {\tt 1},  input files for \HB and \HS are written; default is {\tt 1}
\item[\tt 86] sets the maximal width which is taken as 'invisible' in the output for \HB and \HS; default is {\tt 0.}
\item[\tt 88] sets a maximal mass of particles which are included in loop calculations; default is $10^{16}$~GeV. Note, this option must be turned in \SARAH first
\item[\tt 89] sets the maximal mass for scalars which is treated as numerical zero; default is $10^{-8}$~GeV
\item[\tt 95] if {\tt 1}, mass matrices at one-loop are forced to be real; default is {\tt 0}
\item[\tt 400] fixes initial step-size in numerical derivative for the purely numerical method to calculate two-loop Higgs masses; default is {\tt 0.1}
\item[\tt 401] fixes initial step-size in numerical derivative for the semi-analytical method to calculate two-loop Higgs masses; default is {\tt 0.001}
\item[\tt 510] if {\tt 1}, \SPheno writes solution of tadpole equations at tree-level; default is {\tt 1}. This is needed for \Vevacious.
\item[\tt 515] if {\tt 1}, \SPheno writes all running values at the GUT scale; default is {\tt 0}
\item[\tt 520] if {\tt 1}, \SPheno writes \HB blocks (effective coupling ratios of Higgs particles to SM fields); default is {\tt 1}
\item[\tt 525] if {\tt 1}, \SPheno writes the size of all different contributions to the Higgs diphoton rate; default is {\tt 0}
\item[\tt 530] if {\tt 1}, the tree-level values of the tadpole equations appear in the output instead of the loop corrected ones; default is {\tt 0}
\item[\tt 550] if {\tt 0}, the fine-tuning calculation is skipped; default is {\tt 1}
\item[\tt 551] if {\tt 1}, one-loop corrections to $Z$-mass are included in fine-tuning calculation; default is {\tt 0}
\item[\tt 999] if {\tt 1}, debug information is printed on the screen; default is {\tt 0}
\end{itemize}
\section{Model files for the \BLSSM}
The full model file for the implementation of the \BLSSM in \SARAH is shown. In addition, I summarize all changes in {\tt particles.m} and {\tt parameters.m} compared to the MSSM. This includes definition of new parameters and changed properties of parameters already present in the MSSM. For the particles I restrict myself to the intermediate states and the mass eigenstates after EWSB but skip the gauge eigenstates. Finally, the additional input file to create a \SPheno version for the \BLSSM is given in Appendix~\ref{app:sphenofile}.

\subsection{Model file}
\label{app:BLSSMm}
\begin{lstlisting}[style=file, title=B-L-SSM.m]
Model`Name = "BLSSM";
Model`NameLaTeX ="B-L-SSM";
Model`Authors = "L.Basso, F.Staub";
Model`Date = "2012-09-01";

(* 2013-09-01: changing to new conventions for Superfields, Superpotential and global symmetries *)



(*-------------------------------------------*)
(*   Particle Content*)
(*-------------------------------------------*)

(* Global symmetries *)

Global[[1]] = {Z[2],MParity};
MpM = {-1,-1,1};
MpP = {1,1,-1};

(* Vector Superfields *)

Gauge[[1]]={B,   U[1], hypercharge, g1,  False, MpM};
Gauge[[2]]={WB, SU[2], left,        g2,  True,  MpM};
Gauge[[3]]={G,  SU[3], color,       g3,  False, MpM};
Gauge[[4]]={Bp,  U[1], BminusL,     gBL, False, MpM};

(* Chiral Superfields *)

SuperFields[[1]] = {q, 3, {uL,  dL},    1/6, 2, 3,  1/6, MpM};  
SuperFields[[2]] = {l, 3, {vL,  eL},   -1/2, 2, 1, -1/2, MpM};
SuperFields[[3]] = {Hd,1, {Hd0, Hdm},  -1/2, 2, 1,    0, MpP};
SuperFields[[4]] = {Hu,1, {Hup, Hu0},   1/2, 2, 1,    0, MpP};

SuperFields[[5]] = {d, 3, conj[dR],   1/3, 1, -3, -1/6, MpM};
SuperFields[[6]] = {u, 3, conj[uR],  -2/3, 1, -3, -1/6, MpM};
SuperFields[[7]] = {e, 3, conj[eR],     1, 1,  1,  1/2, MpM};
SuperFields[[8]] = {vR,3, conj[vR],     0, 1,  1,  1/2, MpM};

SuperFields[[9]]  = {C1, 1, C10,  0, 1, 1, -1, MpP};
SuperFields[[10]] = {C2, 1, C20,  0, 1, 1,  1, MpP};


(*------------------------------------------------------*)
(* Superpotential *)
(*------------------------------------------------------*)

SuperPotential = Yu u.q.Hu - Yd d.q.Hd - Ye e.l.Hd + \[Mu] Hu.Hd + Yv vR.l.Hu - MuP C1.C2 + Yn vR.C1.vR;


(*----------------------------------------------*)
(*   ROTATIONS                                  *)
(*----------------------------------------------*)

NameOfStates={GaugeES, EWSB};

(* Dirac Spinors for gauge eigenstates *)
DEFINITION[GaugeES][DiracSpinors]={
  Bino ->{fB, conj[fB]},
  Wino -> {fWB, conj[fWB]},
  Glu -> {fG, conj[fG]},
  H0 -> {FHd0, conj[FHu0]},
  HC -> {FHdm, conj[FHup]},
  Fd1 -> {FdL, 0},
  Fd2 -> {0, FdR},
  Fu1 -> {FuL, 0},
  Fu2 -> {0, FuR},
  Fe1 -> {FeL, 0},
  Fe2 -> {0, FeR},
  Fv1 -> {FvL, 0},
  Fv2 -> {0, FvR},
  FC -> {FC10, conj[FC20]},
  FB -> {fBp, conj[fBp]}
};


(*--- Gauge Sector ---- *)
DEFINITION[EWSB][GaugeSector] =
{ 
  {{VB,VWB[3],VBp},{VP,VZ,VZp},ZZ},
  {{VWB[1],VWB[2]},{VWm,conj[VWm]},ZW},
  {{fWB[1],fWB[2],fWB[3]},{fWm,fWp,fW0},ZfW}
};


     
(*--- VEVs ---- *)
DEFINITION[EWSB][VEVs]= 
{{SHd0, {vd, 1/Sqrt[2]}, {sigmad, I/Sqrt[2]},{phid,1/Sqrt[2]}},
 {SHu0, {vu, 1/Sqrt[2]}, {sigmau, I/Sqrt[2]},{phiu,1/Sqrt[2]}},
 {SvL, {0, 0}, {sigmaL, I/Sqrt[2]},{phiL,1/Sqrt[2]}},
 {SvR, {0, 0}, {sigmaR, I/Sqrt[2]},{phiR,1/Sqrt[2]}},
 {SC10, {x1, 1/Sqrt[2]}, {sigma1, I/Sqrt[2]},{phi1, 1/Sqrt[2]}},
 {SC20, {x2, 1/Sqrt[2]}, {sigma2, I/Sqrt[2]},{phi2, 1/Sqrt[2]}}     
};
 
 
(*--- Matter Sector ---- *)
DEFINITION[EWSB][MatterSector]= 
{    {{SdL, SdR}, {Sd, ZD}},
     {{SuL, SuR}, {Su, ZU}},
     {{SeL, SeR}, {Se, ZE}},
     {{sigmaL,sigmaR}, {SvIm, ZVI}},
     {{phiL,phiR}, {SvRe, ZVR}}, 
     {{phid, phiu,phi1, phi2}, {hh, ZH}}, 
     {{sigmad, sigmau,sigma1,sigma2}, {Ah, ZA}},
     {{SHdm,conj[SHup]},{Hpm,ZP}},
     {{fB, fW0, FHd0, FHu0,fBp,FC10,FC20}, {L0, ZN}}, 
     {{{fWm, FHdm}, {fWp, FHup}}, {{Lm,UM}, {Lp,UP}}},
     {{FvL,conj[FvR]},{Fvm,UV}},
     {{{FeL},{conj[FeR]}},{{FEL,ZEL},{FER,ZER}}},
     {{{FdL},{conj[FdR]}},{{FDL,ZDL},{FDR,ZDR}}},
     {{{FuL},{conj[FuR]}},{{FUL,ZUL},{FUR,ZUR}}}                    \
       }; 


(* Phases *)
DEFINITION[EWSB][Phases]= 
{    {fG, PhaseGlu}
    }; 

(* Dirac Spinors for Mass eigenstates *)    
DEFINITION[EWSB][DiracSpinors]={
 Fd ->{  FDL, conj[FDR]},
 Fe ->{  FEL, conj[FER]},
 Fu ->{  FUL, conj[FUR]},
 Fv ->{  Fvm, conj[Fvm]},
 Chi ->{ L0, conj[L0]},
 Cha ->{ Lm, conj[Lp]},
 Glu ->{ fG, conj[fG]}
};
\end{lstlisting}

\subsection{Parameters files}
\label{app:paramerersfile}
\begin{itemize} 
 \item {\bf New Gauge couplings}.
\begin{lstlisting}[style=file]
{g1BL,    {Description -> "Mixed Gauge Coupling 1",
           LesHouches -> {gauge, 10},
           LaTeX -> "g_{Y B}",
           OutputName -> gYB }},
{gBL1,    {Description -> "Mixed Gauge Coupling 2",
           LesHouches -> {gauge, 11},
           LaTeX -> "g_{B Y}",
           OutputName -> gBY}},
{gBL,     {Description -> "B-L-Coupling", 
           LaTeX -> "g_{B}",
           GUTnormalization -> Sqrt[3/2],
           LesHouches -> {gauge,4},
           OutputName -> gBL }},
\end{lstlisting}

 \item {\bf New Gauge boson mass}.
\begin{lstlisting}[style=file]
{MZp,     {Description -> "Z' mass", 
           LaTeX -> "M_{Z'}",
           Real -> True,
           LesHouches -> None,
           OutputName -> MZp }},
\end{lstlisting}

 \item {\bf New Gaugino masses}.
\begin{lstlisting}[style=file]           
{MassBp,  {Description -> "Bino' Mass",
           LaTeX -> "{M}_{BL}",
           LesHouches -> {BL,31},
           OutputName -> MBp }},
{MassBBp, {Description -> "Mixed Gaugino Mass 1",
           LaTeX -> "{M}_{B B'}",
           LesHouches -> {BL,32},
           OutputName -> MBBp }},
\end{lstlisting}

 \item {\bf New gauge boson mixing angle}.
\begin{lstlisting}[style=file]              
{ThetaWp,  { LaTeX -> "{\\Theta'}_W",
             DependenceNum ->ArcTan[(2 g1BL Sqrt[g1^2+g2^2])
                      /(g1BL^2 + 16 (x1^2+x2^2)/(vd^2+vu^2) -g1^2-g2^2)]/2,
             Real ->True,
             DependenceSPheno -> ArcCos[Abs[ZZ[3,3]]],
             OutputName-> TWp,
             LesHouches -> {ANGLES,10}      }},
\end{lstlisting}

 \item {\bf New angle to give ratio of VEVs}.
\begin{lstlisting}[style=file]
{TBetaP,  { LaTeX -> "\\tan(\\beta')",
             Real ->  True, 
             LesHouches -> None,
             OutputName -> TBp     }},  
\end{lstlisting}

 \item {\bf New superpotential and soft-breaking parameters}.
\begin{lstlisting}[style=file]
{MuP,     {Description -> "Mu' Parameter",
           LaTeX -> "{\\mu_{\\eta}}",
           LesHouches -> {BL,1},
           OutputName -> MuP }},
{B[MuP],  {Description -> "B' Parameter",
           LaTeX -> "B_{\\eta}",
           LesHouches -> {BL,2},
           OutputName -> BMuP}},

{Yv,     {Description -> "Neutrino-Yukawa-Coupling",
          LaTeX -> "Y_\\nu",
          LesHouches -> Yv,
          OutputName-> Yv}},   
{T[Yv],  {Description -> "Trilinear-Neutrino-Coupling",
          LaTeX -> "T_\\nu",
          LesHouches -> Tv,
          OutputName-> Tv}},
{Yn,     {Description -> "Neutrino-X-Yukawa-Coupling",
          LaTeX -> "Y_x",
          OutputName -> Yx,
          LesHouches->Yx }},
{T[Yn],  {Description -> "Trilinear-Neutrino-X-Coupling",
          OutputName -> Tx,
          LaTeX -> "T_x",
          LesHouches->TX}},

{mvR2,  { Description -> "Softbreaking right Sneutrino Mass",
          LaTeX -> "m_{\\nu}^2",
          LesHouches -> mv2,
          OutputName-> mv2 }},
{mC12,   {Description -> "Bilepton 1 Soft-Breaking mass",
          LaTeX->"m_{\\eta}^2",
          LesHouches -> {BL,11} ,
          OutputName -> mC12}},
{mC22,   {Description -> "Bilepton 2 Soft-Breaking mass",
          LaTeX->"m_{\\bar{\\eta}}^2",
          LesHouches -> {BL,12} ,
          OutputName -> mC22}},
\end{lstlisting}

 \item {\bf New VEVs}.
\begin{lstlisting}[style=file]
{x1,   { Description -> "Bilepton 1 VEV",
               LaTeX -> "v_{\\eta}",
             DependenceNum ->  Sin[BetaP]*vX, 
             OutputName -> x1,
             Real -> True,
             LesHouches -> {BL,41}   }}, 
{x2,   {Description -> "Bilepton 2 VEV",
               LaTeX -> "v_{\\bar{\\eta}}",
             DependenceNum ->  Cos[BetaP]*vX, 
             OutputName -> x2,
             Real -> True,
             LesHouches -> {BL,42}   }},
{vX,      {Description -> "Bilepton VEV",
              LaTeX -> "x",
             Dependence ->  None, 
             OutputName -> vX,
             DependenceSPheno -> Sqrt[x1^2 + x2^2],
             Real -> True,
             LesHouches -> {BL,43}   }},
\end{lstlisting}

 \item {\bf New rotation matrices in matter sector}.
\begin{lstlisting}[style=file] 
{ZVR,    { LaTeX -> "Z^R",
           OutputName -> ZVR,
           LesHouches -> SNUMIXR }},
{ZVI,    { LaTeX -> "Z^I",
           OutputName -> ZVI,
           LesHouches -> SNUMIXI }},  

{UV,     {Description ->"Neutrino-Mixing-Matrix", 
          LaTeX -> "U^V",
          LesHouches -> UVMIX,
          OutputName-> UV      }},
\end{lstlisting}

\item {\bf Modified rotation matrix in gauge sector}.
\begin{lstlisting}[style=file]
CW=Cos[ThetaW]; SW=Sin[ThetaW]; CWp=Cos[ThetaWp]; SWp=Sin[ThetaWp];
{ZZ,  {Description -> "Photon-Z-Z' Mixing Matrix",
       Dependence ->   {{CW,-SW CWp,  SW SWp },                             
                        {SW, CW CWp, -CW SW  },
                        {0 , SWp,        CWp }},
       Real ->True,
       LaTeX -> "Z^{\\gamma Z Z'}",
       LesHouches -> None,
       OutputName -> ZZ }},             
\end{lstlisting}

\item {\bf Modified rotation matrix in matter sector}.
\begin{lstlisting}[style=file]
{ZH,   { Description->"Scalar-Mixing-Matrix",
         Dependence->None,
         DependenceNum->None,
         DependenceOptional->None}},
{ZA,   { Description->"Pseudo-Scalar-Mixing-Matrix",
         Dependence->None,
         DependenceNum->None,
         DependenceOptional->None}},
\end{lstlisting}
\end{itemize}

\subsection{Particles files}
\label{app:particlesfile}
\begin{itemize} 
 \item {\bf Intermediate states}. 
\begin{lstlisting}[style=file]
WeylFermionAndIndermediate = {
  ...
  (* Superfields *)
  {vR,   { Description -> "Right Neutrino Superfield" }},
  {C1,   { LaTeX  -> "\\hat{\\eta}" }},
  {C2,   { LaTeX  -> "\\hat{\\bar{\\eta}}" }},

  (* Intermediate fermions *)
  {FC10,   { LaTeX  -> "\\tilde{\\eta}" }},
  {FC20,   { LaTeX  -> "\\tilde{\\bar{\\eta}}" }},
  {fBp,   { LaTeX -> "{\\tilde{B}{}'}"}},  

  (* Intermediate Scalars *)
  {phi1,   { LaTeX  -> "\\phi_{\\eta}" }},
  {phi2,   { LaTeX  -> "\\phi_{\\bar{\\eta}}" }},

  {sigma1,   {   LaTeX -> "\\sigma_{\\eta}"}},
  {sigma2,   {   LaTeX -> "\\sigma_{\\bar{\\eta}}" }},
  
  {SC10,  { LaTeX -> "\\eta"  }},    
  {SC20,  { LaTeX -> "\\bar{\\eta}" }},  

};
\end{lstlisting}
 \item {\bf New mass eigenstates}.
More interesting are the mass eigenstates. The additional information given for those is used in the different output for \SPheno and the MC-tools. We begin with the new states which are not present in the MSSM:
\begin{lstlisting}[style=file]
  {SvRe,  { Description -> "CP-even Sneutrino",
            LaTeX -> "\\nu^R",
            OutputName -> "nR",
            FeynArtsNr -> 41,
            LHPC -> {5, "blue"},
            PDG->{1000012,1000014,1000016,2000012,2000014,2000016},
            PDG.IX ->{200000001,200000002,200000003,
                      200000004,200000005,200000006} }},
  {SvIm, { Description -> "CP-odd Sneutrino",
           LaTeX -> "\\nu^I",
           OutputName -> "nI",
           FeynArtsNr -> 40,
           LHPC -> {5, "turquoise"},
           PDG->{4000012,4000014,4000016,5000012,5000014,5000016},
           PDG.IX ->{202000001,202000002,202000003,
                     202000004,202000005,202000006}}},  
           
  {VZp,  { Description -> "Z'-Boson",
           PDG -> {31},
           PDG.IX -> {122000002},
           Width -> Automatic, 
           Mass -> LesHouches,
           FeynArtsNr -> 10,
           LaTeX -> "{Z'}",
           Goldstone -> Ah[{2}],
           ElectricCharge -> 0,
           OutputName -> "Zp"}} 
  {gZp,   { Description -> "Z'-Ghost",  
            PDG -> 0,
            PDG.IX -> 0,
            Width -> 0, 
            Mass -> Automatic,
            FeynArtsNr -> 10,
            LaTeX -> "\\eta^{Z'}",
            ElectricCharge -> 0,
            OutputName -> "gZp"}},
\end{lstlisting}
Some comments are at place:
\begin{itemize}
 \item {\tt PDG} and {\tt PDG.IX}: \dots
 \item {\tt FeynArtsNr}: \dots
 \item {\tt ElectricCharge}: \dots
 \item {\tt OutputName}: \dots
\end{itemize}
\item {\bf Modified mass eigenstates}.
For other states only more generations appear compared to the MSSM. Therefore, it is only possible to extent the lists for the PDGs
\begin{lstlisting}[style=file]
{hh ,  { Description -> "Higgs",
         PDG -> {25,35,9900025, 9900035},
         PDG.IX->{101000001,101000002,101000003,101000004}  }}, 
{Ah ,  { Description -> "Pseudo-Scalar Higgs",
         PDG -> {0,0,36,9900036},
         PDG.IX->{0,0,102000001,102000002} }},                
{Fv,   { Description -> "Neutrinos",
         Mass -> Automatic,
         Width -> Automatic,
         PDG ->{12,14,16,112,114,116},
         PDG.IX->{111000001,111000002,111000003,
                  111000004,111000005,111000006 } }},    
       
{Chi,  { Description -> "Neutralinos",
         PDG -> {1000022,1000023,1000025,1000035,
                        1000032,1000036,1000039},
         PDG.IX ->{211000001,211000002,211000003,211000004,
                             211000005,211000006,211000007 } }}                     
\end{lstlisting}
\end{itemize}

\subsection{\SPheno file}
\label{app:sphenofile}
\begin{lstlisting}[style=file,title=SPheno.m]
MINPAR={{1,m0},
        {2,m12},
        {3,TanBeta},
        {4,SignumMu},
        {5,Azero},
        {6,SignumMuP}, 
        {7,TBetaP},
        {8,MZp}};

RealParameters = {TanBeta, TBetaP,m0};
ParametersToSolveTadpoles = {B[\[Mu]],B[MuP],\[Mu],MuP};

RenormalizationScaleFirstGuess = m0^2 + 4 m12^2;
RenormalizationScale = MSu[1]*MSu[6];

ConditionGUTscale = (g1*gBL-g1BL*gBL1)/Sqrt[gBL^2+gBL1^2] == g2;

BoundaryHighScale={
{g1,(g1*gBL-g1BL*gBL1)/Sqrt[gBL^2+gBL1^2]},
{g1,Sqrt[(g1^2+g2^2)/2]},
{g2,g1},
{gBL, g1},
{g1BL,0},
{gBL1,0},
{T[Ye], Azero*Ye},
{T[Yd], Azero*Yd},
{T[Yu], Azero*Yu},
{T[Yv], Azero*Yv},
{T[Yn], Azero*Yn},
{mq2, DIAGONAL m0^2},
{ml2, DIAGONAL m0^2},
{md2, DIAGONAL m0^2},
{mu2, DIAGONAL m0^2},
{me2, DIAGONAL m0^2},
{mvR2, DIAGONAL m0^2},
{mHd2, m0^2},
{mHu2, m0^2},
{mC12, m0^2},
{mC22, m0^2},
{MassB, m12},
{MassWB,m12},
{MassG,m12},
{MassBp,m12},
{MassBBp,0}
};


BoundarySUSYScale = {
 {g1T,(g1*gBL-g1BL*gBL1)/Sqrt[gBL^2+gBL1^2]},
 {gBLT, Sqrt[gBL^2+gBL1^2]},
 {g1BLT,(g1BL*gBL+gBL1*g1)/Sqrt[gBL^2+gBL1^2]},
 {g1, g1T},
 {gBL, gBLT},
 {g1BL, g1BLT},
 {gBL1,0},
 {vevP, MZp/gBL},
 {betaP,ArcTan[TBetaP]},
 {x2,vevP*Cos[betaP]},
 {x1,vevP*Sin[betaP]},
 {Yv,  LHInput[Yv]},
 {Yn,  LHInput[Yn]}
};

BoundaryEWSBScale = {
 {g1T,(g1*gBL-g1BL*gBL1)/Sqrt[gBL^2+gBL1^2]},
 {gBLT, Sqrt[gBL^2+gBL1^2]},
 {g1BLT,(g1BL*gBL+gBL1*g1)/Sqrt[gBL^2+gBL1^2]},
 {g1, g1T},
 {gBL, gBLT},
 {g1BL, g1BLT},
 {gBL1,0},
 {vevP, MZp/gBL},
 {betaP,ArcTan[TBetaP]},
 {x2,vevP*Cos[betaP]},
 {x1,vevP*Sin[betaP]}
};

InitializationValues = {
 {gBL, 0.5},
 {g1BL, -0.06},
 {gBL1, -0.06}
 }

BoundaryLowScaleInput={
 {vd,Sqrt[4 mz2/(g1^2+g2^2)]*Cos[ArcTan[TanBeta]]},
 {vu,Sqrt[4 mz2/(g1^2+g2^2)]*Sin[ArcTan[TanBeta]]}
};

ListDecayParticles = Automatic;
ListDecayParticles3B =Automatic;

UseBoundarySUSYatEWSB = True;

(* Example for mSugra input values *)
DefaultInputValues = {m0 -> 1000, m12 -> 1500, TanBeta->20, SignumMu ->1, Azero -> -1500, SignumMuP -> 1, TBetaP -> 1.15, MZp -> 2500, Yn[1,1]->0.37, Yn[2,2]->0.4, Yn[3,3]->0.4};
\end{lstlisting}

\section{Models included in \SARAH}
\label{app:models}
I show here the list of models which are included in the public version of \SARAH. Additional models created and provided by user an also found here
\begin{verbatim}
https://sarah.hepforge.org/trac/wiki 
\end{verbatim}

\subsection{Supersymmetric Models}
\begin{itemize}
 \item Minimal supersymmetric standard model 
    \begin{itemize} 
     \item With general flavour and CP structure ({\tt MSSM})
     \item Without flavour violation ({\tt MSSM/NoFV})
     \item With explicit CP violation in the Higgs sector ({\tt MSSM/CPV})
     \item In SCKM basis ({\tt MSSM/CKM})
    \end{itemize}
   \item Singlet extensions: 
   \begin{itemize}
    \item Next-to-minimal supersymmetric standard model ({\tt NMSSM}, {\tt NMSSM/NoFV}, {\tt NMSSM/CPV}, {\tt NMSSM/CKM})
    \item near-to-minimal supersymmetric standard model  ({\tt near-MSSM}) 
    \item General singlet extended, supersymmetric standard model ({\tt SMSSM})  
    \item Dirac NMSSM ({\tt DiracNMSSM})
    \item Next-to-minimal supersymmetric standard model  with inverse Seesaw ({inverse-Seesaw-NMSSM})
  \end{itemize}
  \item Triplet extensions 
  \begin{itemize} 
    \item Triplet extended MSSM ({\tt TMSSM}) 
    \item Triplet extended NMSSM ({\tt TNMSSM}) 
  \end{itemize}
   \item Models with $R$-parity violation  
  \begin{itemize}
    \item bilinear RpV ({\tt MSSM-RpV/Bi}) 
    \item Lepton number violation ({\tt MSSM-RpV/LnV})
    \item Only trilinear lepton number violation ({\tt MSSM-RpV/TriLnV})
    \item Baryon number violation ({\tt MSSM-RpV/BnV})  
    \item $\mu\nu$SSM ({\tt munuSSM}) 
  \end{itemize}
   \item Additional $U(1)'s$ 
  \begin{itemize}
    \item $U(1)$-extended MSSM ({\tt UMSSM})  
    \item secluded MSSM ({\tt secluded-MSSM}) 
    \item minimal $B-L$ model ({\tt B-L-SSM})  
    \item minimal singlet-extended $B-L$ model ({\tt N-B-L-SSM})
  \end{itemize}
   \item SUSY-scale seesaw extensions
    \begin{itemize}
      \item inverse seesaw ({\tt inverse-Seesaw}) 
      \item linear seesaw ({\tt LinSeesaw}) 
      \item singlet extended inverse seesaw ({\tt inverse-Seesaw-NMSSM}) 
      \item inverse seesaw with $B-L$ gauge group ({\tt B-L-SSM-IS})  
      \item minimal $U(1)_R \times U(1)_{B-L}$ model with inverse seesaw  ({\tt BLRinvSeesaw}) 
\end{itemize}
 \item Models with Dirac Gauginos
   \begin{itemize}
    \item MSSM/NMSSM with Dirac Gauginos ({\tt DiracGauginos}) 
    \item minimal $R$-Symmetric SSM ({\tt MRSSM}) 
    \item minimal Dirac Gaugino SSM ({\tt MDGSSM})
   \end{itemize}
 \item High-scale extensions
\begin{itemize}
 \item Seesaw 1 - 3 ($SU(5)$ version) ,
 ({\tt Seesaw1},{\tt Seesaw2},{\tt Seesaw3}) 
 \item Left/right model ($\Omega$LR) ({\tt Omega})
\end{itemize}
\item Others:
\begin{itemize}
 \item MSSM with non-holomorphic soft-terms ({\tt NHSSM})
 \item MSSM with colour sextets ({\tt MSSM6C})
\end{itemize}

\end{itemize}

\subsection{Non-Supersymmetric Models}
\begin{itemize}
\item Standard Model (SM) ({\tt SM}), Standard model in CKM basis ({\tt SM/CKM}) 
\item inert Higgs doublet model ({\tt Inert}) 
\item B-L extended SM ({\tt B-L-SM})
\item B-L extended SM with inverse seesaw ({\tt B-L-SM-IS})
\item SM extended by a scalar colour octet ({\tt SM-8C})
\item Two Higgs doublet models ({\tt THDM}, {\tt THDM-II}, {\tt THDM-III}, {\tt THDM-LS}, {\tt THDM-Flipped})
\item Singlet extended SM ({\tt SSM})
\item Triplet extended SM ({\tt TSM})
\end{itemize}
\end{appendix}

\bibliographystyle{ArXiv}
\bibliography{lit}

\providecommand{\bysame}{\leavevmode\hbox to3em{\hrulefill}\thinspace}
\begin{thebibliography}{100}

\bibitem{Ramond:1971gb}
P.~Ramond, Phys.Rev. \textbf{D3} (1971), 2415--2418.

\bibitem{Wess:1974tw}
J.~Wess and B.~Zumino, Nucl.Phys. \textbf{B70} (1974), 39--50.

\bibitem{Volkov:1973ix}
D.~Volkov and V.~Akulov, Phys.Lett. \textbf{B46} (1973), 109--110.

\bibitem{Weinberg:1975gm}
S.~Weinberg, Phys.Rev. \textbf{D13} (1976), 974--996.

\bibitem{Weinberg:1979bn}
S.~Weinberg, Phys.Rev. \textbf{D19} (1979), 1277--1280.

\bibitem{Goldberg:1983nd}
H.~Goldberg, Phys.Rev.Lett. \textbf{50} (1983), 1419.

\bibitem{Ellis:1983ew}
J.~R. Ellis, J.~Hagelin, D.~V. Nanopoulos, K.~A. Olive, and M.~Srednicki,
  Nucl.Phys. \textbf{B238} (1984), 453--476.

\bibitem{Drees:1992am}
M.~Drees and M.~M. Nojiri, Phys.Rev. \textbf{D47} (1993), 376--408,
  [hep-ph/9207234].

\bibitem{Dimopoulos:1981yj}
S.~Dimopoulos, S.~Raby, and F.~Wilczek, Phys.Rev. \textbf{D24} (1981),
  1681--1683.

\bibitem{Ibanez:1981yh}
L.~E. Ibanez and G.~G. Ross, Phys.Lett. \textbf{B105} (1981), 439.

\bibitem{Marciano:1981un}
W.~J. Marciano and G.~Senjanovic, Phys.Rev. \textbf{D25} (1982), 3092.

\bibitem{Einhorn:1981sx}
M.~Einhorn and D.~Jones, Nucl.Phys. \textbf{B196} (1982), 475.

\bibitem{Amaldi:1991cn}
U.~Amaldi, W.~de~Boer, and H.~Furstenau, Phys.Lett. \textbf{B260} (1991),
  447--455.

\bibitem{Langacker:1991an}
P.~Langacker and M.-x. Luo, Phys.Rev. \textbf{D44} (1991), 817--822.

\bibitem{Ellis:1990wk}
J.~R. Ellis, S.~Kelley, and D.~V. Nanopoulos, Phys.Lett. \textbf{B260} (1991),
  131--137.

\bibitem{Martin:2001vx}
S.~P. Martin, Phys.Rev. \textbf{D65} (2002), 116003,  [hep-ph/0111209].

\bibitem{Ibanez:1982fr}
L.~E. Ibanez and G.~G. Ross, Phys.Lett. \textbf{B110} (1982), 215--220.

\bibitem{Witten:1981nf}
E.~Witten, Nucl.Phys. \textbf{B188} (1981), 513.

\bibitem{Witten:1982df}
E.~Witten, Nucl.Phys. \textbf{B202} (1982), 253.

\bibitem{Hall:1983iz}
L.~J. Hall, J.~D. Lykken, and S.~Weinberg, Phys.Rev. \textbf{D27} (1983),
  2359--2378.

\bibitem{Nilles:1983ge}
H.~P. Nilles, Phys.Rept. \textbf{110} (1984), 1--162.

\bibitem{AlvarezGaume:1983gj}
L.~Alvarez-Gaume, J.~Polchinski, and M.~B. Wise, Nucl.Phys. \textbf{B221}
  (1983), 495.

\bibitem{Kane:1993td}
G.~L. Kane, C.~F. Kolda, L.~Roszkowski, and J.~D. Wells, Phys.Rev. \textbf{D49}
  (1994), 6173--6210,  [hep-ph/9312272].

\bibitem{Dine:1995ag}
M.~Dine, A.~E. Nelson, Y.~Nir, and Y.~Shirman, Phys.Rev. \textbf{D53} (1996),
  2658--2669,  [hep-ph/9507378].

\bibitem{Dine:1994vc}
M.~Dine, A.~E. Nelson, and Y.~Shirman, Phys.Rev. \textbf{D51} (1995),
  1362--1370,  [hep-ph/9408384].

\bibitem{Giudice:1998bp}
G.~Giudice and R.~Rattazzi, Phys.Rept. \textbf{322} (1999), 419--499,
  [hep-ph/9801271].

\bibitem{Randall:1998uk}
L.~Randall and R.~Sundrum, Nucl.Phys. \textbf{B557} (1999), 79--118,
  [hep-th/9810155].

\bibitem{Craig:2013cxa}
N.~Craig,  (2013),  1309.0528.

\bibitem{Chatrchyan:2012ufa}
CMS Collaboration, S.~Chatrchyan et~al., Phys.Lett. \textbf{B716} (2012),
  30--61,  [1207.7235].

\bibitem{Aad:2012tfa}
ATLAS Collaboration, G.~Aad et~al., Phys.Lett. \textbf{B716} (2012), 1--29,
  [1207.7214].

\bibitem{Ellwanger:2009dp}
U.~Ellwanger, C.~Hugonie, and A.~M. Teixeira, Phys.Rept. \textbf{496} (2010),
  1--77,  [0910.1785].

\bibitem{Ellwanger:2006rm}
U.~Ellwanger and C.~Hugonie, Mod.Phys.Lett. \textbf{A22} (2007), 1581--1590,
  [hep-ph/0612133].

\bibitem{Ma:2011ea}
E.~Ma, Phys.Lett. \textbf{B705} (2011), 320--323,  [1108.4029].

\bibitem{Zhang:2008jm}
Y.~Zhang, H.~An, X.-d. Ji, and R.~N. Mohapatra, Phys.Rev. \textbf{D78} (2008),
  011302,  [0804.0268].

\bibitem{Hirsch:2011hg}
M.~Hirsch, M.~Malinsky, W.~Porod, L.~Reichert, and F.~Staub, JHEP \textbf{1202}
  (2012), 084,  [1110.3037].

\bibitem{Dreiner:2012gx}
H.~K. Dreiner, M.~Kramer, and J.~Tattersall, Europhys.Lett. \textbf{99} (2012),
  61001,  [1207.1613].

\bibitem{Bhattacherjee:2013gr}
B.~Bhattacherjee, J.~L. Evans, M.~Ibe, S.~Matsumoto, and T.~T. Yanagida,
  Phys.Rev. \textbf{D87} (2013), no.~11, 115002,  [1301.2336].

\bibitem{Kim:2014eva}
J.~S. Kim, K.~Rolbiecki, K.~Sakurai, and J.~Tattersall, JHEP \textbf{1412}
  (2014), 010,  [1406.0858].

\bibitem{Giudice:2010wb}
G.~F. Giudice, T.~Han, K.~Wang, and L.-T. Wang, Phys.Rev. \textbf{D81} (2010),
  115011,  [1004.4902].

\bibitem{Gori:2013ala}
S.~Gori, S.~Jung, and L.-T. Wang, JHEP \textbf{1310} (2013), 191,  [1307.5952].

\bibitem{Han:2013usa}
C.~Han, A.~Kobakhidze, N.~Liu, A.~Saavedra, L.~Wu, et~al., JHEP \textbf{1402}
  (2014), 049,  [1310.4274].

\bibitem{Schwaller:2013baa}
P.~Schwaller and J.~Zurita, JHEP \textbf{1403} (2014), 060,  [1312.7350].

\bibitem{Baer:2014cua}
H.~Baer, A.~Mustafayev, and X.~Tata, Phys.Rev. \textbf{D89} (2014), no.~5,
  055007,  [1401.1162].

\bibitem{Han:2014kaa}
Z.~Han, G.~D. Kribs, A.~Martin, and A.~Menon, Phys.Rev. \textbf{D89} (2014),
  no.~7, 075007,  [1401.1235].

\bibitem{Bramante:2014dza}
J.~Bramante, A.~Delgado, F.~Elahi, A.~Martin, and B.~Ostdiek, Phys.Rev.
  \textbf{D90} (2014), no.~9, 095008,  [1408.6530].

\bibitem{Han:2014xoa}
C.~Han, L.~Wu, J.~M. Yang, M.~Zhang, and Y.~Zhang,  (2014),  1409.4533.

\bibitem{Baer:2014kya}
H.~Baer, A.~Mustafayev, and X.~Tata, Phys.Rev. \textbf{D90} (2014), no.~11,
  115007,  [1409.7058].

\bibitem{Gori:2014oua}
S.~Gori, S.~Jung, L.-T. Wang, and J.~D. Wells, JHEP \textbf{1412} (2014), 108,
  [1410.6287].

\bibitem{Bramante:2014tba}
J.~Bramante, P.~J. Fox, A.~Martin, B.~Ostdiek, T.~Plehn, et~al.,  (2014),
  1412.4789.

\bibitem{Han:2015lma}
C.~Han, D.~Kim, S.~Munir, and M.~Park,  (2015),  1502.03734.

\bibitem{Gonzalez-Garcia:2014bfa}
M.~Gonzalez-Garcia, M.~Maltoni, and T.~Schwetz, JHEP \textbf{1411} (2014), 052,
   [1409.5439].

\bibitem{Hisano:1998wn}
J.~Hisano, M.~M. Nojiri, Y.~Shimizu, and M.~Tanaka, Phys.Rev. \textbf{D60}
  (1999), 055008,  [hep-ph/9808410].

\bibitem{Rossi:2002zb}
A.~Rossi, Phys.Rev. \textbf{D66} (2002), 075003,  [hep-ph/0207006].

\bibitem{Buckley:2006nv}
M.~R. Buckley and H.~Murayama, Phys.Rev.Lett. \textbf{97} (2006), 231801,
  [hep-ph/0606088].

\bibitem{Hirsch:2008gh}
M.~Hirsch, S.~Kaneko, and W.~Porod, Phys.Rev. \textbf{D78} (2008), 093004,
  [0806.3361].

\bibitem{Hirsch:2008dy}
M.~Hirsch, J.~Valle, W.~Porod, J.~Romao, and A.~Villanova~del Moral, Phys.Rev.
  \textbf{D78} (2008), 013006,  [0804.4072].

\bibitem{Borzumati:2009hu}
F.~Borzumati and T.~Yamashita, Prog.Theor.Phys. \textbf{124} (2010), 761--868,
  [0903.2793].

\bibitem{Esteves:2009vg}
J.~Esteves, J.~Romao, A.~Villanova~del Moral, M.~Hirsch, J.~Valle, et~al., JHEP
  \textbf{0905} (2009), 003,  [0903.1408].

\bibitem{Esteves:2010ff}
J.~Esteves, J.~Romao, M.~Hirsch, F.~Staub, and W.~Porod, Phys.Rev. \textbf{D83}
  (2011), 013003,  [1010.6000].

\bibitem{Malinsky:2005bi}
M.~Malinsky, J.~Romao, and J.~Valle, Phys.Rev.Lett. \textbf{95} (2005), 161801,
   [hep-ph/0506296].

\bibitem{Abada:2012cq}
A.~Abada, D.~Das, A.~Vicente, and C.~Weiland, JHEP \textbf{1209} (2012), 015,
  [1206.6497].

\bibitem{BhupalDev:2012ru}
P.~Bhupal~Dev, S.~Mondal, B.~Mukhopadhyaya, and S.~Roy, JHEP \textbf{1209}
  (2012), 110,  [1207.6542].

\bibitem{Abada:2014kba}
A.~Abada, M.~E. Krauss, W.~Porod, F.~Staub, A.~Vicente, et~al., JHEP
  \textbf{1411} (2014), 048,  [1408.0138].

\bibitem{Peccei:1977hh}
R.~Peccei and H.~R. Quinn, Phys.Rev.Lett. \textbf{38} (1977), 1440--1443.

\bibitem{Covi:2001nw}
L.~Covi, H.-B. Kim, J.~E. Kim, and L.~Roszkowski, JHEP \textbf{0105} (2001),
  033,  [hep-ph/0101009].

\bibitem{Kim:2001sh}
H.-B. Kim and J.~E. Kim, Phys.Lett. \textbf{B527} (2002), 18--22,
  [hep-ph/0108101].

\bibitem{Covi:2009pq}
L.~Covi and J.~E. Kim, New J.Phys. \textbf{11} (2009), 105003,  [0902.0769].

\bibitem{Choi:2011yf}
K.-Y. Choi, L.~Covi, J.~E. Kim, and L.~Roszkowski, JHEP \textbf{1204} (2012),
  106,  [1108.2282].

\bibitem{Bae:2011iw}
K.~J. Bae, E.~J. Chun, and S.~H. Im, JCAP \textbf{1203} (2012), 013,
  [1111.5962].

\bibitem{Bae:2011jb}
K.~J. Bae, K.~Choi, and S.~H. Im, JHEP \textbf{1108} (2011), 065,  [1106.2452].

\bibitem{Strumia:2010aa}
A.~Strumia, JHEP \textbf{1006} (2010), 036,  [1003.5847].

\bibitem{Dine:1981rt}
M.~Dine, W.~Fischler, and M.~Srednicki, Phys.Lett. \textbf{B104} (1981), 199.

\bibitem{Zhitnitsky:1980tq}
A.~Zhitnitsky, Sov.J.Nucl.Phys. \textbf{31} (1980), 260.

\bibitem{Dreiner:2014eda}
H.~K. Dreiner, F.~Staub, and L.~Ubaldi, Phys.Rev. \textbf{D90} (2014), 055016,
  [1402.5977].

\bibitem{Kim:1983dt}
J.~E. Kim and H.~P. Nilles, Phys.Lett. \textbf{B138} (1984), 150.

\bibitem{Athron:2009ue}
P.~Athron, S.~King, D.~Miller, S.~Moretti, and R.~Nevzorov, Phys.Lett.
  \textbf{B681} (2009), 448--456,  [0901.1192].

\bibitem{Athron:2009bs}
P.~Athron, S.~King, D.~Miller, S.~Moretti, and R.~Nevzorov, Phys.Rev.
  \textbf{D80} (2009), 035009,  [0904.2169].

\bibitem{Arbelaez:2013hr}
C.~Arbelaez, R.~M. Fonseca, M.~Hirsch, and J.~C. Romao, Phys.Rev. \textbf{D87}
  (2013), no.~7, 075010,  [1301.6085].

\bibitem{Arbelaez:2013nga}
C.~Arbeláez, M.~Hirsch, M.~Malinský, and J.~C. Romão, Phys.Rev. \textbf{D89}
  (2014), no.~3, 035002,  [1311.3228].

\bibitem{Fayet:1978qc}
P.~Fayet, Phys.Lett. \textbf{B78} (1978), 417.

\bibitem{Polchinski:1982an}
J.~Polchinski and L.~Susskind, Phys.Rev. \textbf{D26} (1982), 3661.

\bibitem{Hall:1990hq}
L.~Hall and L.~Randall, Nucl.Phys. \textbf{B352} (1991), 289--308.

\bibitem{Fox:2002bu}
P.~J. Fox, A.~E. Nelson, and N.~Weiner, JHEP \textbf{0208} (2002), 035,
  [hep-ph/0206096].

\bibitem{Nelson:2002ca}
A.~E. Nelson, N.~Rius, V.~Sanz, and M.~Unsal, JHEP \textbf{0208} (2002), 039,
  [hep-ph/0206102].

\bibitem{Antoniadis:2005em}
I.~Antoniadis, A.~Delgado, K.~Benakli, M.~Quiros, and M.~Tuckmantel, Phys.Lett.
  \textbf{B634} (2006), 302--306,  [hep-ph/0507192].

\bibitem{Antoniadis:2006uj}
I.~Antoniadis, K.~Benakli, A.~Delgado, and M.~Quiros, Adv.Stud.Theor.Phys.
  \textbf{2} (2008), 645--672,  [hep-ph/0610265].

\bibitem{Amigo:2008rc}
S.~D.~L. Amigo, A.~E. Blechman, P.~J. Fox, and E.~Poppitz, JHEP \textbf{0901}
  (2009), 018,  [0809.1112].

\bibitem{Plehn:2008ae}
T.~Plehn and T.~M. Tait, J.Phys. \textbf{G36} (2009), 075001,  [0810.3919].

\bibitem{Benakli:2008pg}
K.~Benakli and M.~Goodsell, Nucl.Phys. \textbf{B816} (2009), 185--203,
  [0811.4409].

\bibitem{Belanger:2009wf}
G.~Belanger, K.~Benakli, M.~Goodsell, C.~Moura, and A.~Pukhov, JCAP
  \textbf{0908} (2009), 027,  [0905.1043].

\bibitem{Benakli:2009mk}
K.~Benakli and M.~Goodsell, Nucl.Phys. \textbf{B830} (2010), 315--329,
  [0909.0017].

\bibitem{Choi:2009ue}
S.~Choi, J.~Kalinowski, J.~Kim, and E.~Popenda, Acta Phys.Polon. \textbf{B40}
  (2009), 2913--2922,  [0911.1951].

\bibitem{Benakli:2010gi}
K.~Benakli and M.~Goodsell, Nucl.Phys. \textbf{B840} (2010), 1--28,
  [1003.4957].

\bibitem{Choi:2010gc}
S.~Choi, D.~Choudhury, A.~Freitas, J.~Kalinowski, J.~Kim, et~al., JHEP
  \textbf{1008} (2010), 025,  [1005.0818].

\bibitem{Carpenter:2010as}
L.~M. Carpenter, JHEP \textbf{1209} (2012), 102,  [1007.0017].

\bibitem{Kribs:2010md}
G.~D. Kribs, T.~Okui, and T.~S. Roy, Phys.Rev. \textbf{D82} (2010), 115010,
  [1008.1798].

\bibitem{Abel:2011dc}
S.~Abel and M.~Goodsell, JHEP \textbf{1106} (2011), 064,  [1102.0014].

\bibitem{Davies:2011mp}
R.~Davies, J.~March-Russell, and M.~McCullough, JHEP \textbf{1104} (2011), 108,
   [1103.1647].

\bibitem{Benakli:2011vb}
K.~Benakli, Fortsch.Phys. \textbf{59} (2011), 1079--1082,  [1106.1649].

\bibitem{Benakli:2011kz}
K.~Benakli, M.~D. Goodsell, and A.-K. Maier, Nucl.Phys. \textbf{B851} (2011),
  445--461,  [1104.2695].

\bibitem{Kalinowski:2011zz}
J.~Kalinowski, PoS \textbf{EPS-HEP2011} (2011), 265.

\bibitem{Frugiuele:2011mh}
C.~Frugiuele and T.~Gregoire, Phys.Rev. \textbf{D85} (2012), 015016,
  [1107.4634].

\bibitem{Itoyama:2011zi}
H.~Itoyama and N.~Maru, Int.J.Mod.Phys. \textbf{A27} (2012), 1250159,
  [1109.2276].

\bibitem{Rehermann:2011ax}
K.~Rehermann and C.~M. Wells,  (2011),  1111.0008.

\bibitem{Bertuzzo:2012su}
E.~Bertuzzo and C.~Frugiuele, JHEP \textbf{1205} (2012), 100,  [1203.5340].

\bibitem{Davies:2012vu}
R.~Davies, JHEP \textbf{1210} (2012), 010,  [1205.1942].

\bibitem{Argurio:2012cd}
R.~Argurio, M.~Bertolini, L.~Di~Pietro, F.~Porri, and D.~Redigolo, JHEP
  \textbf{1208} (2012), 086,  [1205.4709].

\bibitem{Fok:2012fb}
R.~Fok, G.~D. Kribs, A.~Martin, and Y.~Tsai, Phys.Rev. \textbf{D87} (2013),
  no.~5, 055018,  [1208.2784].

\bibitem{Argurio:2012bi}
R.~Argurio, M.~Bertolini, L.~Di~Pietro, F.~Porri, and D.~Redigolo, JHEP
  \textbf{1210} (2012), 179,  [1208.3615].

\bibitem{Frugiuele:2012pe}
C.~Frugiuele, T.~Gregoire, P.~Kumar, and E.~Ponton, JHEP \textbf{1303} (2013),
  156,  [1210.0541].

\bibitem{Frugiuele:2012kp}
C.~Frugiuele, T.~Gregoire, P.~Kumar, and E.~Ponton, JHEP \textbf{1305} (2013),
  012,  [1210.5257].

\bibitem{Benakli:2012cy}
K.~Benakli, M.~D. Goodsell, and F.~Staub, JHEP \textbf{1306} (2013), 073,
  [1211.0552].

\bibitem{Itoyama:2013sn}
H.~Itoyama and N.~Maru, Phys.Rev. \textbf{D88} (2013), no.~2, 025012,
  [1301.7548].

\bibitem{Chakraborty:2013gea}
S.~Chakraborty and S.~Roy, JHEP \textbf{1401} (2014), 101,  [1309.6538].

\bibitem{Csaki:2013fla}
C.~Csaki, J.~Goodman, R.~Pavesi, and Y.~Shirman, Phys.Rev. \textbf{D89} (2014),
  no.~5, 055005,  [1310.4504].

\bibitem{Itoyama:2013vxa}
H.~Itoyama and N.~Maru,  (2013),  1312.4157.

\bibitem{Beauchesne:2014pra}
H.~Beauchesne and T.~Gregoire, JHEP \textbf{1405} (2014), 051,  [1402.5403].

\bibitem{Benakli:2014daa}
K.~Benakli, EPJ Web Conf. \textbf{71} (2014), 00012,  [1402.4286].

\bibitem{Bertuzzo:2014bwa}
E.~Bertuzzo, C.~Frugiuele, T.~Gregoire, and E.~Ponton,  (2014),  1402.5432.

\bibitem{Alves:2015kia}
D.~S.~M. Alves, J.~Galloway, M.~McCullough, and N.~Weiner,  (2015),
  1502.03819.

\bibitem{Heikinheimo:2011fk}
M.~Heikinheimo, M.~Kellerstein, and V.~Sanz, JHEP \textbf{1204} (2012), 043,
  [1111.4322].

\bibitem{Kribs:2012gx}
G.~D. Kribs and A.~Martin, Phys.Rev. \textbf{D85} (2012), 115014,  [1203.4821].

\bibitem{Alves:2013wra}
D.~S.~M. Alves, J.~Liu, and N.~Weiner,  (2013),  1312.4965.

\bibitem{Kribs:2007ac}
G.~D. Kribs, E.~Poppitz, and N.~Weiner, Phys.Rev. \textbf{D78} (2008), 055010,
  [0712.2039].

\bibitem{Fok:2012me}
R.~Fok,  (2012),  1208.6558.

\bibitem{Dudas:2013gga}
E.~Dudas, M.~Goodsell, L.~Heurtier, and P.~Tziveloglou, Nucl.Phys.
  \textbf{B884} (2014), 632--671,  [1312.2011].

\bibitem{Allanach:2001kg}
B.~Allanach, Comput.Phys.Commun. \textbf{143} (2002), 305--331,
  [hep-ph/0104145].

\bibitem{Allanach:2009bv}
B.~Allanach and M.~Bernhardt, Comput.Phys.Commun. \textbf{181} (2010),
  232--245,  [0903.1805].

\bibitem{Allanach:2014nba}
B.~Allanach, A.~Bednyakov, and R.~Ruiz~de Austri,  (2014),  1407.6130.

\bibitem{Porod:2003um}
W.~Porod, Comput.Phys.Commun. \textbf{153} (2003), 275--315,  [hep-ph/0301101].

\bibitem{Porod:2011nf}
W.~Porod and F.~Staub, Comput.Phys.Commun. \textbf{183} (2012), 2458--2469,
  [1104.1573].

\bibitem{Djouadi:2002ze}
A.~Djouadi, J.-L. Kneur, and G.~Moultaka, Comput.Phys.Commun. \textbf{176}
  (2007), 426--455,  [hep-ph/0211331].

\bibitem{Paige:2003mg}
F.~E. Paige, S.~D. Protopopescu, H.~Baer, and X.~Tata,  (2003),
  hep-ph/0312045.

\bibitem{Baer:2003xc}
H.~Baer, C.~Balazs, A.~Belyaev, R.~Dermisek, A.~Mafi, et~al.,
  Nucl.Instrum.Meth. \textbf{A502} (2003), 560--563.

\bibitem{Baer:1999sp}
H.~Baer, F.~E. Paige, S.~D. Protopopescu, and X.~Tata,  (1999),
  hep-ph/0001086.

\bibitem{Paige:1998xm}
F.~E. Paige, S.~D. Proto~pescu, H.~Baer, and X.~Tata,  (1998),  hep-ph/9810440.

\bibitem{Paige:1998ux}
F.~E. Paige, S.~D. Protopopescu, H.~Baer, and X.~Tata,  (1998),
  hep-ph/9804321.

\bibitem{Baer:1993ae}
H.~Baer, F.~E. Paige, S.~D. Protopopescu, and X.~Tata,  (1993),
  hep-ph/9305342.

\bibitem{Heinemeyer:1998yj}
S.~Heinemeyer, W.~Hollik, and G.~Weiglein, Comput.Phys.Commun. \textbf{124}
  (2000), 76--89,  [hep-ph/9812320].

\bibitem{Hahn:2009zz}
T.~Hahn, S.~Heinemeyer, W.~Hollik, H.~Rzehak, and G.~Weiglein,
  Comput.Phys.Commun. \textbf{180} (2009), 1426--1427.

\bibitem{Staub:2008uz}
F.~Staub,  (2008),  0806.0538.

\bibitem{Staub:2009bi}
F.~Staub, Comput.Phys.Commun. \textbf{181} (2010), 1077--1086,  [0909.2863].

\bibitem{Staub:2010jh}
F.~Staub, Comput.Phys.Commun. \textbf{182} (2011), 808--833,  [1002.0840].

\bibitem{Staub:2012pb}
F.~Staub, Computer Physics Communications \textbf{184} (2013), pp. 1792--1809,
  [1207.0906].

\bibitem{Staub:2013tta}
F.~Staub, Comput.Phys.Commun. \textbf{185} (2014), 1773--1790,  [1309.7223].

\bibitem{Porod:2014xia}
W.~Porod, F.~Staub, and A.~Vicente, Eur.Phys.J. \textbf{C74} (2014), no.~8,
  2992,  [1405.1434].

\bibitem{Goodsell:2014bna}
M.~D. Goodsell, K.~Nickel, and F.~Staub,  (2014),  1411.0675.

\bibitem{Goodsell:2015ira}
M.~Goodsell, K.~Nickel, and F.~Staub,  (2015),  1503.03098.

\bibitem{Pukhov:2004ca}
A.~Pukhov,  (2004),  hep-ph/0412191.

\bibitem{Boos:1994xb}
E.~Boos, M.~Dubinin, V.~Ilyin, A.~Pukhov, and V.~Savrin,  (1994),
  hep-ph/9503280.

\bibitem{Hahn:2000kx}
T.~Hahn, Comput.Phys.Commun. \textbf{140} (2001), 418--431,  [hep-ph/0012260].

\bibitem{Hahn:2009bf}
T.~Hahn, PoS \textbf{ACAT08} (2008), 121,  [0901.1528].

\bibitem{Kilian:2007gr}
W.~Kilian, T.~Ohl, and J.~Reuter, Eur.Phys.J. \textbf{C71} (2011), 1742,
  [0708.4233].

\bibitem{Moretti:2001zz}
M.~Moretti, T.~Ohl, and J.~Reuter,  (2001),  hep-ph/0102195.

\bibitem{Degrande:2011ua}
C.~Degrande, C.~Duhr, B.~Fuks, D.~Grellscheid, O.~Mattelaer, et~al.,
  Comput.Phys.Commun. \textbf{183} (2012), 1201--1214,  [1108.2040].

\bibitem{Alwall:2011uj}
J.~Alwall, M.~Herquet, F.~Maltoni, O.~Mattelaer, and T.~Stelzer, JHEP
  \textbf{1106} (2011), 128,  [1106.0522].

\bibitem{Cullen:2011ac}
G.~Cullen, N.~Greiner, G.~Heinrich, G.~Luisoni, P.~Mastrolia, et~al.,
  Eur.Phys.J. \textbf{C72} (2012), 1889,  [1111.2034].

\bibitem{Gieseke:2003hm}
S.~Gieseke, A.~Ribon, M.~H. Seymour, P.~Stephens, and B.~Webber, JHEP
  \textbf{0402} (2004), 005,  [hep-ph/0311208].

\bibitem{Gieseke:2006ga}
S.~Gieseke, D.~Grellscheid, K.~Hamilton, A.~Ribon, P.~Richardson, et~al.,
  (2006),  hep-ph/0609306.

\bibitem{Bellm:2013lba}
J.~Bellm, S.~Gieseke, D.~Grellscheid, A.~Papaefstathiou, S.~Platzer, et~al.,
  (2013),  1310.6877.

\bibitem{Gleisberg:2003xi}
T.~Gleisberg, S.~Hoeche, F.~Krauss, A.~Schalicke, S.~Schumann, et~al., JHEP
  \textbf{0402} (2004), 056,  [hep-ph/0311263].

\bibitem{Gleisberg:2008ta}
T.~Gleisberg, S.~Hoeche, F.~Krauss, M.~Schonherr, S.~Schumann, et~al., JHEP
  \textbf{0902} (2009), 007,  [0811.4622].

\bibitem{Hoche:2014kca}
S.~Höche, S.~Kuttimalai, S.~Schumann, and F.~Siegert,  (2014),  1412.6478.

\bibitem{Bechtle:2008jh}
P.~Bechtle, O.~Brein, S.~Heinemeyer, G.~Weiglein, and K.~E. Williams,
  Comput.Phys.Commun. \textbf{181} (2010), 138--167,  [0811.4169].

\bibitem{Bechtle:2011sb}
P.~Bechtle, O.~Brein, S.~Heinemeyer, G.~Weiglein, and K.~E. Williams,
  Comput.Phys.Commun. \textbf{182} (2011), 2605--2631,  [1102.1898].

\bibitem{Bechtle:2013xfa}
P.~Bechtle, S.~Heinemeyer, O.~Stål, T.~Stefaniak, and G.~Weiglein,  (2013),
  1305.1933.

\bibitem{Athron:2014yba}
P.~Athron, J.-h. Park, D.~Stöckinger, and A.~Voigt,  (2014),  1406.2319.

\bibitem{Camargo-Molina:2013qva}
J.~Camargo-Molina, B.~O'Leary, W.~Porod, and F.~Staub,  (2013),  1307.1477.

\bibitem{Stal:2011cz}
O.~Stal and G.~Weiglein, JHEP \textbf{1201} (2012), 071,  [1108.0595].

\bibitem{Ender:2011qh}
K.~Ender, T.~Graf, M.~Muhlleitner, and H.~Rzehak, Phys.Rev. \textbf{D85}
  (2012), 075024,  [1111.4952].

\bibitem{Aparicio:2012vk}
L.~Aparicio, P.~Camara, D.~Cerdeno, L.~Ibanez, and I.~Valenzuela, JHEP
  \textbf{1302} (2013), 084,  [1212.4808].

\bibitem{Graf:2012hh}
T.~Graf, R.~Grober, M.~Muhlleitner, H.~Rzehak, and K.~Walz, JHEP \textbf{1210}
  (2012), 122,  [1206.6806].

\bibitem{SchmidtHoberg:2012yy}
K.~Schmidt-Hoberg and F.~Staub, JHEP \textbf{1210} (2012), 195,  [1208.1683].

\bibitem{SchmidtHoberg:2012ip}
K.~Schmidt-Hoberg, F.~Staub, and M.~W. Winkler, JHEP \textbf{1301} (2013), 124,
   [1211.2835].

\bibitem{Ross:2012nr}
G.~G. Ross, K.~Schmidt-Hoberg, and F.~Staub, JHEP \textbf{1208} (2012), 074,
  [1205.1509].

\bibitem{Kaminska:2013mya}
A.~Kaminska, G.~G. Ross, and K.~Schmidt-Hoberg,  (2013),  1308.4168.

\bibitem{Binjonaid:2014oga}
M.~Y. Binjonaid and S.~F. King, Phys.Rev. \textbf{D90} (2014), no.~7, 055020,
  [1403.2088].

\bibitem{Kaminska:2014wia}
A.~Kaminska, G.~G. Ross, K.~Schmidt-Hoberg, and F.~Staub, JHEP \textbf{1406}
  (2014), 153,  [1401.1816].

\bibitem{Muhlleitner:2014vsa}
M.~Muhlleitner, D.~T. Nhung, H.~Rzehak, and K.~Walz,  (2014),  1412.0918.

\bibitem{Arina:2014xya}
C.~Arina, V.~Martin-Lozano, and G.~Nardini, JHEP \textbf{1408} (2014), 015,
  [1403.6434].

\bibitem{Bandyopadhyay:2014vma}
P.~Bandyopadhyay, K.~Huitu, and A.~S. Keceli,  (2014),  1412.7359.

\bibitem{List:2013dga}
J.~List and B.~Vormwald,  (2013),  1307.4074.

\bibitem{Dreiner:2012mx}
H.~Dreiner, K.~Nickel, F.~Staub, and A.~Vicente, Phys.Rev. \textbf{D86} (2012),
  015003,  [1204.5925].

\bibitem{Dreiner:2013jta}
H.~Dreiner, K.~Nickel, and F.~Staub,  (2013),  1309.1735.

\bibitem{Biswas:2014gga}
S.~Biswas, D.~Chowdhury, S.~Han, and S.~J. Lee,  (2014),  1409.0882.

\bibitem{Allanach:2014lca}
B.~Allanach, S.~Biswas, S.~Mondal, and M.~Mitra,  (2014),  1408.5439.

\bibitem{Chamoun:2014eda}
N.~Chamoun, H.~Dreiner, F.~Staub, and T.~Stefaniak, JHEP \textbf{1408} (2014),
  142,  [1407.2248].

\bibitem{Dreiner:2014lqa}
H.~K. Dreiner, K.~Nickel, and F.~Staub,  (2014),  1411.3731.

\bibitem{Abada:2011mg}
A.~Abada, A.~Figueiredo, J.~Romao, and A.~Teixeira, JHEP \textbf{1108} (2011),
  099,  [1104.3962].

\bibitem{Hirsch:2012yv}
M.~Hirsch, W.~Porod, C.~Weiss, and F.~Staub, Phys.Rev. \textbf{D87} (2013),
  013010,  [1211.0289].

\bibitem{Banerjee:2013fga}
S.~Banerjee, P.~S.~B. Dev, S.~Mondal, B.~Mukhopadhyaya, and S.~Roy,  (2013),
  1306.2143.

\bibitem{Krauss:2013gya}
M.~E. Krauss, W.~Porod, F.~Staub, A.~Abada, A.~Vicente, et~al., Phys.Rev.
  \textbf{D90} (2014), 013008,  [1312.5318].

\bibitem{DeRomeri:2012qd}
V.~De~Romeri and M.~Hirsch, JHEP \textbf{1212} (2012), 106,  [1209.3891].

\bibitem{Boucenna:2015zwa}
S.~M. Boucenna, J.~W.~F. Valle, and A.~Vicente,  (2015),  1502.07546.

\bibitem{Esteves:2010si}
J.~Esteves, J.~Romao, M.~Hirsch, A.~Vicente, W.~Porod, et~al., JHEP
  \textbf{1012} (2010), 077,  [1011.0348].

\bibitem{Esteves:2011gk}
J.~Esteves, J.~Romao, M.~Hirsch, W.~Porod, F.~Staub, et~al., JHEP \textbf{1201}
  (2012), 095,  [1109.6478].

\bibitem{DeRomeri:2011ie}
V.~De~Romeri, M.~Hirsch, and M.~Malinsky, Phys.Rev. \textbf{D84} (2011),
  053012,  [1107.3412].

\bibitem{Krauss:2013jva}
M.~E. Krauss, W.~Porod, and F.~Staub, Phys.Rev. \textbf{D88} (2013), 015014,
  [1304.0769].

\bibitem{O'Leary:2011yq}
B.~O'Leary, W.~Porod, and F.~Staub, JHEP \textbf{1205} (2012), 042,
  [1112.4600].

\bibitem{Hirsch:2012kv}
M.~Hirsch, W.~Porod, L.~Reichert, and F.~Staub, Phys.Rev. \textbf{D86} (2012),
  093018,  [1206.3516].

\bibitem{Frank:2014bma}
M.~Frank and S.~Mondal, Phys.Rev. \textbf{D90} (2014), no.~7, 075013,
  [1408.2223].

\bibitem{Brooijmans:2014eja}
G.~Brooijmans, R.~Contino, B.~Fuks, F.~Moortgat, P.~Richardson, et~al.,
  (2014),  1405.1617.

\bibitem{Corcella:2014lha}
G.~Corcella,  (2014),  1412.6831.

\bibitem{Busbridge:2014sha}
D.~Busbridge,  (2014),  1408.4605.

\bibitem{Benakli:2014cia}
K.~Benakli, M.~Goodsell, F.~Staub, and W.~Porod, Phys.Rev. \textbf{D90} (2014),
  no.~4, 045017,  [1403.5122].

\bibitem{Diessner:2014ksa}
P.~Dießner, J.~Kalinowski, W.~Kotlarski, and D.~Stöckinger, JHEP
  \textbf{2014} (2014), 124,  [1410.4791].

\bibitem{Lalak:2015xea}
Z.~Lalak, M.~Lewicki, and J.~D. Wells,  (2015),  1502.05702.

\bibitem{Athron:2012sq}
P.~Athron, S.~King, D.~Miller, S.~Moretti, and R.~Nevzorov, Phys.Rev.
  \textbf{D86} (2012), 095003,  [1206.5028].

\bibitem{Alves:2012fx}
D.~S. Alves, P.~J. Fox, and N.~Weiner,  (2012),  1207.5522.

\bibitem{Bharucha:2013ela}
A.~Bharucha, A.~Goudelis, and M.~McGarrie,  (2013),  1310.4500.

\bibitem{Ding:2015wma}
R.~Ding, T.~Li, F.~Staub, C.~Tian, and B.~Zhu,  (2015),  1502.03614.

\bibitem{Brummer:2013upa}
F.~Brümmer, M.~McGarrie, and A.~Weiler, JHEP \textbf{1404} (2014), 078,
  [1312.0935].

\bibitem{Ding:2013pya}
R.~Ding, T.~Li, F.~Staub, and B.~Zhu, JHEP \textbf{1403} (2014), 130,
  [1312.5407].

\bibitem{Louis:2014pia}
J.~Louis, K.~Schmidt-Hoberg, and L.~Zarate,  (2014),  1402.2977.

\bibitem{Ding:2014bqa}
R.~Ding, L.~Wang, and B.~Zhu, Phys.Lett. \textbf{B733} (2014), 373--379,
  [1403.3908].

\bibitem{Kyae:2014aka}
B.~Kyae and C.~S. Shin, Phys.Rev. \textbf{D90} (2014), 035023,  [1403.6527].

\bibitem{Abel:2014fka}
S.~Abel and M.~McGarrie, JHEP \textbf{1407} (2014), 145,  [1404.1318].

\bibitem{Un:2014afa}
C.~S. Ün, Å.~H. Tanyıldızı, S.~Kerman, and L.~Solmaz,  (2014),  1412.1440.

\bibitem{Fichet:2015oha}
S.~Fichet, B.~Herrmann, and Y.~Stoll,  (2015),  1501.05307.

\bibitem{Camargo-Molina:2014pwa}
J.~Camargo-Molina, B.~Garbrecht, B.~O'Leary, W.~Porod, and F.~Staub, Phys.Lett.
  \textbf{B737} (2014), 156--161,  [1405.7376].

\bibitem{Chattopadhyay:2014gfa}
U.~Chattopadhyay and A.~Dey, JHEP \textbf{1411} (2014), 161,  [1409.0611].

\bibitem{Ellwanger:2006rn}
U.~Ellwanger and C.~Hugonie, Comput.Phys.Commun. \textbf{177} (2007), 399--407,
   [hep-ph/0612134].

\bibitem{Allanach:2013kza}
B.~Allanach, P.~Athron, L.~C. Tunstall, A.~Voigt, and A.~Williams,
  Comput.Phys.Commun. \textbf{185} (2014), 2322--2339,  [1311.7659].

\bibitem{Baglio:2013iia}
J.~Baglio, R.~Gröber, M.~Mühlleitner, D.~Nhung, H.~Rzehak, et~al.,
  Comput.Phys.Commun. \textbf{185} (2014), no.~12, 3372--3391,  [1312.4788].

\bibitem{Goodsell:2014pla}
M.~D. Goodsell, K.~Nickel, and F.~Staub,  (2014),  1411.4665.

\bibitem{Staub:2010ty}
F.~Staub, W.~Porod, and B.~Herrmann, JHEP \textbf{1010} (2010), 040,
  [1007.4049].

\bibitem{Ibanez:1991hv}
L.~E. Ibanez and G.~G. Ross, Phys.Lett. \textbf{B260} (1991), 291--295.

\bibitem{Ibanez:1991pr}
L.~E. Ibanez and G.~G. Ross, Nucl.Phys. \textbf{B368} (1992), 3--37.

\bibitem{Banks:1991xj}
T.~Banks and M.~Dine, Phys.Rev. \textbf{D45} (1992), 1424--1427,
  [hep-th/9109045].

\bibitem{Dreiner:2005rd}
H.~K. Dreiner, C.~Luhn, and M.~Thormeier, Phys.Rev. \textbf{D73} (2006),
  075007,  [hep-ph/0512163].

\bibitem{Dreiner:2006xw}
H.~K. Dreiner, C.~Luhn, H.~Murayama, and M.~Thormeier, Nucl.Phys. \textbf{B774}
  (2007), 127--167,  [hep-ph/0610026].

\bibitem{Fonseca:2011sy}
R.~M. Fonseca, Comput.Phys.Commun. \textbf{183} (2012), 2298--2306,
  [1106.5016].

\bibitem{Martin:1997ns}
S.~P. Martin,  (1997),  hep-ph/9709356.

\bibitem{Fayet:1974jb}
P.~Fayet and J.~Iliopoulos, Phys.Lett. \textbf{B51} (1974), 461--464.

\bibitem{Holdom:1985ag}
B.~Holdom, Phys.Lett. \textbf{B166} (1986), 196.

\bibitem{Goodsell:2012fm}
M.~D. Goodsell, JHEP \textbf{1301} (2013), 066,  [1206.6697].

\bibitem{Hall:1980kf}
L.~J. Hall, Nucl.Phys. \textbf{B178} (1981), 75.

\bibitem{Weinberg:1979sa}
S.~Weinberg, Phys.Rev.Lett. \textbf{43} (1979), 1566--1570.

\bibitem{Weinberg:1980bf}
S.~Weinberg, Phys.Rev. \textbf{D22} (1980), 1694.

\bibitem{Adler:1969er}
S.~L. Adler and W.~A. Bardeen, Phys.Rev. \textbf{182} (1969), 1517--1536.

\bibitem{Witten:1982fp}
E.~Witten, Phys.Lett. \textbf{B117} (1982), 324--328.

\bibitem{Martin:1993zk}
S.~P. Martin and M.~T. Vaughn, Phys.Rev. \textbf{D50} (1994), 2282,
  [hep-ph/9311340].

\bibitem{Yamada:1994id}
Y.~Yamada, Phys.Rev. \textbf{D50} (1994), 3537--3545,  [hep-ph/9401241].

\bibitem{Jack:1997eh}
I.~Jack, D.~Jones, and A.~Pickering, Phys.Lett. \textbf{B426} (1998), 73--77,
  [hep-ph/9712542].

\bibitem{Jack:1999zs}
I.~Jack and D.~Jones, Phys.Lett. \textbf{B473} (2000), 102--108,
  [hep-ph/9911491].

\bibitem{Jack:2000nm}
I.~Jack, D.~Jones, and S.~Parsons, Phys.Rev. \textbf{D62} (2000), 125022,
  [hep-ph/0007291].

\bibitem{Jack:1997pa}
I.~Jack and D.~Jones, Phys.Lett. \textbf{B415} (1997), 383--389,
  [hep-ph/9709364].

\bibitem{Fonseca:2011vn}
R.~M. Fonseca, M.~Malinsky, W.~Porod, and F.~Staub, Nucl.Phys. \textbf{B854}
  (2012), 28--53,  [1107.2670].

\bibitem{Braam:2011xh}
F.~Braam and J.~Reuter, Eur.Phys.J. \textbf{C72} (2012), 1885,  [1107.2806].

\bibitem{Sperling:2013eva}
M.~Sperling, D.~Stöckinger, and A.~Voigt, JHEP \textbf{1307} (2013), 132,
  [1305.1548].

\bibitem{Sperling:2013xqa}
M.~Sperling, D.~Stöckinger, and A.~Voigt, JHEP \textbf{1401} (2014), 068,
  [1310.7629].

\bibitem{Machacek:1983tz}
M.~E. Machacek and M.~T. Vaughn, Nucl. Phys. \textbf{B222} (1983), 83.

\bibitem{Machacek:1983fi}
M.~E. Machacek and M.~T. Vaughn, Nucl. Phys. \textbf{B236} (1984), 221.

\bibitem{Machacek:1984zw}
M.~E. Machacek and M.~T. Vaughn, Nucl. Phys. \textbf{B249} (1985), 70.

\bibitem{Luo:2002ti}
M.-x. Luo, H.-w. Wang, and Y.~Xiao, Phys. Rev. \textbf{D67} (2003), 065019,
  [hep-ph/0211440].

\bibitem{Fonseca:2013bua}
R.~M. Fonseca, M.~Malinsky, and F.~Staub,  (2013),  1308.1674.

\bibitem{Pierce:1996zz}
D.~M. Pierce, J.~A. Bagger, K.~T. Matchev, and R.-j. Zhang, Nucl.Phys.
  \textbf{B491} (1997), 3--67,  [hep-ph/9606211].

\bibitem{Martin:2002wn}
S.~P. Martin, Phys.Rev. \textbf{D67} (2003), 095012,  [hep-ph/0211366].

\bibitem{Martin:2005eg}
S.~P. Martin, Phys.Rev. \textbf{D71} (2005), 116004,  [hep-ph/0502168].

\bibitem{Brignole:2001jy}
A.~Brignole, G.~Degrassi, P.~Slavich, and F.~Zwirner, Nucl.Phys. \textbf{B631}
  (2002), 195--218,  [hep-ph/0112177].

\bibitem{Degrassi:2001yf}
G.~Degrassi, P.~Slavich, and F.~Zwirner, Nucl.Phys. \textbf{B611} (2001),
  403--422,  [hep-ph/0105096].

\bibitem{Brignole:2002bz}
A.~Brignole, G.~Degrassi, P.~Slavich, and F.~Zwirner, Nucl.Phys. \textbf{B643}
  (2002), 79--92,  [hep-ph/0206101].

\bibitem{Dedes:2002dy}
A.~Dedes and P.~Slavich, Nucl.Phys. \textbf{B657} (2003), 333--354,
  [hep-ph/0212132].

\bibitem{Dedes:2003km}
A.~Dedes, G.~Degrassi, and P.~Slavich, Nucl.Phys. \textbf{B672} (2003),
  144--162,  [hep-ph/0305127].

\bibitem{Degrassi:2009yq}
G.~Degrassi and P.~Slavich, Nucl.Phys. \textbf{B825} (2010), 119--150,
  [0907.4682].

\bibitem{Avdeev:1997sz}
L.~Avdeev and M.~Y. Kalmykov, Nucl.Phys. \textbf{B502} (1997), 419--435,
  [hep-ph/9701308].

\bibitem{Bednyakov:2002sf}
A.~Bednyakov, A.~Onishchenko, V.~Velizhanin, and O.~Veretin, Eur.Phys.J.
  \textbf{C29} (2003), 87--101,  [hep-ph/0210258].

\bibitem{Spira:1995rr}
M.~Spira, A.~Djouadi, D.~Graudenz, and P.~Zerwas, Nucl.Phys. \textbf{B453}
  (1995), 17--82,  [hep-ph/9504378].

\bibitem{Dreiner:2012dh}
H.~Dreiner, K.~Nickel, W.~Porod, and F.~Staub, Comput.Phys.Commun. \textbf{184}
  (2013), 2604--2617,  [1212.5074].

\bibitem{Ellis:1986yg}
J.~R. Ellis, K.~Enqvist, D.~V. Nanopoulos, and F.~Zwirner, Mod.Phys.Lett.
  \textbf{A1} (1986), 57.

\bibitem{Barbieri:1987fn}
R.~Barbieri and G.~Giudice, Nucl.Phys. \textbf{B306} (1988), 63.

\bibitem{Ghilencea:2012qk}
D.~Ghilencea and G.~Ross, Nucl.Phys. \textbf{B868} (2013), 65--74,
  [1208.0837].

\bibitem{Khalil:2007dr}
S.~Khalil and A.~Masiero, Phys.Lett. \textbf{B665} (2008), 374--377,
  [0710.3525].

\bibitem{Barger:2008wn}
V.~Barger, P.~Fileviez~Perez, and S.~Spinner, Phys.Rev.Lett. \textbf{102}
  (2009), 181802,  [0812.3661].

\bibitem{FileviezPerez:2010ek}
P.~Fileviez~Perez and S.~Spinner, Phys.Rev. \textbf{D83} (2011), 035004,
  [1005.4930].

\bibitem{CamargoMolina:2012hv}
J.~Camargo-Molina, B.~O'Leary, W.~Porod, and F.~Staub, Phys.Rev. \textbf{D88}
  (2013), 015033,  [1212.4146].

\bibitem{Marshall:2014kea}
Z.~Marshall, B.~A. Ovrut, A.~Purves, and S.~Spinner, Phys.Lett. \textbf{B732}
  (2014), 325--329,  [1401.7989].

\bibitem{Elsayed:2012ec}
A.~Elsayed, S.~Khalil, S.~Moretti, and A.~Moursy, Phys.Rev. \textbf{D87}
  (2013), no.~5, 053010,  [1211.0644].

\bibitem{Basso:2012gz}
L.~Basso, B.~O'Leary, W.~Porod, and F.~Staub, JHEP \textbf{1209} (2012), 054,
  [1207.0507].

\bibitem{Basso:2012tr}
L.~Basso and F.~Staub, Phys.Rev. \textbf{D87} (2013), 015011,  [1210.7946].

\bibitem{FileviezPerez:2011kd}
P.~Fileviez~Perez, S.~Spinner, and M.~K. Trenkel, Phys.Rev. \textbf{D84}
  (2011), 095028,  [1103.5504].

\bibitem{Khalil:2012gs}
S.~Khalil and S.~Moretti, J.Mod.Phys. \textbf{4} (2013), 7--10,  [1207.1590].

\bibitem{Krauss:2012ku}
M.~E. Krauss, B.~O'Leary, W.~Porod, and F.~Staub, Phys.Rev. \textbf{D86}
  (2012), 055017,  [1206.3513].

\bibitem{Perez:2013kla}
P.~Fileviez~Perez and S.~Spinner, Phys.Lett. \textbf{B728} (2014), 489--495,
  [1308.0524].

\bibitem{Chankowski:2006jk}
P.~H. Chankowski, S.~Pokorski, and J.~Wagner, Eur.Phys.J. \textbf{C47} (2006),
  187--205,  [hep-ph/0601097].

\bibitem{Basso:2010jm}
L.~Basso, S.~Moretti, and G.~M. Pruna, Phys.Rev. \textbf{D82} (2010), 055018,
  [1004.3039].

\bibitem{Staub:2011dp}
F.~Staub, T.~Ohl, W.~Porod, and C.~Speckner, Comput.Phys.Commun. \textbf{183}
  (2012), 2165--2206,  [1109.5147].

\bibitem{Dev:2009aw}
P.~B. Dev and R.~Mohapatra, Phys.Rev. \textbf{D81} (2010), 013001,
  [0910.3924].

\bibitem{BhupalDev:2010he}
P.~Bhupal~Dev and R.~Mohapatra, Phys.Rev. \textbf{D82} (2010), 035014,
  [1003.6102].

\bibitem{Brooijmans:2012yi}
G.~Brooijmans, B.~Gripaios, F.~Moortgat, J.~Santiago, P.~Skands, et~al.,
  (2012),  1203.1488.

\bibitem{Basso:2012ew}
L.~Basso, A.~Belyaev, D.~Chowdhury, M.~Hirsch, S.~Khalil, et~al.,
  Comput.Phys.Commun. \textbf{184} (2013), 698--719,  [1206.4563].

\bibitem{Ohl:FeynMF}
T.~Ohl,  (1997),  hep-ph/0301101.

\bibitem{Skands:2003cj}
P.~Z. Skands, B.~Allanach, H.~Baer, C.~Balazs, G.~Belanger, et~al., JHEP
  \textbf{0407} (2004), 036,  [hep-ph/0311123].

\bibitem{Allanach:2008qq}
B.~Allanach, C.~Balazs, G.~Belanger, M.~Bernhardt, F.~Boudjema, et~al.,
  Comput.Phys.Commun. \textbf{180} (2009), 8--25,  [0801.0045].

\bibitem{Adam:2013mnn}
MEG Collaboration, J.~Adam et~al., Phys.Rev.Lett. \textbf{110} (2013), 201801,
  [1303.0754].

\bibitem{Aubert:2009ag}
BaBar Collaboration, B.~Aubert et~al., Phys.Rev.Lett. \textbf{104} (2010),
  021802,  [0908.2381].

\bibitem{Bellgardt:1987du}
SINDRUM Collaboration, U.~Bellgardt et~al., Nucl.Phys. \textbf{B299} (1988), 1.

\bibitem{Hayasaka:2010np}
K.~Hayasaka, K.~Inami, Y.~Miyazaki, K.~Arinstein, V.~Aulchenko, et~al.,
  Phys.Lett. \textbf{B687} (2010), 139--143,  [1001.3221].

\bibitem{Dohmen:1993mp}
SINDRUM II Collaboration., C.~Dohmen et~al., Phys.Lett. \textbf{B317} (1993),
  631--636.

\bibitem{Bertl:2006up}
SINDRUM II Collaboration, W.~H. Bertl et~al., Eur.Phys.J. \textbf{C47} (2006),
  337--346.

\bibitem{Liu:2004qw}
J.~J. Liu, C.~S. Li, L.~L. Yang, and L.~G. Jin, Phys.Lett. \textbf{B599}
  (2004), 92--101,  [hep-ph/0406155].

\bibitem{AguilarSaavedra:2002ns}
J.~Aguilar-Saavedra and B.~Nobre, Phys.Lett. \textbf{B553} (2003), 251--260,
  [hep-ph/0210360].

\bibitem{Abe:1997fz}
CDF Collaboration, F.~Abe et~al., Phys.Rev.Lett. \textbf{80} (1998),
  2525--2530.

\bibitem{Nilles:1982dy}
H.~P. Nilles, M.~Srednicki, and D.~Wyler, Phys.Lett. \textbf{B120} (1983), 346.

\bibitem{Derendinger:1983bz}
J.~Derendinger and C.~A. Savoy, Nucl.Phys. \textbf{B237} (1984), 307.

\bibitem{Claudson:1983et}
M.~Claudson, L.~J. Hall, and I.~Hinchliffe, Nucl.Phys. \textbf{B228} (1983),
  501.

\bibitem{Kounnas:1983td}
C.~Kounnas, A.~Lahanas, D.~V. Nanopoulos, and M.~Quiros, Nucl.Phys.
  \textbf{B236} (1984), 438.

\bibitem{Drees:1985ie}
M.~Drees, M.~Gluck, and K.~Grassie, Phys.Lett. \textbf{B157} (1985), 164.

\bibitem{Gunion:1987qv}
J.~Gunion, H.~Haber, and M.~Sher, Nucl.Phys. \textbf{B306} (1988), 1.

\bibitem{Komatsu:1988mt}
H.~Komatsu, Phys.Lett. \textbf{B215} (1988), 323.

\bibitem{Langacker:1994bc}
P.~Langacker and N.~Polonsky, Phys.Rev. \textbf{D50} (1994), 2199--2217,
  [hep-ph/9403306].

\bibitem{Casas:1995pd}
J.~Casas, A.~Lleyda, and C.~Munoz, Nucl.Phys. \textbf{B471} (1996), 3--58,
  [hep-ph/9507294].

\bibitem{Casas:1996de}
J.~Casas and S.~Dimopoulos, Phys.Lett. \textbf{B387} (1996), 107--112,
  [hep-ph/9606237].

\bibitem{Camargo-Molina:2013sta}
J.~Camargo-Molina, B.~O'Leary, W.~Porod, and F.~Staub,  (2013),  1309.7212.

\bibitem{lee2008hom4ps}
T.~Lee, T.~Li, and C.~Tsai, Computing \textbf{83} (2008), no.~2, 109--133.

\bibitem{James:1975dr}
F.~James and M.~Roos, Comput.Phys.Commun. \textbf{10} (1975), 343--367.

\bibitem{Wainwright:2011kj}
C.~L. Wainwright, Comput.Phys.Commun. \textbf{183} (2012), 2006--2013,
  [1109.4189].

\bibitem{Belanger:2010st}
G.~Belanger, N.~D. Christensen, A.~Pukhov, and A.~Semenov, Comput.Phys.Commun.
  \textbf{182} (2011), 763--774,  [1008.0181].

\bibitem{Ade:2013zuv}
Planck, P.~Ade et~al., Astron.Astrophys. \textbf{571} (2014), A16,
  [1303.5076].

\bibitem{Conte:2012fm}
E.~Conte, B.~Fuks, and G.~Serret, Comput.Phys.Commun. \textbf{184} (2013),
  222--256,  [1206.1599].

\end{thebibliography}

\end{document}